\documentclass[a4paper,11pt]{article}
\usepackage[latin1]{inputenc}
\usepackage{graphicx}
\usepackage{indentfirst}
\usepackage{amsmath}
\usepackage{amssymb}
\usepackage{amsfonts}
\usepackage{latexsym}
\usepackage{epsfig}
\usepackage{psfig}
\usepackage{float}
\begin{document}
\title{\bf Seismic motion in urban sites consisting  of blocks in welded contact with a soft
layer overlying a hard half space: I. Finite set of blocks\rm\\}
\author{Jean-Philippe Groby\thanks{Laboratorium voor Akoestieke en Thermische Fysica, KULeuven,
Celestineslaan 200D, 3001 Heverlee, Belgium ({\tt JeanPhilippe.Groby@fys.kuleuvne.be})} and
Armand Wirgin\thanks{LMA/CNRS, 31 chemin Joseph Aiguier, 13402
Marseille cedex 20, France, ({\tt wirgin@lma.cnrs-mrs.fr})} }
\date{\today}
\maketitle
\begin{abstract}
We address the problem of the response to a seismic wave of an
urban site consisting of $N$ non-identical, non-equispaced blocks
overlying a soft layer underlain by a hard substratum. The results
of a theoretical analysis, appealing to a space-frequency
mode-matching (MM) technique, are compared to those
 obtained by a space-time  finite element
(FE) technique. The two methods are shown to give rise to the same
prediction of the seismic response for $N=1$ and $N=2$ blocks. The
mechanism of the interaction between blocks and the ground, as
well as that of the mutual interaction between blocks, are
studied. It is shown that the presence of a small number of blocks
modifies the seismic disturbance in a manner which evokes
qualitatively, but not quantitatively,  what was observed during
the 1985 Michoacan earthquake in Mexico City. Disturbances at a
much greater level, induced by a large number of blocks (in fact,
a periodic set) are studied in the companion paper.
\end{abstract}
\newline
\newline
Abbreviated title: Seismic response in  urban sites
\newline
\newline
Corresponding author: Armand Wirgin, tel.: 33 4 91 16 40 50, fax:
33 4 91 16 42 70\\ e-mail: wirgin@lma.cnrs-mrs.fr
\newpage
\tableofcontents
\newpage
\newpage
\section{Introduction}\label{intro}
The Michoacan earthquake that struck Mexico City in 1985 presented
some particular characteristics which have since been encountered
at various other locations
\cite{savage,semblat,iwata,maeda,haghshenas}, but at a lower level
of intensity. Other than the fact that the response in downtown
Mexico varied considerably in a spatial sense \cite{flores}, was
quite intense and of very long duration  at certain locations (as
much as $\approx$3min \cite{Perezrocha}), and often took the form
of a quasi-monochromatic signal with beatings \cite{mateos}, a
remarkable feature of this earthquake (studied in
\cite{Furumuraetkennet1998,Cardenassato,grobyetwirgin2005,grobyetwirgin2005II})
was that such strong motion could be caused by a seismic source so
far from the city (the epicenter was located in the subduction
zone off the Pacific coast, approximately 350km from Mexico City).
It is now recognized \cite{Cardenassato,cardenas} that the
characteristics of the abnormal response recorded in downtown
Mexico were partially present in the waves entering into the
 city (notably $60$km from the city as recorded by the authors of
\cite{Furumuraetkennet1998}) after having accomplished their
voyage
 from the source, this being thought to be due to the excitation
 of Love and generalized-Rayleigh modes by the irregularities of the
 crust
\cite{Cardenassato,chavezgarciasalazar,Furumuraetkennet1998}).

In the present investigation (as well as in the companion paper),
we focus on the influence of the presence of the built features of
the urban site as a complementary explanation of the abnormal
response: the so-called {\textit{city-site effect}}. A building or
a group of buildings over a hard half-space, solicited by a plane
incident SH wave, has been shown to modify the seismic waves on
the ground near the building \cite{trifunac,luco}, the
modification being larger when more buildings are taken into
account because of multiple-interaction: i.e., the so-called
{\textit{structure-soil-structure interaction}}. For models of the
geophysical structure involving only a hard half-space, the
stress-free base block mode appears to be the main cause of the
modification \cite{luco}.

The studies that deal with a geophysical structure involving, in
addition, a soft-layer overlying the hard-half space, have been
mainly concerned either with an infinite set of
periodically-arranged
\cite{wirginetbard,Boutinroussillon,Boutinrous} or randomly-
arranged \cite{lombert,clouteau} buildings on, or partially
imbedded in, the ground. In \cite{wirginetbard}, the authors
suggest that the large duration and amplitude are strongly linked
to resonant phenomena of the soft-layer associated with waves
whose structure is close to that of Love waves. The solicitation
being of the form of a plane incident wave, such modes cannot be
excited in the absence of buildings \cite{grobyetwirgin2005}.

In \cite{levander,hill}, it was shown that the modes of a soft
layer/hard half space can be excited when the interface between
the subtratum and the layer present some irregularities. These
effects were qualified as "vertical and lateral interferences" in
a previous numerical study \cite{akla70}. The question of the
excitation of modes, via surface irregularities constituted by the
set of buildings on the ground, was subsequently addressed in
\cite{wirgintsogkagroby}. In \cite{wirgingroby,wirginandgroby2006}
it was found that the excitation of vibration modes associated
with a periodically-modulated surface impedance, modeling  a
periodic distribution of blocks emerging from a flat ground, can
lead to enhanced durations and amplifications of the cumulative
displacement and velocity as compared to what is found for a flat
stress-free or constant surface impedance surface. The authors of
\cite{wirginandgroby2006} show that these modes manifest
themselves by amplified evanescent waves in the substratum.

The contributions \cite{Boutinroussillon,Boutinrous} employ
homogeneized models of a periodic city, but the fact that these
models  are restricted to low frequencies may explain why they  do
not account for the amplifications obtained in
\cite{wirgintsogkagroby} \cite{wirginandgroby2006}. In a host of
other  numerical studies
\cite{lombert,Guegen,chavezgarciabard,Boutinroussillon,Boutinrous,clouteau,fah,febi06},
the presence of buildings is found either to hardly modify, or to
de-amplify, the seismic disturbance, in contradiction with what is
shown in  \cite{semblat2003,wirgintsogka,grobythese}.

In \cite{rial89},  the spatial variability of damage to structures
on the ground was attributed to the variability of the resonance
frequencies of the buildings and of the soil structure beneath
each building, with the implication that the most dangerous
situation is when the natural frequency of the building (often
treated as a one degree (or several degrees) of freedom oscillator
\cite{jebi73,gu00,clouteau,Boutinrous,ro06}) is coincident with
that (obtained by a 1D analysis) of the substructure below the
base of the building (a well-known paradigm in the civil
engineering community known as the {\it double resonance}).

Another point of view is to consider the building as a seismic
source, either when it is solicited artificially by a vibrator
located on its roof \cite{jennings,wongtrif}, or when it re-emits
vibrations received from the incident seismic disturbance (or
other form of solicitation such as that coming from   an
underground nuclear explosion \cite{shaw,doby}). It is not
unreasonable to think that the presence of one or more buildings
on the ground enables the excitation of the (Love, Rayleigh) modes
of the underground system.  This is known to be possible when a
flat stress-free surface overlying a soft layer in welded contact
with a hard substratum is solicited by a source located in the
layer  or substratum \cite{grobyetwirgin2005} and should therefore
also occur when the source (i.e., the building) is on the free
surface.

The present work originated in the observation that no
satisfactory {\it theoretical} explanation has been given until
now of the influence of buildings on anomalous seismic response in
urban environments with soft layers, or large basins, overlying a
hard substratum. The principal reason for this knowledge gap
probably lies in the complexity of the sites examined in previous
studies and in the complexity of the phenomena. Thus, it appears
to be opportune to develop a theoretical model which is as
complete and as simple as possible, on an idealized, although
rather representative urban site, in order to address the
following questions:

(i) how should one account for the principal features of the
seismic response in the cases of a relatively small, and then,
large number of blocks?

(ii) what are the modes of the global structures (i.e. the
superstructure plus the geophysical structure) and what are the
mechanisms of their excitation and interaction?

(iii) what are the repercussions of resonant phenomena on the
seismic response?

(iv) what are the differences in seismic response between
configurations with a small and a large number of blocks?

The investigation herein focuses on the seismic response of one
and two blocks (the case of a periodic set of blocks is considered
in the companion parper) in welded contact with a soft layer
overlying a hard half-space. The modal analysis of the whole
configuration, backed up by extensive numerical computations,
shows that: i) the presence of one block induces the excitation of
two types of modes, the first
 whose structure is close to that of a mode of the
geophysical structure, and the second whose structure is close to
that of a mode of the superstructure (i.e., the set of blocks,
each of which is formed of one or several buildings), ii) the
presence of more than one block gives rise to {\it coupled modes}
resulting from a combination of the two other types of modes.

We uncover the mechanism of the (so-called {\it soil-structure})
interaction between the superstructure and the geophysical
substructure. Despite the fact that differences are noticed
between the computed displacements for a configuration with, and
in the absence of, buildings, mainly consisting in a longer
duration and
 a larger displacement in the building than at the same
 location in the absence of the building, and in a modification, due
 to the presence of the block(s) of
the structure of the waves traveling in the layer, no very
pronounced effects are apparent in the case of only a few (one or
two) buildings. This could mean that the \textit{city-site effect}
is important only when a large number of buildings is accounted
for. This case is investigated theoretically and numerically in
the companion paper for a periodic arrangement of identical
blocks. \clearpage
\section{Description of the configuration}\label{config}
\begin{figure}[ptb]
\begin{center}
\includegraphics[width=12cm] {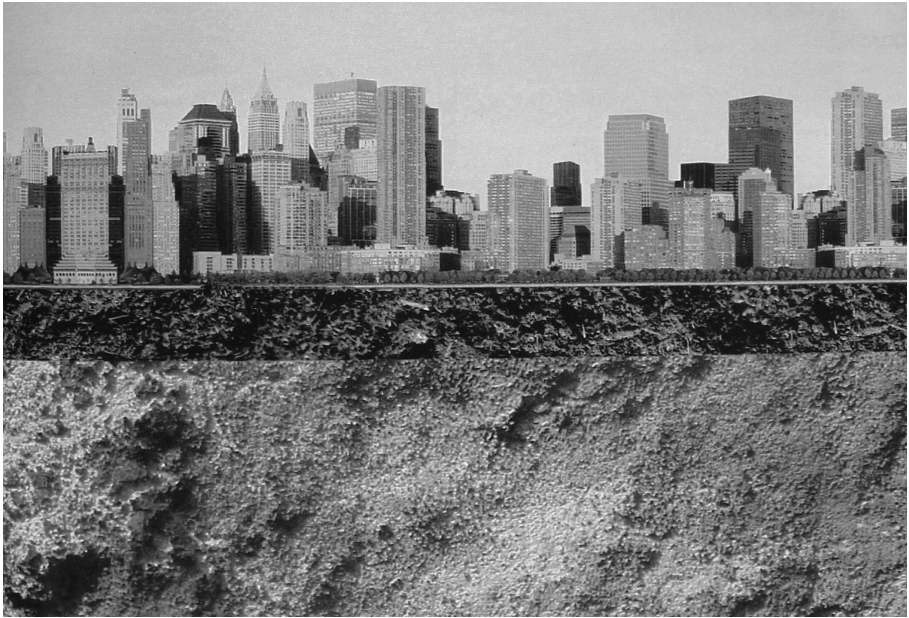}
\caption{View of a modern city with the underground.}
\label{ville}
\end{center}
\end{figure}

\begin{figure}[ptb]
\begin{center}
\includegraphics[scale=0.3] {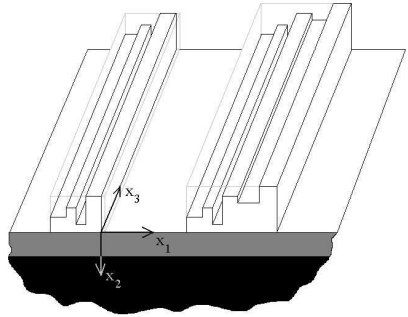}
\end{center}
\caption{View of the 2D city (only two of the blocks are
represented).} \label{fig1}
\end{figure}

We focus on a portion of a modern city consisting of a
 set of  blocks (see e.g., fig. \ref{ville}
in which it will be noted that the blocks (i.e., buildings or
groups of buildings) are not generally identical, nor arranged in
periodic manner). The city has 2D geometry, with $x_{3}$ the
ignorable coordinate of a $Ox_{1}x_{2}x_{3}$ cartesian coordinate
system (see Fig. \ref{fig1}).
The buildings are assumed to be in welded contact, across the flat
ground surface, with the substructure. The latter is composed of a
horizontal  soft layer underlain by a hard half space (see fig.
\ref{ville}). Each block is characterized by two constants, its
height $b_{j}$ and width $w_{j}$, and all blocks have the same
rectangular geometry (but not the same sizes)  and composition.
Let $d_{j}$ be the $x_{1}$ coordinate of the center of the base
segment of the $j-$th block. The distance between the blocks $j$
and $i$ is denoted by $d_{ji}=|d_{j}-d_{i}|$ and is not
necessarily constant between successive pairs of blocks.

\begin{figure}[ptb]
\begin{center}
\includegraphics[scale=0.3] {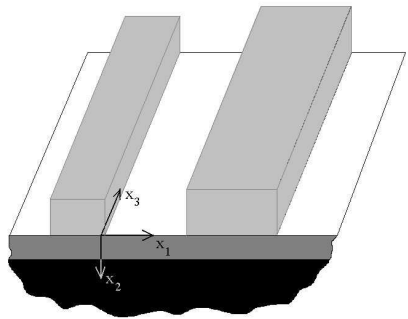}
\caption{View of the 2D city with homogenized blocks (only two of
the blocks are represented).}
\label{fig2}
\end{center}
\end{figure}

For the purpose of analysis, each block is homogenized (this {\it
does not mean} that the set of blocks is reduced to a single
horizontal, homogeneous layer, as in
\cite{Boutinroussillon,Boutinrous}), so that the final aspect of
the city is as in Fig. \ref{fig2}.
Let $\mathbb{B}\in \mathbb{Z}$ denote the set of indices by which
the blocks are identified (e.g., for three blocks: $\{1,2,3\}$ or
$\{-1,0,1\}$). The cardinal of $\mathbb{B}$ is designated by $N$
(i.e., $N$ denotes the number of blocks in the configuration, and
this number will either be finite (in the following analysis) or
infinite (as in the companion paper).

$\Gamma_{f}$ is the stress-free surface composed of a ground
portion $\Gamma_{g}$, assumed to be flat and horizontal, and a
portion $\Gamma_{ag}$, constituting the reunion of the
above-ground-level boundaries $\Gamma_{ag}^{j}~;~j\in \mathbb{B}$
of the blocks. The ground $\Gamma_{G}$ is flat and horizontal, and
is the reunion of $\Gamma_{g}$ and the base segments
$\Gamma_{bs}^{j}~;~j\in \mathbb{B}$ joining the blocks to the
underground.

The medium in contact with, and above, $\Gamma_{f}$ is air,
assumed to be the vacumn (which is why $\Gamma_{f}$ is
stress-free). The medium in contact with, and below $\Gamma_{G}$
is the mechanically-soft layer occupying the domain $\Omega_{1}$,
which is laterally-infinite and of thickness $h$, and whose lower
boundary is $\Gamma_{h}$, also assumed to be flat and horizontal.
The soft material in the layer is in welded contact across
$\Gamma_{h}$ with the mechanically-hard material in the
semi-infinite domain (substratum) $\Omega_{0}$.

The domain of the $j$-th block is denoted by $\Omega_{2}^{j}$ and
the reunion of all the $\Omega_{2}^{j}~;~j\in \mathbb{B}$ is
denoted by $\Omega_{2}$. The material in each block is in welded
contact with the material in the soft layer across the base
segments $\Gamma_{bs}^{j}~;~j\in \mathbb{B}$.
\begin{figure}[ptb]
\begin{center}
\includegraphics[scale=0.3] {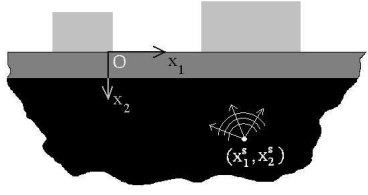}
\caption{Sagittal ($x_{1}0x_{2}$) plane view of the 2D city with
homogenized blocks (only two of the blocks are represented)
solicited by a cylindrical wave radiated by a line source located
at $(x_{1}^{s},x_{2}^{s})\in\Omega_{0}$.} \label{fig3b}
\end{center}
\end{figure}
\begin{figure}[ptb]
\begin{center}
\includegraphics[scale=0.3] {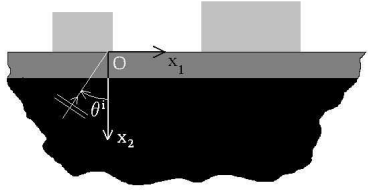}
\caption{Sagittal ($x_{1}0x_{2}$) plane view of the 2D city with
homogenized blocks (only two of the blocks are represented)
solicited by a plane wave with incident angle $\theta^{i}$.}
\label{fig3c}
\end{center}
\end{figure}

The origin $O$ of the cartesian coordinate system is on the
ground, $x_{2}$ increases with depth and $x_{3}$ is perpendicular
to the (sagittal) plane of the figs. \ref{fig3b}-\ref{fig3c}. With
$\mathbf{i}_{j}$ the unit vector along the positive $x_{j}$ axis,
we note that the unit vectors normal to $\Gamma_{G}$ and
$\Gamma_{h}$ are $-\mathbf{i}_{2}$.

The media filling $\Omega_{0}$, $\Omega_{1}$ and $\bigcup_{j\in
\mathbb{B}}\Omega_{2}^{j}$ are $M^{0}$, $M^{1}$ and $M^{2}$
respectively and the latter are assumed to be initially
stress-free, linear, isotropic and homogeneous (thus, each block,
which is generally inhomogeneous, is assumed to be homogenized in
our analysis). We assume that $M^{0}$ is non-dissipative whereas
$M^{1}$ and $M^{2}$ are dissipative, described by a constant
quality factor $Q^{j}$ in the frequency range of excitation.

The seismic disturbance is delivered to the site in the form of a
shear-horizontal (SH) cylindrical wave (radiated by a line source
parallel to the $x_{3}$ axis and located in $\Omega_{0}$; see fig.
\ref{fig3b}) or a plane wave (with incident angle $\theta^{i}$;
see fig. \ref{fig3c}), propagating initially in $\Omega_{0}$ (this
meaning, that in the absence of the layer, the city, and the air,
the total field is precisely that associated with this cylindrical
or plane wave). The SH nature of the {\it incident wave}
(indicated by the superscript $i$ in the following) means that the
motion associated with it is strictly transverse (i.e., in the
$x_{3}$ direction and independent of the $x_{3}$ coordinate). Both
the SH polarization and the invariance of the incident wave with
respect to $x_{3}$ are communicated to the fields that are
generated at the site in response to the incident wave. Thus, our
analysis deals with the propagation of 2D SH waves (i.e., waves
that depend exclusively on the two cartesian coordinates
$(x_{1},x_{2})$ and that are associated with motion in the $x_{3}$
direction only).

We shall be concerned with a description of the elastodynamic
wavefield on the free surface (i.e., on $\Gamma_{f}$) resulting
from the cylindrical or plane seismic wave sollicitation of the
site.
\section{Governing equations}\label{goveqs}
%
\subsection{Space-time framework wave equations}\label{xtframe}
In a generally-inhomogeneous, isotropic elastic or viscoelastic
medium $M$ occupying $\mathbb{R}^{3}$, the space-time framework
wave equation for SH waves is:
\begin{equation}\label{wgoveqs.1}
 \nabla\cdot (\mu(\mathbf{x},\omega)\nabla u(\mathbf{x},t))-
 \rho(\mathbf{x})\partial_{t}^{2}u(\mathbf{x},t)=-\rho(\mathbf{x})
 f(\mathbf{x},t) ~,
\end{equation}
wherein $u$ is the displacement component in the $\mathbf{i}_{3}$
direction,  $f$ the component of applied force density in the
$\mathbf{i}_{3}$ direction, $\mu$ the Lam\'e descriptor of
rigidity, $\rho$ the mass density, $t$ the time variable, $\omega$
the angular frequency, $\partial^{n}_{t}$ the $n-$th partial
derivative with respect to $t$, and $\mathbf{x}=(x_{1},x_{2})$.
Since our configuration involves three homogeneous media, and the
applied force is assumed to be non vanishing only in $\Omega_{0}$,
we have
\begin{equation}\label{wgoveqs.2}
 \left ( c^{m}(\omega)\right ) ^{2}\nabla\cdot\nabla u^{m}(\mathbf{x},t)-
 \partial_{t}^{2}u^{m}(\mathbf{x},t)= -f(\mathbf{x},t)\delta_{m0}
 ~~;~~\mathbf{x}\in\Omega_{m}~;~m=0,1,2~,
\end{equation}
wherein superscripts  $m$ designate the medium (0 for $M^{0}$,
etc.), $\delta_{m0}=1$ for $m=0$, $\delta_{m0}=0$ for $m\neq 0$,
and $c^{m}$ is the generally-complex velocity of shear body waves
in $M^{m}$, related to the density and rigidity by
\begin{equation}\label{wgoveqs.3}
  \left ( c^{m}(\omega)\right ) ^{2}=\frac{\mu^{m}(\omega)}{\rho^{m}}~,
\end{equation}
it being understood that $\rho^{m},~\mu^{m}(\omega)~;~m=0,1,2$ are
constants with respect to $\mathbf{x}$. In addition, the densities
are positive real and we assume that substratum is a
dissipation-free solid so that the rigidity therein is a positive
real constant with respect to $\omega$, i.e.,
$\mu^{0}(\omega)=\mu^{0}>0$.
\subsection{Space-frequency framework wave
equations}\label{xomegaframe}
The space-frequency framework versions of the wave equations are
obtained by expanding the force density and displacement in
Fourier integrals:
\begin{equation}\label{xomegaframe.1}
  f(\mathbf{x},t)=\int_{-\infty}^{\infty}f(\mathbf{x},\omega)e^{-\mbox{i}\omega t}
  d\omega~~,~~u^{m}(\mathbf{x},t)=\int_{-\infty}^{\infty}
  u^{m}(\mathbf{x},\omega)e^{-\mbox{i}\omega t}d\omega~,\forall t\in
  \mathbb{R}~,
\end{equation}
so as to give rise to the Helmholtz equations
\begin{equation}\label{xomegaframe.2}
 \nabla\cdot\nabla
  u^{m}(\mathbf{x},\omega)+\left ( k^{m}(\omega)\right ) ^{2}
  u^{m}(\mathbf{x},\omega)=
  -f(\mathbf{x},\omega)\delta_{m0}~~;~~\forall\mathbf{x}\in
  \Omega_{m}~~;~~m=0,1~,
\end{equation}
wherein
\begin{equation}\label{xomegaframe.3}
  k^{m}(\omega):=\frac{\omega}{c^{m}(\omega)}=
  \omega\sqrt{\frac{\rho^{m}}{\mu^{m}(\omega)}}~.
\end{equation}
is the generally-complex wavenumber in $M^{m}$. Actually, due to
the assumptions made in sects. \ref{config} and \ref{xomegaframe}:
\begin{equation}\label{xomegaframe.4}
  k^{0}(\omega):=\frac{\omega}{c^{0}}=
  \omega\sqrt{\frac{\rho^{0}}{\mu^{0}}}~,
\end{equation}
(i.e., $k^{0}$ is a positive real quantity which depends linearly
on $\omega$ ).

As mentioned above, we shall be concerned with cylindrical or
plane wave excitation of the city. Plane waves correspond to $f=0$
and cylindrical waves  to $f\neq 0$.

The incident field is chosen to take the form of a pseudo
Ricker-type pulse in the time domain.
\subsection{Space-frequency framework expression of the driving
force for cylindrical wave excitation}\label{xomegaexcit}
The space-frequency framework expression of the driving force
density for a cylindrical wave radiated from a line source located
at $\mathbf{x}^{s}:=(x_{1}^{s},x_{2}^{s})\in\Omega_{0}$ is
\begin{equation}\label{xomegaexcit.1}
  f(\mathbf{x},\omega)=S(\omega)
  \delta(\mathbf{x}-\mathbf{x^{s}})~,
\end{equation}
wherein $S(\omega)$ is the spectrum of the incident pulse and
is chosen to be a time derivative of a Ricker pulse and
$\delta(.)$ is the Dirac distribution.
 The
amplitude spectrum $S(\omega)$ is given by
\begin{equation}\label{xomegaexcit.2}
   S(\omega)=\frac{12\pi\alpha^{2} \omega^{2}}{\sqrt{\pi}}\frac{\omega^{2}}{4\alpha^{3}}
   \exp\left (\mbox{i}t_{s}\omega-\frac{\omega^{2}}{4\alpha^{2}}\right) ~,
\end{equation}
to which corresponds the temporal variation (Fourier inverse of
$S(\omega)$):
\begin{equation}\label{xomegaexcit.3}
   S(t)=-24\pi\alpha^{4}\left[-3(t_{s}-t)+ 2\alpha^{2}(t_{s}-t)^{3}\right]
   \exp\left [ -\alpha^{2}(t_{s}-t)^{2}\right] ~,
\end{equation}
wherein $\alpha=\pi/t_{p}$, $t_{p}$ is the characteristic period
of the pulse, and $t_{s}$ the time at which the pulse attains its
maximal value. In the remainder of this paper, we shall take
$t_{s}=t_{p}=2$ sec.

The (incident) wave associated with this driving force is
\begin{equation}\label{xomegaexcit.4}
u^{i}(\mathbf{x},\omega)=\int_{\mathbb{R}^{2}}G^{0}(\|\mathbf{x}-\mathbf{y}\|,\omega)
f(\mathbf{y},\omega)d\varpi(\mathbf{y})~,
\end{equation}
wherein $\mathbf{y}:=(y_{1},y_{2})$ is an integration point in the
sagittal plane, $d\varpi(\mathbf{y})$ the differential area
element at point $\mathbf{y}$ and
$G^{0}(k^{0}\|\mathbf{x}-\mathbf{y}\|)$  the 2D free-space Green's
function which satisfies:
\begin{equation}\label{xomegaexcit.5}
  \left[ \Delta+\left ( k^{0}\right ) ^{2}\right]
  G^{0}(\|\mathbf{x}-\mathbf{y}\|,\omega)=
  -\delta(\mathbf{x}-\mathbf{y})~;~\forall\mathbf{x}
  \in\mathbb{R}^{2}~,
\end{equation}
(with $\delta(~.~)$ the Dirac delta distribution) and the
(outgoing wave) radiation condition:
\begin{equation}\label{xomegaexcit.6}
G^{0}(\|\mathbf{x}-\mathbf{y}\|,\omega)\sim ~ \text{ outgoing
waves}~~;~~\|\mathbf{x}-\mathbf{y}\| \rightarrow \infty~.
\end{equation}
The free-space Green's function is given by \cite{MF53}:
\begin{equation}\label{xomegaexcit.7}
G^{0}(\|\mathbf{x}-\mathbf{y}\|,\omega)=\frac{\mbox{i}}{4}H_{0}^{(1)}
(k^{0}\|\mathbf{x}-\mathbf{y}\|)=\frac{\mbox{i}}{4\pi}\int_{-\infty}^{\infty}
\exp\{i[k_{1}(x_{1}-y_{1})+k_{2}^{0}|x_{2}-y_{2}|]\}\frac{dk_{1}}{k_{2}^{0}}~,
\end{equation}
with $H_{0}^{(1)}(.)$ the Hankel function of the first kind and
order $0$, and
\begin{equation}\label{xomegaexcit.7a}
k_{2}^{j}=\sqrt{(k^{j})^{2}-(k_{1})^{2}}~~;~~\Re k_{2}^{j}\geq
0~~,~~\Im k_{2}^{j}\geq 0 ~~\text{for}~\omega\geq 0~.
\end{equation}
 Introducing (\ref{xomegaexcit.7}) and
(\ref{xomegaexcit.2}) into (\ref{xomegaexcit.4}) results in
\begin{equation}\label{xomegaexcit.8}
u^{i}(\mathbf{x},\omega)=S(\omega)G^{0}(\|\mathbf{x}-\mathbf{x}^{s}\|,\omega)=
S(\omega)\frac{\mbox{i}}{4}H_{0}^{(1)} (k^{0}
\|\mathbf{x}-\mathbf{x}^{s}\|)~,
\end{equation}
which is the space-frequency expression of a cylindrical wave.
\subsection{Material constants in a dissipative
medium}\label{dissmat}
A word is now in order concerning the dissipative nature of the
layer and blocks. In seismological applications involving
viscoelastic media, the quality factor is usually considered to be
either constant or a weakly-varying function of frequency
\cite{fah} in the bandwith of the source. We shall therefore
assume that $Q^{j}(\omega)=Q^{j}$, with $Q^{j}$ constants,
$j=1,2$. It can be shown \cite{kj79} that this implies
\begin{equation}\label{dissmat.16}
 \mu^{j}(\omega)=\mu^{j}_{ref} \left( \frac{-\mbox{i} \omega}
 {\omega_{ref}} \right) ^{\frac{2}{\pi}
 \arctan\left ( \frac{1}{Q^{j}}\right ) }~;~j=1,2~,
\end{equation}
wherein: $\omega_{ref}$ is a reference angular frequency, chosen
herein to be equal to $9\times 10^{-2}$Hz. Hence
\begin{equation}\label{dissmat.17}
 c^{j}(\omega)=c^{j}_{ref} \left( \frac{-\mbox{i} \omega}
 {\omega_{ref}} \right) ^{\frac{1}{\pi}
 \arctan\left ( \frac{1}{Q^{j}}\right ) }~;~j=1,2~,
\end{equation}
with $c^{j}_{ref}:=\sqrt\frac{\mu^{j}_{ref}}{\rho^{j}}$. Note
should be taken of the fact that even though $Q^{j}~,~j=1,2$ are
non-dispersive (i.e., do not depend on $\omega$) under the present
assumption, the phase velocities $c^{j}~;~j=1,2$ {\it are}
dispersive.
\subsection{Boundary and radiation conditions in the
space-frequency framework}\label{boundrad}
The translation of the stress-free (i.e., vanishing traction)
nature of $\Gamma_{f}=\Gamma_{g}\bigcup\Gamma_{ag}$, with
$\Gamma_{ag}:=\bigcup_{j\in\mathbb{B}}\Gamma_{ag}^{j}$, is:
\begin{equation}\label{boundrad.1}
\mu^{1}(\omega)\partial_{n}
u^{1}(\mathbf{x},\omega)=0~;~\mathbf{x} \in \Gamma_{g},
\end{equation}
\begin{equation}\label{boundrad.2}
\mu^{2}(\omega)\partial_{n}
u^{2(j)}(\mathbf{x},\omega)=0~;~\mathbf{x} \in \Gamma_{ag}^{j}~,~j\in\mathbb{B}
\end{equation}
wherein $\mathbf{n}$ denotes the generic unit vector normal to a
boundary and $\partial_{n}$ designates the operator
$\partial_{n}=\mathbf{n}\cdot\nabla$.

That $M^{1}$ and $M^{2}$ are in welded contact across
$\Gamma_{bs}:=\bigcup_{j\in\mathbb{B}}\Gamma_{bs}^{j}$ is
translated by the fact that the displacement and traction are
continuous across $\Gamma_{bs}$:
\begin{equation}\label{boundrad.3}
u^{1}(\mathbf{x},\omega)-u^{2(j)}(\mathbf{x},\omega)=0~;~\mathbf{x}
\in \Gamma_{bs}^{j}~,~j\in\mathbb{B}
\end{equation}
\begin{equation}\label{boundrad.4}
\mu^{1}(\omega)\partial_{n}
u^{1}(\mathbf{x},\omega)-\mu^{2}(\omega)\partial_{n}
u^{2(j)}(\mathbf{x},\omega)=0~;~\mathbf{x} \in \Gamma_{bs}^{j}~,~j\in\mathbb{B}.
\end{equation}
That $M^{1}$ and $M^{0}$ are in welded contact across $\Gamma_{h}$
is translated by the fact that the displacement and traction are
continuous across this interface:
\begin{equation}\label{boundrad.5}
u^{1}(\mathbf{x},\omega)-u^{0}(\mathbf{x},\omega)~;~\mathbf{x} \in
\Gamma_{h}~,
\end{equation}
\begin{equation}\label{boundrad.6}
\mu^{1}(\omega)\partial_{n}
u^{1}(\mathbf{x},\omega)-\mu^{0}(\omega)\partial_{n}
u^{0}(\mathbf{x},\omega)~;~\mathbf{x} \in \Gamma_{h}.
\end{equation}
The uniqueness of the solution to the forward-scattering problem
is assured by the radiation condition in the substratum:
\begin{equation}\label{boundrad.7}
u^{0}(\mathbf{x},\omega)-u^{i}(\mathbf{x},\omega)\sim ~ \text{
outgoing waves}~~;~~ \|\mathbf{x}\|\rightarrow \infty,~~x_{2}>h~.
\end{equation}
\subsection{Boundary and radiation conditions in the
space-time framework}\label{xtboundrad}
Since our finite element method
\cite{grobyettsogka,wirgintsogkagroby,grobythese} for solving the
wave equation in a heterogeneous medium $M$ (in our case,
involving three homogeneous components, $M^{0}$, $M^{1}$ and
$M^{2}$) relies on the assumption that $M$ be a continuum, it does
not appeal to any boundary conditions except on $\Gamma_{f}$ where
the vanishing traction condition is invoked (fictitious domain
method). Furthermore, since the essentially unbounded nature of
the geometry of the city cannot be implemented numerically, we
take this geometry to be finite and surround it (except on the
$\Gamma_{f}$ portion) by a  perfectly-matched layer (PML)
\cite{collinoettsogka} which enables closure of the computational
domain without generating unphysical reflected waves (from the PML
layer). In a sense, this replaces the radiation condition of the
unbounded domain. The stress-free boundary condition on $\Gamma_f$
is modeled with the help of the fictitious domain method
\cite{becachejolyettsogka}, which allows us to account for the
diffraction of waves by a boundary of complicated geometry, not
necessarily matching the volumic mesh.
\subsection{Statement of the boundary-value (forward - scattering)
problem in the space-time framework}\label{xtboundvalpb}
The problem is to determine the time record of the displacement
fields $u^{1}(\mathbf{x},t)$ on $\Gamma_{g}$ and
$u^{2(j)}(\mathbf{x},t)$ on $\Gamma_{ag}^{j}$, $j\in\mathbb{B}$.
\subsection{Recovery of the  space-frequency displacements from the space-time displacements}\label{xomegaboundvalpb}
The spectra of the displacements are obtained from the time
records of the displacements by Fourier inversion, i.e.,
\begin{equation}\label{xomegaboundvalpb.1}
 u^{j}(\mathbf{x},\omega)= \frac{1}{2\pi}\int_{-\infty}^{\infty}u^{j}(\mathbf{x},t)
 e^{\mbox{i}\omega t}
  dt~;~j=1,2~.
\end{equation}
\section{Field representations in the space-frequency framework
for $N<\infty$} \label{xtfieldrep}
%
\subsection{Field in $\Omega_{0}$}
\label{fieldrepomega0}
It is useful to consider the boundary $\partial\Omega_{0}$ of
$\Omega_{0}$ to be composed of $\Gamma_{h}$ plus a semi-circle of
infinite radius $\Gamma_{\infty}$ joining $\Gamma_{h}$ at
$\mathbf{x}=(-\infty,h)$ and $\mathbf{x}=(-\infty,h)$. The unit
vector $\mathbf{n}$ normal to $\partial\Omega_{0}$ is taken to be
outward with respect to $\Omega_{0}$, so that it is equal to
$-\mathbf{i}_{2}$ on $\Gamma_{h}$.

We seek the field representation in $\Omega_{0}$. Applying Green's
second identity to $u^{0}$ and $G^{0}$ in $\Omega_{0}$ and making
use of the radiation condition at infinity relative to these two
functions, gives
\begin{multline}\label{fieldrepomega0.1}
  \mathcal{H}_{\Omega_{0}}(\mathbf{x})u^{0}(\mathbf{x},\omega)=
u^{i}(\mathbf{x},\omega)+
\\
\int_{\Gamma_{h}} \left [ G^{0}(\|\mathbf{x}-\mathbf{y}\|,\omega)
\partial_{n}u^{0}(\mathbf{y},\omega)-
u^{0}(\mathbf{y},\omega)
\partial_{n}G^{0}(\|\mathbf{x}-\mathbf{y}\|,\omega)
\right] d\gamma(\mathbf{y})~,
\end{multline}
wherein $d\gamma(\mathbf{y})$ is the infinitesimal arc length
along $\Gamma_{h}$ and
\begin{equation}\label{fieldrepomega0.2}
  \mathcal{H}_{\Omega_{0}}(\mathbf{x})=\left \{
\begin{array}{cc}
1 & ~;~\mathbf{x}\in \Omega_{0}
\\
0 & ~~~~~~~~;~\mathbf{y}\in \mathbb{R}^{2}\setminus
\overline{\Omega}_{0}
\\
1/2 & ;~\mathbf{y}\in \Gamma_{h}
\end{array}\right.
~.
\end{equation}
Introducing  the  cartesian coordinate integral representation of
the Green's function (\ref{xomegaexcit.7}) into the boundary
integral representation of the field (\ref{fieldrepomega0.1}),
while paying attention to the absolute values, leads to the
following result:
\begin{equation}\label{fieldrepomega0.3}
u^{0}(\mathbf{x},\omega)=
u^{i}(\mathbf{x},\omega)+
\int_{-\infty}^{\infty}B^{0}(k_{1},\omega)
\exp\left\{\mbox{i}\left[k_{1}x_{1}+k_{2}^{0}(x_{2}-h)\right]\right\}
  \frac{dk_{1}}{k_{2}^0}~;~\mathbf{x}\in \Omega_{0}~,
\end{equation}
wherein:
\begin{equation}\label{fieldrepomega0.4}
B^{0}(k_{1},\omega)=\frac{i}{4\pi}\int_{-\infty}^{\infty}
\left\{\partial_{y_{1}}u^{0}(y_{1},h,\omega)+
ik_{2}^{0}u^{0}(y_{1},h,\omega)\right\}\exp\left( -i
k_{1}y_{1}\right) dy_{1}~,
\end{equation}
At this point, we must distinguish between plane wave excitation
(briefly alluded-to in a subsequent section) and cylindrical wave
excitation (to which all the following numerical results apply).

In the case of  {\it plane-wave excitation} we can write
\begin{equation}\label{w2.5.9}
u^{i}(\mathbf{x},\omega)=
\int_{-\infty}^{\infty}A^{0-}(k_{1},\omega)
\exp\left\{i\left[k_{1}x_{1}-k_{2}^{0}x_{2}\right]\right\}
 \frac{ dk_{1}}{k_{2}^{0}}~;~\mathbf{x}\in \mathbb{R}^{2}~,
\end{equation}
wherein
\begin{equation}\label{w2.5.10}
A^{0-}(k_{1},\omega)=S(\omega)k_{2}^{0}\delta(k_{1}-k_{1}^{i})~,
\end{equation}
with $k_{1}^{i}=k^{0}\sin\theta^{i}$, and $\theta^{i}$ the angle
of incidence,  so that
\begin{equation}\label{w2.5.11}
u^{i}(\mathbf{x},\omega)=S(\omega) \exp\left\{
\mbox{i}\left[k_{1}^{i}x_{1}-k_{2}^{i}x_{2}\right]\right\}
 ~;~\mathbf{x}\in \mathbb{R}^{2}~,
\end{equation}
wherein $k_{2}^{i}=k^{0}\cos\theta^{i}$.

 In the case of
{\it cylindrical wave excitation}, we have, on account of
(\ref{xomegaexcit.7}) and (\ref{xomegaexcit.8}):
\begin{equation}\label{fieldrepomega0.5}
u^{i}(\mathbf{x},\omega)= \left\{
\begin{array}{cc}
\displaystyle \int_{-\infty}^{\infty}A^{0+}(k_{1},\omega)
\exp\left\{\mbox{i}\left[k_{1}x_{1}+k_{2}^{0}x_{2}\right]\right\}
  \frac{dk_{1}}{k_{2}^{0}} & ~;~\mathbf{x}\in \Omega_{0}^{+}\\
\displaystyle  \int_{-\infty}^{\infty}A^{0-}(k_{1},\omega)
\exp\left\{\mbox{i}\left[k_{1}x_{1}-k_{2}^{0}x_{2}\right]\right\}
   \frac{dk_{1}}{k_{2}^{0}} & ~;~\mathbf{x}\in \Omega_{0}^{-}
  \end{array}
  \right. ~,
\end{equation}
wherein
\begin{equation}\label{fieldrepomega0.6}
 \Omega_{0}^{+}=\{\forall x_{1}\in\mathbb{R}~;~x_{2}>x_{2}^{s}\}
  ~,
\end{equation}
\begin{equation}\label{fieldrepomega0.7}
 \Omega_{0}^{-}=\{\forall x_{1}\in\mathbb{R}~;~h<x_{2}<x_{2}^{s}\}
  ~,
\end{equation}
\begin{equation}\label{fieldrepomega0.8}
A^{0+}(k_{1},\omega)=S(\omega)\frac{\mbox{i}}{4\pi}\exp
\left\{-\mbox{i}\left[k_{1}x_{1}^{s}+k_{2}^{0}x_{2}^{s}\right]\right\}
 ~,
\end{equation}
\begin{equation}\label{fieldrepomega0.9}
A^{0-}(k_{1},\omega)=S(\omega)\frac{\mbox{i}}{4\pi}\exp
\left\{-\mbox{i}\left[k_{1}x_{1}^{s}-k_{2}^{0}x_{2}^{s}\right]\right\}
 ~.
\end{equation}
Since we shall henceforth be interested only in the field in the
subdomain $\Omega_{0}^{-}$ of $\Omega_{0}$, we can write
\begin{multline}\label{fieldrepomega0.10}
u^{0}(\mathbf{x},\omega)=
\int_{-\infty}^{\infty}A^{0-}(k_{1},\omega)
\exp\left\{\mbox{i}\left[k_{1}x_{1}-k_{2}^{0}x_{2}\right]\right\}
  \frac{dk_{1}}{k_{2}^{0}}+
\\
\int_{-\infty}^{\infty}B^{0}(k_{1},\omega)
\exp\left\{\mbox{i}\left[k_{1}x_{1}+k_{2}^{0}(x_{2}-h)\right]\right\}
  \frac{dk_{1}}{k_{2}^{0}}~;~\mathbf{x}\in \Omega_{0}^{-}~,
\end{multline}
with the understanding that: i) $S(\omega)$ is known a priori and
given by its expression in (\ref{xomegaexcit.2}), ii)
$A^{0\pm}(k_{1},\omega)$ are known a priori, iii)
$B^{0}(k_{1},\omega)$ is an unknown function, iv)
$A^{0\pm}(k_{1},\omega)$ and $B^{0}(k_{1},\omega)$ have units of
(length) since $u^{0}$ (and, in general, all displacements), have
units of length, v) $u^{0}$ is expressed as a sum of incoming and
outgoing plane (bulk and evanescent) waves.
\subsection{Field in $\Omega_{1}$}\label{fieldrepomega1}
By proceeding in the same manner as previously we find
\begin{multline}\label{fieldrepomega1.1}
u^{1}(\mathbf{x},\omega)=
\int_{-\infty}^{\infty}A^{1}(k_{1},\omega)
\exp\left\{\mbox{i}\left[k_{1}x_{1}-k_{2}^{1}x_{2}\right]\right\}
  \frac{dk_{1}}{k_{2}^{1}}+
\\
\int_{-\infty}^{\infty}B^{1}(k_{1},\omega)
\exp\left\{\mbox{i}\left[k_{1}x_{1}+k_{2}^{1}x_{2}\right]\right\}
  \frac{dk_{1}}{k_{2}^{1}}~;~\mathbf{x}\in \Omega_{0}~,
\end{multline}
with the understanding that: i) now both $A^{1}(k_{1},\omega)$ and
$B^{1}(k_{1},\omega)$ are unknown functions,  ii) $u^{1}$ is
expressed as a sum of incoming and outgoing plane (bulk and
evanescent) waves.

\subsection{Field in the $j$-th block}\label{fieldrepblock}
The task is here to obtain a suitable representation of the field
in the generic block $j$ (of height $b_j$ and width $w_j$)
occupying the domain $\Omega_{2}^{j}$. The boundary of this domain
is $\partial\Omega_{2}^{j}=\Gamma_{ag}^{j}\bigcup\Gamma_{bs}^{j}$.
It should be recalled that the field satisfies a Neumann boundary
condition on the emerged boundary $\Gamma_{ag}^{j}$ of the block.
No boundary condition of the Neumann or Dirichlet type is
available on the segment $\Gamma_{bs}^{j}$ so that, strictly
speaking, we are not searching for a modal representation of the
field in the block domain, but rather for a {\it quasi-modal}
representation, the latter satisfying a priori the boundary
condition on $\Gamma_{ag}^{j}$, but no particular boundary
condition on $\Gamma_{bs}^{j}$.

Let $O^{j}x_{1}^{j}x_{2}^{j}x_{3}^{j}$ be the (local) cartesian
coordinate system attached to  $\Omega_{2}^{j}$  such that the
origin $O^{j}$ is located on, and at the center of, the segment
$\Gamma_{bs}^{j}$. We note that
\begin{equation}\label{fieldrepblock.1}
x_{1}=d_{j}+x_{1}^{j}~~,~~x_{2}=x_{2}^{j}~~;~~\forall j\in
\mathbb{B}~,
\end{equation}
wherein it should be recalled that $d_{j}$ is the $x_{1}$
coordinate of the center of the base segment of the $j-$th block.

We apply the separation of variables technique and the boundary
conditions on $\partial\Omega_{2}^{j}$ to obtain
\begin{equation}\label{fieldrepblock.2}
u^{2(j)}(\mathbf{x},\omega)=\sum_{m=0}^{\infty}B_{m}^{2(j)}(\omega)\cos\left[
k^{2(j)}_{1m}\left( x_{1}^{j}+\frac{w_j}{2}\right) \right] \cos\left[
k^{2(j)}_{2m}\left( x_{2}^{j}+b_j\right) \right]
\\
~~;~~ \mathbf{x}\in\Omega_{2}^{j}~,~\forall j\in \mathbb{B}~,
\end{equation}
wherein
\begin{equation}\label{fieldrepblock.3}
k^{2(j)}_{1m}=\frac{m\pi}{w_{j}}~;~ k^{2(j)}_{2m}=\sqrt{\left(
k^{2}\right) ^2-\left( k^{2(j)}_{1m}\right) ^2}~~;~~\Re\left(
k^{2(j)}_{2m}\right) \geq 0~~,~~\Im\left( k^{2(j)}_{2m}\right)
\geq 0~~\text{for}~\omega\geq 0~,
\end{equation}
and $B_{m}^{2(j)}(\omega)$ has units of length. On account of (\ref{fieldrepblock.1}) we finally obtain
\begin{multline}\label{fieldrepblock.5}
u^{2(j)}(\mathbf{x},\omega)=\sum_{m=0}^{\infty}B_{m}^{2(j)}(\omega)\cos\left[
k^{2(j)}_{1m}\left( x_{1}-d_{j}+\frac{w_{j}}{2}\right) \right] \cos\left[
k^{2(j)}_{2m}\left( x_{2}+b_j\right) \right]
\\
~~;~~ \mathbf{x}\in\Omega_{2}^{j}~,~\forall j\in \mathbb{B}~,
\end{multline}
it being understood that  the
$\mathbf{B}^{2(j)}:=\{B_{m}^{2(j)}(\omega)~;~m\in\mathbb{Z}\}~~,~~j\in\mathbb{B}$
are all unknown vectors.

\section{Determination of the various unknown coefficients by
application of boundary and continuity conditions on $\Gamma_{G}$
and  $\Gamma_{h }$ for the case $N<\infty$}\label{coeffs}
\subsection{Application of the  boundary and continuity conditions
concerning the traction on $\Gamma_{G}$}\label{stressgamg}
From (\ref{boundrad.2}) and (\ref{boundrad.3}) we obtain
\begin{multline}\label{stressgamg.1}
\mu^{1}\int_{-\infty}^{\infty}\partial_{x_{2}}u^{1}(x_{1},0,\omega)\exp(-\mbox {i}K_{1}x_{1})dx_{1}-
\\
\mu^{2}\sum_{j\in\mathbb{B}}\int_{d_{j}-w/2}^{d_{j}+w/2}\partial_{x_{2}}u^{2(j)}(x_{1},0,\omega)
\exp(-\mbox{i}K_{1}x_{1})dx_{1}=0
~~;~~ \forall K_{1}\in\mathbb{R}~.
\end{multline}
Introducing the appropriate field representations therein and
making use of the orthogonality condition
\begin{equation}\label{stressgamg.2}
\int_{-\infty}^{\infty}\exp[-\mbox{i}(k_{1}-K_{1})x_{1}]dx_{1}=2\pi\delta(k_{1}-K_{1})
~~;~~ \forall k_{1}~,~K_{1}\in\mathbb{R}~,
\end{equation}
gives rise to
\begin{multline}\label{stressgamg.3}
A^{1}(k_{1},\omega)-B^{1}(k_{1},\omega)=
\\
\frac{1}{2\pi
\mbox{i}}\sum_{j\in\mathbb{B}}
e^{-\mbox{i}k_{1}(d_{j}-w_{j}/2)}\sum_{m=0}^{\infty}B_{m}^{2(j)}(\omega)\frac{\mu^{2}k_{2m}^{2(j)}w_{j}}{\mu^{1}}
I_{m}^{(j)-}(k_{1},\omega)\sin(k_{2m}^{2(j)}b_{j})
 ~~;~~ \forall
k_{1}\in\mathbb{R}~,
\end{multline}
wherein
\begin{equation}\label{stressgamg.4}
I_{m}^{(j)\pm}(k_{1},\omega):=\int_{0}^{1}\exp(\pm
\mbox{i}k_{1}w_{j}\eta)\cos(k_{1m}^{2(j)}w_{j}\eta)d\eta~.
\end{equation}
\subsection{Application of the continuity conditions
concerning the displacement on $\Gamma_{G}$}\label{dispgammag}
From (\ref{boundrad.4})  we obtain
\begin{multline}\label{dispgammag.1}
\int_{d_{l}-\frac{w}{2}}^{d_{l}+\frac{w}{2}}u^{1}(x_{1},0,\omega)\cos\left[
k_{1n}^{2(l)}(x_{1}-d_{l}+w_{l}/2)\right] dx_{1}-
\\
\int_{d_{l}-\frac{w}{2}}^{d_{l}+\frac{w}{2}}u^{2(l)}(x_{1},0,\omega)\cos\left[
k_{1n}^{2(l)}(x_{1}-d_{l}+w_{l}/2)\right] dx_{1}=0 ~~;~~ \forall
l\in\mathbb{B}~.
\end{multline}
Introducing the appropriate field representations therein, and
making use of the orthogonality condition
\begin{multline}\label{dispgammag.2}
\int_{d_{l}-\frac{w}{2}}^{d_{l}+\frac{w}{2}}\cos\left[
k_{1m}^{2(l)}(x_{1}-d_{l}+w_{l}/2)\right] \cos\left[
k_{1n}^{2(l)}(x_{1}-d_{l}+w_{l}/2)\right]
dx_{1}=\frac{w_{l}}{\epsilon_{m}}\delta_{mn} \\
 ~~;~~ \forall
m~,~n=0,1,2,....~,
\end{multline}
wherein $\delta_{mn}$ is the Kronecker symbol and $\epsilon_{m}$
the Neumann symbol ($\epsilon_{m}=1$ for $m=0$, $\epsilon_{m}=2$
for $m>0$), gives rise to
\begin{multline}\label{dispgammag.3}
B_{m}^{2(l)}(\omega)=\frac{\epsilon_{m}}{\cos(k_{2m}^{2(l)}b_{l})}\int_{-\infty}^{\infty}
\left[ A^{1}(k_{1},\omega)+B^{1}(k_{1},\omega)\right]
I_{n}^{(l)+}(k_{1},\omega)e^{\mbox{i}k_{1}(d_{l}-w_{l}/2)}\frac{dk_{1}}{k_{2}^{1}}\\
 ~~;~~
\forall m=0,1,2,....~~,~~l\in\mathbb{B}~.
\end{multline}
\subsection{Application of the continuity conditions
concerning the traction on $\Gamma_{h}$}\label{stressgammah}
From (\ref{boundrad.5})  we obtain
\begin{equation}\label{stressgammah.1}
\mu^{0}\int_{-\infty}^{\infty}\partial_{x_{2}}u^{0}(x_{1},h,\omega)
\exp(-\mbox{i}K_{1}x_{1})dx_{1}-
\mu^{1}\int_{-\infty}^{\infty}\partial_{x_{2}}u^{1}(x_{1},h,\omega)
\exp(-\mbox{i}K_{1}x_{1})dx_{1}=0 ~~;~~ \forall
K_{1}\in\mathbb{R}~.
\end{equation}
Introducing the appropriate field representations therein, and
making use of the orthogonality relation (\ref{stressgamg.2}), gives
rise to
\begin{equation}\label{stressgammah.2}
-\mu^{0}A^{0-}(k_{1},\omega)e^{-\mbox{i}k_{2}^{0}h}+\mu^{0}B^{0}(k_{1},\omega)+
\mu^{1}A^{1}(k_{1},\omega)e^{-\mbox{i}k_{2}^{1}h}-\mu^{1}B^{1}(k_{1},\omega)e^{\mbox{i}k_{2}^{1}h}=0
~~;~~ \forall k_{1}\in\mathbb{R}~.
\end{equation}
\subsection{Application of  the continuity conditions
concerning the displacement on $\Gamma_{h}$}\label{displgammah}
From (\ref{boundrad.6})  we obtain
\begin{equation}\label{displgammah.1}
\int_{-\infty}^{\infty}u^{0}(x_{1},h,\omega)\exp(-\mbox{i}K_{1}x_{1})dx_{1}-
\int_{-\infty}^{\infty}u^{1}(x_{1},h,\omega)\exp(-\mbox{i}K_{1}x_{1})dx_{1}=0
~~;~~ \forall K_{1}\in\mathbb{R}~.
\end{equation}
Introducing the appropriate field representations therein, and
making use of the orthogonality relation (\ref{stressgamg.2}), gives
rise to
\begin{equation}\label{displgammah.2}
\frac{A^{0-}(k_{1},\omega)}{k_{2}^{0}}e^{-\mbox{i}k_{2}^{0}h}
+\frac{B^{0}(k_{1},\omega)}{k_{2}^{0}}-
\\
\frac{A^{1}(k_{1},\omega)e^{-\mbox{i}k_{2}^{1}h}}{k_{2}^{1}}-\frac{B^{1}(k_{1},\omega)e^{\mbox{i}k_{2}^{1}h}}{k_{2}^{1}}=0
~~;~~ \forall k_{1}\in\mathbb{R}~.
\end{equation}
\subsection{Determination of the various unknowns}
\subsubsection{Elimination ~of ~$B_{m}^{2j}(\omega)$ ~to
~obtain ~an ~integral equation for
$B^{0}(k_{1},\omega)$}\label{inteqb0}
 After a series of substitutions, the following integral equation for
$B^{0}(k_{1},\omega)$ is obtained (wherein $K_{1}$ and $K_{2}^{j}$
play  the same roles, and are related to each other in the same
manner, as $k_{1}$ and $k_{2}^{j}$ respectively):
\begin{equation}\label{inteqb0.2}
C(k_{1},\omega)B^{0}(k_{1},\omega)-\int_{-\infty}^{\infty}
D(k_{1},K_{1},\omega)B^{0}(K_{1},\omega)dK_{1}=F(k_{1},\omega)
~~;~~ \forall k_{1}\in\mathbb{R}~,
\end{equation}
wherein
\begin{equation}\label{inteqb0.3}
 C(k_{1},\omega)=\cos(k_{2}^{1}h)-\mbox{i}\frac{\mu^1 k_2^1}{\mu^0 k_2^0}\sin(k_{2}^{1}h)~,
\end{equation}
\begin{multline}\label{inteqb0.4}
D(k_{1},K_{1},\omega)= \left[
\cos(K_{2}^{1}h)-\mbox{i}\frac{\mu^1 K_2^1}{\mu^0 K_2^0}\sin(K_{2}^{1}h)\right] \times
\\
\frac{i}{2\pi}
\sum_{j\in\mathbb{B}}e^{\mbox{i}(K_{1}-k_{1})(d_{j}-w_{j}/2)}\sum_{m=0}^{\infty}\epsilon_{m}
\frac{\mu^{2}k_{2m}^{2(j)}w_{j}}{\mu^{0}K_{2}^{0}}
\tan(k_{2m}^{2(j)}b_j)I_{m}^{(j)-}(k_{1},\omega)I_{m}^{(j)+}(K_{1},\omega)
\\
 ~~;~~ \forall
k_{1},~K_{1}\in\mathbb{R}~,
\end{multline}
and
\begin{multline}\label{inteqb0.5}
F(k_{1},\omega)= A^{0-}(k_{1},\omega)
e^{-\mbox{i}k_{2}^{0}h}\left[ \cos(k_{2}^{1}h)+\mbox{i}\frac{\mu^1
k_2^1}{\mu^0 k_2^0}\sin(k_{2}^{1}h)\right] +
\\
\int_{-\infty}^{\infty}\frac{A^{0-}(K_{1},\omega)}{i}
e^{-\mbox{i}K_{2}^{0}h} \left[ \cos(K_{2}^{1}h)+\mbox{i}
\frac{\mu^0 K_2^0}{\mu^1 K_2^1}\sin(K_{2}^{1}h)\right] \times
\\
\frac{1}{2\pi}
\sum_{j\in\mathbb{B}}e^{\mbox{i}(K_{1}-k_{1})(d_{j}-w_{j}/2)}\sum_{m=0}^{\infty}\epsilon_{m}
\frac{\mu^{2}k_{2m}^{2(j)}w_{j}}{\mu^{0}K_{2}^{0}}
\tan(k_{2m}^{2(j)}b_j)I_{m}^{(j)-}(k_{1},\omega)I_{m}^{(j)+}(K_{1},\omega)dK_{1}
\\
 ~~;~~ \forall
k_{1}\in\mathbb{R}~,
\end{multline}
Eq.(\ref{inteqb0.2}) is a Fredholm integral equation of the second kind for
the unknown function $\{B^{0}(k_{1},\omega)~;~k_{1}\in
\mathbb{R}\}$.

By means of the changes of variables $k_{1}=k^{0}\sigma_{1}$,
$K_{1}=k^{0}S_{1}$ (note that $\sigma_{1}$ and $S^{1}$ are
dimensionless), we can cast (\ref{inteqb0.2}) into the form
\begin{equation}\label{resinteq.1}
C(\sigma_{1},\omega)B^{0}(\sigma_{1},\omega)-\int_{-\infty}^{\infty}
E(\sigma_{1},S_{1},\omega)B^{0}(S_{1},\omega)dS_{1}=F(\sigma_{1},\omega)
~~;~~ \forall \sigma_{1}\in\mathbb{R}~,
\end{equation}
wherein $C(\sigma_{1},\omega)$, $B^{0}(\sigma_{1},\omega)$ and
$F(\sigma_{1},\omega)$ are, by definition, $C(k_{1},\omega)$,
$B^{0}(k_{1},\omega)$ and $F(k_{1},\omega)$ in which $k_{1}$ is
replaced by $k^{0}\sigma_{1}$, $D(\sigma_{1},S_{1},\omega)$ is
$D(k_{1},K_{1},\omega)$ in which $k_{1}$, $K_{1}$ are replaced by
$k^{0}\sigma_{1}$, $K^{0}S_{1}$ respectively, and
\begin{equation}\label{resinteq.2}
E(\sigma_{1},S_{1},\omega):=k^{0}D(\sigma_{1},S_{1},\omega)~.
\end{equation}
Adding and subtracting the same term on the left side of the
integral equation gives
\begin{multline}\label{resinteq.3}
\left[ C(\sigma_{1},\omega)-E(\sigma_{1},\sigma_{1},\omega)\right]
B^{0}(\sigma_{1},\omega)-
\\
\int_{-\infty}^{\infty} E(\sigma_{1},S_{1},\omega)\left[
1-\delta(S_{1}-\sigma_{1})\right]
B^{0}(S_{1},\omega)dS_{1}=F(\sigma_{1},\omega) ~~;~~ \forall
\sigma_{1}\in\mathbb{R}~,
\end{multline}
from which we obtain
\begin{multline}\label{resinteq.4}
 B^{0}(\sigma_{1},\omega)=
 \frac{F(\sigma_{1},\omega)+
\int_{-\infty}^{\infty} E(\sigma_{1},S_{1},\omega)\left[
1-\delta(S_{1}-\sigma_{1})\right]
B^{0}(S_{1},\omega)dS_{1}}{C(\sigma_{1},\omega)-E(\sigma_{1},\sigma_{1},\omega)}
 ~~;~~ \forall
\sigma_{1}\in\mathbb{R}~.
\end{multline}
An iterative approach for solving this integral equation consists
in computing successively:
\begin{equation}\label{resinteq.5}
 B^{0(0)}(\sigma_{1},\omega)=
\frac{F(\sigma_{1},\omega)}{C(\sigma_{1},\omega)-E(\sigma_{1},\sigma_{1},\omega)}
~~;~~ \forall \sigma_{1}\in\mathbb{R}~,
\end{equation}
\begin{multline}\label{resinteq.6}
 B^{0(1)}(\sigma_{1},\omega)= B^{(0)}(\sigma_{1},\omega)+
 \\
 \frac{\int_{-\infty}^{\infty} E(k_{1},S_{1},\omega)\left[
1-\delta(S_{1}-\sigma_{1})\right]
B^{0(0)}(S_{1},\omega)dS_{1}}{C(\sigma_{1},\omega)-E(\sigma_{1},\sigma_{1},\omega)}
 ~~;~~ \forall
\sigma_{1}\in\mathbb{R}~,
\end{multline}
and so on.
\subsubsection{Elimination ~of ~$B^{0}(k_{1},\omega)$ ~to
~obtain ~a ~linear system of equations for
$B_{m}^{2j}(\omega)$}\label{bmbis}
The procedure is again to make a series of substitutions which now
lead to the linear system of equations for $B_{n}^{2(l)}(\omega)$,
$\forall l \in \mathbb{B}$, $\forall n \in \mathbb{N}$ :
\begin{equation}\label{bmbis.3}
C_{n}(\omega)B_{n}^{2(l)}(\omega)=F_{n}^{(l)}(\omega)+\sum_{j\in\mathbb{B}}
\sum_{m\in\mathbb{N}}D_{nm}^{(lj)}(\omega)B_{m}^{2(j)}(\omega)
~~;~~ \forall l\in\mathbb{B}~;~\forall n\in\mathbb{N}~,
\end{equation}
wherein
\begin{equation}\label{bmbis.4}
C_{n}(\omega)=\cot(k_{2n}^{2}b)~;~\forall n\in\mathbb{N}~,
\end{equation}
\begin{multline}\label{bmbis.5}
F_{n}^{(l)}(\omega)= \frac{2\epsilon_{n}}{
\sin(k_{2n}^{2(l)}b_{l})}
\int_{-\infty}^{\infty}A^{0-}(k_{1},\omega)e^{-\mbox{i}k_{2}^{0}h}\left[
\frac{I_{n}^{(l)+}(k_{1},\omega)
e^{ik_{1}(d_{l}-w_{l}/2)}}{\cos(k_{2}^{1}h)-i\frac{\mu^{1}k_{2}^{1}}{\mu^{0}k_{2}^{0}}\sin(k_{2}^{1}h)}\right]
\frac{dk_{1}}{k_{2}^{0}}
\\
  ~~;~~ \forall l\in\mathbb{B}~;~\forall
n\in\mathbb{N}~,
\end{multline}
and
\begin{multline}\label{bmbis.6}
D_{nm}^{(lj)}(\omega)=
 \frac{\mbox{i}k_{2m}^{2(j)}\mu^{2}w_{j}
\epsilon_{n}\sin(k_{2m}^{2(j)}b_{j})}
{2 \pi \sin(k_{2n}^{2(l)}b_{l})\mu^{0}}\times
\\
\int_{-\infty}^{\infty}I_{n}^{(l)+}(k_{1},\omega)I_{m}^{(j)-}(k_{1},\omega)
\left[
\frac{\cos(k_{2}^{1}h)-\mbox{i}\frac{\mu^{0}k_{2}^{0}}{\mu^{1}k_{2}^{1}} \sin(k_{2}^{1}h)}
{\cos(k_{2}^{1}h)-\mbox{i}\frac{\mu^{1}k_{2}^{1}}{\mu^{0}k_{2}^{0}}\sin(k_{2}^{1}h)
}\right]
e^{\mbox{i}k_{1}\left(\left(d_{l}-d_{j}\right)-\frac{w_{l}-w_{j}}{2}\right)}\frac{dk_{1}}{k_{2}^{0}}
\\
~~;~~ \forall l,~j\in\mathbb{B}~;~\forall n,~m\in\mathbb{N}~.
\end{multline}

Eq. (\ref{bmbis.3}) can be written as:
\begin{multline}\label{bmbis.8}
B_{n}^{2(l)}(\omega)=\frac{F_{n}^{(l)}(\omega)+\sum_{j\in\mathbb{B}}\sum_{m=0}^{\infty}
D_{nm}^{(lj)}(\omega)\left( 1-\delta_{nm}\delta_{lj}\right)
B_{m}^{2(j)}(\omega)}{C_{n}(\omega)-D_{nn}^{(ll)}(\omega)}
\\
~~;~~ \forall l\in\mathbb{B}~;~\forall n\in\mathbb{N}~,
\end{multline}

An iterative procedure for solving this linear set of equations is
as follows:
\begin{equation}\label{bmbis.9}
\left(B_{n}^{2(l)}(\omega)\right)^{(0)}=\frac{F_{n}^{(l)}(\omega)}{C_{n}(\omega)-D_{nn}^{ll}(\omega)}
~~;~~ \forall l\in\mathbb{B}~;~\forall n\in\mathbb{N}~,
\end{equation}
\begin{multline}\label{bmbis.10}
\left(B_{n}^{2(l)}(\omega)\right)^{(p)}=\frac{F_{n}^{(l)}(\omega)+\sum_{j\in\mathbb{B}}\sum_{m=0}^{\infty}
D_{nm}^{(lj)}(\omega)\left( 1-\delta_{nm}\delta_{lj}\right)
\left(B_{m}^{2(j)}(\omega)\right)^{(p-1)}}{C_{n}(\omega)-D_{nn}^{(ll)}(\omega)}
\\ ~~;~~p=1,2,...~;~ \forall l\in\mathbb{B}~;~\forall
n\in\mathbb{N}~.
\end{multline}
\section{Modal analysis}
\subsection{The emergence of the natural modes of the configuration from
the iterative solution of the integral equation for
$B^{0}(k_{1},\omega)$}\label{natmodesinteq}
Eqs. (\ref{resinteq.5}), (\ref{resinteq.6}), etc. show that the $n$-th order
iterative approximation of the solution to the integral equation
(\ref{inteqb0.2}) is of the form
\begin{equation}\label{natmodesinteq.1}
 B^{0(n)}(\sigma_{1},\omega)= \frac{\mathcal{N}^{(n)}(\sigma_{1},\omega)}
 {C(\sigma_{1},\omega)-E(\sigma_{1},\sigma_{1},\omega)}:=
 \frac{\mathcal{N}^{(n)}(\sigma_{1},\omega)}{\mathcal{D}(\sigma_{1},\omega)}
  ~;~n=1,2,... ~,
\end{equation}
wherein
\begin{equation}\label{natmodesinteq.2}
 \mathcal{N}^{(0)}(\sigma_{1},\omega)= F(\sigma_{1},\omega)~,
\end{equation}
\begin{equation}\label{natmodesinteq.3}
 \mathcal{N}^{(n>0)}(\sigma_{1},\omega)=
 F(\sigma_{1},\omega)+\int_{-\infty}^{\infty} E(\sigma_{1},S_{1},\omega)\left[
1-\delta(S_{1}-\sigma_{1})\right] B^{0(n-1)}(S_{1},\omega)dS_{1}~,
\end{equation}
from which it becomes apparent that the solution
$B^{0(n)}(\sigma_{1},\omega)$, {\it to any order of
approximation}, is expressed as a fraction, the denominator
$\mathcal{D}(\sigma_{1},\omega)$ of which (not depending on the
order of approximation), can become small for certain values of
$\sigma_{1}$ and $\omega$ so as to make
$B^{0(n)}(\sigma_{1},\omega)$, and (possibly) the field in the
substratum, large for these values. When this happens, a {\it
natural mode of the configuration}, comprising the blocks, the
soft layer and the hard substratum, is excited, this taking the
form of a {\it resonance}
 with respect to
$B^{0(n)}(\sigma_{1},\omega)$, i.e., with respect to a plane wave
component of the field in the substratum. As $B^{0}(k_{1},\omega)$
is related to $A^{1}(\sigma_{1},\omega)$ and $B^{1}(\sigma_{1})$
via (\ref{stressgammah.2})-(\ref{displgammah.2}), the structural
resonance manifests itself for the same $\sigma_{1}$ and $\omega$
as concerns the field in the layer.

We say that $B^{0(n)}(\sigma_{1},\omega)$,  and the fields in the
layer and substratum, can become {\it possibly} large at resonance
because until now we have not taken into account the numerator
$\mathcal{N}^{(n)}(\sigma_{1},\omega)$, which might be small when
the denominator is small, or such as to prevent, by other means,
the fields in the layer and substratum from becoming large.
Moreover, since the field is expressed as a sum of plane waves,
the fact that $B^{0}(k_{1},\omega)$ may become large for some
$k_{1}^{\star}$, does not necessarily mean that the sum of plane
waves (including waves whose horizontal wavenumber $k_{1}\neq
k_{1}^{\star}$), and therefore the field, will be large at a
resonance frequency.
\subsection{$B^{0}(k_{1},\omega)$ in the absence of the
blocks}\label{b0outblocks}
The easiest way to account for the absence of blocks is to take
$b=0$. Then:
\begin{equation}\label{b0outblocks.1}
F(k_{1},\omega)= A^{0-}(k_{1},\omega)e^{-\mbox{i}k_{2}^{0}h}\left[
\cos(k_{2}^{1}h)+\mbox{i}\frac{\mu^1 k_2^1}{\mu^0
k_2^0}\sin(k_{2}^{1}h)\right]
 ~~;~~ \forall
k_{1}\in\mathbb{R}~,
\end{equation}
\begin{equation}\label{b0outblocks.2}
D(k_{1},K_{1},\omega)=0
 ~~;~~ \forall
k_{1}, K_{1}\in\mathbb{R}~,
\end{equation}
so that (\ref{inteqb0.2}) yields
\begin{multline}\label{b0outblocks.3}
B^{0}(k_{1},\omega)=\frac{F(k_{1},\omega)}{C(k_{1},\omega)}=
\frac{\mbox{i}S(\omega)}{4\pi}e^{-\mbox{i}\left(k_{1}x_{1}^{s}+k_{2}^{0}(h-x_{2}^{s})\right)}\left[\frac{
\cos(k_{2}^{1}h)+\mbox{i}\frac{\mu^1 k_2^1}{\mu^0 k_2^0}\sin(k_{2}^{1}h)}{
\cos(k_{2}^{1}h)-\mbox{i}\frac{\mu^1 k_2^1}{\mu^0 k_2^0}\sin(k_{2}^{1}h)}\right]
 ~~;~~ \forall
k_{1}\in\mathbb{R}~.
\end{multline}

First consider the case of {\it bulk plane wave excitation} at incidence angle $\theta^i$.
Using the property of the Dirac delta distribution
$\delta(x-y)F(y)=\delta(x-y)F(x)$, we obtain
\begin{multline}\label{b0outblocks.4}
B^{0}(k_{1},\omega)=\frac{\mbox{i}S(\omega)}{4\pi}e^{-\mbox{i}k_{2}^{0,i}h}\left[\frac{
\cos(k_{2}^{1,i}h)+\mbox{i}\frac{\mu^1 k_2^{1,i}}{\mu^0 k_2^{0,i}}\sin(k_{2}^{1,i}h)}{
\cos(k_{2}^{1,i}h)-\mbox{i}\frac{\mu^1 k_2^{1,i}}{\mu^0 k_2^{0,i}}\sin(k_{2}^{1,i}h) }\right]
 \delta(k_{1}-k_{1}^{i})
 ~~;~~ \forall
k_{1}\in\mathbb{R}~,
\end{multline}
wherein
\begin{equation}\label{b0outblocks.5}
k_{1}^{i}:=k^{0}\sin\theta^{i}~,~~k_{2}^{i}=k_{2}^{0,i}~,~~k_{2}^{j,i}:=\sqrt{\left(
k^{j}\right) ^{2}-\left( k_{1}^{i}\right) ^{2}}~;~~\Re\left(
k_{2}^{j,i}\right) \geq 0~,~\Im\left( k_{2}^{j,i}\right) \geq
0~;~j=0,1,2~.
\end{equation}
What this result means is that the amplitude function
$B^{0}(k_{1},\omega)$ vanishes for all $k_{1}$ except
$k_{1}=k_{1}^{i}$, and since (if we assume that there is no
dissipation in the layer and substratum), $\mu^1 k_2^{1,i}/\mu^0
k_2^{0,i} $, $k_{2}^{0,i}$, $k_{2}^{1,i}$, $\sin(k_{2}^{1,i}h)$,
$\cos(k_{2}^{1,i}h)$ are real, the denominator in
(\ref{b0outblocks.4}) cannot vanish (neither be small if there is
a reasonable amount of dissipation in the layer and/or the
substratum), which is another way of saying that {\it no
structural resonances can exist when the configuration (without
blocks) is excited by a bulk plane wave}.

In fact, it is easy to ascertain that the existence of structural
resonances, in the case of the configuration without  blocks, is
tied up (when there is no dissipation in the layer and substrate)
with the possibility of $k_{2}^{0}$ becoming pure imaginary,
because then the denominator can effectively vanish (or become
very small for a reasonable amount of dissipation in the layer
and/or substratum). This possibility is connected with the
excitation of a {\it Love mode}, characterized by the simultaneous
existence of a surface  wave (associated with pure imaginary
$k_{2}^{0}$) in the substratum, and a standing bulk wave
(associated with pure real $k_{2}^{1}$) in the layer.

To make this more palpable, we introduce (\ref{b0outblocks.4}) into
(\ref{fieldrepomega0.10}) so as to find
\begin{multline}\label{b0outblocks.6}
u^{0}(\mathbf{x},\omega)=S(\omega)
\exp\left\{\mbox{i}\left[k_{1}^{i}x_{1}-k_{2}^{0,i}x_{2}\right]\right\}+
\\
S(\omega)\frac{\cos(k_{2}^{1,i}h)
+\mbox{i}\frac{\mu^{1}k_{2}^{1,i}}{\mu^{0}k_{2}^{0,i}}\sin(k_{2}^{1,i}h)
}{
\cos(k_{2}^{1,i}h)-\mbox{i}\frac{\mu^{1}k_{2}^{1,i}}{\mu^{0}k_{2}^{0,i}}\sin(k_{2}^{1,i}h)
}
\exp\left\{i\left[k_{1}^{i}x_{1}+k_{2}^{0,i}(x_{2}-2h)\right]\right\}
~~;~~\forall\mathbf{x}\in\Omega_{0} ~,
\end{multline}
which expresses the fact that the field in $\Omega_{0}$ is the sum
of the incident plane wave and a reflected plane  wave for all
$k^{0}$. Both of these waves are {\it bulk waves} (i.e., both of
the cartesian components of their wavevectors are {\it real} for
non-dissipative media) as opposed to the requirement of there
being a {\it surface wave} (i.e., whose $x_{2}$ component is
imaginary for non-dissipative media) in $\Omega_{0}$ when a Love
mode is excited. This means that {\it it is impossible to excite a
Love mode, in a configuration without blocks consisting of a soft
layer overlying a hard halfspace, when the incident wave is a
plane bulk wave}.

However, as shown in \cite{grobyetwirgin2005,grobyetwirgin2005II},
{\it it is possible to excite a Love mode, in a configuration
without blocks consisting of a soft layer overlying a hard
halfspace, when the incident wave is a cylindrical
 wave radiated by a shallow source}.

In the next section, we shall show that it is possible to excite
something like a Love mode in a configuration consisting of a set
of buildings in welded contact with a soft layer overlying a hard
halfspace even when the incident wave is a plane bulk wave, the
same being true when the solicitation is a cylindrical wave.
\subsection{Quasi Love modes}\label{quasilove}
Let us return to the denominator of the expression of $B^{0(n)}$,
which, for convenience, we re-write in terms of $k_{1}$:
\begin{multline}\label{quasilove.1}
\mathcal{D}(k_{1},\omega)=C(k_{1},\omega)-k^{0}D(k_{1},k_{1},\omega)
\\
=\cos(k_{2}^{1}h)-\mbox{i}\frac{\mu^1 k_2^1}{\mu^0 k_2^0}\sin(k_{2}^{1}h) -\left[
\cos(k_{2}^{1}h)-\mbox{i}\frac{\mu^0 k_2^0}{\mu^1 k_2^1} \sin(k_{2}^{1}h)\right] \times
\\
\frac{\mbox{i}}{2\pi} \sum_{m=0}^{\infty}\epsilon_{m}
\frac{\mu^{2}k_{2m}^{2(j)}w_{j}}{\mu^{0}k_{2}^{0}}
\tan(k_{2m}^{2(j)}b_{j})I_{m}^{(j)-}(k_{1},\omega)I_{m}^{(j)+}(k_{1},\omega)
 ~~;~~ \forall
k_{1}\in\mathbb{R}~,
\end{multline}
wherein it is easy to show that
\begin{equation}\label{w6.2.8}
  I_{m}^{(j)\pm}(k_{1},\omega)=\frac{\mbox{i}^{m}}{2}e^{\pm \mbox{i}k_{1}\frac{w_j}{2}}\left[
  \text{sinc}\left( \frac{\pm k_{1}w_j}{2}+\frac{m\pi}{2}\right) +
  (-1)^{m}\text{sinc}\left( \frac{\pm k_{1}w_j}{2}-\frac{m\pi}{2}\right)
  \right] ~,
\end{equation}
with
\begin{equation}\label{w6.2.9}
  \text{sinc}(\zeta):=\frac{\sin(\zeta)}{\zeta}~.
\end{equation}
Since $\mathcal{C}(k_{1},\omega):=C(k_{1},\omega)=0$ is the
dispersion relation for (i.e., providing the means of determining
the $(k_{1},\omega)$ couples leading to a possible resonance
associated with the excitation of) Love modes, we can say that
(\ref{quasilove.1}) is the dispersion relation for {\it quasi Love
modes}.
\newline
\newline
{\it Remark 1}\newline Quasi Love modes are different from Love
modes which, at present, means that the $(k_{1}, \omega)$ couples
for which $\mathcal{D}(k_{1},\omega)=0$ are not identical to the
$(k_{1}, \omega)$ couples for which $\mathcal{C}(k_{1},\omega)=0$.
\newline
\newline
{\it Remark 2}\newline When $b\rightarrow 0$, the dispersion
relation for quasi Love modes becomes the dispersion relation for
Love modes.
\newline
\newline
{\it Remark 3}\newline For small $k_{20}^{2(j)}b$, the quasi Love
modes are a small perturbation of Love modes, which means here
that the $(k_{1}, \omega)$ couples for which
$\mathcal{D}(k_{1},\omega)=0$ are close to the $(k_{1}, \omega)$
couples for which $\mathcal{C}(k_{1},\omega)=0$.
\newline
\newline
{\it Remark 4}\newline For small $\mu^{2}/\mu^{1}$, the quasi Love
modes are a small perturbation of Love modes, which means here
that the $k_{1}, \omega$ couples for which
$\mathcal{D}(k_{1},\omega)=0$ are close to the $k_{1}, \omega$
couples for which $\mathcal{C}(k_{1},\omega)=0$.
\newline
\newline
{\it Remark 5}\newline For small $k^{2}w$ the quasi Love modes are
a small perturbation of Love modes, which means here that the
$k_{1}, \omega$ couples for which $\mathcal{D}(k_{1},\omega)=0$
are close to the $k_{1}, \omega$ couples for which
$\mathcal{C}(k_{1},\omega)=0$.
\newline
\newline
{\it Remark 6}\newline The dispersion relation for quasi Love
modes is independent of the number of blocks (provided this number
is greater than 0).
\newline
\newline
To substantiate these remarks (when necessary) and obtain a more
detailed picture of the features of the quasi Love modes as
compared to those of the Love modes, we must analyze more closely
(\ref{quasilove.1}). For $m=0$ we have
\begin{equation}\label{w6.2.10}
  I_{0}^{(j)\pm}(k_{1},\omega)=
  e^{\pm \mbox{i}k_{1}\frac{w_j}{2}}
  \text{sinc}\left( \frac{k_{1}w_j}{2}\right)
  ~,
\end{equation}
and for $k_{1}w\neq\pm m\pi$, we have
\begin{equation}\label{w6.2.11}
  I_{m}^{(j)\pm}(k_{1},\omega)=\frac{\pm 2k_{1}w_j \mbox{i}^{m}}{
 \left( k_{1}w_j \right) ^{2}-\left( m\pi \right)
  ^{2}}
  e^{\pm \mbox{i}k_{1}\frac{w}{2}}
  \sin\left( \frac{\pm k_{1}w_j}{2}+\frac{m\pi}{2}\right)
  ~.
\end{equation}
To make a complex issue relatively simple, we assume that $k_{1}w$
is effectively such as to be different from $m\pi$ for
$m=0,1,2,...$. Then
\begin{multline}\label{w6.2.12}
\mathcal{D}(k_{1},\omega)=
\cos(k_{2}^{1}h) -\mbox{i} \frac{\mu^1 k_2^1}{\mu^0 k_2^0}\sin(k_{2}^{1}h) -\left[\frac{\mu^1 k_2^1}{\mu^0 k_2^0}
\cos(k_{2}^{1}h)-\mbox{i}\sin(k_{2}^{1}h)\right] \times
\\
\frac{\mbox{i}}{2\pi}\frac{\mu^{2}}{\mu^{1}}\Bigg\{
\frac{k^{2}}{k_{2}^{1}}\tan(k^{2}b_j)\text{sinc}^{2}\left(
\frac{k_{1}w_j}{2}\right)+2\sum_{m=1}^{\infty}\frac{k_{2m}^{2}}{k_{2}^{1}}\frac{
4(k_{1}w_j)^{2}}{
 \left[ \left( k_{1}w \right) ^{2}-\left( m\pi \right)
  ^{2}\right] ^{2}}\times
\\
  \sin\left( \frac{k_{1}w_j}{2}+\frac{m\pi}{2}\right)
  \sin\left( -\frac{k_{1}w_j}{2}+\frac{m\pi}{2}\right)
  \tan(k_{2m}^{2(j)}b_j)\Bigg\}
~,
\end{multline}
which rather clearly substantiates the aforementioned remarks.

Consider the first term in $\{..\}$. This term is significant only
for small $k_{1}w_j/2$ due to the sinc function whose modulus
decays rapidly as its argument increases. Another feature of this
term is that it vanishes when $k^{2}b_j=l\pi~;~l=0,1,2,...$, which
occurs when the zeroth-order quasi-mode in a block encounters a
stress-free boundary condition at the base of the block (i.e.
$\left.u^{2(j)}_{,2}\right|_{x_2=0}=0$), in which case the latter
is disconnected from the underground (since no wave can penetrate
into the layer) from the point of view of the fundamental block
quasi mode. This corresponds to the \textit{stress-free base block
mode}. It is then logical that this quasi mode of the block should
not perturb the dispersion characteristics of the mode of the
whole configuration.

The analysis of the series term in $\{..\}$ is more difficult. It
is clear that a few terms of this series should be retained,
unless $k^{2}b$ and/or $k_{1}w$ are very small. The subsequent
terms of the series become rapidly small (with $m$) due to the
fact that $\tan(k_{2m}^{2(j)}b_j)\sim \mbox{i}\tanh(m\pi
b_j/w_j)\rightarrow \mbox{i}~;~m\rightarrow\infty$ and
$k_{2m}^{2(j)}\frac{ 4(k_{1}w_j)^{2}\mbox{i}}{ \left[ \left(
k_{1}w_j \right) ^{2}-\left( m\pi \right) ^{2}\right]
^{2}}=O(m^{-3})~;~m\rightarrow \infty$.

In any case, it seems legitimate to adopt the following picture of
what is going on: the base of a given block is a location where
diffraction is produced resulting in incident (from either the
block into the underground or from the underground into the block)
bulk waves being transformed into diffracted bulk and surface
waves, as is testified by the presence of the terms $I_{m}^{(j)+}$
and $I_{m}^{(j)-}$ in the series, and by the fact that these
diffraction effects disappear as the width of the base segment
goes to zero. Naturally, the presence of these locally-produced
diffracted waves perturbs the overall wave structure (with respect
to what it was in the absence of the blocks) and therefore results
in a modification of the characteristics of the modes in the layer
and substratum (which were Love modes when the blocks were
absent). This picture is consistent with the observation that the
diffracted waves are more difficult to produce when the base
segment of a block appears as a stress-free surface due  to the
fact that either $k^{2}b_j=l\pi~;~l=0,1,2,...$ or
$\mu^{2}/\mu^{1}<<1$.

Beyond this, it is necessary to carry out a numerical study in
order to see how the different parameters involved in the problem,
notably those of the block, modify the dispersion characteristics
of the modes of the configuration with respect to what these modes
were (i.e., Love modes) in the absence of the blocks. The
numerical study should also seek to evaluate the modification of
the response (notably on the ground) of the configuration due to
the presence of blocks.

\subsection{The emergence of the natural modes of the configuration from the
linear system of equations for $B_{m}^{2(l)}$: quasi
displacement-free base block modes}\label{quasistfrbablmo}
Eqs.(\ref{bmbis.9})-(\ref{bmbis.10}) signify that
$\left(B_{n}^{2(l)}\right)^{(p)}$ becomes large when $C_{n}-D_{nn}^{(ll)}$ becomes
small, and that this occurs at all orders $p$ of approximation.
The fact that $\left(B_{n}^{2(l)}\right)^{(p)}$ becomes inordinately large is
associated with the excitation of a natural mode of the
configuration. The equation $C_{n}(\omega)-D_{nn}^{(ll)}(\omega)=0$
is the dispersion relation of the $n$-th natural mode of the
configuration.

Let us examine this relation in more detail.
\begin{multline}\label{quasistfrbablmo.1}
\cot(k_{2n}^{2(l)}b_{l})-\frac{\mbox{i} \mu^{2}k_{2n}^{2(l)}w_{l}}{2\pi \mu^{0}}
 \int_{-\infty}^{\infty}I_{n}^{(l)+}(k_{1},\omega)I_{n}^{(l)-}(k_{1},\omega) \left[
\frac{\cos(k_{2}^{1}h)-\mbox{i}\frac{\mu^{0}k_{2}^{0}}{\mu^{1}k_{2}^{1}}
\sin(k_{2}^{1}h)}{\cos(k_{2}^{1}h)-\mbox{i}\frac{\mu^{1}k_{2}^{1}}{\mu^{0}k_{2}^{0}}\sin(k_{2}^{1}h)
}\right] \frac{dk_{1}}{k_{2}^{0}}=0~,
\end{multline}
which shows that the natural modes of the configuration result
from the {\it interaction} of the fields in two substructures: the
superstructure (i.e., block(s) above the ground), associated with
the term
\begin{equation}\label{quasistfrbablmo.2}
 \mathcal{F}_{1n}^{(l)}(b,\mu^{2},\rho^{2},\omega)= \mathcal{F}_{1n}^{(l)}=\cot(k_{2n}^{2(l)}b_{l})~,
\end{equation}
and the substructure (i.e., soft layer plus hard half space below
the ground), associated with the term
\begin{multline}\label{quasistfrbablmo.3}
 \mathcal{F}_{2n}^{(l)}(h,\mu^{0},\rho^{0},\mu^{1},\rho^{1},\mu^{2},\rho^{2},w,\omega)=\mathcal{F}_{2n}^{(l)}=
\\
\frac{\mbox{i} \mu^{2}k_{2n}^{2(l)}w_{l}}{2\pi \mu^{0}}
 \int_{-\infty}^{\infty}I_{n}^{(l)+}(k_{1},\omega)I_{n}^{(l)-}(k_{1},\omega) \left[
\frac{\cos(k_{2}^{1}h)-\mbox{i}\frac{\mu^{0}k_{2}^{0}}{\mu^{1}k_{2}^{1}}
\sin(k_{2}^{1}h)}{\cos(k_{2}^{1}h)-\mbox{i}\frac{\mu^{1}k_{2}^{1}}{\mu^{0}k_{2}^{0}}\sin(k_{2}^{1}h)
}\right]
\frac{dk_{1}}{k_{2}^{0}}
\end{multline}
Each of these two substructures possesses its own natural modes,
i.e., arising from $\mathcal{F}_{1n}(b,\mu^{2},\rho^{2},\omega)=0$
for the superstructure, and
$\mathcal{F}_{2n}(h,\mu^{0},\rho^{0},\mu^{1},\rho^{1},\mu^{2},\rho^{2},w,\omega)=0$
for the substructure,  but {\it the natural modes of the complete
structure (superstructure plus substructure) are neither the modes
of the superstructure nor those of the substructure} since they
are defined by
\begin{equation}\label{quasistfrbablmo.6}
 \mathcal{F}_{1n}^{(l)}(b,\mu^{2},\rho^{2},\omega)-
  \mathcal{F}_{2n}^{(l)}(h,\mu^{0},\rho^{0},\mu^{1},\rho^{1},\mu^{2},\rho^{2},w,\omega)=0
~,
\end{equation}
which emphasizes the fact that the natural modes of the complete
structure result from the {\it interaction} of the natural modes
of the superstructure with those of the substructure. Note that
the solutions of (\ref{quasistfrbablmo.6}) can be quite different
from those of either/both $\mathcal{F}_{1n}=0$ and
$\mathcal{F}_{2n}=0$.

In order to establish the natural frequencies of the complete
structure, we first analyze the natural frequencies of each
substructure and assume that all media in the structure are
non-dissipative (i.e., elastic).

The solutions of the dispersion relation for the superstructure are:
\begin{multline}\label{quasistfrbablmo.7}
 \cot\left( k_{2n}^{2(l)}b_{l}\right) =0~~\Leftrightarrow~~k_{2n}^{2(l)}b_{l}=\left(2m+1\right)\pi~~
 \Leftrightarrow~~\omega=
 \omega_{nm}^{l}=c^{2}\sqrt{\left( \frac{\left(2m+1\right)\pi}{b_{l}}\right) ^{2}+
 \left( \frac{n\pi}{w_{l}}\right) ^{2}}
 \\
 ~;~m,n=0,1,2,...~,
\end{multline}
which are the natural frequencies of  vibration of a block with a
{\it displacement-free base} (i.e., at these natural frequencies,
$\left.u^{2(l)}\right|_{x_2=0}$ vanishes on the base segment of
the block). This relation shows that there corresponds to the
$n$-th block mode an infinite set of sub-modes, identified by the
index $m$. The same can be said about the natural modes of the
entire configuration.

Now consider the dispersion relation for the substructure of the
entire structure. Due to the fact that the integrand is an even
function of $k_{1}$, it becomes
\begin{multline}\label{quasistfrbablmo.8}
\mathcal{F}_{2n}^{(l)}= \frac{\mbox{i}
\mu^{2}k_{2n}^{2(l)}w_{l}}{\pi \mu^0}\int_{0}^{\infty}
I_{n}^{(l)+}(k_{1},\omega)I_{n}^{(l)-}(k_{1},\omega) \left[
\frac{\cos(k_{2}^{1}h)-\mbox{i}\frac{\mu^{0}k_{2}^{0}}{\mu^{1}k_{2}^{1}}
\sin(k_{2}^{1}h)}{\cos(k_{2}^{1}h)-\mbox{i}\frac{\mu^{1}k_{2}^{1}}{\mu^{0}k_{2}^{0}}\sin(k_{2}^{1}h)
}\right] \frac{dk_{1}}{k_{2}^{0}}=
\\
\frac{\mbox{i} \mu^{2}k_{2n}^{2(l)}w_{l}}{\pi \mu^1}\int_{0}^{\infty}
I_{n}^{(l)+}(k_{1},\omega)I_{n}^{(l)-}(k_{1},\omega) \left[
\frac{\mu^{1}k_{2}^{1}\cos(k_{2}^{1}h)-\mbox{i}\mu^{0}k_{2}^{0}
\sin(k_{2}^{1}h)}{\mu^{0}k_{2}^{0}\cos(k_{2}^{1}h)-\mbox{i}\mu^{1}k_{2}^{1}\sin(k_{2}^{1}h)
}\right]
\frac{dk_{1}}{k_{2}^{1}}=0~.
\\
\end{multline}
A few preliminary remarks are in order:
\newline
\newline
{\it Remark 1}\newline Since the term $\mathcal{F}_{2n}$ in the
dispersion relation is absent in the absence of the
infrastructure, we can say that the natural modes of the complete
configuration are {\it quasi displacement-free base block modes}.
Quasi displacement-free base block modes are different from
displacement-free base block modes which, at present, means that
the $(n, \omega)$ couples for which
$\mathcal{F}_{1n}(b,\mu^{2},\rho^{2},\omega)=0$ are not identical
to the $(n, \omega)$ couples for which
$\mathcal{F}_{2n}(h,\mu^{0},\rho^{0},\mu^{1},\rho^{1},\mu^{2},
\rho^{2},w,\omega)=0$.
\newline
\newline
{\it Remark 2}\newline For small $\mu^{2}/\mu^{1}$, the quasi
displacement- free base block modes are a small perturbation of
displace-ment- free base block modes, which means here ~that the
$(n, \omega)$ couples ~for which
$\mathcal{F}_{1n}(b,\mu^{2},\rho^{2},\omega)=0$ are close to the
$(n, \omega)$ couples for which
$\mathcal{F}_{2n}(h,\mu^{0},\rho^{0},\mu^{1},$$\rho^{1},\mu^{2},
\rho^{2},w,\omega)=0$. This is a relatively- logical result in
that when  $\mu^{2}/\mu^{1}$ is small, the waves coming from the
infrastructure have more trouble penetrating into the blocks and
modifying therein the modal structure.
\newline
\newline
{\it Remark 3}\newline The dispersion relation for quasi
displacement- free base block modes is independent of the number
of blocks (provided this number is greater than 0). This is a
somewhat surprising result related to the choice of the iteration
method for solving the linear system of equations for
$B_{n}^{2(l)}$, since a more accurate choice of method (one of
which is described in sect. \ref{anotherquasisfb}) can be shown to
lead to a somewhat different (although much more complicated)
dispersion relation which depends on the number of blocks in the
configuration.
\newline
\newline
We now analyze in more detail $\mathcal{F}_{2n}$, and, in
particular, $\mathcal{F}_{20}$. Recalling (\ref{w6.2.10}), we get:
\begin{equation}\label{w6.3.22}
\mathcal{F}_{20}^{(l)}= \frac{\mbox{i}k^{2}w_l}{\pi }\frac{\mu^{2}}{\mu^{0}}\int_{0}^{\infty}
\text{sinc}^{2}\left( \frac{k_{1}w_l}{2}\right) \left[
\frac{\cos(k_{2}^{1}h)-\mbox{i}\frac{\mu^{0}k_{2}^{0}}{\mu^{1}k_{2}^{1}}
\sin(k_{2}^{1}h)}{\cos(k_{2}^{1}h)-\mbox{i}\frac{\mu^{1}k_{2}^{1}}{\mu^{0}k_{2}^{0}}\sin(k_{2}^{1}h)
}\right]
\frac{dk_{1}}{k_{2}^{0}}~.
\end{equation}
Proceeding as \cite{grobyetwirgin2005,grobyetwirgin2005II}, we decompose the integral into
three parts (under the assumption $k^{1}>k^{0}>0$) so as to
obtain:
\begin{equation}\label{w6.3.23}
\mathcal{F}_{20}^{(l)}(h,\mu^{0},\rho^{0},\mu^{1},\rho^{1},\mu^{2},\rho^{2},w,\omega)=
\mathcal{F}_{20}^{1(l)}+\mathcal{F}_{20}^{2(l)}+\mathcal{F}_{20}^{3(l)}
\end{equation}
wherein
\begin{equation}\label{w6.3.24}
\mathcal{F}_{20}^{1(l)}=
\frac{ik^{2}w}{\pi }\frac{\mu^{2}}{\mu^{0}}\int_{0}^{k^{0}}
\text{sinc}^{2}\left( \frac{K_{1}w}{2}\right) \left[
\frac{\cos(K_{2}^{1}h)-\mbox{i}\frac{\mu^{0}K_{2}^{0}}{\mu^{1}K_{2}^{1}}
\sin(K_{2}^{1}h)}{\cos(K_{2}^{1}h)-\mbox{i}\frac{\mu^{1}K_{2}^{1}}{\mu^{0}K_{2}^{0}}\sin(K_{2}^{1}h)
}\right]
\frac{dk_{1}}{K_{2}^{0}}~,
\end{equation}
\begin{equation}\label{w6.3.25}
\mathcal{F}_{20}^{2(l)}=
\frac{k^{2}w}{\pi}\frac{\mu^{2}}{\mu^{0}}\int_{k^{0}}^{k^{1}}
\text{sinc}^{2} \left( \frac{K_{1}w}{2}\right) \left[
\frac{\cos(K_{2}^{1}h)+\frac{\mu^{0}\kappa_{2}^{0}}{\mu^{1}k_{2}^{1}}
\sin(k_{2}^{1}h)}{\cos(k_{2}^{1}h)-\frac{\mu^{1}k_{2}^{1}}{\mu^{0}\kappa_{2}^{0}}\sin(k_{2}^{1}h)
}\right]
\frac{dk_{1}}{\kappa_{2}^{0}}~,
\end{equation}
\begin{equation}\label{w6.3.26}
\mathcal{F}_{20}^{3(l)}= \frac{k^{2}w}{\pi}\frac{\mu^{2}}{\mu^{0}}\int_{k^{1}}^{\infty}
\text{sinc}^{2} \left( \frac{K_{1}w}{2}\right) \left[
\frac{\cosh(\kappa_{2}^{1}h)+\frac{\mu^{0}\kappa_{2}^{0}}{\mu^{1}\kappa_{2}^{1}}
\sinh(\kappa_{2}^{1}h)}{\cosh(\kappa_{2}^{1}h)+\frac{\mu^{1}\kappa_{2}^{1}}{\mu^{0}\kappa_{2}^{0}}\sinh(\kappa_{2}^{1}h)
}\right]
\frac{dk_{1}}{\kappa_{2}^{0}}~,
\end{equation}
with
\begin{equation}\label{w6.3.27}
k_{2}^{j}(\omega)=K_{2}^{j}(\omega):=\bigg| \sqrt{\left(
k^{j}(\omega)\right) ^{2}-\left( k_{1}\right) ^{2}}\bigg|
~~;~~k_{1}\leq k_{2}^{j}(\omega)~~;~~\omega\geq 0~,
\end{equation}
\begin{equation}\label{w6.3.28}
k_{2}^{j}(\omega)=i\kappa_{2}^{j}:=\bigg| \sqrt{\left(
k_{1}\right) ^{2}-\left( k^{j}(\omega)\right) ^{2}}\bigg|
~~;~~k_{1}\geq k^{j}(\omega)~~;~~\omega\geq 0.
\end{equation}
As shown in \cite{grobyetwirgin2005,grobyetwirgin2005II},
$\mathcal{F}_{20}^{2}$ usually dominates the other two terms, and
this is due to the fact that the denominator in the integrand of
$\mathcal{F}_{20}^{2}$ can vanish for $k_{1}$ over a large portion
of the interval of integration for frequencies at which  Love
modes are excited in the infrastructure, this occurring near the
Haskell frequencies
\begin{equation}\label{w6.3.29}
\nu_{l}^{LOVE}\approx\nu_{m}^{HASK}=\frac{2m+1}{2}\frac{c^{1}}{2h}~~;~~m\in\mathbb{N}
~,
\end{equation}
which corresponds to
\begin{equation}\label{w6.3.30}
k_{m}^{0~LOVE}\approx
k_{m}^{0~HASK}=\frac{2\pi\nu_{m}^{HASK}}{c^{0}}=
\frac{2m+1}{2}\frac{\pi c^{1}}{h c^{0}}~~;~~m\in\mathbb{N}~.
\end{equation}
\begin{equation}\label{w6.3.30a}
k_{m}^{1~LOVE}\approx k_{l}^{1~
HASK}=\frac{2\pi\nu_{m}^{HASK}}{c^{1}}=\frac{2m+1}{2}\frac{\pi}{h}~~;~~m\in\mathbb{N}
~.
\end{equation}
The sinc$^{2}$ function in the integrand of $\mathcal{F}_{20}^{2}$
is significantly large only in the interval $[0, 2\pi/w]$, so that
a minimal requirement for capturing most of the contribution of
the Love modes in $\mathcal{F}_{20}^{2}$ at their frequencies of
resonance is that
\begin{equation}\label{w6.3.31}
k_{1}^{LOVE}<\frac{2\pi}{w} ~.
\end{equation}

Of course, there exist other terms in the dispersion relation of
the $n=0$ mode, i.e., $\mathcal{F}_{20}^{1}$,
$\mathcal{F}_{20}^{3}$,  and the contributions of the
higher-than-fundamental Love modes to $\mathcal{F}_{20}^{2}$. As
concerns $\mathcal{F}_{20}^{1}$ and $\mathcal{F}_{20}^{3}$, we
notice that the former is complex and the latter is pure
imaginary, whereas $\mathcal{F}_{10}^{1}$, is real, so that we
would expect  to have $\mathcal{F}_{20}^{1}$ and
$\mathcal{F}_{20}^{3}$ to contribute less to the dispersion
relation than $\mathcal{F}_{20}^{2}$. The contribution to the
$n-th$ order natural mode of higher-than-fundamental Love modes to
$\mathcal{F}_{20}^{2}$ is empirically found to be always  less
than that of the zeroth order Love mode.

The analysis of the $n>0$ natural modes of the complete structure
proceeds in the same manner as previously, and will not be given
here.
\subsection{Another look at quasi displacement-free base block
modes}\label{anotherquasisfb}
The system of linear equations (\ref{bmbis.3}) can be written as
\begin{equation}\label{anotherquasisfb.1}
B_{n}^{2(l)}(\omega)=P_{n}^{(l)}(\omega)+
\sum_{j\in\mathbb{B}}\sum_{m=0}^{\infty}Q_{nm}^{(lj)}(\omega)B_{m}^{2(j)}(\omega)
~~;~~ \forall l\in\mathbb{B}~;~\forall n\in\mathbb{Z}~,
\end{equation}
wherein
\begin{multline}\label{anotherquasisfb.2}
P_{n}^{(l)}(\omega)= \frac{2\epsilon_{n}}{\cos(k_{2n}^{2(l)}b_l)}
\int_{-\infty}^{\infty}A^{0-}(k_{1},\omega)e^{-ik_{2}^{0}h}\left[
\frac{I_{n}^{(l)+}(k_{1},\omega)
e^{ik_{1}(d_l-w_l/2)}}{\cos(k_{2}^{1}h)-
\mbox{i}\frac{\mu^{1}k_{2}^{1}}{\mu^{0}k_{2}^{0}}i\sin(k_{2}^{1}h)
}\right] \frac{dk_{1}}{k_2^0}
\\
 ~~;~~ \forall l\in\mathbb{B}~;~\forall
n\in\mathbb{Z}~,
\end{multline}
and
\begin{multline}\label{anotherquasisfb.3}
Q_{nm}^{(lj)}(\omega)=
 \frac{\mbox{i}w_j}{2\pi
}\frac{\epsilon_{n}}{\cos(k_{2n}^{2(j)}b_j)}
\frac{\mu^{2}k_{2m}^{2}}{\mu^{0}}\sin(k_{2m}^{2(l)}b_l)\times
\\
\int_{-\infty}^{\infty}I_{n}^{(l)+}(k_{1},\omega)I_{m}^{(j)-}(k_{1},\omega)
\left[
\frac{\cos(k_{2}^{1}h)-i\frac{\mu^{0}k_{2}^{0}}{\mu^{1}k_{2}^{1}}
\sin(k_{2}^{1}h)}{\cos(k_{2}^{1}h)}-i\frac{\mu^{1}k_{2}^{1}}{\mu^{0}k_{2}^{0}}\sin(k_{2}^{1}h)
\right]
e^{ik_{1}\left((d_l-d_j)-\frac{w_l-w_j}{2}\right)}\frac{dk_{1}}{k_{2}^{0}}
\\
~~;~~ \forall l,~j\in\mathbb{B}~;~\forall n,~m\in\mathbb{Z}~.
\end{multline}
{\it Remark 1}\newline If $d_l=d_j$, then $Q_{nm}^{(lj)}=0$ for
$m+n=2p+1$, $\forall p \in \mathbb{N}$.
\newline
\newline
{\it Remark 2}\newline If $d_l=0$, then $P_{n}{(l)}=0$ for
$n=2p+1$, $\forall p \in \mathbb{N}$.
\newline
\newline
{\it Remark 3}\newline According to remarks 1 and 2,
$B_{2p+1}^{2(l)}=0$, $\forall p \in \mathbb{N}$, when the source
is located on the vertical line passing through the center of the
base of the $l$th bloc. This block then only admits displacement
via the even quasi-modes.
\newline
\newline
Let $\mathbb{B}:=\{1,2,3,...,N\}$ and
\begin{equation}\label{anotherquasisfb.4}
\mathbf{b}:= \left(
\mathbf{b}^{2(1)}~\mathbf{b}^{2(2)}~....~\mathbf{b}^{2(N)}\right) ^{T}
~,
\end{equation}
where $T$ is the transpose operator, and
\begin{equation}\label{anotherquasisfb.5}
\mathbf{b}^{2(l)}:= \left(
B^{2(l)}_{0}~B^{2(l)}_{1}~B^{2(l)}_{2}~....~\right) ^{T} ~.
\end{equation}
Similarly, let
\begin{equation}\label{anotherquasisfb.6}
\mathbf{p}:= \left(
\mathbf{p}^{(1)}~\mathbf{p}^{(1)}~....~\mathbf{p}^{(N)}\right) ^{T} ~,
\end{equation}
\begin{equation}\label{anotherquasisfb.7}
\mathbf{p}^{(l)}:= \left( P^{(l)}_{0}~P^{(l)}_{1}~P^{(l)}_{2}~....~\right)
^{T} ~,
\end{equation}
\begin{equation}\label{anotherquasisfb.8}
\mathbf{Q}:= \left(
\begin{matrix}
\mathbf{Q}^{(11)} & \mathbf{Q}^{(12)} & .... & \mathbf{Q}^{(1N)}
\\
\mathbf{Q}^{(21)} & \mathbf{Q}^{(22)} & .... & \mathbf{Q}^{(2N)}
\\ .            &                 &      &   .
\\ .            &                 &      &   .
\\ .            &                 &      &   .
\\ \mathbf{Q}^{(N1)} & \mathbf{Q}^{(N2)} & .... & \mathbf{Q}^{(NN)}
\end{matrix}
\right),
\end{equation}
\begin{equation}\label{anotherquasisfb.9}
\mathbf{Q}^{(lj)}:= \left(
\begin{matrix}
Q^{(lj)}_{00} & Q^{(lj)}_{01} & Q^{(lj)}_{02} &........
\\
Q^{(lj)}_{10} & Q^{(lj)}_{11} & Q^{(lj)}_{12} &........
\\
Q^{(lj)}_{20} & Q^{(lj)}_{21} & Q^{(lj)}_{22} &........
\\ .            &                 &    .
\\ .            &                 &    .
\\ .            &                 &    .
\\ .            &                 &    .
\\ .            &                 &    .
\\ .            &                 &    .
\end{matrix}
\right),
\end{equation}
so that (\ref{anotherquasisfb.1}) can be written as
\begin{equation}\label{anotherquasisfb.10}
  \mathbf{b}=\mathbf{p}+\mathbf{Q}\mathbf{b}~,
\end{equation}

This is a matrix equation enabling the determination of the set
$\mathbf{b}$ of unknown vectors
$\{\mathbf{b}^{2(l)}~;~l\in\mathbb{B}\}$; it can be written as
\begin{equation}\label{anotherquasisfb.11}
  (\mathbf{I}-\mathbf{Q})\mathbf{b}=\mathbf{p}~,
\end{equation}
wherein $\mathbf{I}$ is the identity matrix (i.e., the matrix
having the same dimensions as $\mathbf{Q}$ with one's on the
diagonal and zeros elsewhere).

The {\it natural modes} of the city are obtained by turning off
the excitation, embodied in the vector $\mathbf{p}$. Thus:
\begin{equation}\label{anotherquasisfb.12}
  (\mathbf{I}-\mathbf{Q})\mathbf{b}=0~,
\end{equation}
which possesses a non-trivial solution only if
\begin{equation}\label{anotherquasisfb.13}
  \text{det}(\mathbf{I}-\mathbf{Q})=0~,
\end{equation}
wherein det$(\mathbf{M})$ signifies the determinant of the matrix
$\mathbf{M}$.

To get a grip on this relation, consider the case in which there
is only one block in the city. Then (\ref{anotherquasisfb.13}) becomes
\begin{equation}\label{anotherquasisfb.14}
  \text{det} \left(
  \begin{matrix}
  1-Q_{00}^{(11)} & -Q_{01}^{(11)} & -Q_{02}^{(11)} & .......\\
   -Q_{10}^{(11)} & 1-Q_{11}^{(11)} & -Q_{12}^{(11)} & .......\\
   -Q_{20}^{(11)} & -Q_{21}^{(11)} & 1-Q_{22}^{(11)} & .......\\
   .            &              &               &        \\
   .            &              &               &        \\
   .            &              &               &        \\
   .            &              &               &        \\
\end{matrix}
\right) =0~,
\end{equation}
A procedure, called the {\it partition method}, for solving this
equation (as well as the equation (\ref{anotherquasisfb.11}) for
the response), particularly appropriate if the off-diagonal
elements of the matrix are small compared to the diagonal
elements, is first to consider the matrix to have one row and one
column,
\begin{equation}\label{anotherquasisfb.16}
  1-Q_{00}^{(11)} =0~,
\end{equation}
then to consider the matrix to have two rows and two columns,
\begin{equation}\label{anotherquasisfb.17}
  \text{det} \left(
  \begin{matrix}
  1-Q_{00}^{(11)} & -Q_{01}^{(11)}\\
   -Q_{10}^{(11)} & 1-Q_{11}^{(11)}\\
\end{matrix}
\right) =0~,
\end{equation}
or
\begin{equation}\label{anotherquasisfb.18}
  \left( 1-Q_{00}^{(11)}\right) \left( 1-Q_{11}^{(11)}\right) -Q_{01}^{(11)}Q_{10}^{(11)}=0~,
\end{equation}
then to consider the matrix to have three rows and three columns,
\begin{equation}\label{anotherquasisfb.19}
  \text{det} \left(
  \begin{matrix}
  1-Q_{00}^{(11)} & -Q_{01}^{(11)} & -Q_{02}^{(11)} \\
   -Q_{10}^{(11)} & 1-Q_{11}^{(11)} & -Q_{12}^{(11)} \\
   -Q_{20}^{(11)} & -Q_{21}^{(11)} & 1-Q_{22}^{(11)}\\
\end{matrix}
\right) =0~,
\end{equation}
or
\begin{multline}\label{anotherquasisfb.20}
  \left( 1-Q_{00}^{(11)}\right) \left( 1-Q_{11}^{(11)}\right)
  \left( 1-Q_{22}^{(11)}\right)-
  \\
  Q_{10}^{(11)}Q_{21}^{(11)}Q_{02}^{(11)}-
  Q_{20}^{(11)}Q_{12}^{(11)}Q_{01}^{(11)}-
  \left( 1-Q_{11}^{(11)}\right) Q_{20}^{(11)}Q_{02}^{(11)}-
   \left( 1-Q_{22}^{(11)}\right) Q_{10}^{(11)}Q_{01}^{(11)}=0~,
\end{multline}
etc.

If the off-diagonal elements of the matrix are considered to be
negligible compared to the diagonal elements, the dispersion
relation (\ref{anotherquasisfb.14}) is simply that
 the product of the diagonal elements of the
matrix should be nil, which means that any diagonal element of the
matrix should be nil.

Consider the case in which it is $1-Q_{00}^{(11)}$ that vanishes.
This is equivalent to
\begin{multline}\label{anotherquasisfb.22}
 0=\cot(k^{2}b)-
 \frac{\mbox{i}k^{2}w_1}{2\pi}
\frac{\mu^{2}}{\mu^{0}}
\int_{-\infty}^{\infty}I_{0}^{(1)+}(k_{1},\omega)I_{0}^{(1)-}(k_{1},\omega)
\left[
\frac{\cos(k_{2}^{1}h)-\mbox{i}\frac{\mu^{0}k_{2}^{0}}{\mu^{1}k_{2}^{1}}
\sin(k_{2}^{1}h)}{\cos(k_{2}^{1}h)-
\mbox{i}\frac{\mu^{1}k_{2}^{1}}{\mu^{0}k_{2}^{0}}\sin(k_{2}^{1}h)}\right]
\frac{dk_{1}}{k_{2}^{0}}~,
\end{multline}
which is the same as the dispersion relation of the zeroth-order
quasi displacement-free base block mode obtained in the previous
section. Thus, there is no apparent gain
in adopting the analysis of the present section over that of the
previous section in an attempt to resolve the difficulty mentioned
in {\it Remark} 3 in the previous section, namely the fact that
the dispersion relations do not depend on the number $N$ of blocks
in the city.

To resolve this problem, we must take into account the
off-diagonal elements of the matrix because  these elements
contain the information on the number of blocks. Unfortunately,
the inclusion of these off-diagonal elements makes the dispersion
relation increasingly complicated and difficult to analyze as the
order of approximation of the partition method (which consists in
solving increases (it is already quite complicated at first
order). The only way to solve these dispersion relations (which
consist in equating determinants of increasing rank to zero) is by
numerical computation.

When more than one block (for example 2 blocks) are present, the
dispersion relations becomes (if the off-diagonal elements of the
matrix for each block considered independently can be neglected):
\begin{equation}\label{2b}
\mbox{det}
\left(
\begin{array}{ll}
1-Q_{00}^{(11)}& -Q_{00}^{(12)}\\
-Q_{00}^{(21)}& 1-Q_{00}^{(22)}
\end{array}
\right)=
\left(1-Q_{00}^{(11)} \right)\left(1-Q_{00}^{(22)} \right)-Q_{00}^{(12)}Q_{00}^{(21)}=0
\end{equation}
which is quite different from the zeroth-order quasi
displacement-free base block dispersion relation because the
\textit{coupling term} $Q_{00}^{(12)}Q_{00}^{(21)}$ does not
vanish and cannot be neglected. This term corresponds to the
so-called \textit{structure-soil-structure interaction} and
accounts for the distance separating the two buildings. Its form
is close to that of the term representative of the action of the
geophysical structure. Nevertheless, due to its complexity, it is
difficult to carry out an analytical study of this relation.

The {\it partition method} emphasizes the role of the global
superstructure, while the {\it iteration method}, described in
sect. \ref{quasistfrbablmo}, emphasizes the role of only one
component (i.e. one block) of the superstructure. The partition
method
 also accounts for all the possible
interactions between blocks (the term
$\cot\left(k_{20}b_l\right)$) and the geophysical structure
through the terms $Q_{nm}^{(ll)}$, and the interaction between
blocks through the terms $Q_{nm}^{(lj)}$ and $Q_{nm}^{(jl)}$.

Ultimately, the choice of method reduces to determining which one
gives the best results, i.e., results that are closest to reality.
This can  be determined only by full-blown numerical studies, the
results of which will have to be compared to those of the FE
method (employed as a reference). In fact, we find that the
partition method gives the best results, and is therefore employed
in all the subsequent numerical computations.
\section{Expression of the fields $u^{2(j)}(\mathbf{x},\omega)$,
$u^{1}(\mathbf{x},\omega)$ and $u^{0}(\mathbf{x},\omega)$ for line
source excitation}\label{expu1u0}
Once the quasi-modal coefficients $B_{m}^{2(l)}$, $\forall m \in
\mathbb{N}$, $\forall l \in \mathbb{B}$ are obtained from the
matrix equation (\ref{anotherquasisfb.11}), the field in the block
domain $\Omega^{2(l)}$ is computed via (\ref{fieldrepblock.5}).
This field vanishes on the ground at the frequency of occurrence
of the {\textit{displacement-free base mode}} of the block.
\newline
\newline
{\textit{Remark 1}} \newline If higher-than-the- zeroth-order
quasi modal coefficients   can be neglected, the field in the
block $l\in\mathbb{B}$ takes the form
\begin{equation}
u^{2(l)}\approx B_{0}^{2(l)}(\omega)\cos\left(k^{2}\left(x_2+b_l \right) \right)
\end{equation}
which indicates that the displacement field is independent of
$x_1$  and takes the form of a standing wave in the block.

Let us next consider the field in the layer. Combining
(\ref{stressgamg.3}),  (\ref{dispgammag.3}),
(\ref{stressgammah.2}) and (\ref{displgammah.2}), leads, via
(\ref{fieldrepomega1.1}), to:
\begin{multline}\label{expu1u0.1}
u^{1}(\mathbf{x},\omega)=
\\
\frac{\mbox{i}S(\omega)}{2\pi}\int_{-\infty}^{\infty}
\frac{\cos\left(k_{2}^{1}x_{2}\right) e^{\mbox{i}\left(
k_{1}(x_{1}-x_{1}^{s})+k_{2}^{0}(x_{2}^{s}-h)\right)}}
{\cos\left(k_{2}^{1}h
\right)-\mbox{i}\frac{\mu^{1}k_{2}^{1}}{\mu^{0}k_{2}^{0}}\sin\left(k_{2}^{1}h\right)}
\frac{dk_{1}}{k_{2}^{0}}+
\frac{\mbox{i}}{2\pi}\sum_{l\in\mathbb{B}}\sum_{n=0}^{\infty}\frac{\mu^{2}}{\mu^{0}}B_{n}^{2(l)}k_{2n}^{2(l)}w_{l}
\sin\left(k_{2n}^{2(l)}b_{l}\right)\times
\\
\int_{\infty}^{\infty}\frac{\cos\left(k_{2}^{1}(x_{2}-h)
\right)+\mbox{i}\frac{\mu^{0}k_{2}^{0}}{\mu^{1}k_{2}^{1}}\sin\left(k_{2}^{1}(x_{2}-h)\right)}{\cos\left(k_{2}^{1}h
\right)-\mbox{i}\frac{\mu^{1}k_{2}^{1}}{\mu^{0}k_{2}^{0}}\sin\left(k_{2}^{1}h\right)}
I_{n}^{(l)-}(k_{1},\omega)e^{\mbox{i}k_{1}(x_{1}-(d_{l}-\frac{w_{l}}{2}))}\frac{dk_{1}}{k_{2}^{0}}~,
\end{multline}
which can be cast into the form
\begin{equation}\label{expu1u0.2}
u^{1}(\mathbf{x},\omega)=u_{c}^{1}(\mathbf{x},\omega)+\sum_{l\in\mathbb{B}}
u_{\mathbb{B}}^{1(l)}(\mathbf{x},\omega)~,
\end{equation}
with
\begin{equation}\label{expu1u0.3}
u_{c}^{1}(\mathbf{x},\omega)=\frac{\mbox{i}S(\omega)}{2\pi}
\int_{-\infty}^{\infty}\frac{\cos\left(k_{2}^{1}x_{2}\right)
e^{\mbox{i}\left(
k_{1}(x_{1}-x_{1}^{s})+k_{2}^{0}(x_{2}^{s}-h)\right)}}
{\cos\left(k_{2}^{1}h
\right)-\mbox{i}\frac{\mu^{1}k_{2}^{1}}{\mu^{0}k_{2}^{0}}
\sin\left(k_{2}^{1}h\right)}\frac{dk_{1}}{k_{2}^{0}}~,
\end{equation}
and
\begin{multline}\label{expu1u0.4}
u_{\mathbb{B}}^{1(l)}(\mathbf{x},\omega)=
\frac{\mbox{i}}{2\pi}\sum_{n=0}^{\infty}
\frac{\mu^{2}}{\mu^{0}}B_{n}^{2(l)}k_{2n}^{2(l)}w_{l}
\sin\left(k_{2n}^{2(l)}b_{l}\right)
\times
\\
\int_{\infty}^{\infty}\frac{\cos\left(k_{2}^{1}(x_{2}-h)
\right)+\mbox{i}\frac{\mu^{0}k_{2}^{0}}{\mu^{1}k_{2}^{1}}
\sin\left(k_{2}^{1}(x_{2}-h)\right)}{\cos\left(k_{2}^{1}h
\right)-\mbox{i}\frac{\mu^{1}k_{2}^{1}}{\mu^{0}k_{2}^{0}}
\sin\left(k_{2}^{1}h\right)}
I_{n}^{(l)-}(k_{1},\omega)e^{ik_{1}(x_{1}-(d_{l}-\frac{w_{l}}{2}))}
\frac{dk_{1}}{k_{2}^{0}}~.
\end{multline}
This expression indicates that the field in the  layer is composed
of: i) the field obtained in the absence of blocks and induced by
the incident cylindrical wave radiated, ii) the fields induced by
the presence of each block. The displacement field
$u^{1}(\mathbf{x},\omega)-u_{c}^{1}(\mathbf{x},\omega)$ appears as
a sum of block fields which  are strongly linked together, since
each coefficient $B_{n}^{2(j)}$ is calculated by taking into
account the presence of the other blocks via
(\ref{anotherquasisfb.11}).

Let us finally consider the field in the substratum. Combining
(\ref{stressgamg.3}),  (\ref{dispgammag.3}),
(\ref{stressgammah.2}) and (\ref{displgammah.2}), leads, via
(\ref{fieldrepomega0.10}), to:
\begin{multline}\label{expu1u0.5}
u^{0}(\mathbf{x},\omega)=u^{i}(\mathbf{x},\omega)+
\frac{\mbox{i}S(\omega)}
{4\pi}\int_{-\infty}^{\infty}\frac{\cos\left(k_{2}^{1}h \right)+
\mbox{i}\frac{\mu^{1}k_{2}^{1}}{\mu^{0}k_{2}^{0}}\sin\left(k_{2}^{1}h\right)}
{\cos\left(k_{2}^{1}h
\right)-\mbox{i}\frac{\mu^{1}k_{2}^{1}}{\mu^{0}k_{2}^{0}}
\sin\left(k_{2}^{1}h\right)}
e^{\mbox{i}(k_{1}(x_{1}-x_{1}^{s})+k_{2}^{0}(x_{2}+x_{2}^{s}-2h))}
\frac{dk_{1}}{k_{2}^{0}}+
\\
\frac{\mbox{i}}{2\pi}\sum_{l\in\mathbb{B}}\sum_{n=0}^{\infty}\frac{\mu^{2}}
{\mu^{0}}B_{n}^{2(l)}k_{2n}^{2(l)}w_{l}
\sin\left(k_{2n}^{2(l)}b_{l}\right)
\int_{\infty}^{\infty}\frac{I_{n}^{(l)-}(k_{1},\omega)e^{\mbox{i}(k_{1}(x_{1}-(d_{l}-\frac{w_{l}}{2})+k_{2}^{0}(x_{2}-h))
)}}{\cos\left(k_{2}^{1}h
\right)-\mbox{i}\frac{\mu^{1}k_{2}^{1}}{\mu^{0}k_{2}^{0}}\sin\left(k_{2}^{1}h\right)}\frac{dk_{1}}{k_{2}^{0}}~,
\end{multline}
which can be cast into the form
\begin{equation}\label{expu1u0.6}
u^{0}(\mathbf{x},\omega)=\left\{u^{i}(\mathbf{x},\omega)+u_{c}^{d0}(\mathbf{x},\omega)\right\}+
\sum_{j\in\mathbb{B}}u_{\mathbb{B}}^{0(l)}(\mathbf{x},\omega)
\end{equation}
with
\begin{equation}\label{expu1u0.7}
u_{c}^{d0}(\mathbf{x},\omega)=\frac{\mbox{i}S(\omega)}{4\pi}\int_{-\infty}^{\infty}
\frac{\cos\left(k_{2}^{1}h
\right)+\mbox{i}\frac{\mu^{1}k_{2}^{1}}{\mu^{0}k_{2}^{0}}
\sin\left(k_{2}^{1}h\right)}{\cos\left(k_{2}^{1}h
\right)-\mbox{i}\frac{\mu^{1}k_{2}^{1}}
{\mu^{0}k_{2}^{0}}\sin\left(k_{2}^{1}h\right)}
e^{\mbox{i}(k_{1}(x_{1}-x_{1}^{s})+k_{2}^{0}(x_{2}+
x_{2}^{s}-2h))}\frac{dk_{1}}{k_{2}^{0}}~,
\end{equation}
and
\begin{multline}\label{expu1u0.8}
u_{\mathbb{B}}^{0(l)}(\mathbf{x},\omega)=\frac{\mbox{i}}{2\pi}\sum_{n=0}^{\infty}
\frac{\mu^{2}}{\mu^{0}}B_{n}^{2(l)}k_{2n}^{2(l)}w_{l}
\sin\left(k_{2n}^{2(l)}b_{l}\right)\times
\\
\int_{\infty}^{\infty}\frac{I_{n}^{(l)-}(k_{1},\omega)e^{i(k_{1}(x_{1}-(d_{l}-
\frac{w_{l}}{2})+k_{2}^{0}(x_{2}-h)) )}}{\cos\left(k_{2}^{1}h
\right)-\mbox{i}
\frac{\mu^{1}k_{2}^{1}}{\mu^{0}k_{2}^{0}}\sin\left(k_{2}^{1}h\right)}\frac{dk_{1}}{k_{2}^{0}}~.
\end{multline}
This expression indicates that the field in the substratum is
composed of:  i) the field obtained in absence of blocks,
including the incident  plus  diffracted fields, the latter being
induced  by the incident cylindrical wave, ii) the fields induced
by the presence of each block; the latter takes the form of a sum
of block fields which  are strongly linked together, since each
coefficient $B_{n}^{2(j)}$ is calculated by taking into account
the presence of the other blocks via (\ref{anotherquasisfb.11}).
\subsection{Interpretation of the fields $u_{\mathbb{B}}^{0(j)}$ and
$u_{\mathbb{B}}^{1(j)}$}\label{intub1ub0}
If the leading term of the quasi modal representation in each
block $l\in\mathbb{B}$ is dominant (i.e., the higher-order terms
can be neglected), the fields
$u_{\mathbb{B}}^{0(j)}(\mathbf{x},\omega)$ in (\ref{expu1u0.4})
and $u_{\mathbb{B}}^{1(j)}(\mathbf{x},\omega)$ in
(\ref{expu1u0.8}) reduce to:
\begin{equation}\label{intub1ub0.1}
u_{\mathbb{B}}^{1(j)}(\mathbf{x},\omega)=\frac{\mbox{i}}{2\pi}\frac{\mu^{2}}{\mu^{0}}B_{0}^{2(j)}k^{2}w_{j}
\sin\left(k^{2}b_{j}\right)
\int_{\infty}^{\infty}\frac{\mbox{sinc}\left(
k_{1}\frac{w_{j}}{2}\right)e^{\mbox{i}(k_{1}(x_{1}-d_{j})+k_{2}^{0}(x_{2}-h))
)}}{\cos\left(k_{2}^{1}h
\right)-\mbox{i}\frac{\mu^{1}k_{2}^{1}}{\mu^{0}k_{2}^{0}}\sin\left(k_{2}^{1}h\right)}\frac{dk_{1}}{k_{2}^{0}}~,
\end{equation}
\begin{equation}\label{intub1ub0.2}
u_{\mathbb{B}}^{0(j)}(\mathbf{x},\omega)=\frac{\mbox{i}}{2\pi}\frac{\mu^{2}}{\mu^{0}}B_{0}^{2(j)}k^{2}w_{j}
\sin\left(k^{2}b_{j}\right)
\int_{\infty}^{\infty}\frac{\mbox{sinc}\left(
k_{1}\frac{w_{j}}{2}\right)e^{\mbox{i}(k_{1}(x_{1}-d_{j})+k_{2}^{0}(x_{2}-h))
)}}{\cos\left(k_{2}^{1}h
\right)-\mbox{i}\frac{\mu^{1}k_{2}^{1}}{\mu^{0}k_{2}^{0}}\sin\left(k_{2}^{1}h\right)}\frac{dk_{1}}{k_{2}^{0}}~.
\end{equation}
With the help of the material in appendix (\ref{aux3}), each
 $u_{\mathbb{B}}^{0(j)}(\mathbf{x},\omega)$ can be
interpreted as the field radiated by ribbon source of width
$w_{j}$, located at the base of each block $j\in\mathbb{B}$, and
whose amplitude is of the form $\displaystyle
\frac{\mu^{1}}{\mu^{2}}B_{0}^{2(j)}k^{2})w_{j}
\sin\left(k^{2}b_{j}\right)$. These sources are induced sources,
i.e.,  they do not introduce energy into the system, but each of
them induces a modification of the repartition of the energy over
the excitation frequency bandwidth. These sources are located at
the top of the layer and should excite quasi-Love waves, as shown
in \cite{grobyetwirgin2005,grobyetwirgin2005II,grobyandwirgin2003}
(for applied sources).

If higher-order quasi modes are relevant, the even-order modes
corresponds to ribbon sources of width $w_l$, located at $d_l$,
$l\in\mathbb{B}$, while the odd-order modes correspond to line
sources located at the edges of the blocks. The amplitudes of both
of these types of induced sources  depend on the order of the
quasi modes and on the characteristics of the corresponding block.
\section{Numerical results for one block in a Mexico City-like site}
All intervals over the which integration is performed in the
calculation of the quasi-modal coefficients, through the linear
system (\ref{anotherquasisfb.11}), i.e. in the calculation of
$P_n^{(l)}$ and of $Q_{nm}^{(jl)}$, $(n,m)\in \mathbb{N}^2$,
$(j,l)\in \mathbb{B}^2$, and in the calculation of the
displacement fields $\displaystyle
u^{1}=u_c^{1}+\sum_{l\in\mathbb{B}}u_{\mathbb{B}}^{1(l)}$ and
$\displaystyle
u^{0}=u^i+u_c^{0d}+\sum_{l\in\mathbb{B}}u_{\mathbb{B}}^{0(l)}$,
are first reduced to $\displaystyle \int_{0}^{+\infty}$. These
intervals are then separated into
$\mathcal{I}_1=[0,\Re\left(k^0\right)]$ (interference of
propagative waves),
$\mathcal{I}_2=[\Re\left(k^0\right),\Re\left(k^1\right)]$
(excitation of quasi-Love waves) and
$\mathcal{I}_3=[\Re\left(k^1\right),+\infty]$ (evanescent waves in
the layer) in order to point out the different type of waves
associated with the different possible phenomena in the
geophysical structure. The numerical evaluation of these integrals
is carried out by the the procedure described in
\cite{grobyetwirgin2005II}.

The results in this section apply to a single block whose base
segment center is located at (0m,0m).

We compute the response inside the block and on the ground near
the block.

The latter is supposed to be situated in a Mexico City-like site
wherein: $\rho^{0}=2000$ kg/m$^{3}$, $c^{0}$=600 m/s,
$Q^{0}=\infty$, $\rho^{1}=1300$ kg/m$^{3}$, $c^{1}$=60 m/s,
$Q^{1}=30$, with the soft layer thickness being $h=50$m. In
addition, the material constants of the block are: $\rho^{2}=325$
kg/m$^{3}$, $c^{2}$=100m/s, $Q^{2}=100$.

The source is placed  consecutively at (0m, 3000m) or (-65m,
3000m), which are deep locations for  which Love modes can hardly
be excited in the absence of the block, and at (-3000m, 100m), a
shallow location at the which Loves modes can easily be excited in
the absence of the block,
\cite{grobyetwirgin2005,grobyetwirgin2005II}.

The eigenfrequencies of the block displacement-free base block are
$\displaystyle \nu_{0m}^{FB}=\frac{c^{2}(2m+1)}{2b}$, and the
Haskell frequencies are $\displaystyle
\nu_{m}^{HASK}=\frac{2m+1}{2}\frac{c^{1}}{2h}$, wherein
$m=0,1,2,...$. Thus, the Haskell frequencies are 0.3, 0.9, 1.5
...Hz. The fundamental displacement-free base block eigenfrequency
(whose value is supposed to be close to the one of the
corresponding quasi-mode), depends on the choice of the mechanical
parameters of the medium filling $\Omega_2$, and occurs at
$\displaystyle \nu_{00}^{DFB}=\frac{c^{2}}{2b}\approx
\frac{25}{b}$. This expression agrees with the empirical one:
$\displaystyle \nu_{00}^{FB}\approx \frac{30}{b}$ employed in
\cite{Boutinroussillon}.

If the zeroth-order quasi-mode is dominant, the dispersion
relation (\ref{anotherquasisfb.22}) takes the form:
\begin{equation}
0= \cot(k^{2}b)-
 \frac{\mbox{i}k^{2}w}{2\pi}
\frac{\mu^{2}}{\mu^{0}}
\int_{-\infty}^{\infty}\left[
\frac{\cos(k_{2}^{1}h)-\mbox{i}\frac{\mu^{0}k_{2}^{0}}{\mu^{1}k_{2}^{1}}
\sin(k_{2}^{1}h)}{\cos(k_{2}^{1}h)-
\mbox{i}\frac{\mu^{1}k_{2}^{1}}{\mu^{0}k_{2}^{0}}\sin(k_{2}^{1}h)}\right]
\mbox{sinc}^{2}\left(k_1 \frac{w}{2} \right)\frac{dk_{1}}{k_{2}^{0}}= \mathcal{F}_1-\mathcal{F}_2
\label{disrel1b}
\end{equation}
\subsection{Results relative to one $40m\,\times\, 40m$
block}\label{onebloc4040}
The block is $40m$ high and $40m$ wide. The displacement-free base
block eigenfrequencies are then 0.625, 1.875 Hz....
\begin{figure}[ptb]
\centering\includegraphics[width=6.3cm]
{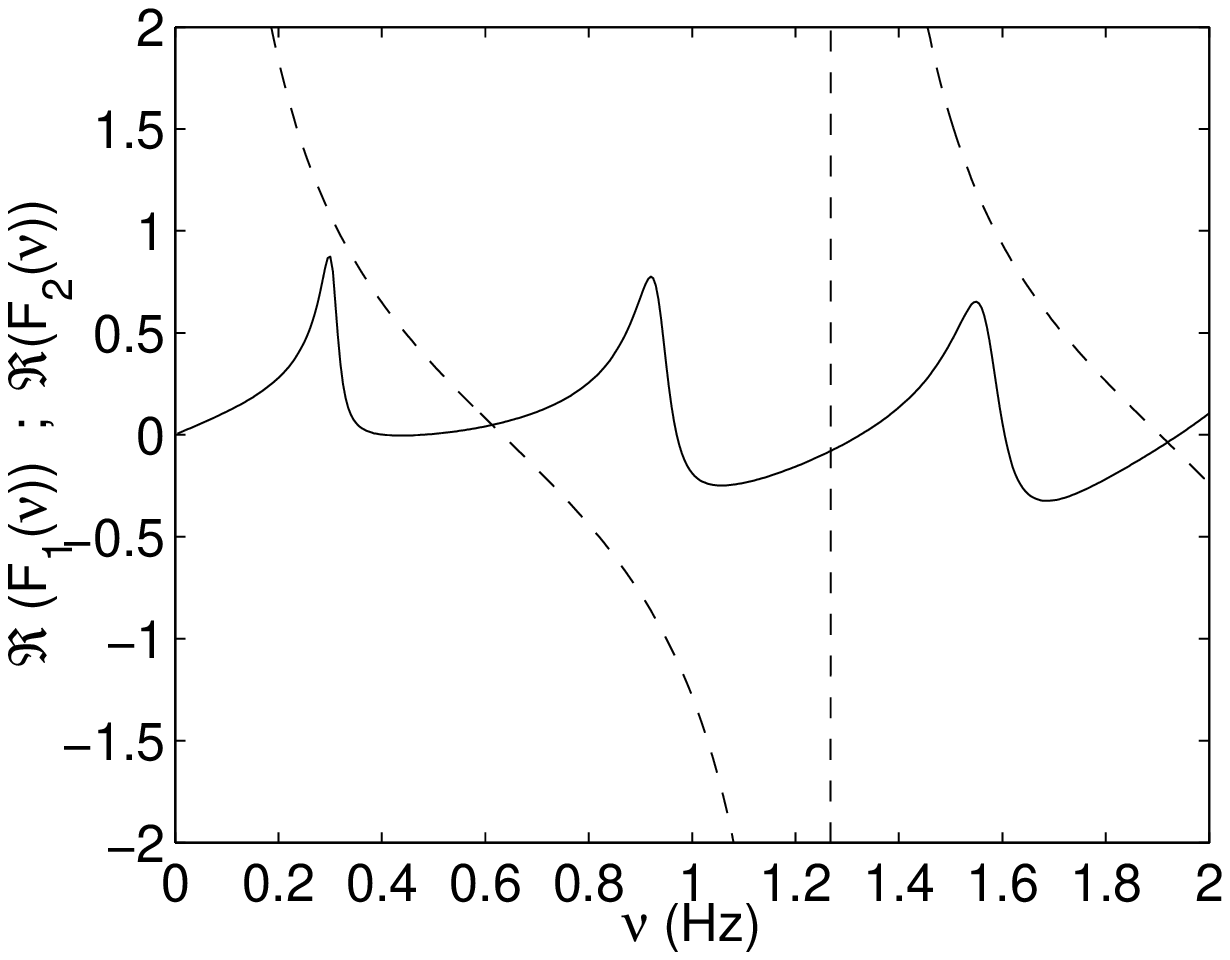}
\centering\includegraphics[width=6.3cm]
{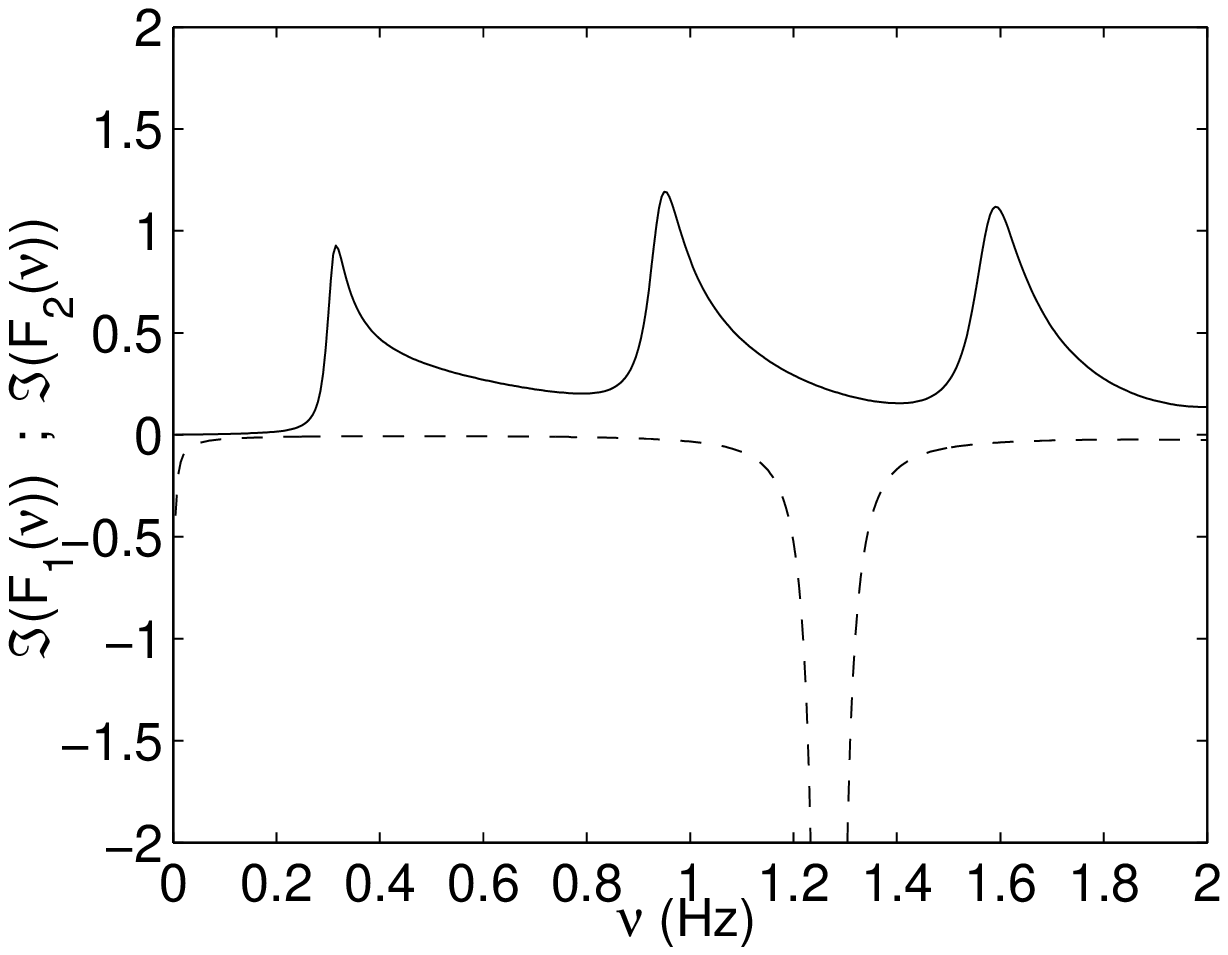} \caption{Indications
concerning the solution of the dispersion relation
$\mathcal{F}=\mathcal{F}_1-\mathcal{F}_2=0$ for a single
$40m\,\times\, 40m$ block in a Mexico City-like site. In the left
panel the solid curve describes $\Re\left(\mathcal{F}_{1}\right)$
versus frequency ($\nu=\omega/2\pi$ in Hz) and the dashed curve
describes $\Re\left(\mathcal{F}_{2}\right)$ versus frequency. The
right panel again contains two curves: the solid one describes
$\Im\left(\mathcal{F}_{1}\right)$ versus frequency and the dashed
curve describes $\Im\left(\mathcal{F}_{2}\right)$ versus frequency
(Hz).} \label{reim4040}
\end{figure}

Fig. \ref{reim4040} gives indications on the frequencies of
occurrence of the modes of the entire (superstructure +
substructure) built site. An eigenfrequency is a frequency for
which $\mathcal{F}=\mathcal{F}_1-\mathcal{F}_2=0$, i.e., for which
$\Re\left(\mathcal{F}_{1}\right)=\Re\left(\mathcal{F}_{2}\right)$
at the least. One notes that this occurs for $\nu\approx
0.3\mbox{, }0.63\mbox{, }1.35\mbox{, }1.92$ Hz in the frequency
range of the figure. An indication of the attenuation associated
with a particular mode (at a frequency $\nu^{\star}$) is the
quantity
$|\Im\left(\mathcal{F}_{1}(\nu^{\star})\right)-\Im\left(\mathcal{F}_{2}(\nu^{\star})\right)|$.
The quasi-Love mode at $\nu\approx 0.3 Hz$ and $\approx 0.9 Hz$,
and the quasi displacement-free base mode block at
$\nu^{QDFB}\approx 0.63 Hz$ and $\approx 1.9 Hz$ are modes with
relatively-small attenuation. On the contrary, the quasi
stress-free base block (close to stress-free base block modes,
corresponding to $\left.u_{2}^{2}\right|_{x_2=0}=0$ and satisfying
$\tan\left(k^{2}b\right)=0$) at $\nu_{01}^{QSFB}\approx 1.35 Hz$
is a mode with  very large attenuation. This third type of mode
(quasi stress-free base block modes) should therefore have a small
influence on the response at the site, as was mentioned in section
\ref{quasilove}.

\begin{figure}[ptb]
\begin{center}
\includegraphics[width=6.0cm] {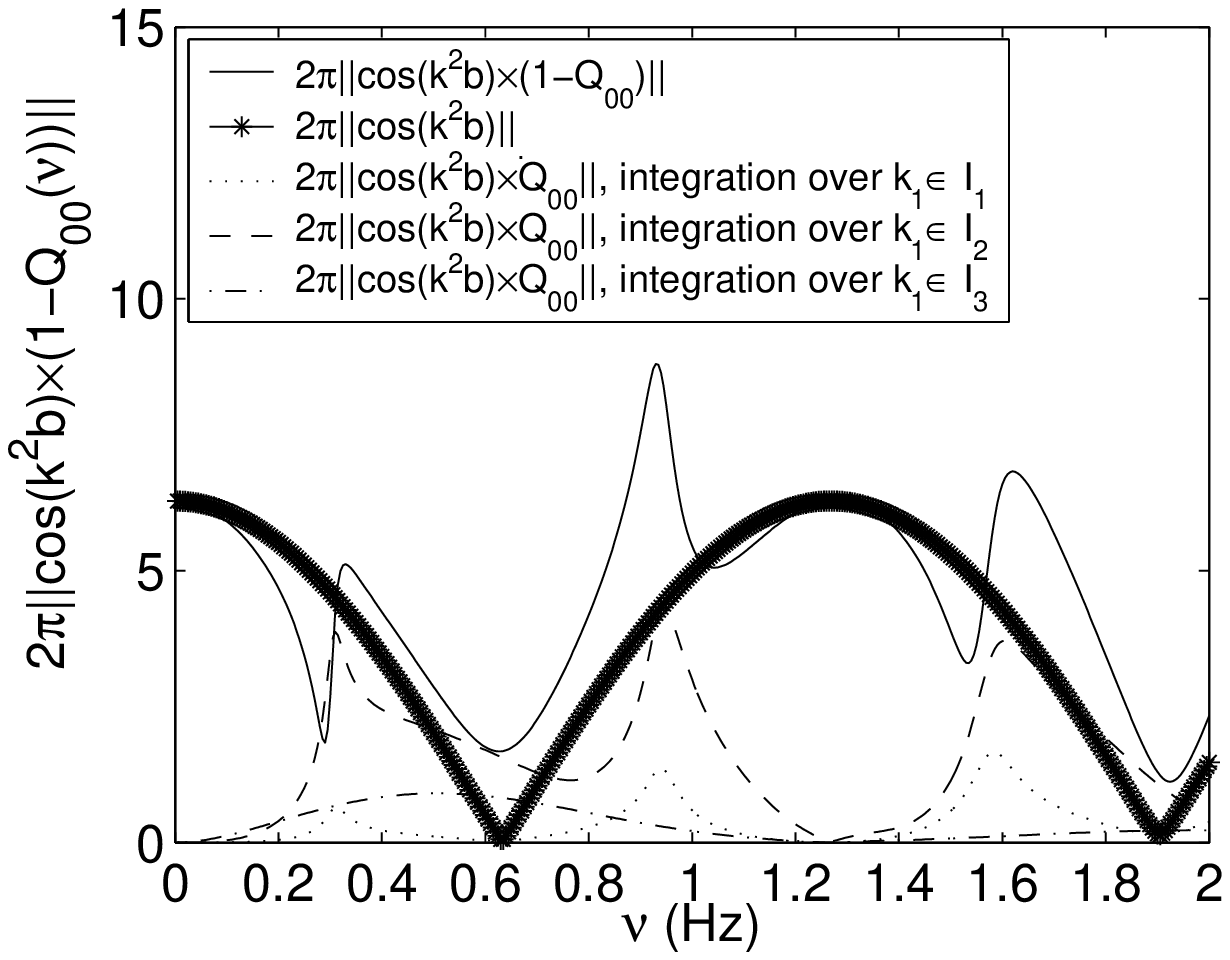}
\includegraphics[width=6.3cm] {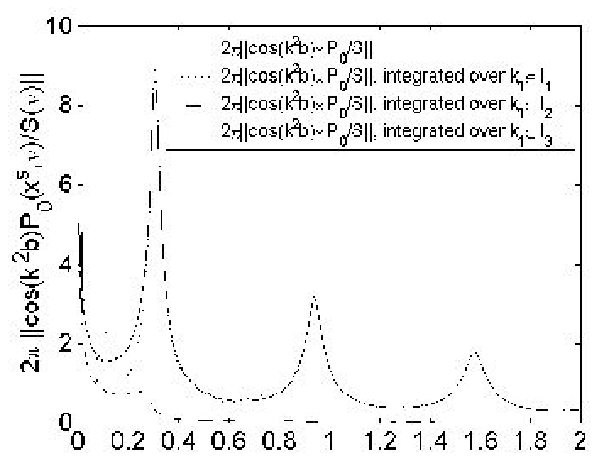}
\end{center}
\caption{$2\pi \|\cos(k^{2}b)(1-Q_{00})\|$ (left panel) and the driving agent
$2\pi\|\cos(k^{2}b)P_{0}/S(\omega)\|$ (right panel).}
\label{E00G04040}
\end{figure}

To give an indication on how the block is excited, and how it
re-radiates the field, we  depict in figure \ref{E00G04040} the
absolute values of the zeroth-order terms, i.e. $(1-Q_{00})$ and
$P_{0}$, involved in the resolution of the linear system
(\ref{anotherquasisfb.11}). The integration interval is divided
into three subintervals: $\mathcal{I}_{k}$, $k=1,2,3$. The source
term $2\pi\|\cos(k^{2}b)P_{0}(\omega)/S(\omega)\|$ is similar to
the transfer function in the absence of the block in
\cite{grobyetwirgin2005II}. The term
$I_{0}^{(j)-}(k_{1},\omega)=\mbox{sinc}\left(k_{1}\frac{w}{2}\right)$
exerts a small influence on $P_{0}(\omega)$. The solicitation of
the block takes  a form close to waves traveling in the layer in
the absence of the block. The deep source being located close to a
line going straight down from the block, the main component of
$P_{0}(\omega)$ comes from the integration over $\mathcal{I}_{1}$
(i.e. interference of propagative waves in the substructure),
while the main component of $Q_{00}(\omega)$, which is independent
of the source location, comes from the integration over
$\mathcal{I}_{2}$ (i.e. Love mode excitation).
\subsubsection{Displacement field on the top and bottom segments of
the block for deep line source solicitation}
We now examine the displacement field on  the horizontal
boundaries of the block.
\begin{figure}[ptb]
\begin{center}
\includegraphics[width=6.0cm] {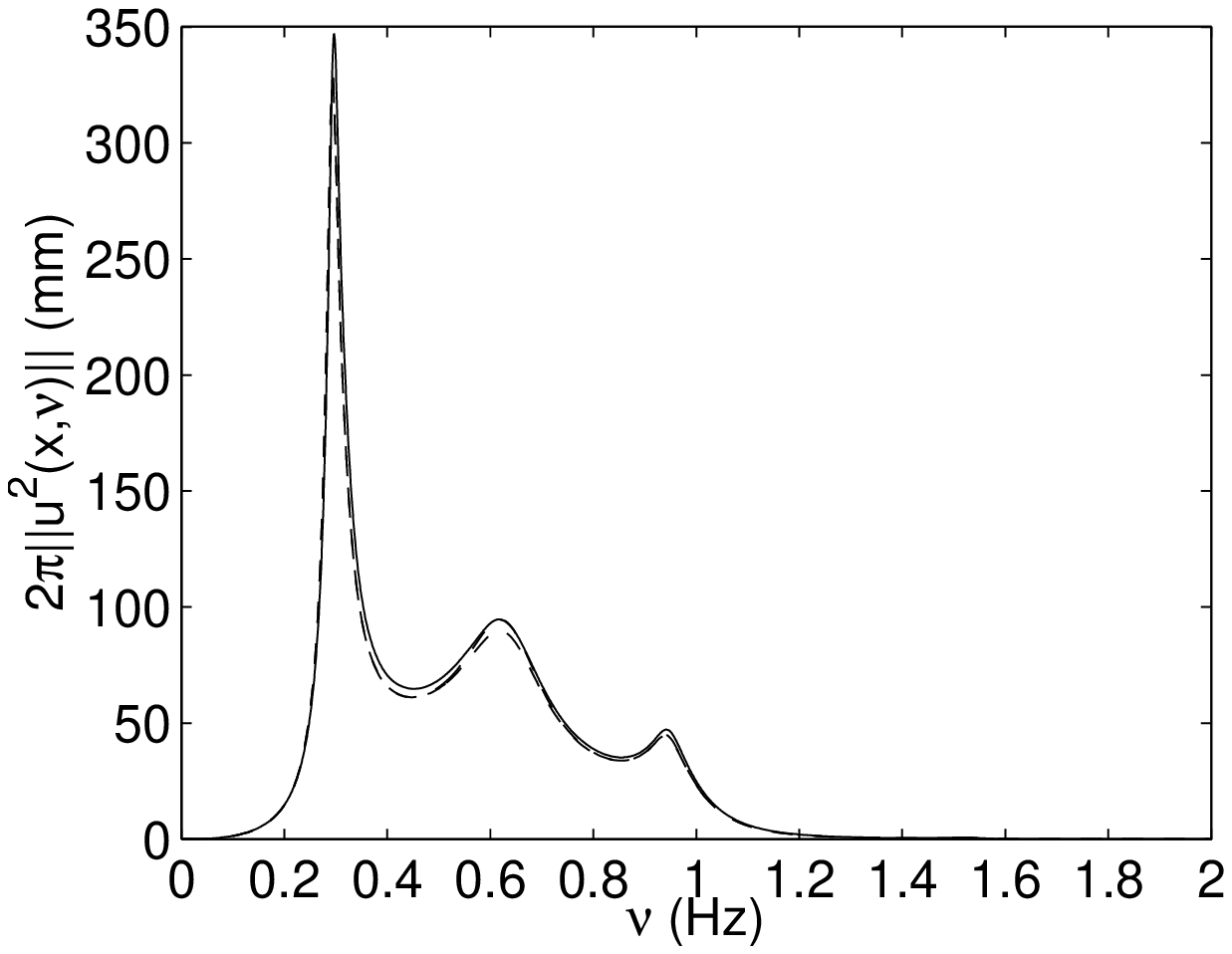}
\includegraphics[width=6.0cm] {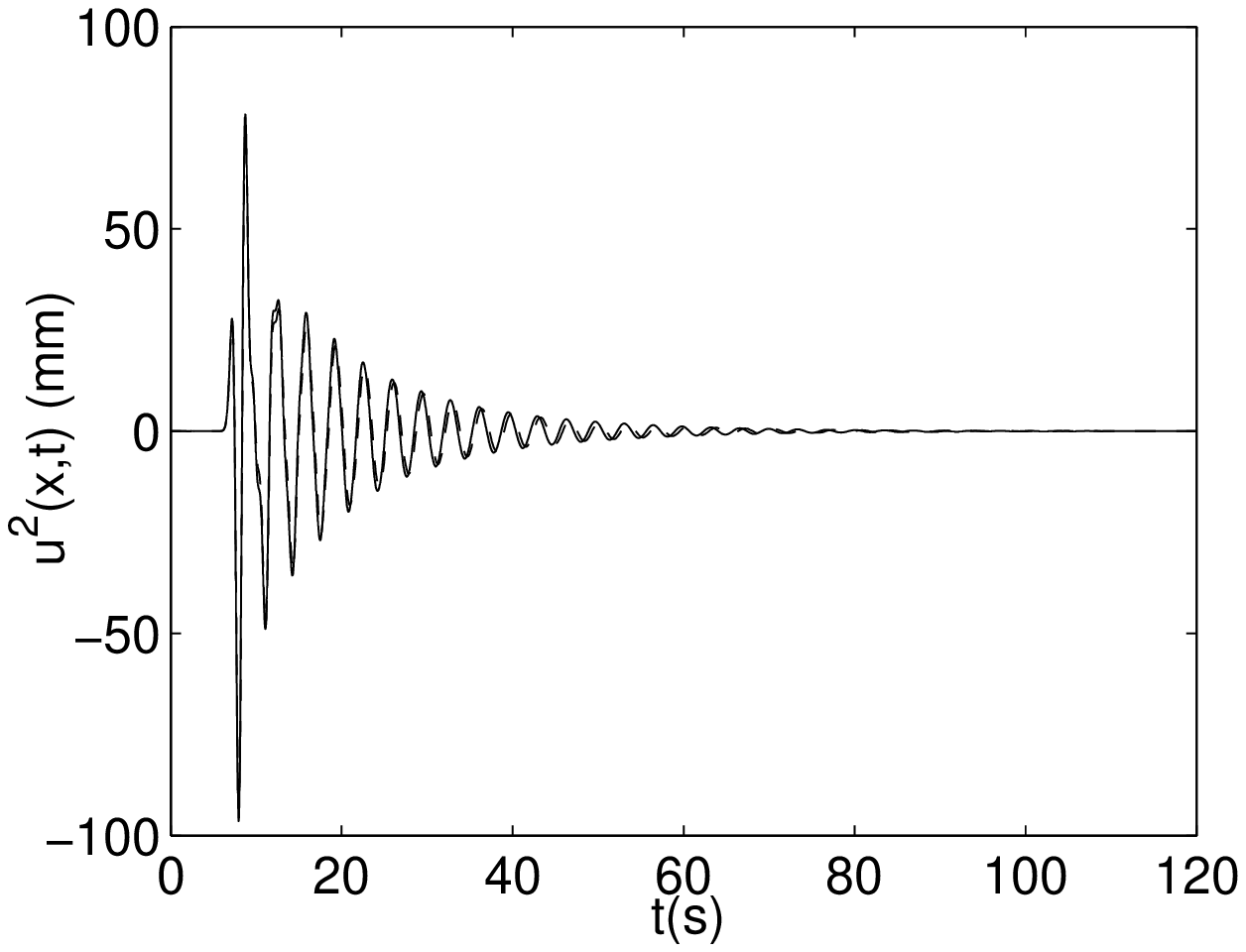}
\end{center}
\caption{$2\pi$ times the spectrum (left panel) and time history
(right panel) of  the total displacement response to the
cylindrical wave (radiated by a deep source located at
$\mathbf{x^{s}}=(-65m,3000m)$) at the center of the summit segment
of a single $40m\times 40m$ block. The dashed curves correspond to
the semi-analytical (mode-matching, one mode) result  and the
solid curves to the numerical (finite-element) result.}
\label{specttimesum4040}
\end{figure}

Figure \ref{specttimesum4040}, depicts  the spectrum and time
history of the {\it total displacement at the center of the summit
segment of the block} as computed by the mode-matching method
(with account taken of one quasi-mode) and the finite-element
method for a deep source. No noticeable differences are found
between the results of the two methods of computation. The neglect
of the quasi-modes of order larger than $0$ is valid for this
block width. The block acts as a ribbon source of width $w$
located at the base segment.
\begin{figure}[ptb]
\begin{center}
\includegraphics[width=6.0cm] {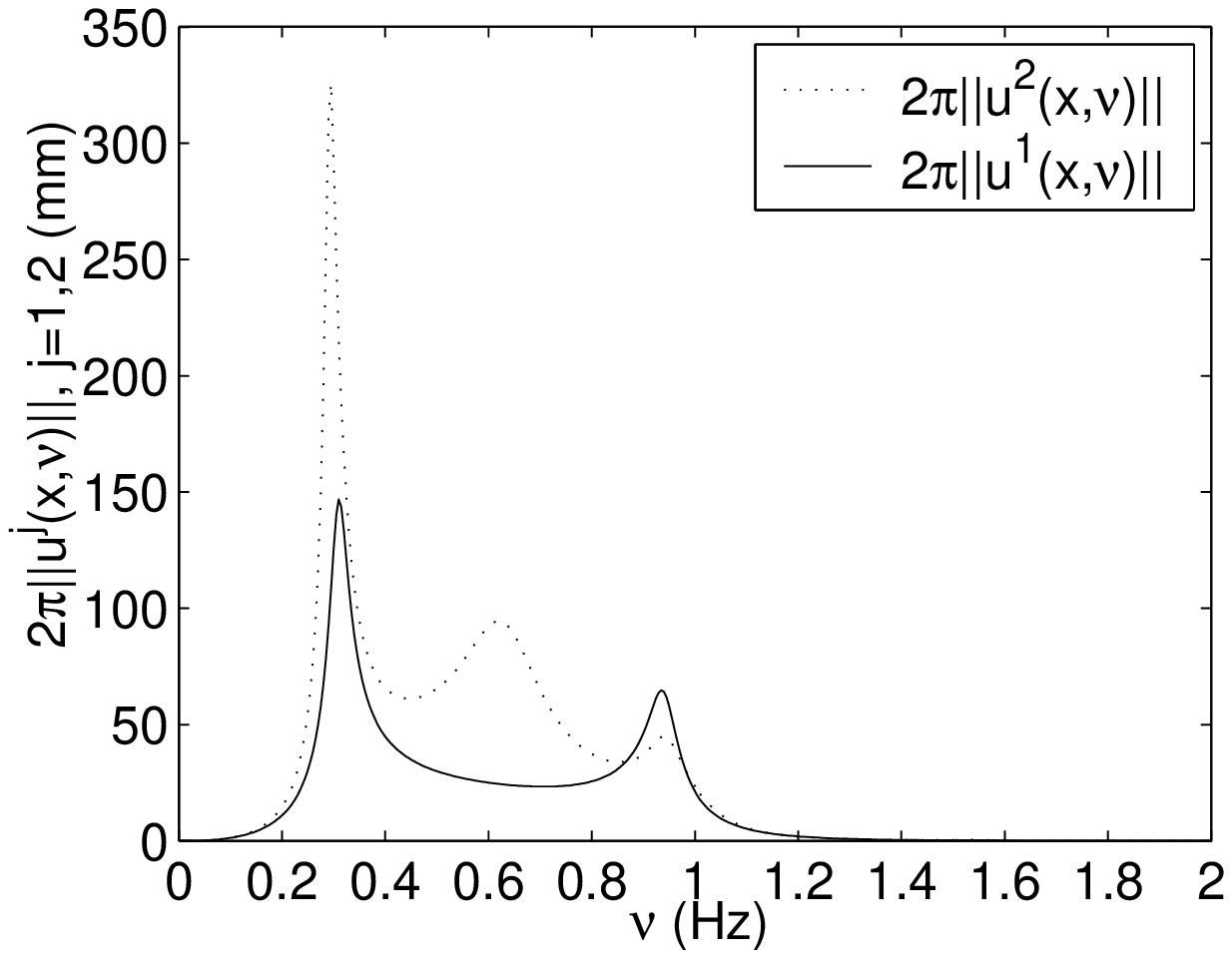}
\includegraphics[width=6.0cm] {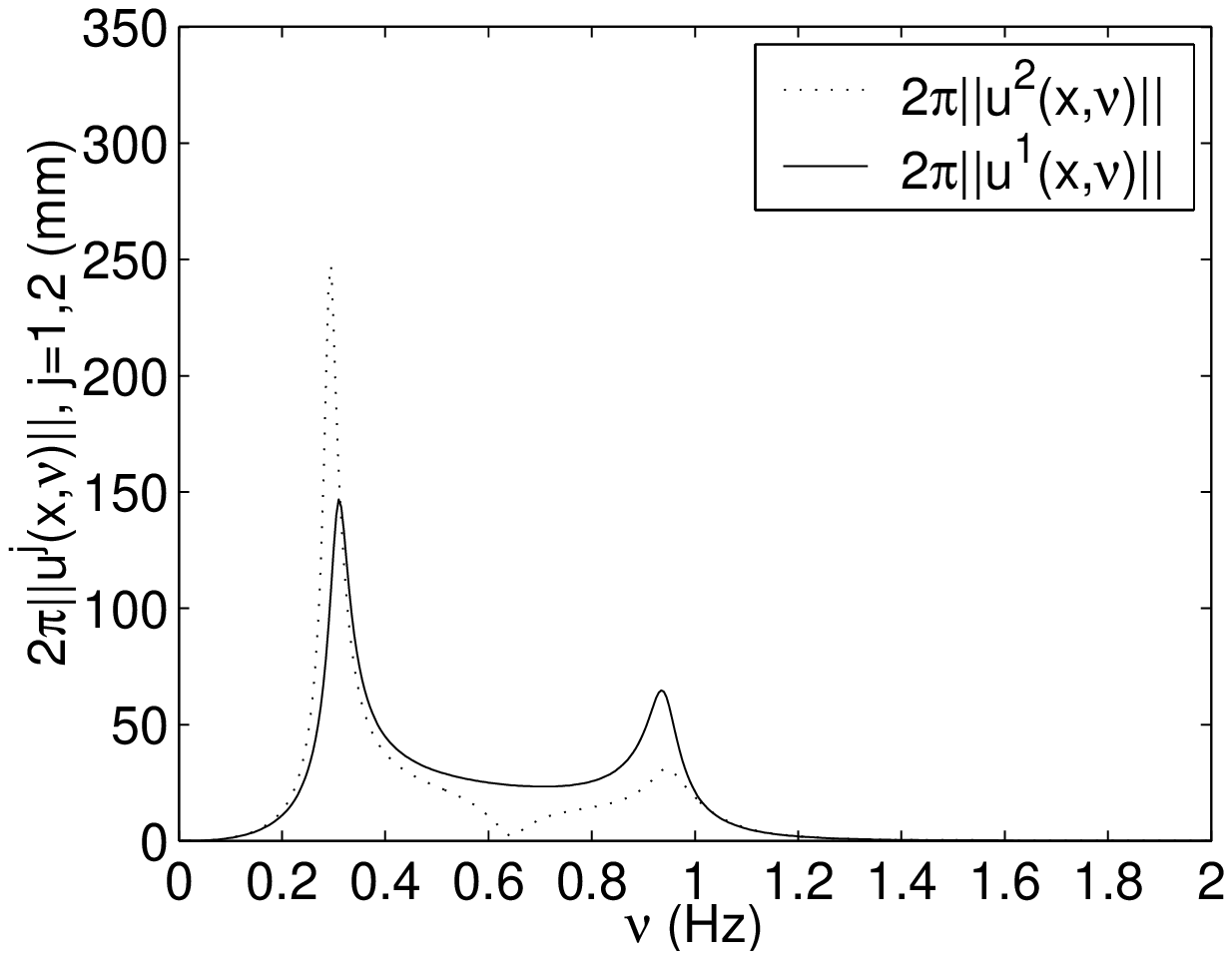}
\includegraphics[width=6.0cm] {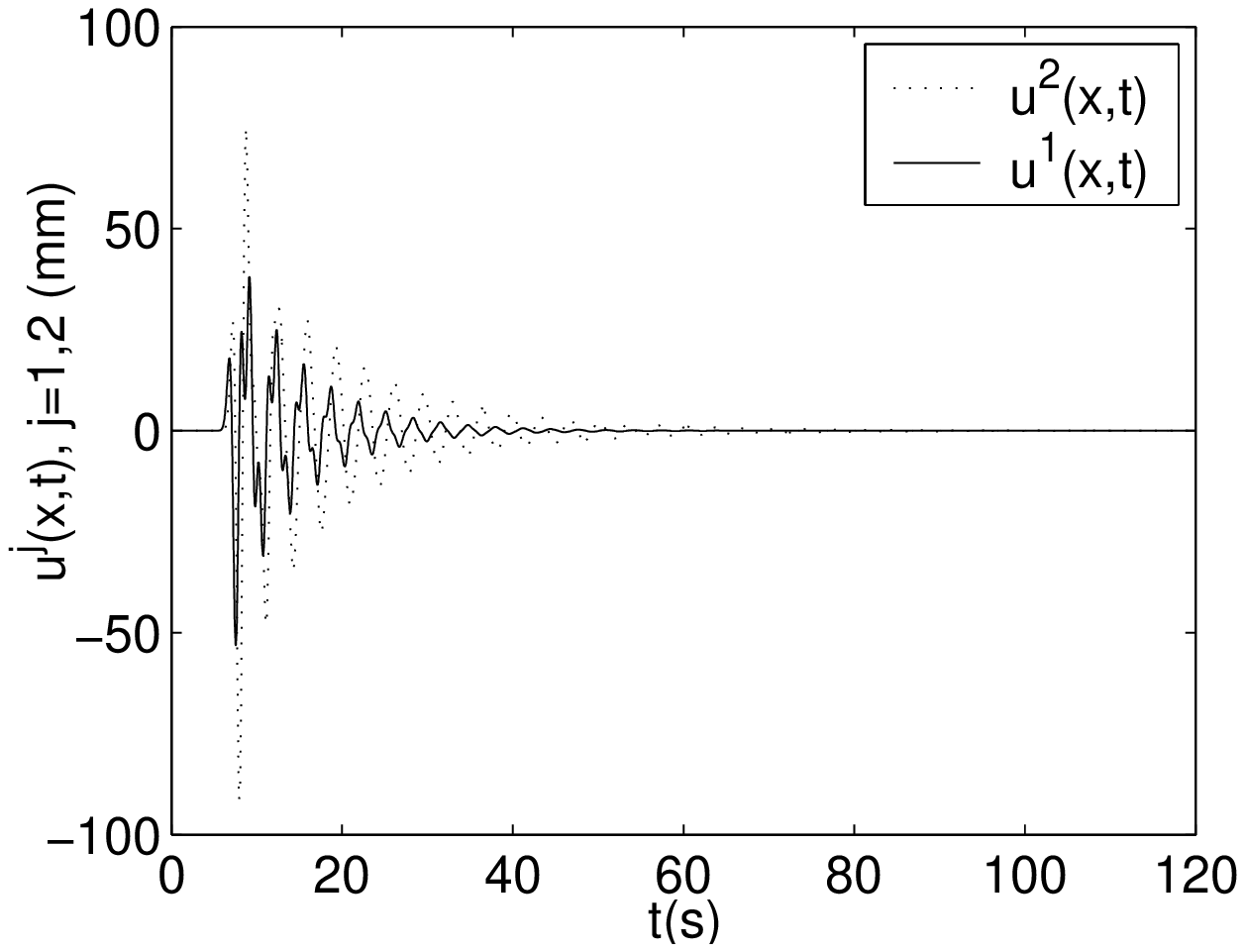}
\includegraphics[width=6.0cm] {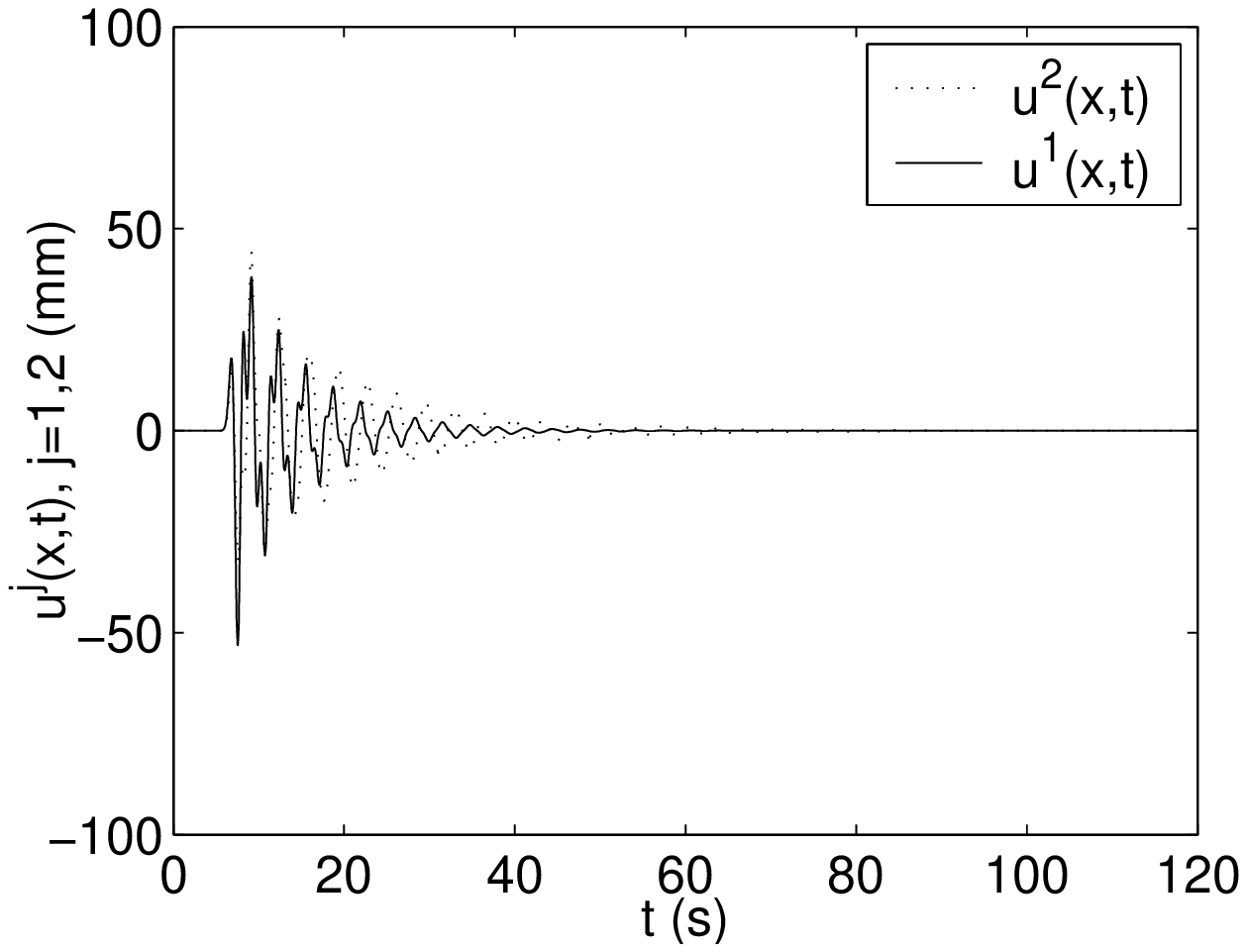}
\caption{Comparison of $2\pi$ times the spectrum (top panels) and
time history  (bottom panels) of the total displacement on the
ground in absence of block (solid curves) with the total
displacement (dotted curves) at the center of the top segment
(left panel) and at the center of the base segment (right panel)
of a single $40m\times 40m$ block, for a deep source located at
$\mathbf{x^{s}}=(-65m,3000m)$.} \label{compspecttime4040}
\end{center}
\end{figure}

Figure \ref{compspecttime4040} depicts the spectrum and  the time
history of {\it the total displacement on the  ground in the
absence of the block} as well as {\it the total displacement at
the centers of the top and base segments of the block} for a deep
source. In the time domain, a small increase of the duration, and
 a fairly-large increase of the peak amplitude, can be noticed, particularly
 on the top segment of the block, where  the quasi
 displacement-free block mode makes
itself felt. In the frequency domain, there is an increase of the
maximum amplitude and the sharpness of the first resonance peak of
the substructure at around $0.3Hz$, as noticed previously, to some
extent, in \cite{lombert,clouteau}, and is a characteristic
feature of the so-called {\textit{soil-structure interaction}}. A
similar phenomenon was mentioned in \ref{quasilove}, and is due to
the ability of shallow sources to excite Love modes in the
configuration without blocks \cite{grobyetwirgin2005}. Here, a
quasi-Love mode can be excited due to the presence of the  block,
which acts like a shallow (actually located on the ground) source.
A similar fact was already noticed in \cite{grobyetwirgin2005II},
where the spectra of the displacement on the ground for a deep
source (provoking only interference phenomena) was compared to
that for a shallow source (provoking Love mode excitation).

The so-called {\textit{soil-structure interaction}} consists (in
the configuration with blocks) in a modification of the phenomenon
from a state where interference  (when a soft layer is present) or
reflection (when the soft layer is absent) phenomena  are dominant
to a state where a quasi-Love mode is excited. This mode is
excited because of the presence of the block in a configuration.
In the absence of the latter, no mode can be excited.

Another point of view is that the block takes the form of an
induced source at the top of the layer (see section
\ref{intub1ub0}), and it is now known that Love modes are excited
when such a configuration is solicited by a line source near the
boundaries of the layer
 \cite{grobyandwirgin2003}.

A particular feature of the spectrum of the displacement when the
block is present, (top-right subfigure of fig.
\ref{compspecttime4040}) is that it vanishes  at the base segment
for  $\displaystyle \nu_{00}^{DFB}\approx \frac{c_{2}}{2b}$, which
is the fundamental displacement-free base block eigenfrequency.
This is made evident in the field representation in the block and
can be understood to be a geometrical effect which, in the case of
a single block, corresponds to the excitation of a quasi-mode.
\subsection{Results relative to one $50m\,\times\, 30m$
block}\label{onebloc5030}
The block is $50 m$ high and $30 m$ wide. The displacement-free
base block eigenfrequencies are then $0.5$, $1.5 Hz$, .....
\begin{figure}[ptb]
\begin{center}
\includegraphics[width=6.0cm] {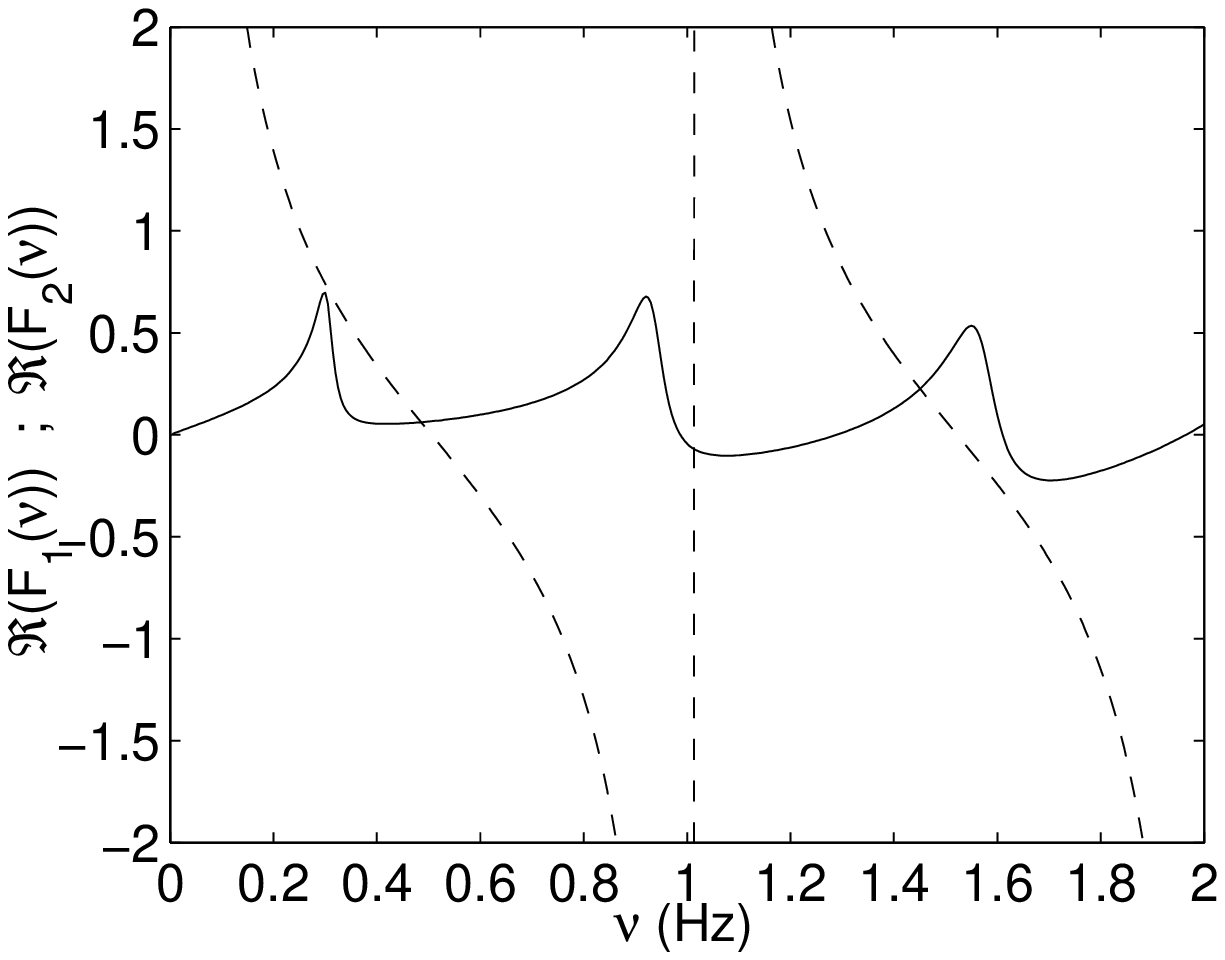}
\includegraphics[width=6.0cm] {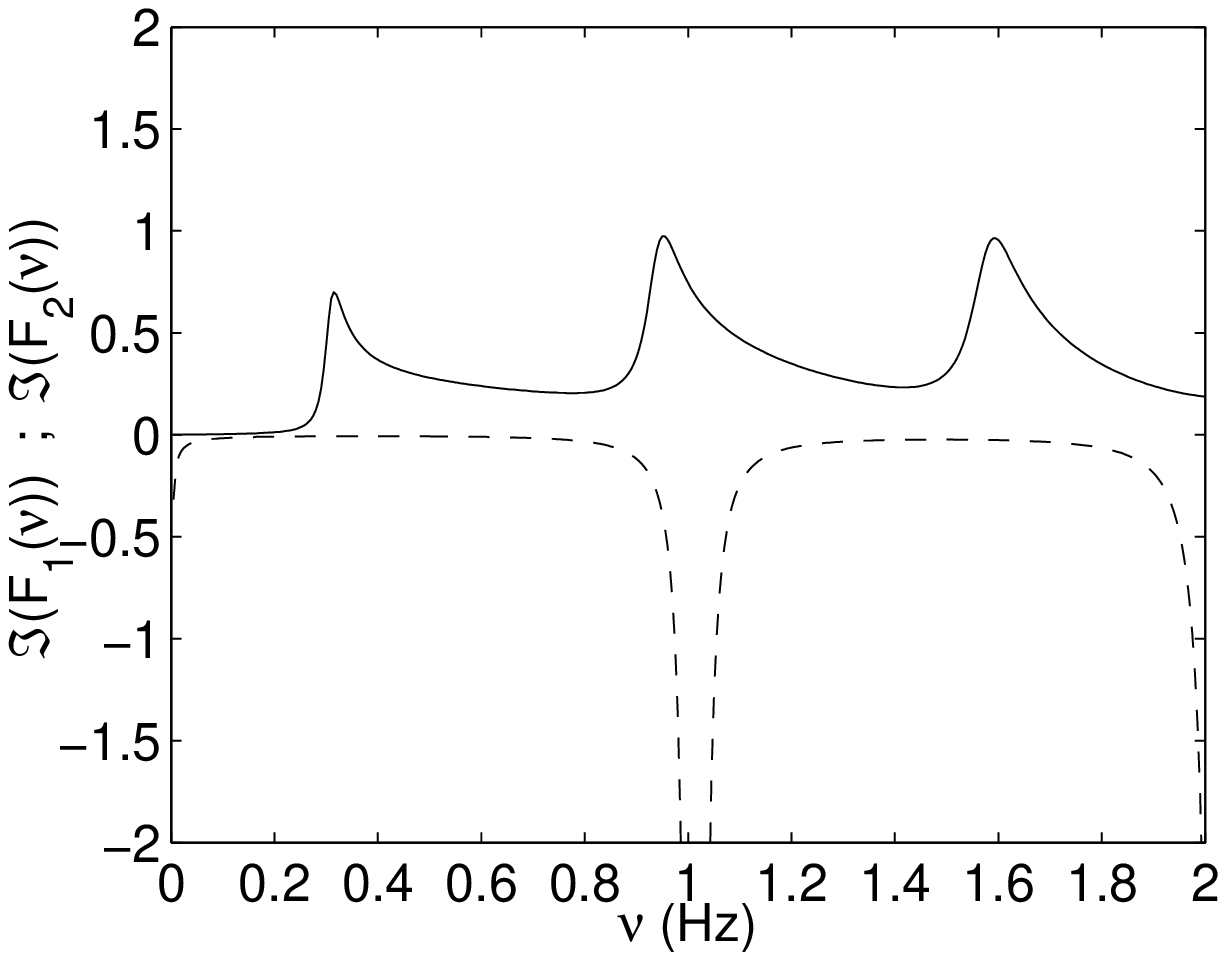}
\end{center}
\caption{Indications concerning the solution of the dispersion
relation $\mathcal{F}=\mathcal{F}_1-\mathcal{F}_2=0$ for a single
$50m\,\times\, 30m$ block in a Mexico City-like site. In the left
panel, the solid curve describes $\Re\left(\mathcal{F}_{1}\right)$
versus frequency ($\nu$ in Hz) and the dashed curve describes
$\Re\left(\mathcal{F}_{2}\right)$ versus frequency. The right
panel again contains two curves: the solid one describes
$\Im\left(\mathcal{F}_{1}\right)$ versus frequency and the dashed
curve describes $\Im\left(\mathcal{F}_{2}\right)$ versus frequency
.} \label{reim5030}
\end{figure}
Figure \ref{reim5030} gives an indication of the modes  of the
configuration. One notes that the eigenfrequencies (frequencies at
which $\Re(\mathcal{F}_{1})=\Re(\mathcal{F}_{2})$, at the least)
are $\nu\approx 0.3\mbox{, }0.5\mbox{, }1.0\mbox{, }1.53\mbox{,
}1.75\mbox{, }1.93$. The attenuations of the quasi-Love mode at
$\nu\approx 0.3$ and $0.9$ Hz, and of the quasi displacement-free
base block at $\nu\approx 0.5$ and $1.5Hz$ are relatively-small.
As previously, the attenuation of the quasi stress-free base block
(close to the zeros of $\tan\left(k^{2}b\right)$) at
$\nu_{0,1}^{QSFB}\approx 1Hz$ is relatively-large.

\begin{figure}[ptb]
\begin{center}
\includegraphics[width=6.3cm] {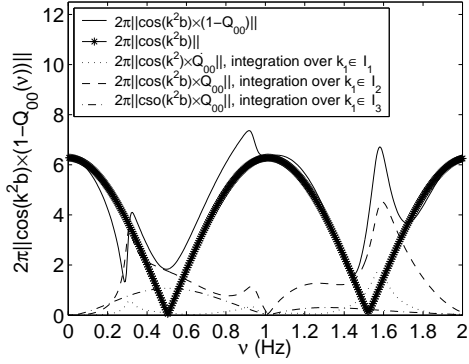}
\end{center}
\caption{$2\pi\|E_{00}\|$.}\label{E00G05030}
\end{figure}

\begin{figure}[ptb]
\begin{center}
\includegraphics[width=6.0cm] {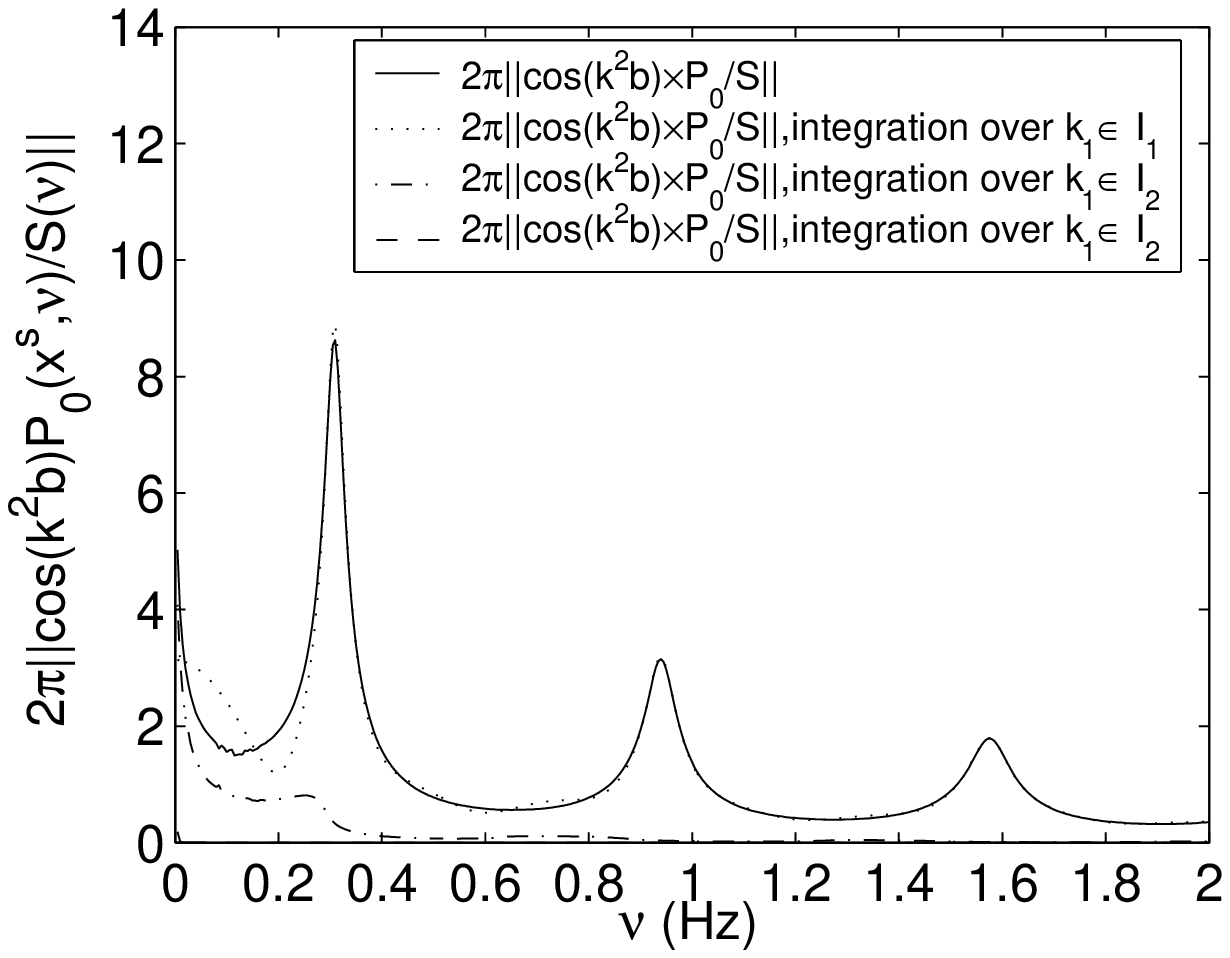}
\includegraphics[width=6.0cm] {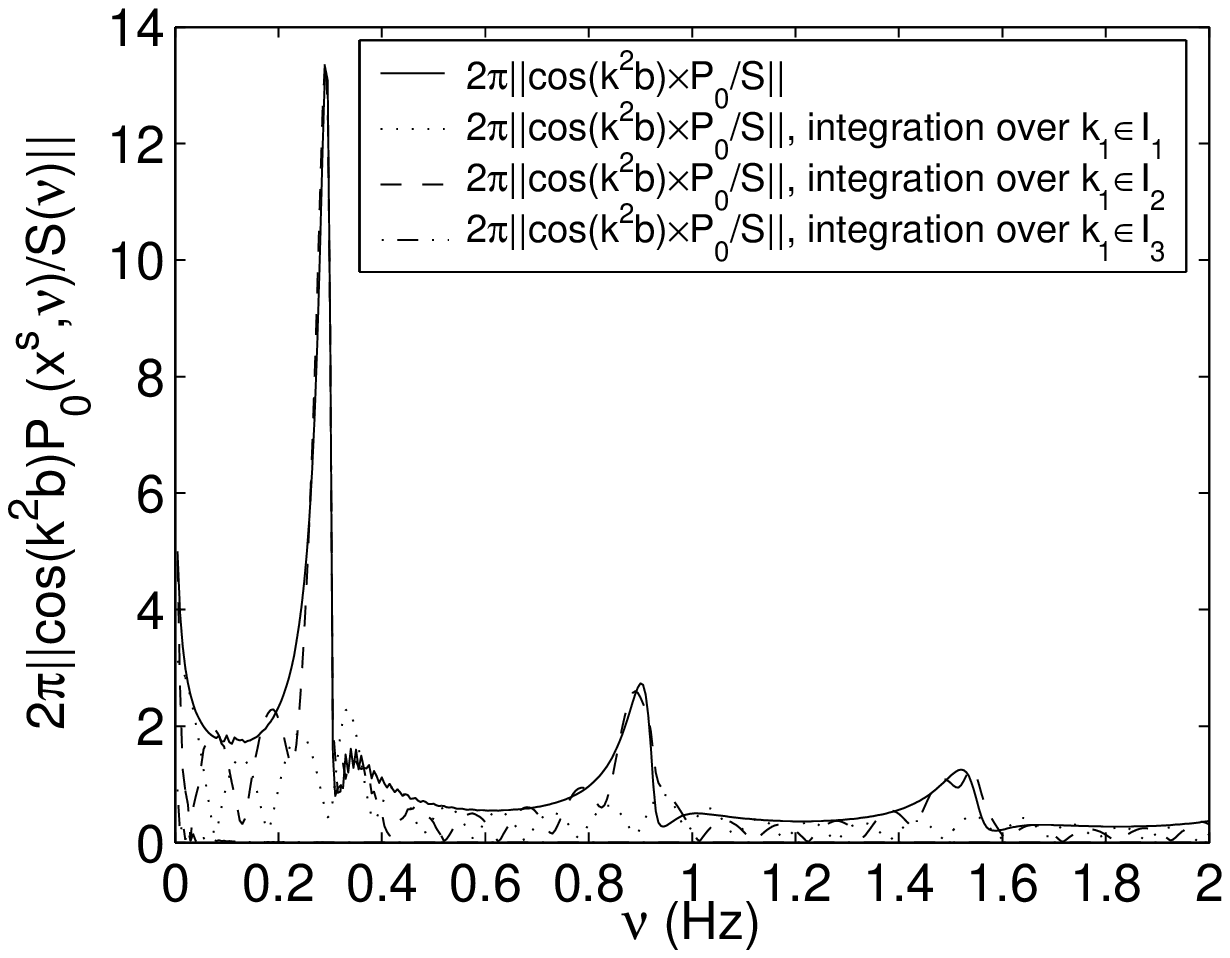}
\end{center}
\caption{Driving agent $2\pi\|\cos(k^{2}b)P_{0}/S(\omega)\|$ for a
deep source located at $\mathbf{x^{s}}=(0m,3000m)$ (left panel)
and for a shallow source located at $\mathbf{x^{s}}=(-3000m,100m)$
(right panel).} \label{G01b5030}
\end{figure}

In figs. \ref{E00G05030} and \ref{G01b5030}, we plot the absolute
values of the zeroth-order terms involved in the resolution of the
linear system (\ref{anotherquasisfb.11}). The source terms
$2\pi\|\cos(k^{2}b)P_{0}(\omega)/S(\omega)\|$ are once again close
to the transfer functions calculated without the  block for both
locations of the source $\mathbf{x^{s}}=(0m,3000m)$ and
$\mathbf{x^{s}}=(-3000m,100m)$.

The term
$I_{0}^{(j)-}(k_{1},\omega)=\mbox{sinc}\left(k_{1}\frac{w}{2}\right)$
again has a small influence on $P_{0}(\omega)$. For a deep source,
the main component of $P_{0}(\omega)$ comes from the integration
over $\mathcal{I}_{1}$ (i.e. the solicitation of the block takes
the form of interfering propagative waves traveling in the layer
associated with bulk waves in the substratum) while for a shallow
source, the main component of $P_{0}(\omega)$ comes from the
integration over $\mathcal{I}_{2}$ (i.e. the solicitation of the
block takes the form of propagative waves traveling in the layer
associated with evanescent waves in the substratum, i.e., Love
modes). The integral over $\mathcal{I}_{2}$ again dominates
$Q_{00}(\omega)$.
\subsubsection{Displacement field on the top and bottom segments of
the block for deep line source solicitation}
We now examine the displacement field on  the horizontal
boundaries of the block.

We first consider that the seismic disturbance is delivered to the
site by a {\it deep line source} located at
$\mathbf{x^{s}}=(0,3000m)$. This means that in absence of the
block, the displacement field is mainly composed of propagative
waves in the substratum and interfering propagative waves in the
layer.
\begin{figure}[ptb]
\begin{center}
\includegraphics[width=6.0cm] {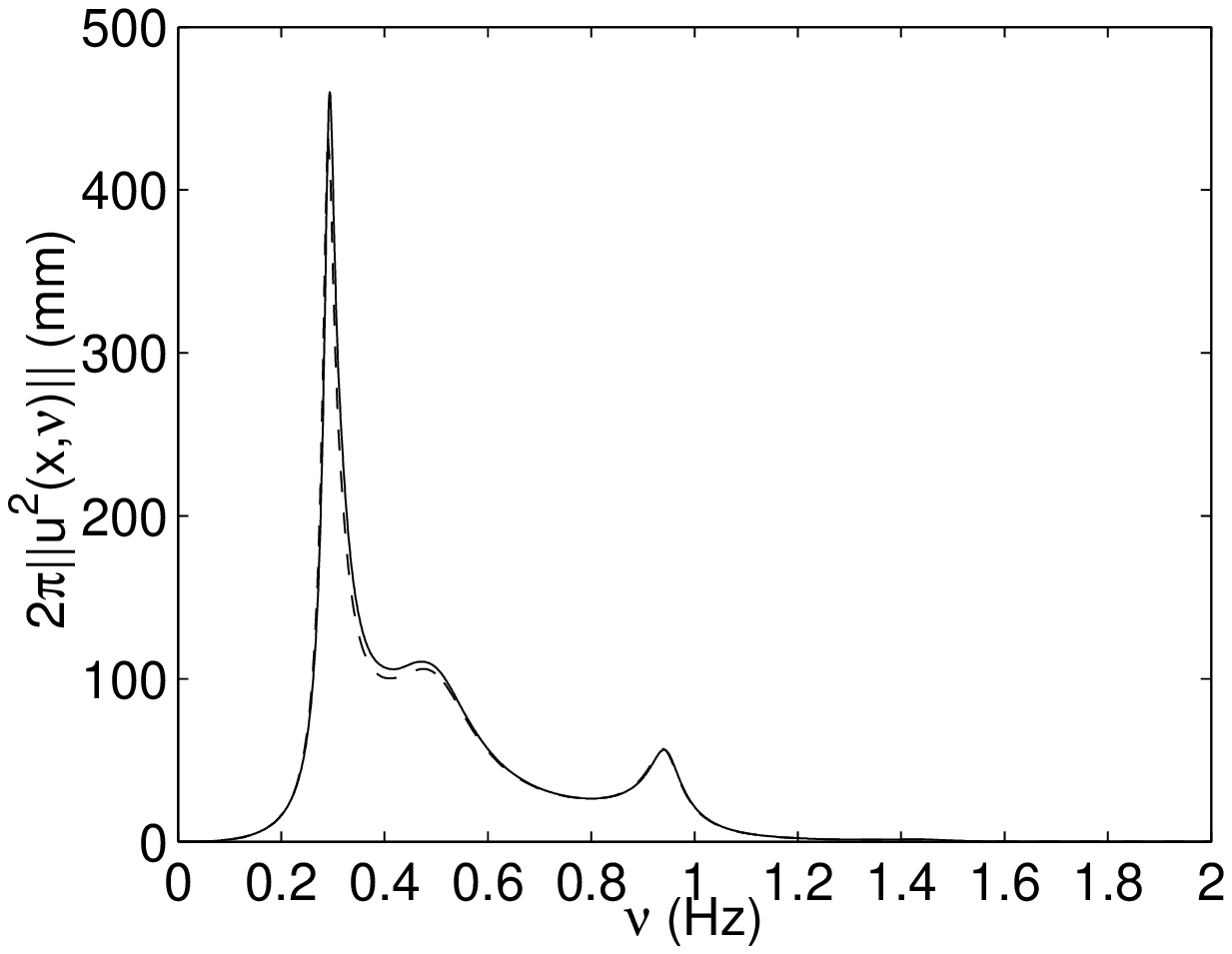}
\includegraphics[width=6.0cm] {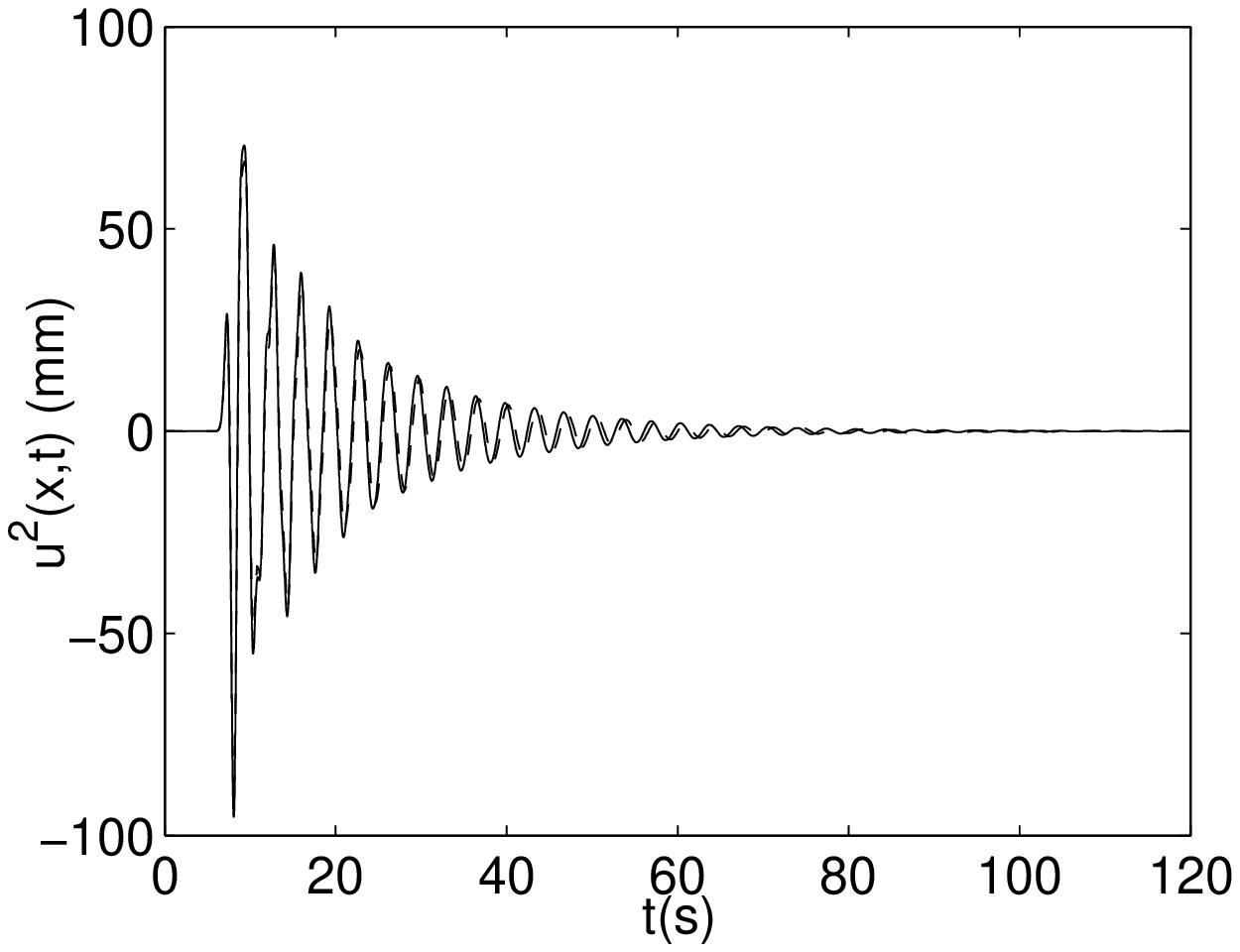}
\includegraphics[width=6.0cm] {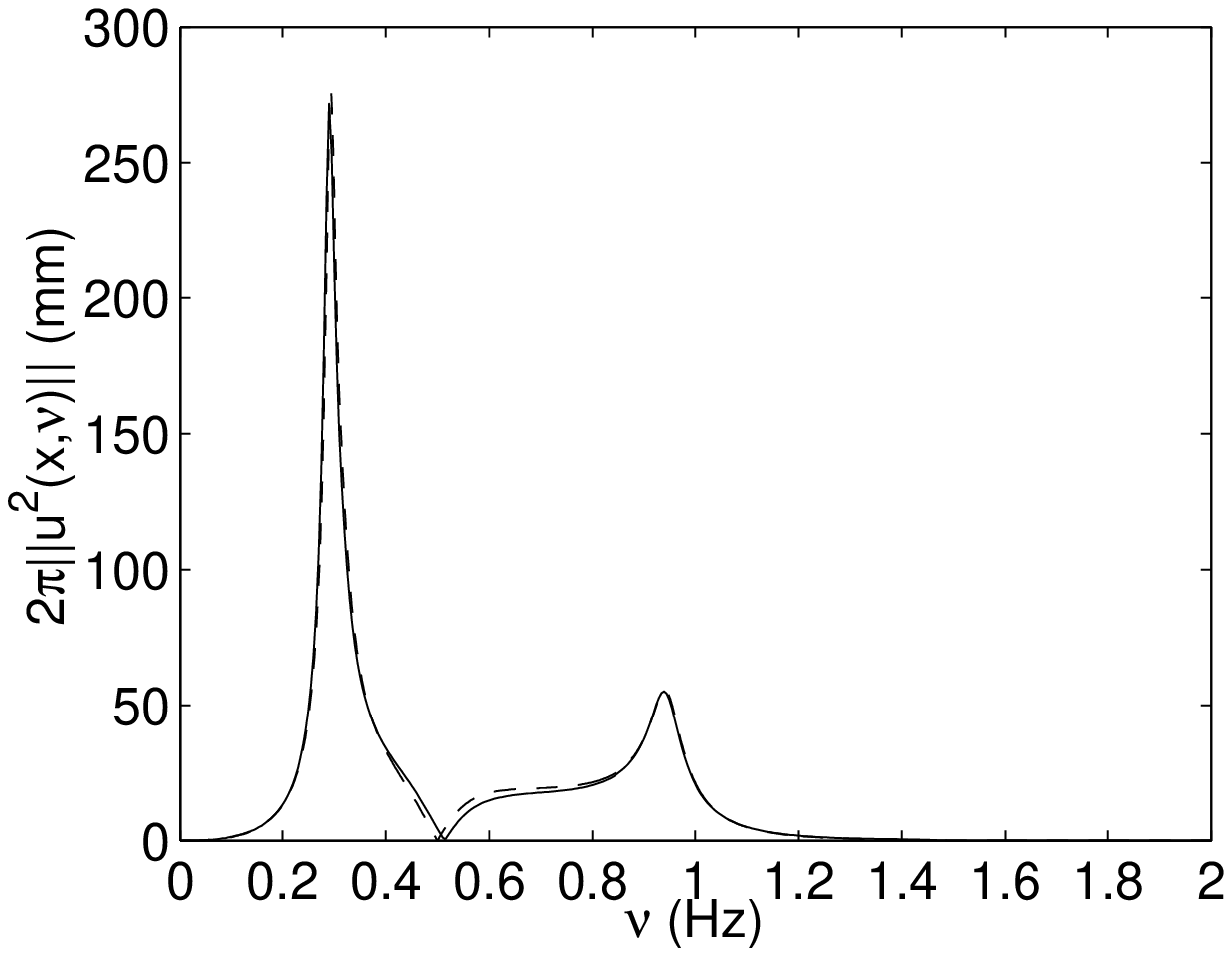}
\includegraphics[width=6.0cm] {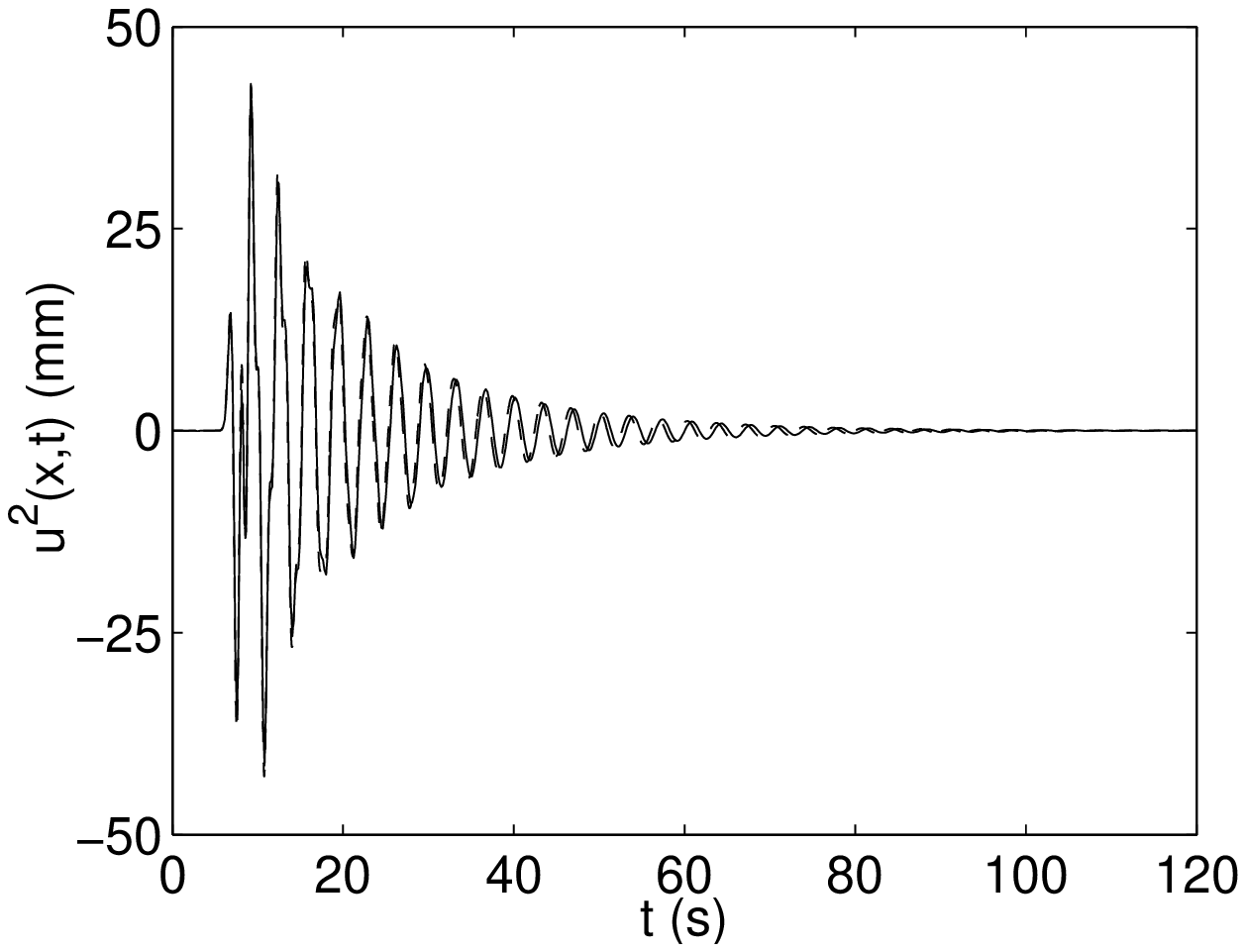}
\end{center}
\caption{$2\pi$ times the spectra (left panels) and time histories
(right panels) of  the total displacement response to a
cylindrical wave (radiated by a deep source located at
$\mathbf{x^{s}}=(0m,3000m)$) at the center of the summit segment
(top panels) and at the center of the base segment (bottom panels)
of a single $50m\times 30m$ block. The dashed curves correspond to
the semi-analytical (mode-matching, one mode) result, and the
solid curves to the numerical (finite-element) result.}
\label{specttimesum5030}
\end{figure}

Figure \ref{specttimesum5030}, depicts  the spectra and time
histories of the {\it total displacement at the center of the top
and bottom segments of the block}, as computed by the
mode-matching method (with account taken of one quasi-mode) and
the finite-element method, for a deep source. No noticeable
differences are found between the results of the two methods of
computation. The neglect of the quasi-modes of order larger than
$0$ is valid for this block width. The block acts as a ribbon
source of width $w$ located at the base segment.
\begin{figure}[ptb]
\begin{center}
\includegraphics[width=6.0cm] {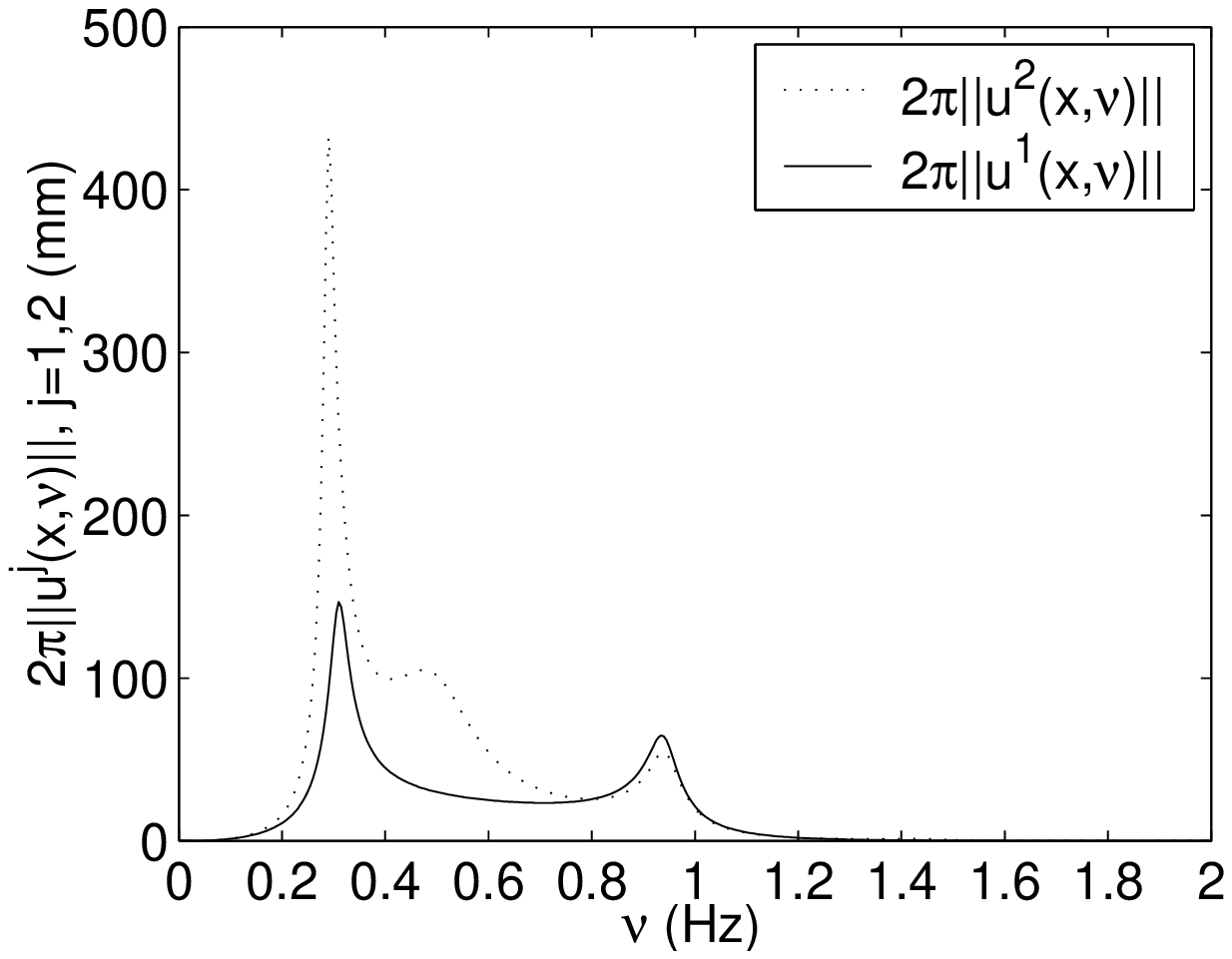}
\includegraphics[width=6.0cm] {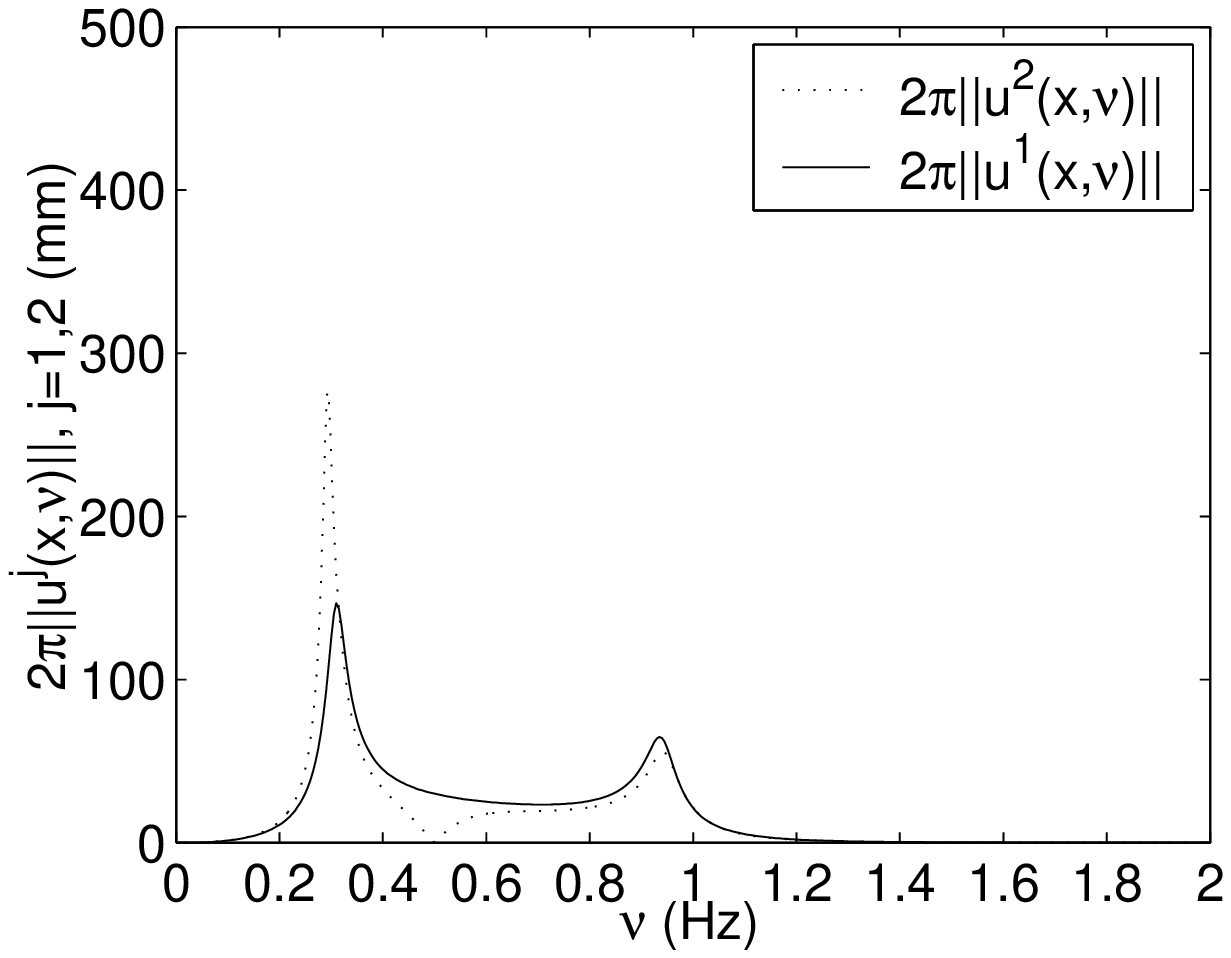}
\includegraphics[width=6.0cm] {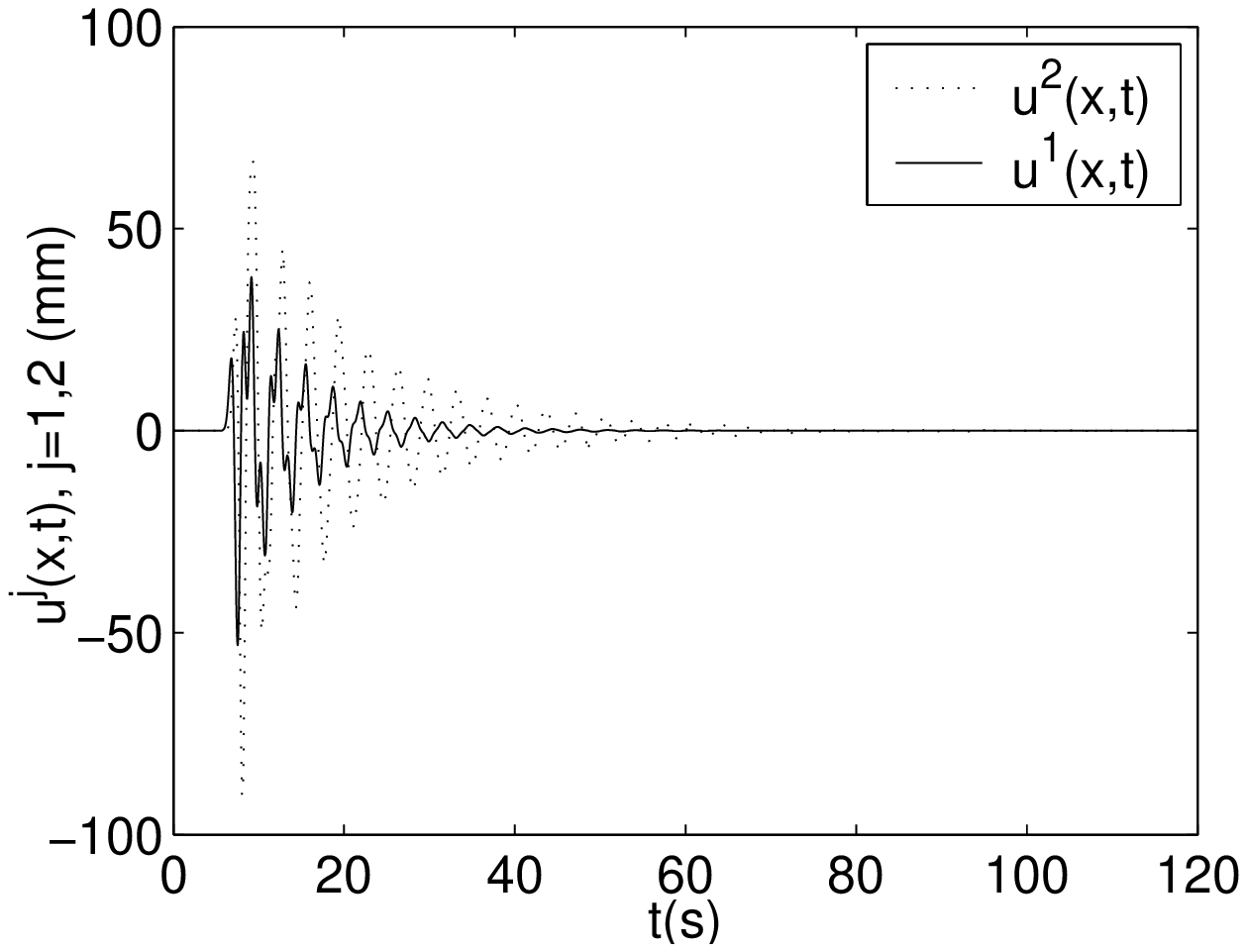}
\includegraphics[width=6.0cm] {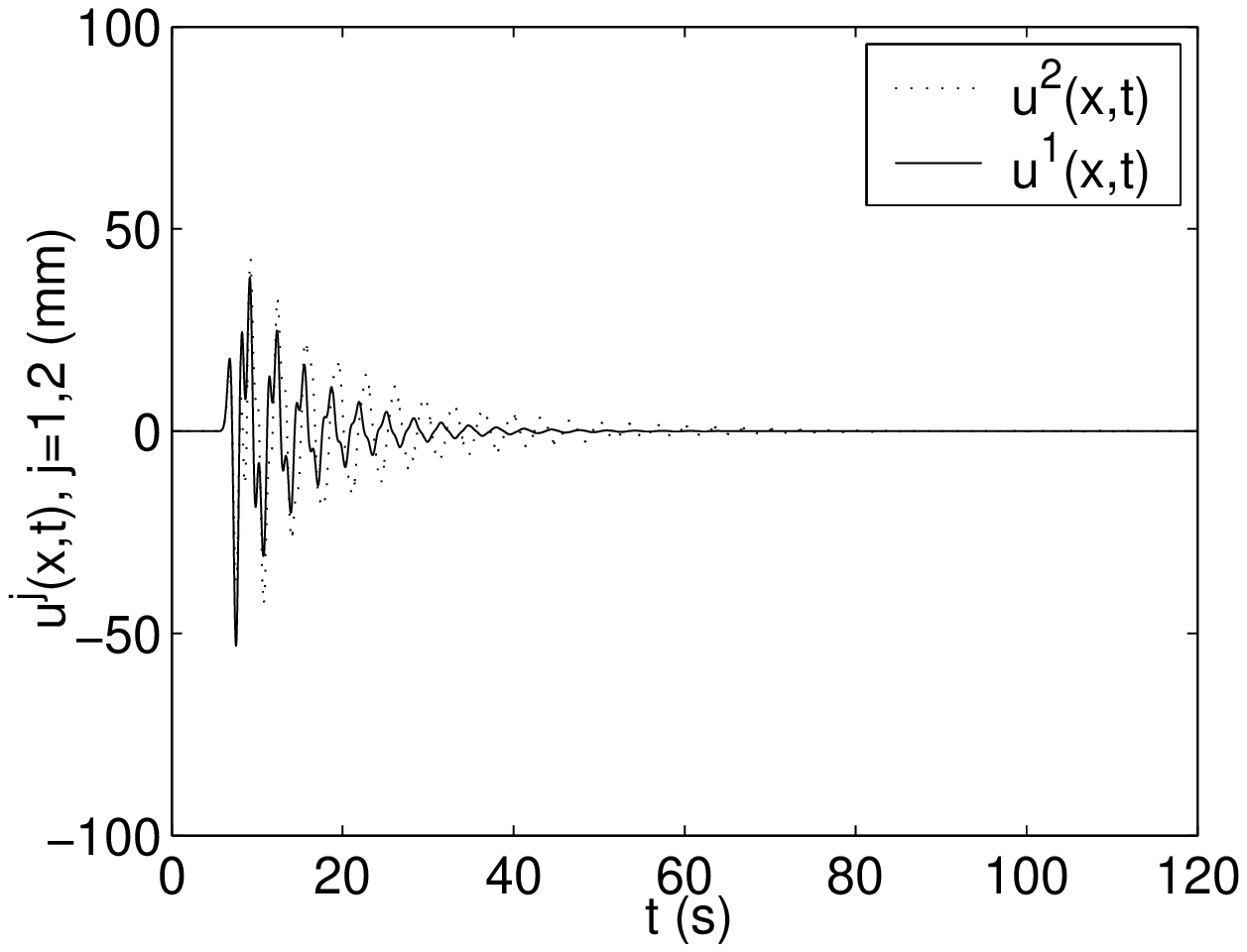}
\caption{Comparison of $2\pi$ times the spectrum (on the top) and
time history  (on the bottom) of the total displacement on the
ground in absence of the block (solid curves) with the total
displacement (dotted curves) at the center of the top segment
(left panel) and at the center of the base segment (right panel)
of a single $50m\times 30m$ block, for a deep source located at
$\mathbf{x^{s}}=(0m,3000m)$.} \label{compspecttime5030}
\end{center}
\end{figure}

Figure \ref{compspecttime5030} depicts the spectra and  the time
histories of {\it the total displacement on the  ground in the
absence of the block} as well as {\it the total displacement at
the centers of the top and base segments of the block} for a deep
source. In the time domain, a small increase of the duration
(although larger than in the $40m\times 40m$ block case), and
 a fairly-large increase of the peak amplitude, can be noticed, particularly
 on the top segment of the block, where  the quasi
 displacement-free block mode makes itself felt.

In the frequency domain, there is an increase of the maximum
amplitude and of the sharpness of the first resonance peak of the
substructure, which is a characteristic feature of the
{\textit{soil-structure interaction}}. Again, the displacement
field vanishes at the base segment for a frequency close to
$\nu_{00}^{DF}$.
\subsubsection{Displacement field on the top and bottom segments of
the block for shallow line source solicitation}
We now consider what happens when the seismic disturbance is
delivered to the site by a {\it shallow line source} located at
$\mathbf{x^{s}}=(-3000m,100m)$. This means that, in the absence of
the block, the displacement field in the substructure is that of
Love modes at the resonance frequencies of these modes.

\begin{figure}[ptb]
\begin{center}
\includegraphics[width=6.0cm] {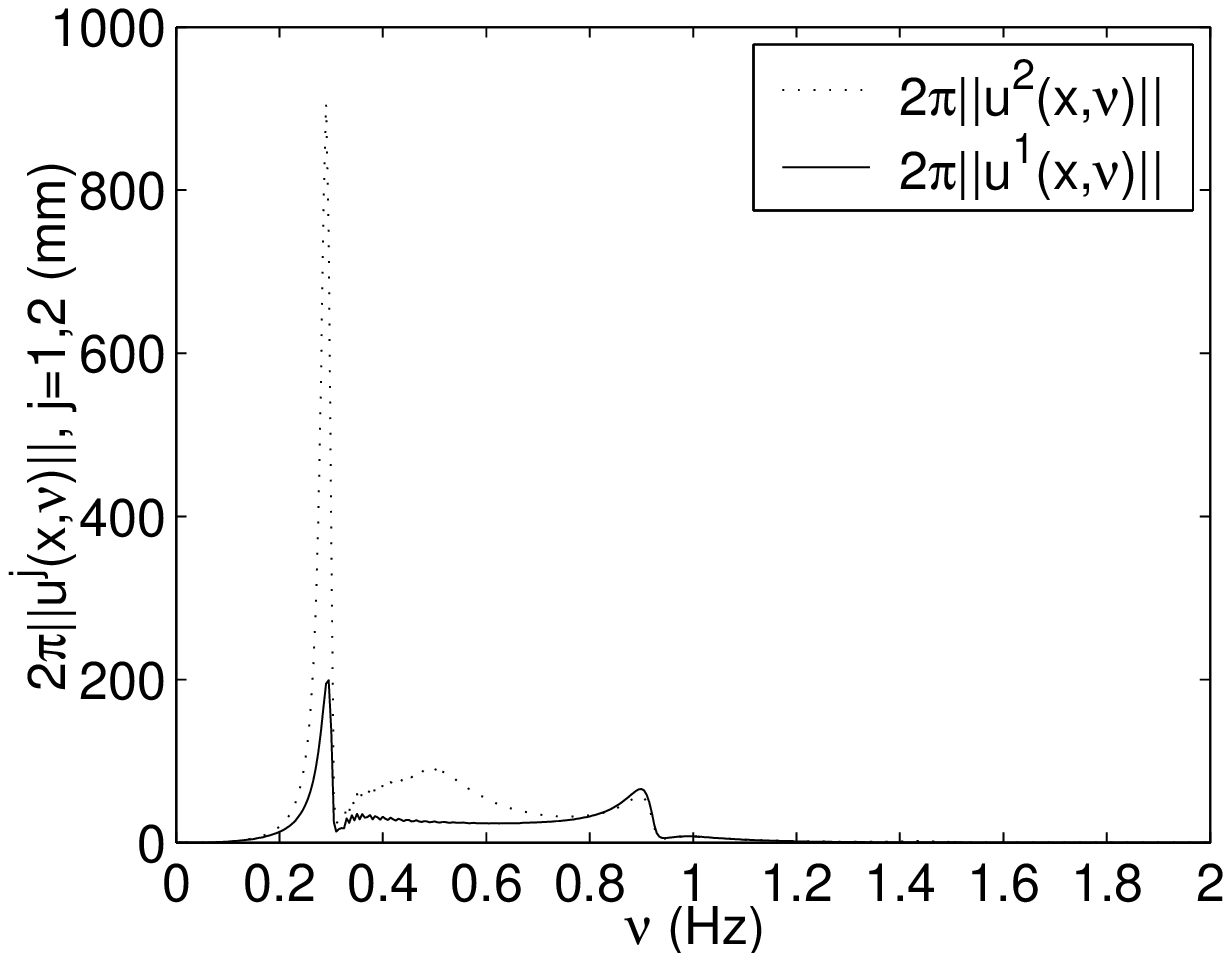}
\includegraphics[width=6.0cm] {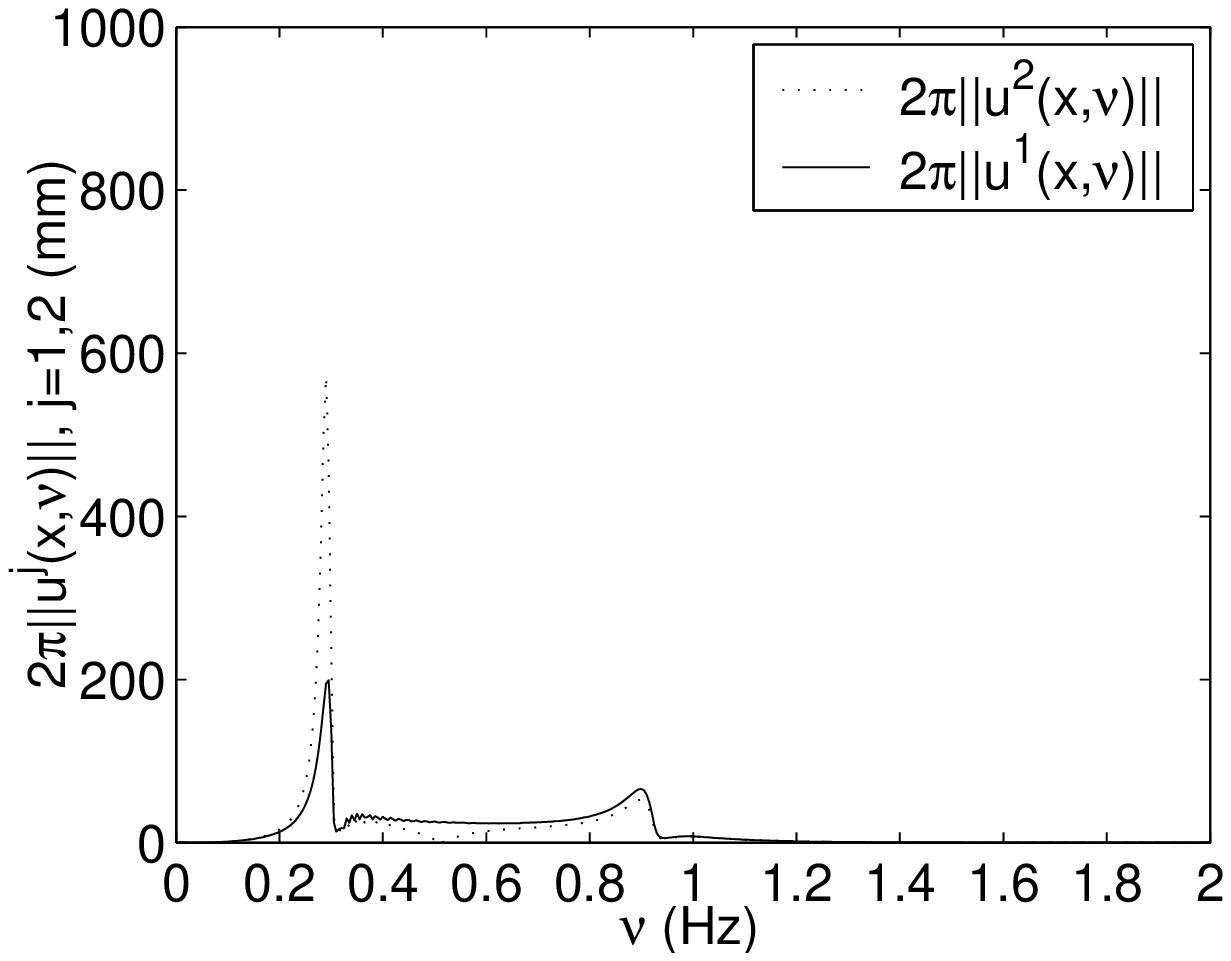}
\includegraphics[width=6.0cm] {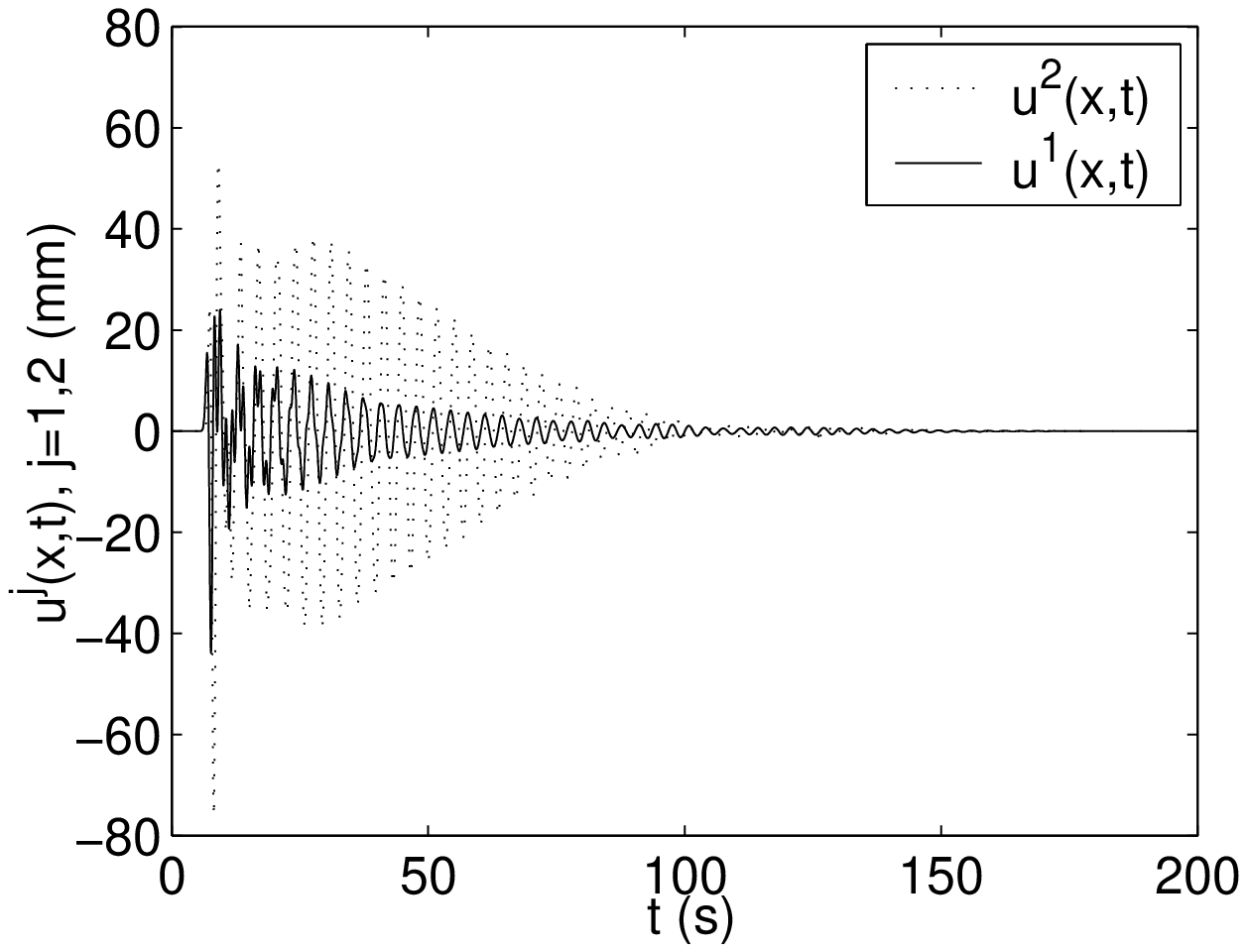}
\includegraphics[width=6.0cm] {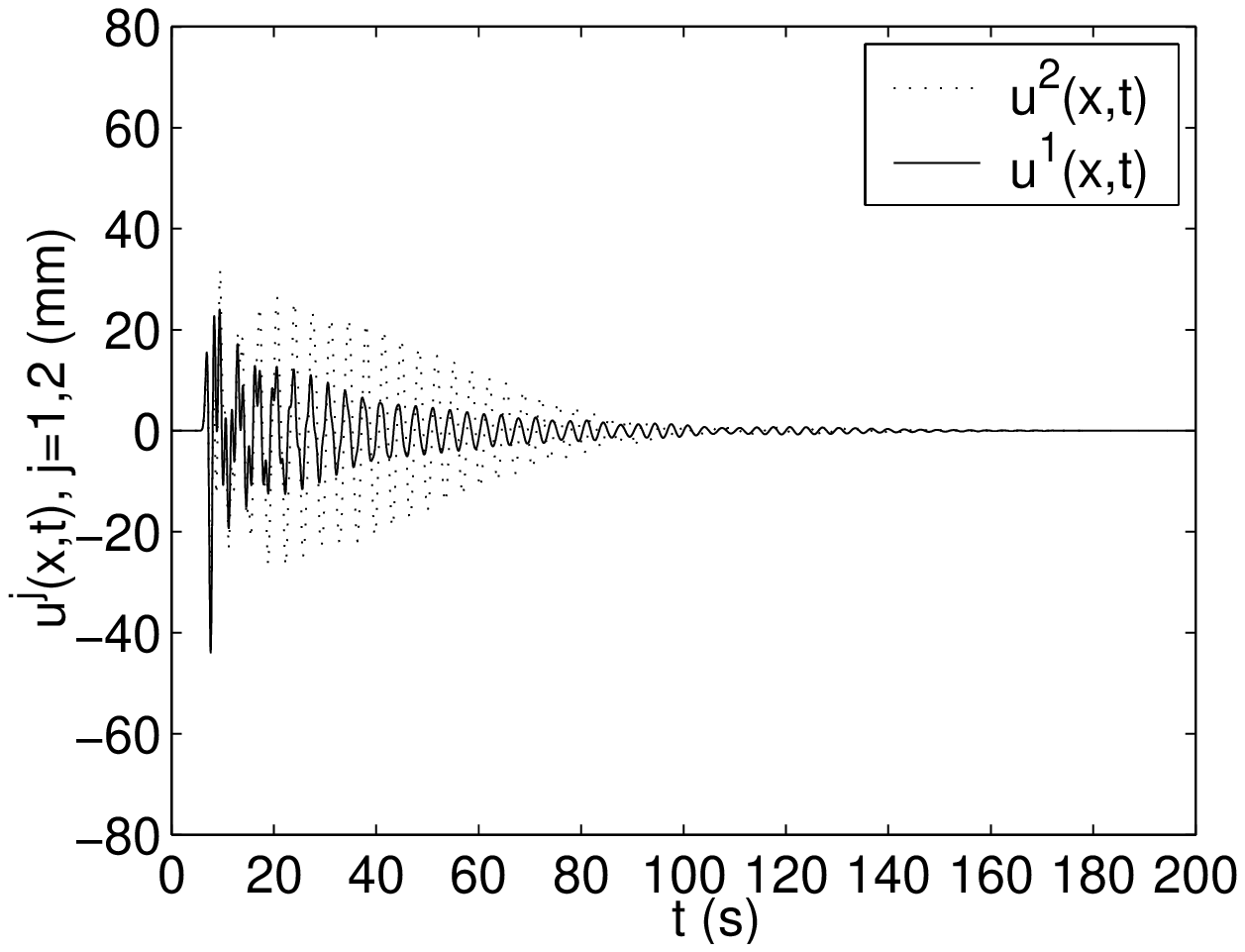}
\caption{Comparison of $2\pi$ times the spectrum (on the top), and
time  history (on the bottom) of the total displacement on the
ground in the absence of the block (solid curves) with the total
displacement, (dotted curves) at the center of the top segment
(left panel) and of the bottom segment (right panel) of a single
$50m\times 30m$ block, for a shallow line source located at
$\mathbf{x^{s}}=(-3000m,100m)$.} \label{compspecttime5030ys100}
\end{center}
\end{figure}

Figure \ref{compspecttime5030ys100} compares  the displacement on
the ground in the absence of the block to that in the block (on
the  top and bottom segments thereof) for a shallow source. In the
time domain, the durations are substantially the same when the
block is present or absent, but the peak and cumulative amplitudes
are larger in the presence of the block.  In the frequency domain,
both the sharpness and amplitude of the first peak, corresponding
to the first Love mode in absence of the blocks, increase. This
means that the  \textit{soil-structure interaction} obtained for a
deep source is also found for a shallow source. The fact that the
position of this peak is hardly shifted means that the structure
of the quasi-Love mode is very nearly that of the Love mode
existing in the absence of the block. However, the presence of the
block enables this mode to be excited more efficiently than in its
absence, i.e.,  when the site is solicited solely by the shallow
source and not by waves re-emitted by the block.
\subsubsection{Displacement on the ground on one side of the block
for deep line source solicitation}\label{onebloc5030ongroun}
To obtain  a more complete picture of the modification  of the
phenomena due to the presence of a block, we now focus our
attention on the ground motion outside of the block.

The {\it deep seismic line source} is located at
$\mathbf{x^{s}}=(0m,3000m)$.
\begin{figure}[ptb]
\begin{center}
\includegraphics[width=6.0cm] {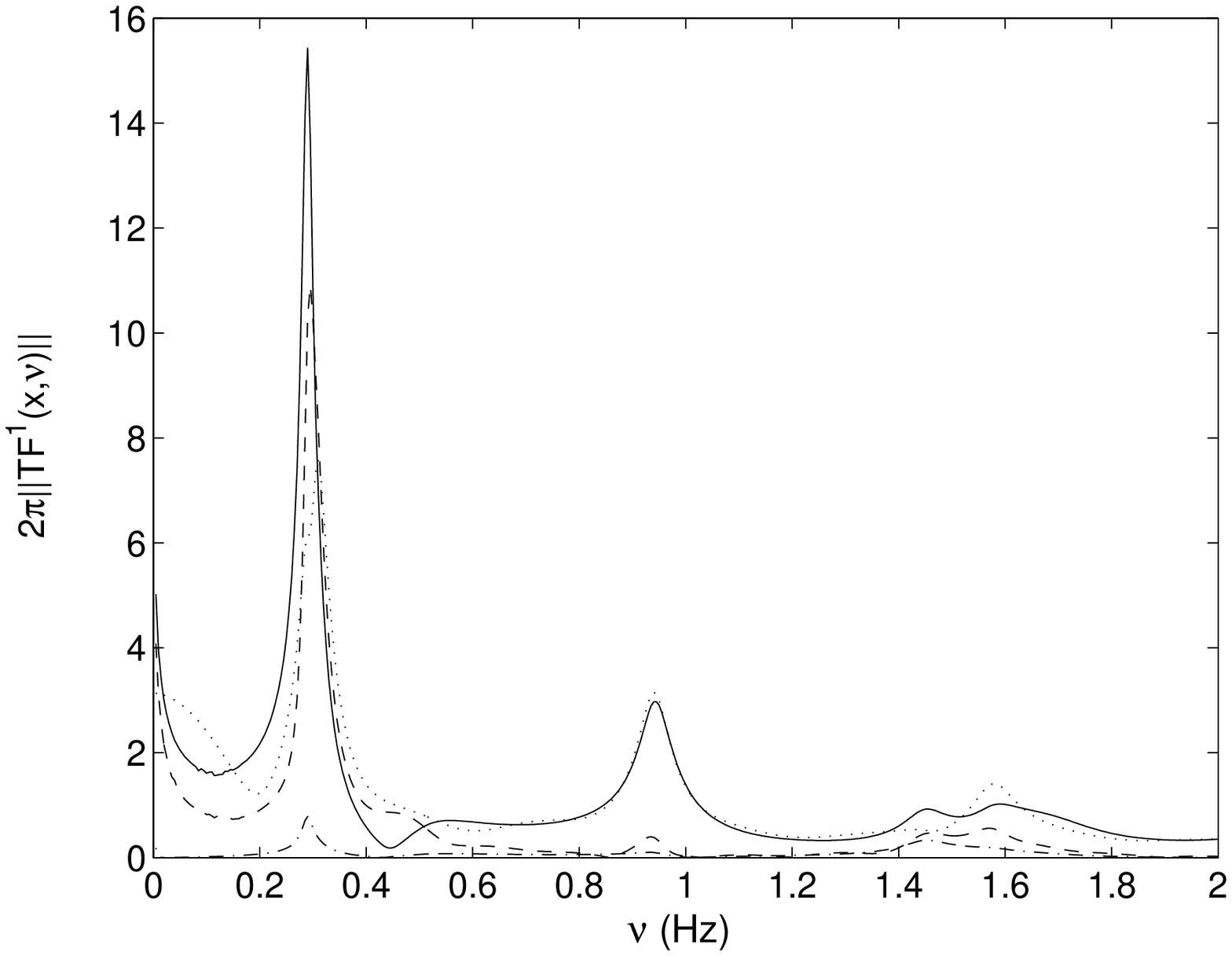}
\includegraphics[width=6.0cm] {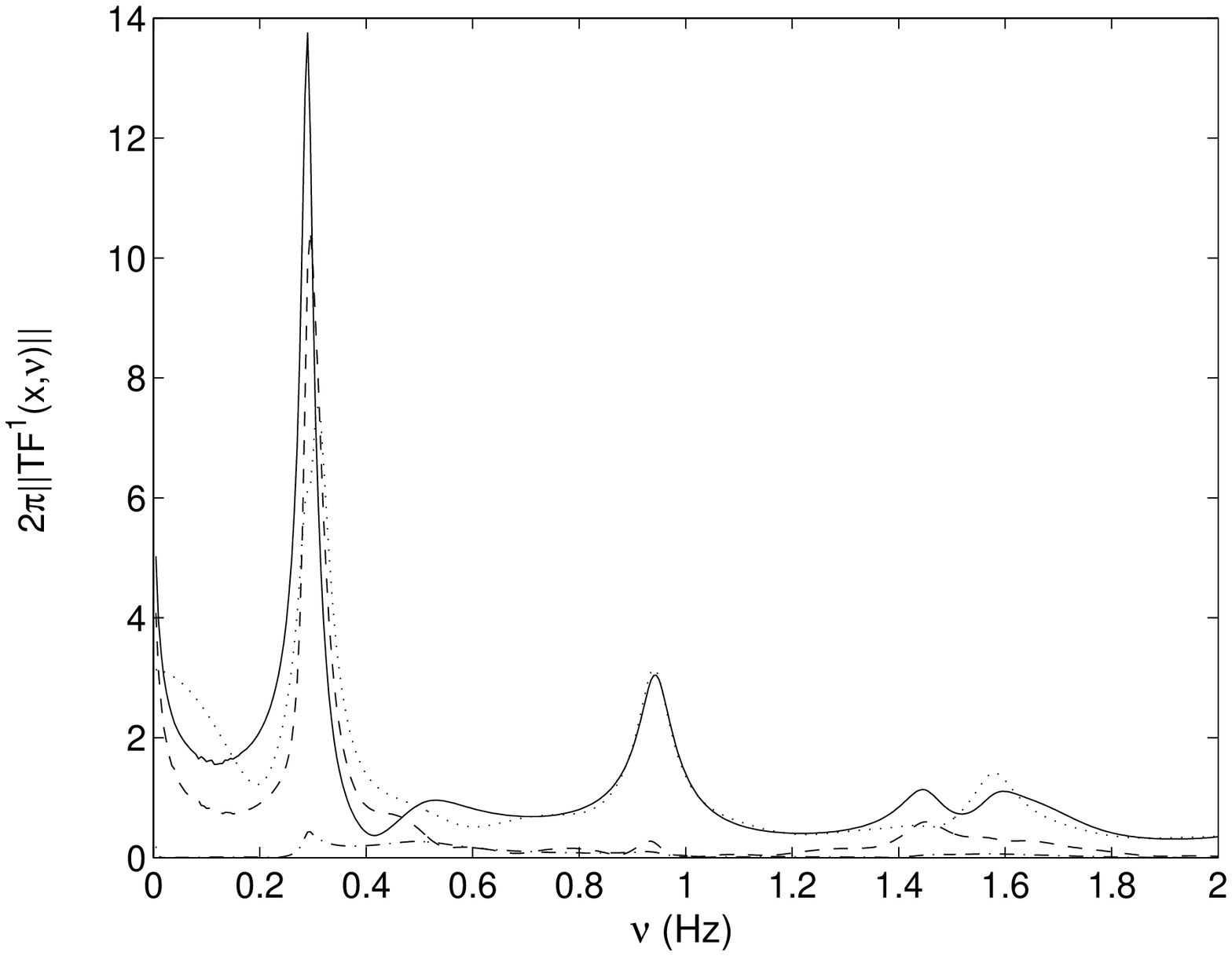}
\includegraphics[width=6.0cm] {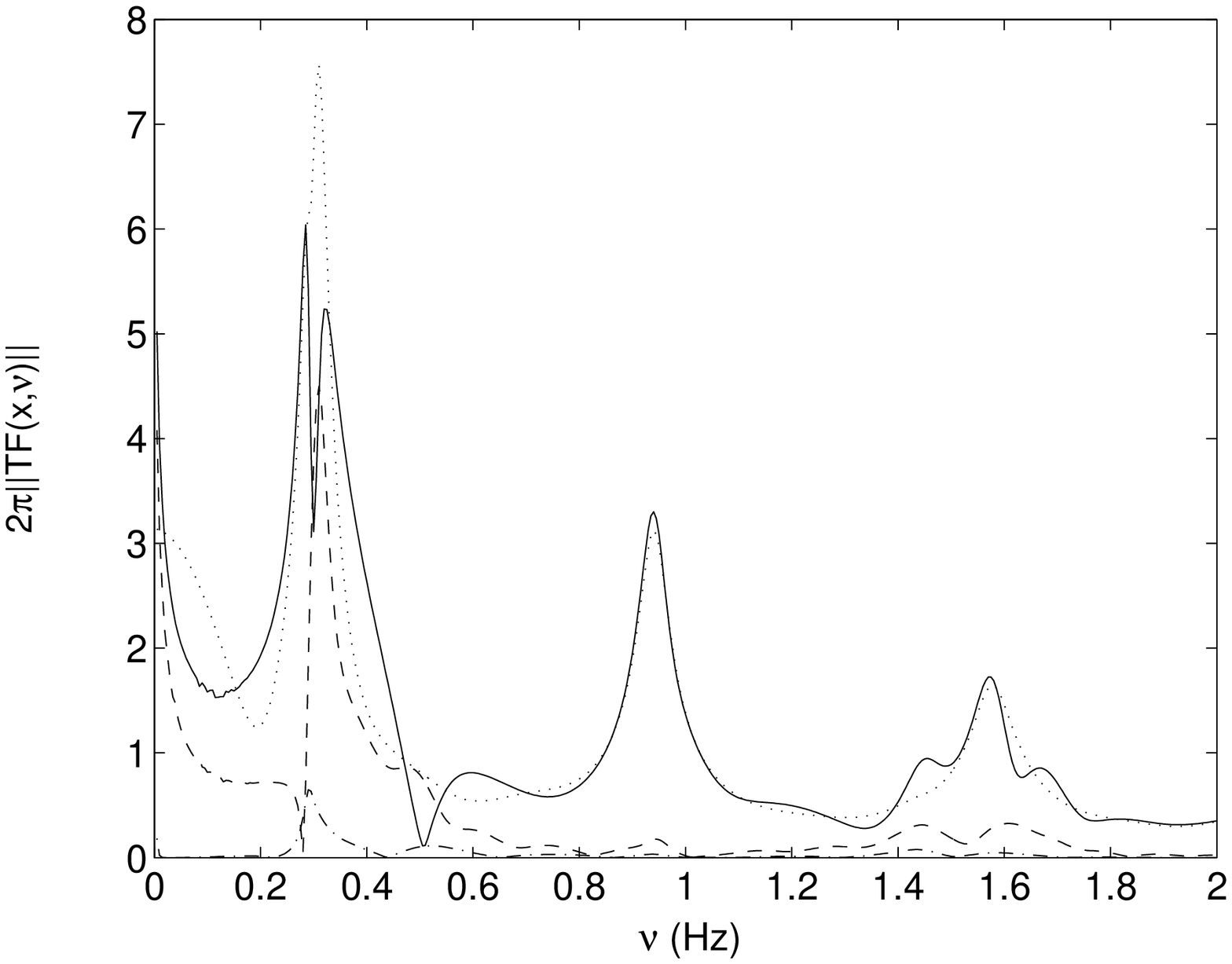}
\includegraphics[width=6.0cm] {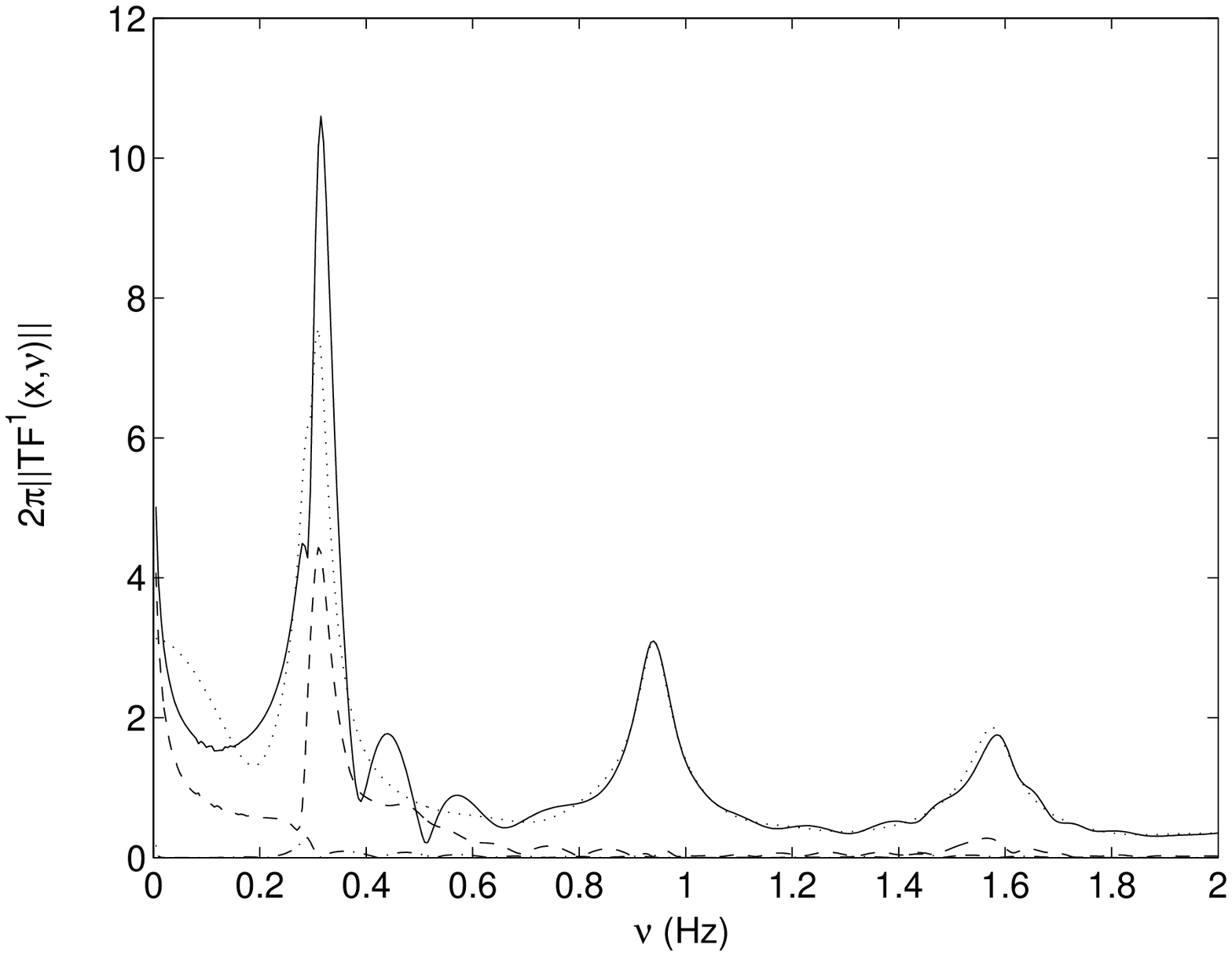}
\caption{$2\pi$ times the transfer function  (solid curves) of the
total displacement on the ground at $\mathbf{x}=(20m,0m)$
(top-left), $\mathbf{x}=(30m,0m)$ (top-right),
$\mathbf{x}=(150m,0m)$ (bottom-left), and $\mathbf{x}=(300m,0m)$
(bottom-right). The deep line source is located at
$\mathbf{x^{s}}=(0m,3000m)$. The dotted curve represents the
contribution along the interval  $\mathcal{I}_{1}$, the dashed
curve the contribution along the interval $\mathcal{I}_{2}$, and
the dot-dashed curve  the contribution along the interval
$\mathcal{I}_{3}$.} \label{ftransground}
\end{center}
\end{figure}
Figures \ref{ftransground} represents $2\pi$ times  the transfer
function $u^{1}(\mathbf{x},\omega)/S(\omega)$ at various locations
on the ground outside of the block, computed by the mode-matching
method by taking into account only the zeroth-order quasi-mode.
From the fact that the deep source  is  close to the vertical line
passing through these ground locations,  in the absence of the
block, the first peak of the transfer function is principally due
to the integral over $\mathcal{I}_{1}$
\cite{grobyetwirgin2005,grobyetwirgin2005II}. The presence of the
block induces: i) a double dependency of the first peak on the
integrals over $\mathcal{I}_{1}$ and over $\mathcal{I}_{2}$ which
leads to splitting of the first peak  at some locations (as seen
in the results at $\mathbf{x_{g}}=(150m,0m)$), ii) surface waves
in the layer (represented by the contribution of the integration
over $\mathcal{I}_{3}$) which normally are not excited in the case
of a deep source, \cite{grobyetwirgin2005,grobyetwirgin2005II}.
The presence of the block allows the excitation of a quasi-Love
mode as is testified by the importance of the integration over
$\mathcal{I}_2$.

\begin{figure}[ptb]
\begin{center}
\includegraphics[width=6.0cm] {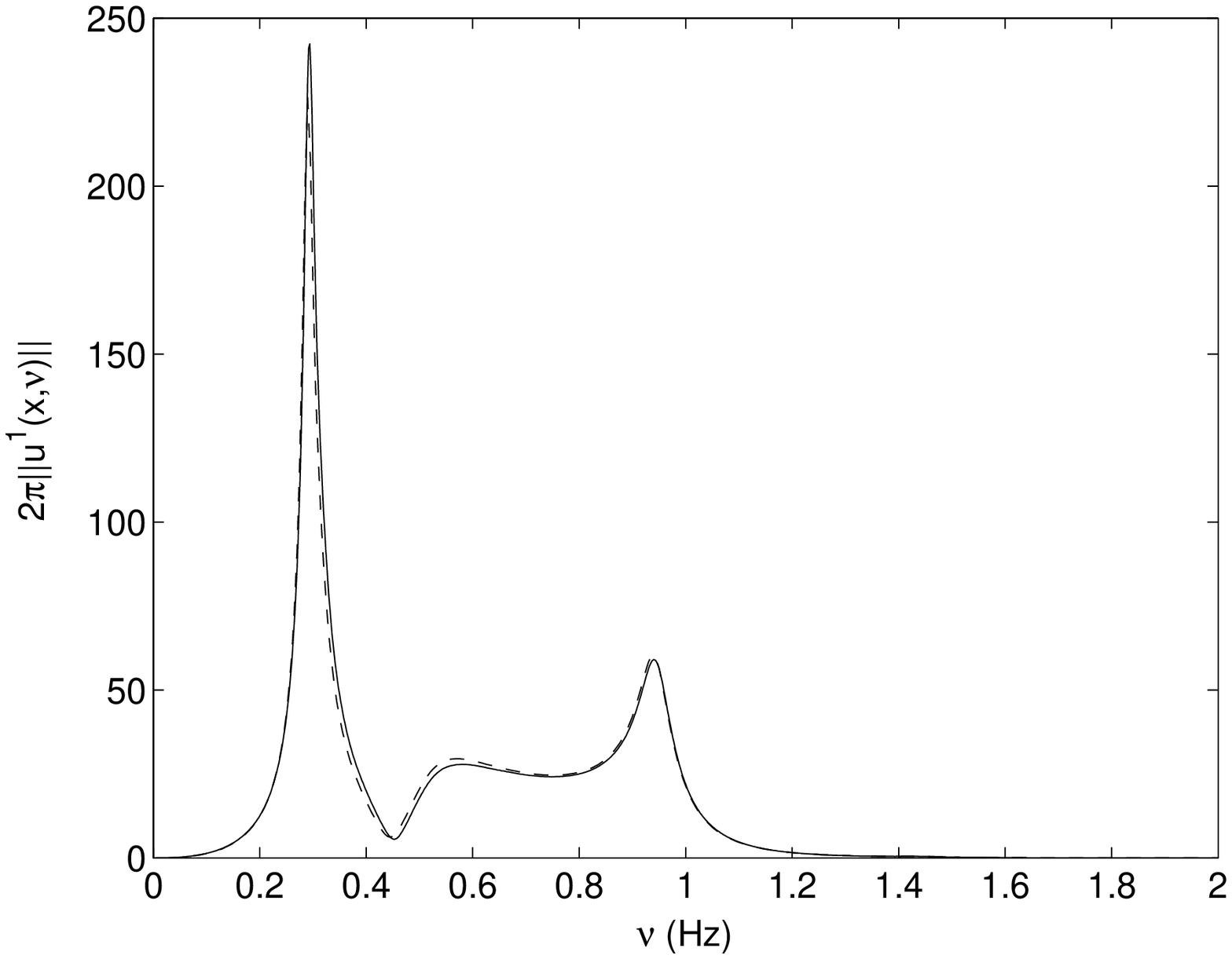}
\includegraphics[width=6.0cm] {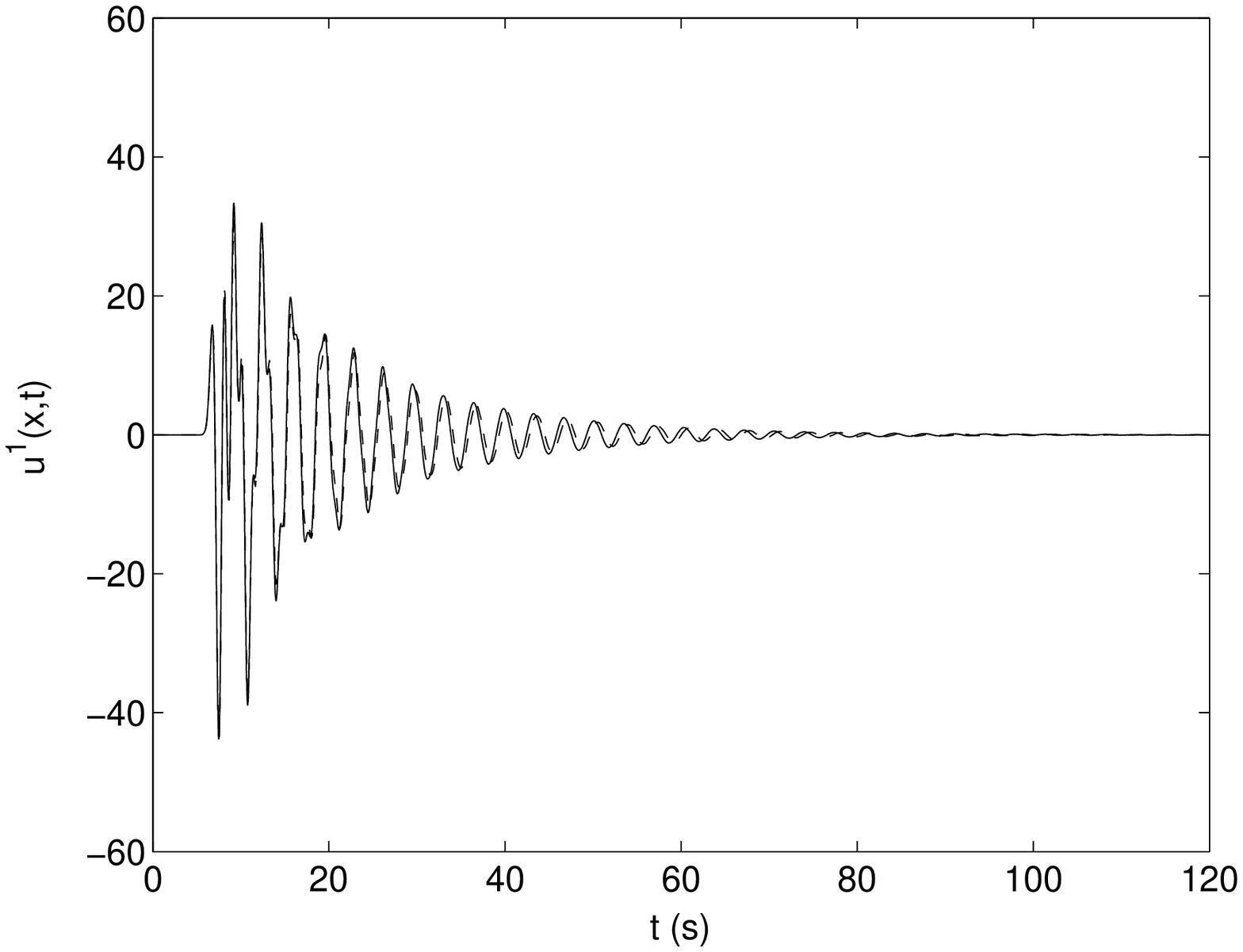}
\includegraphics[width=6.0cm] {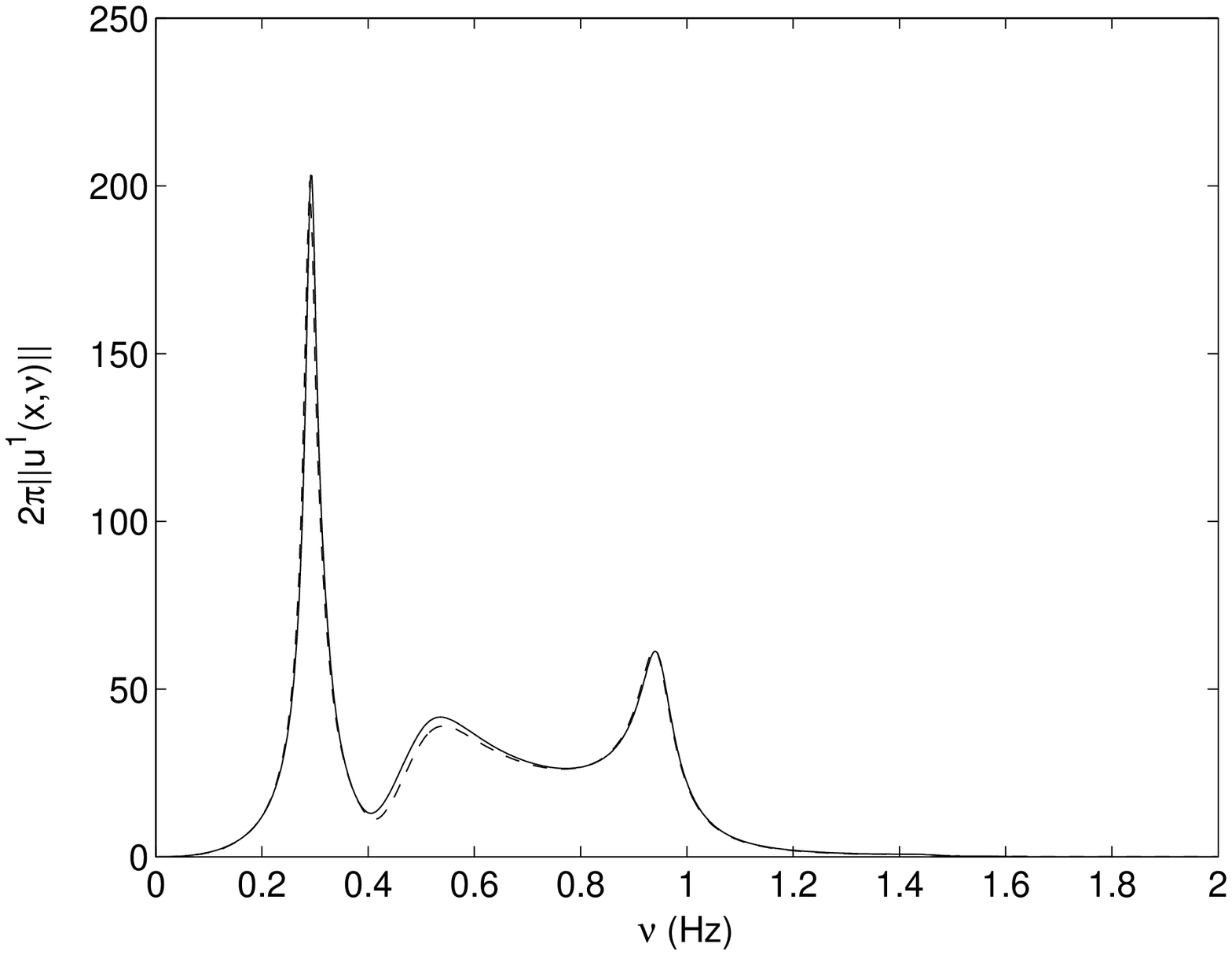}
\includegraphics[width=6.0cm] {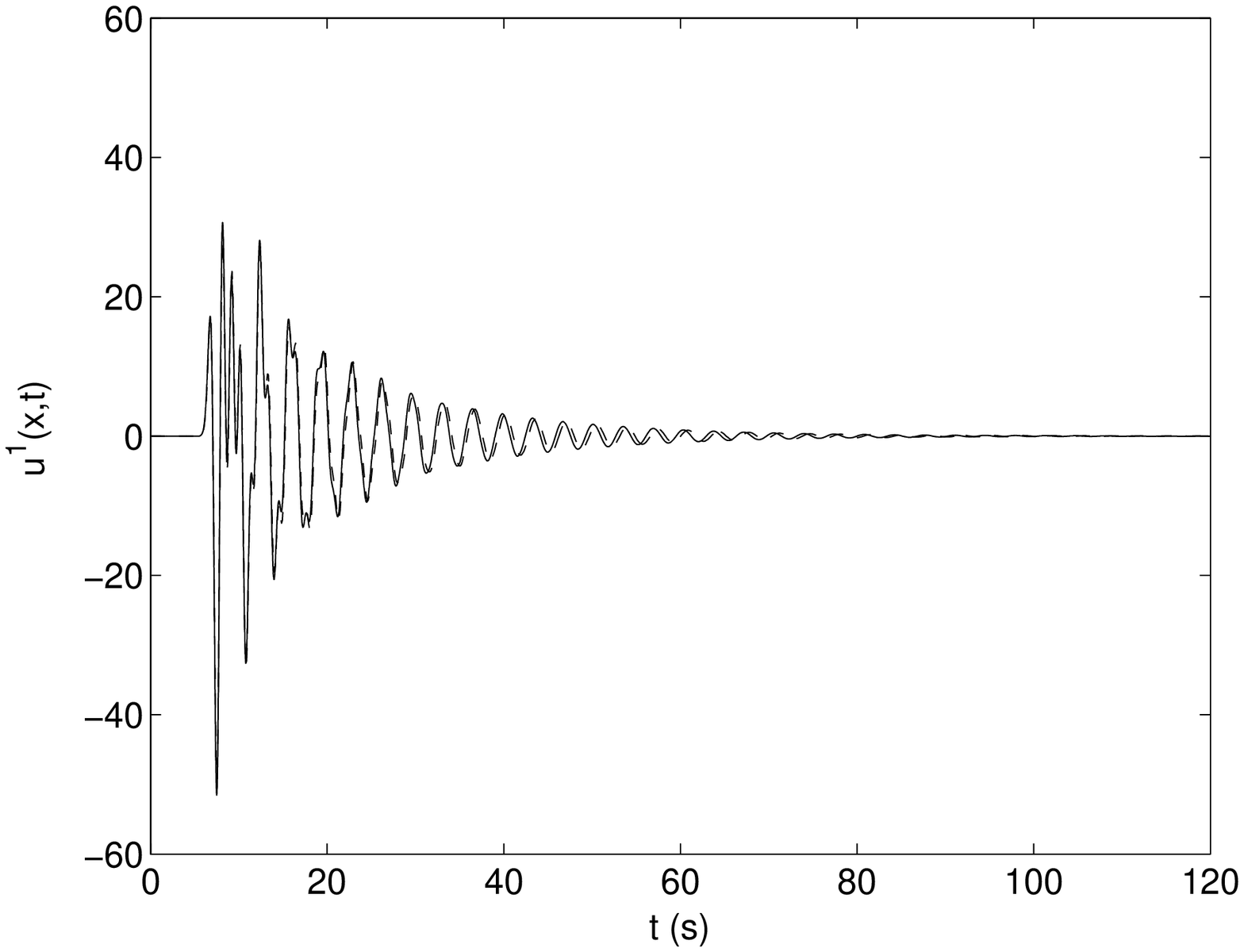}
\includegraphics[width=6.0cm] {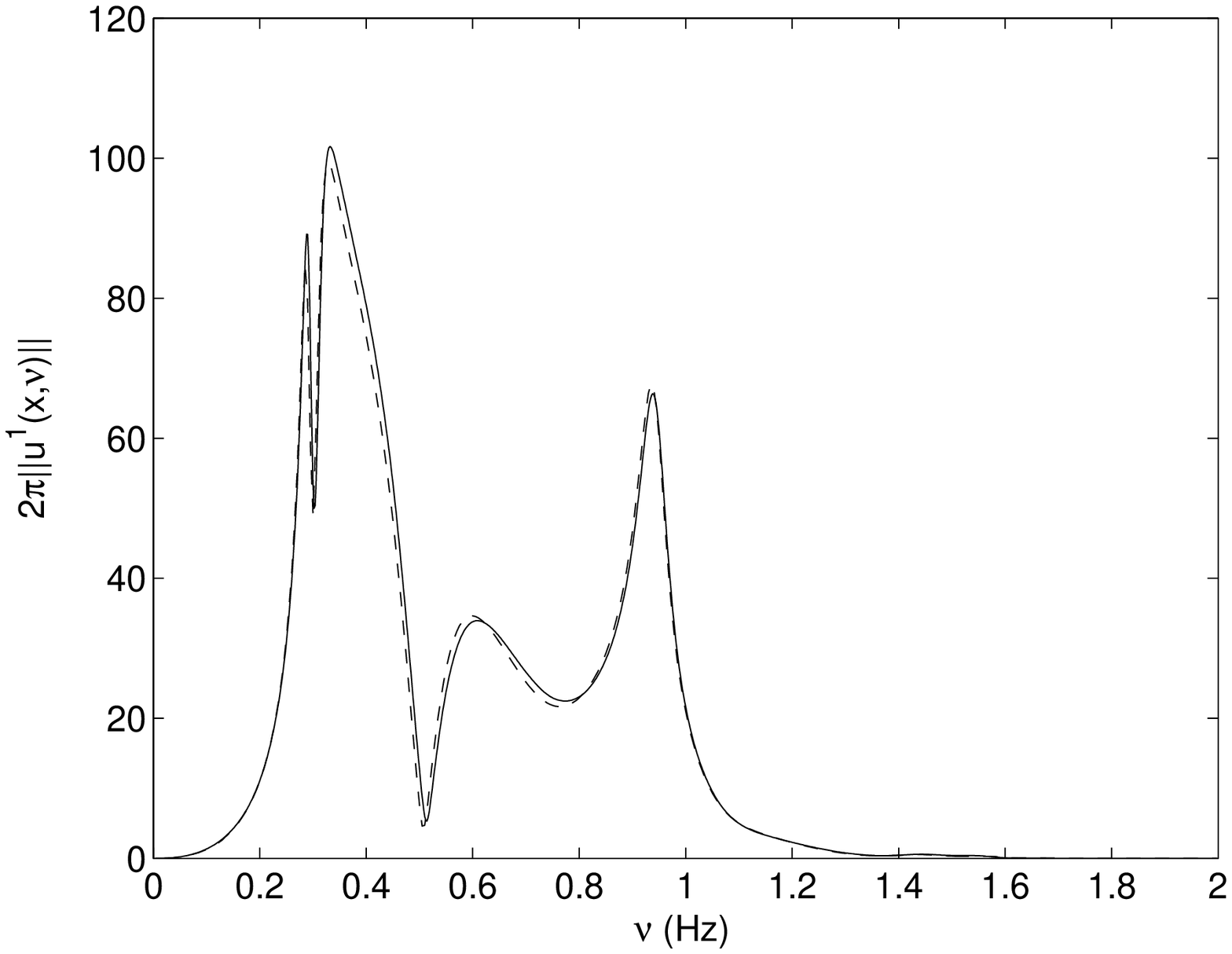}
\includegraphics[width=6.0cm] {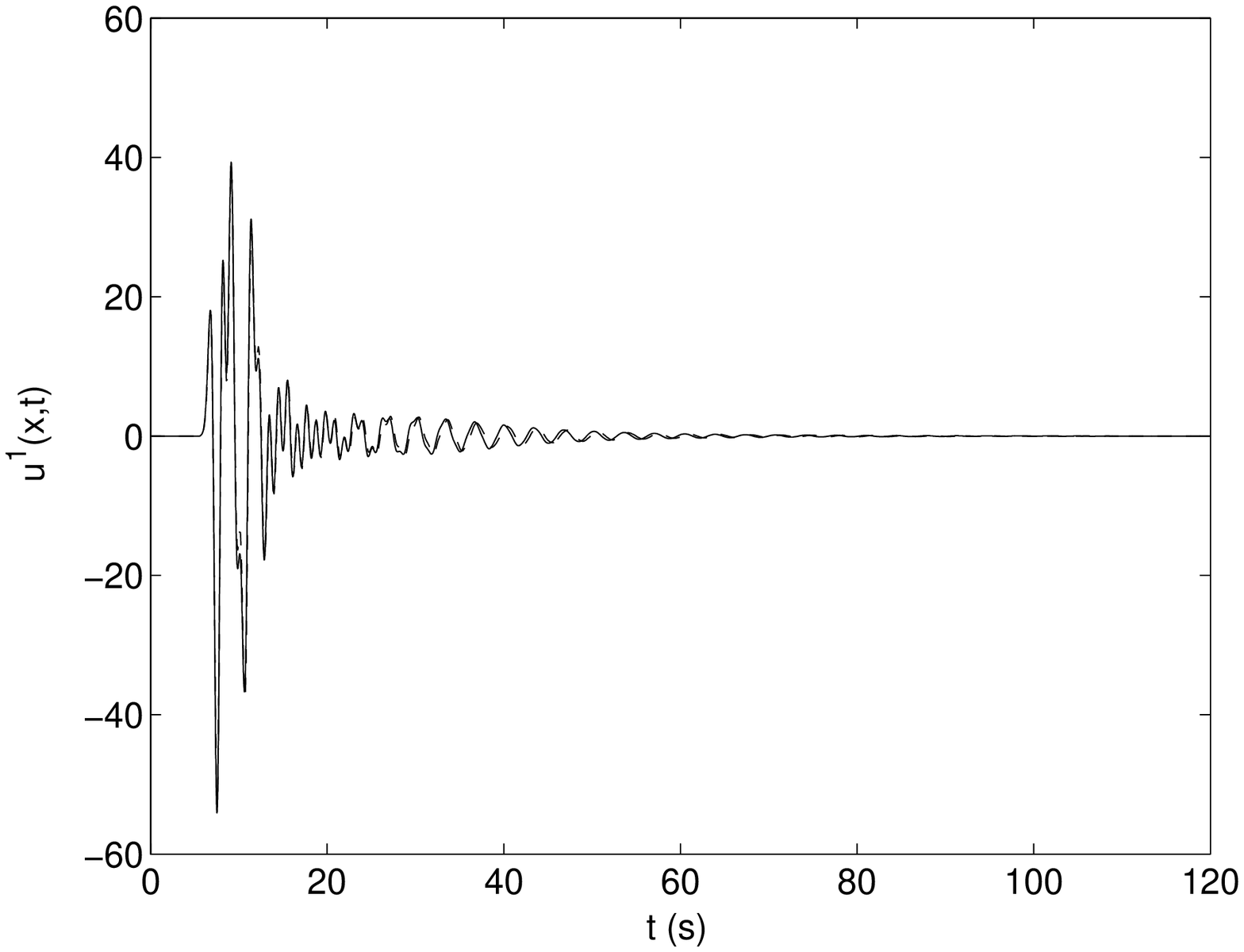}
\includegraphics[width=6.0cm] {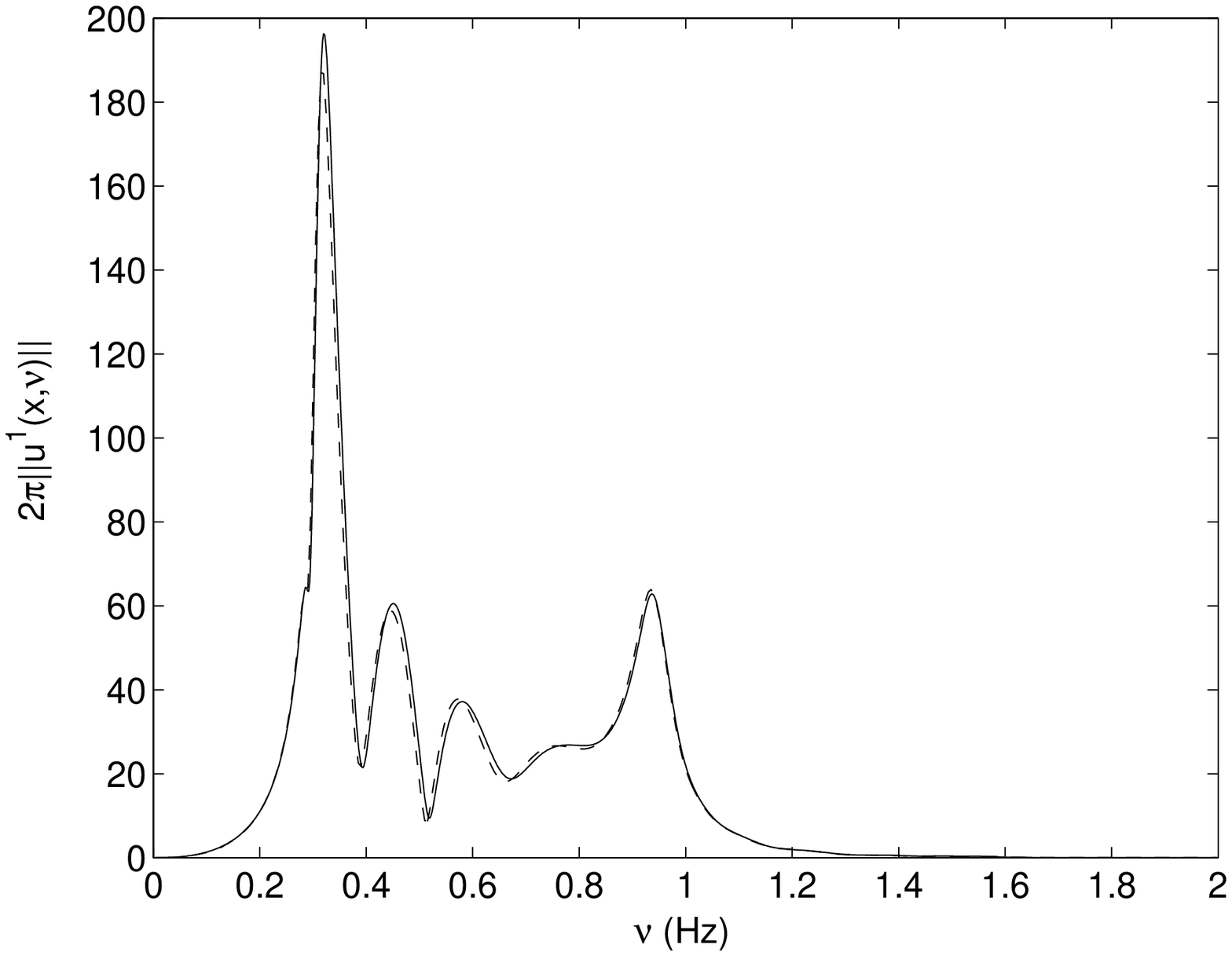}
\includegraphics[width=6.0cm] {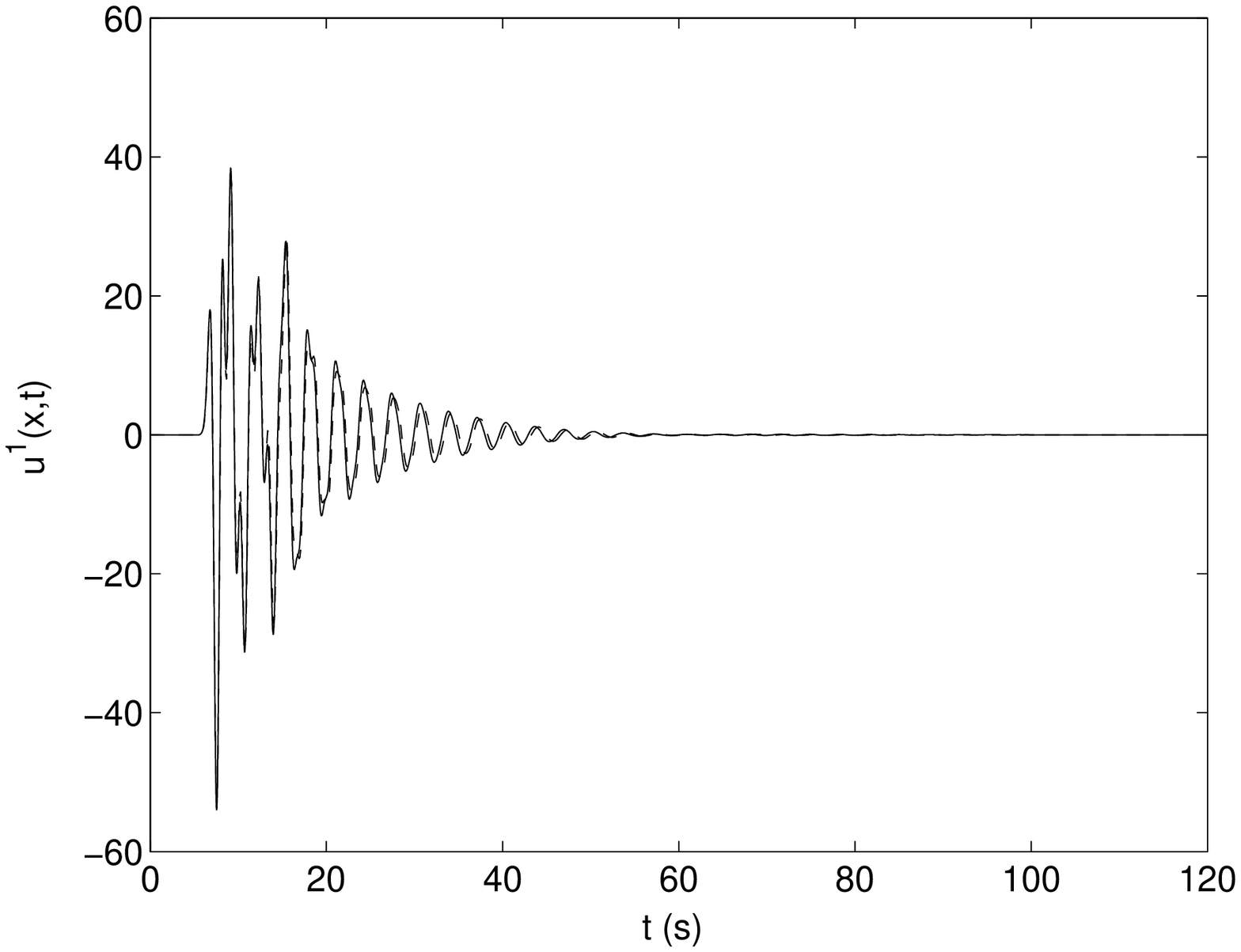}
\end{center}
\caption{$2\pi$ times the spectrum (left panel) and time history
(right panel)  of the total displacement  at various locations on
the ground: (from the top to  bottom) at 20m, 30m, 150, and 300m
from the center of a single $50m\times 30m$ block. The dashed
curves correspond to the semi-analytical (mode-matching) result
(with only the fundamental quasi-mode taken into account), and the
solid curves to the finite element result.}
\label{specttimeground} \label{specttimebasxg}
\end{figure}

Fig. \ref{specttimebasxg} shows once again that there is a  good
agreement between the results of the mode-matching method (with
only the fundamental quasi-mode taken into account) and the finite
element method.

\begin{figure}[ptb]
\begin{center}
\includegraphics[width=6.0cm] {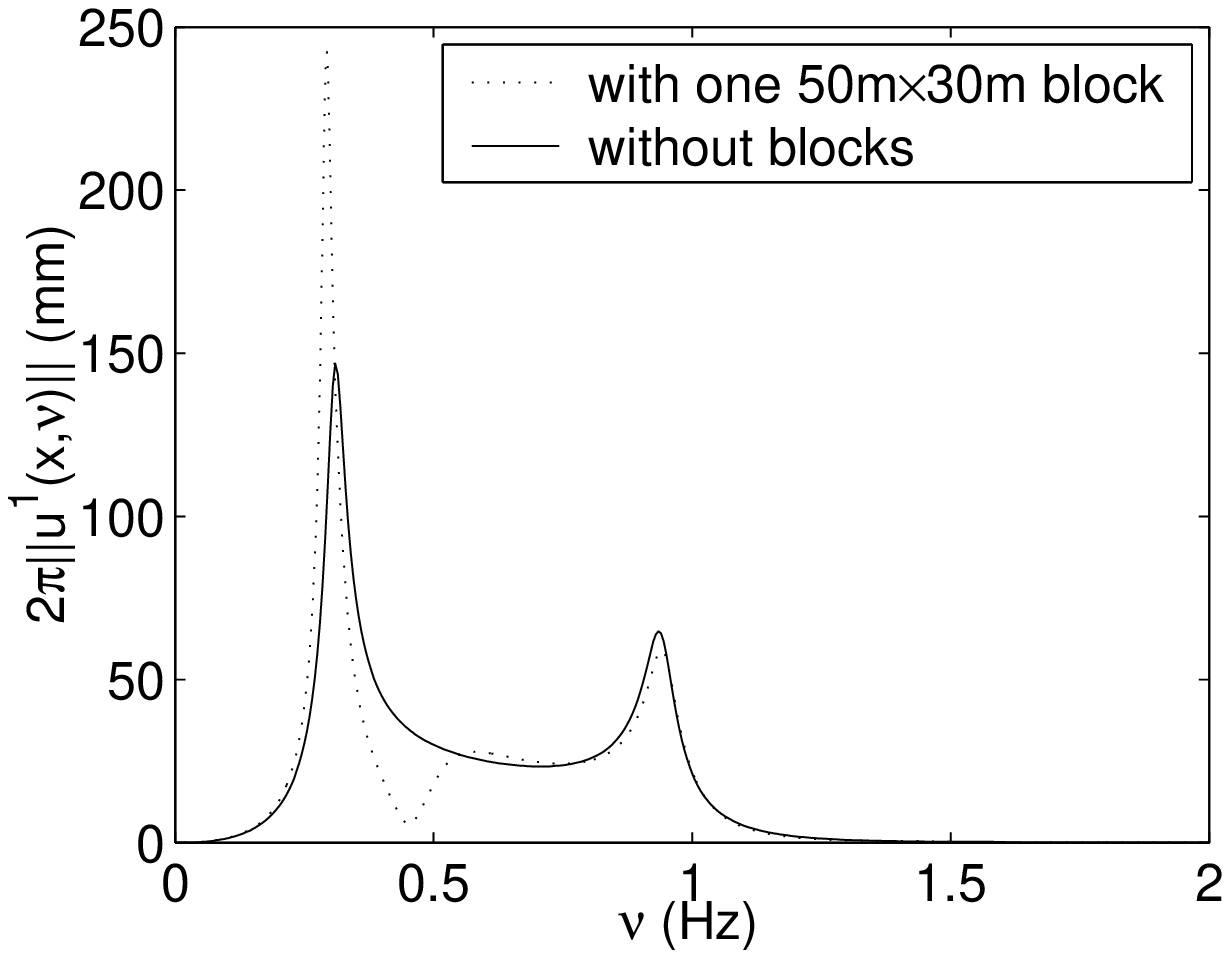}
\includegraphics[width=6.0cm] {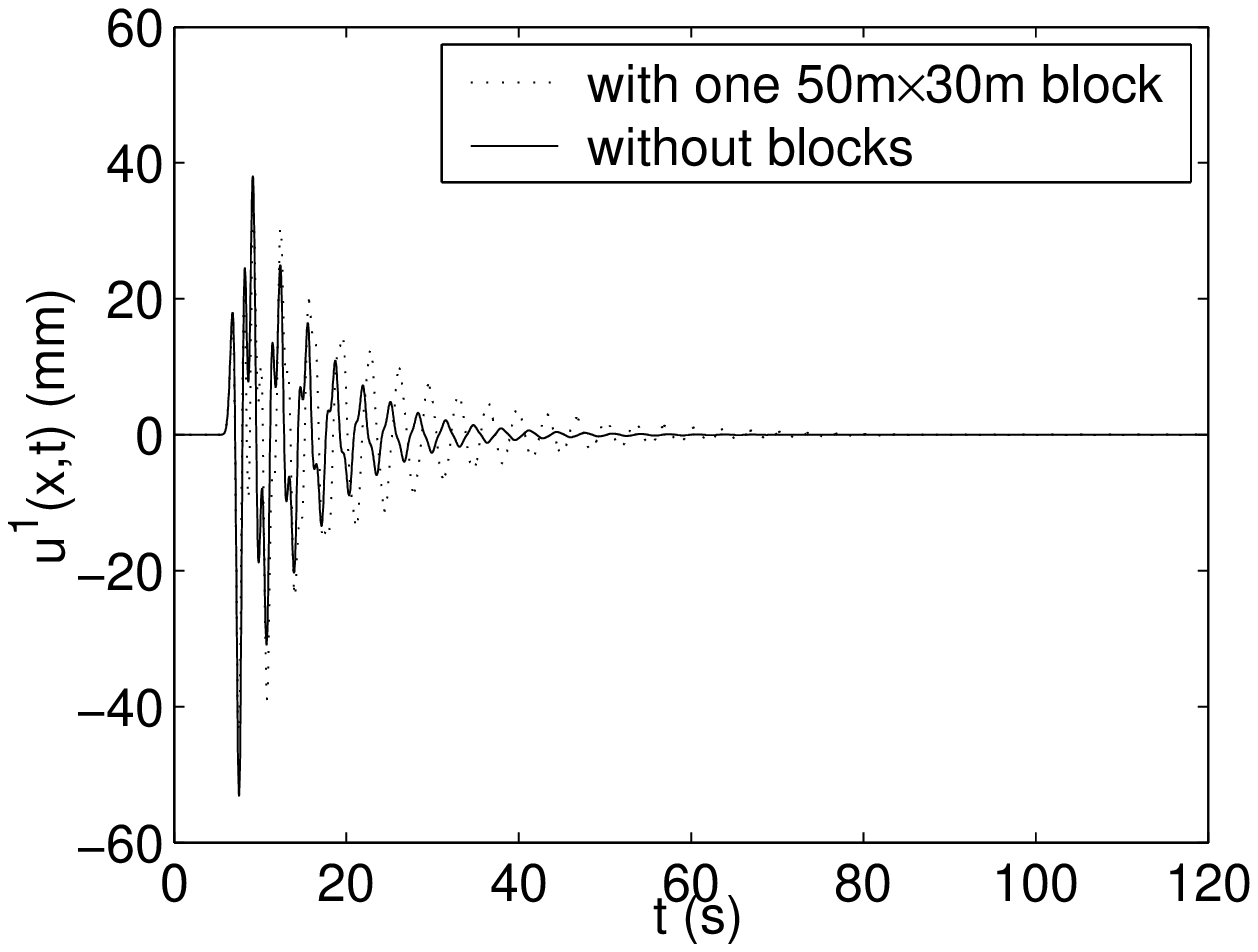}
\includegraphics[width=6.0cm] {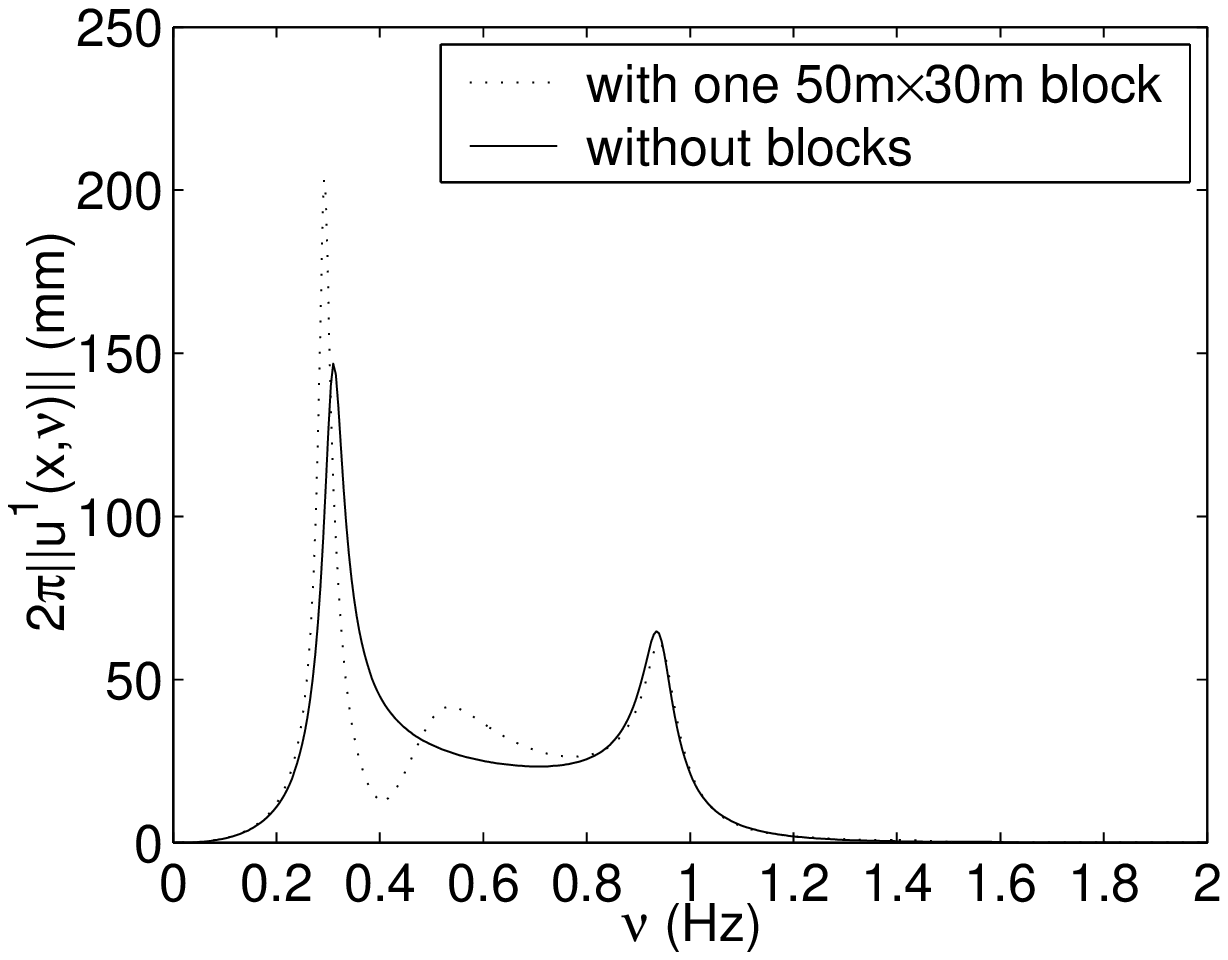}
\includegraphics[width=6.0cm] {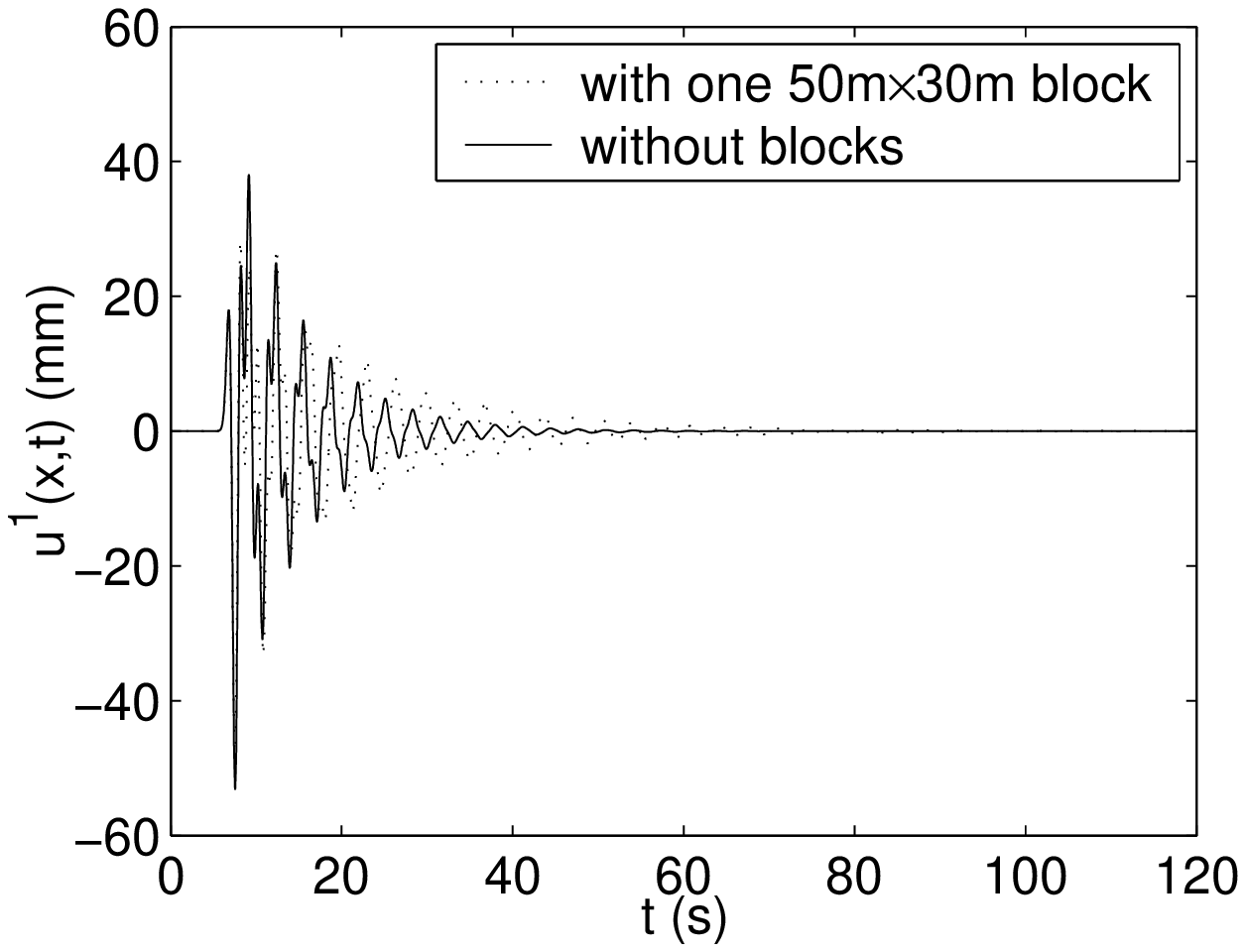}
\includegraphics[width=6.0cm] {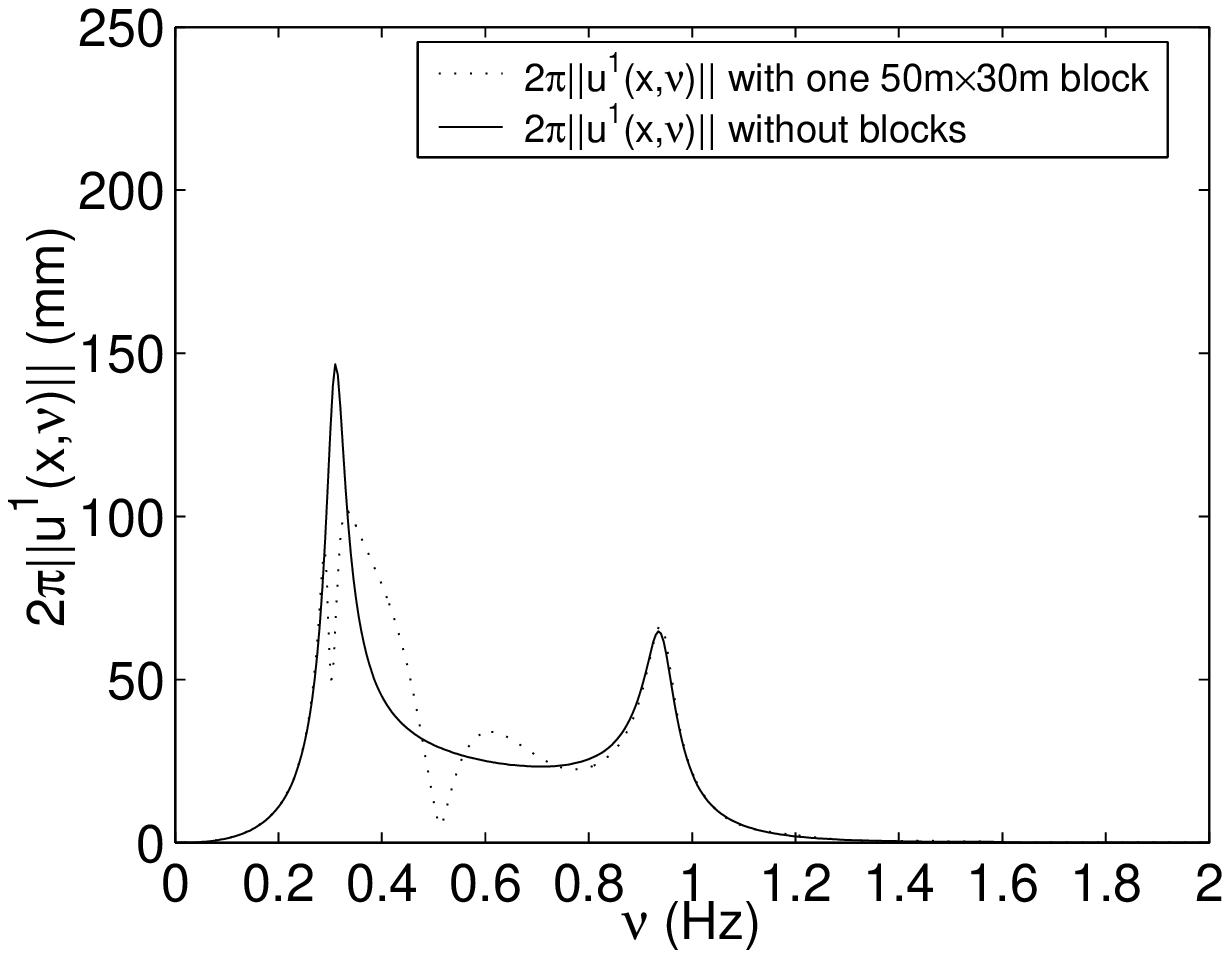}
\includegraphics[width=6.0cm] {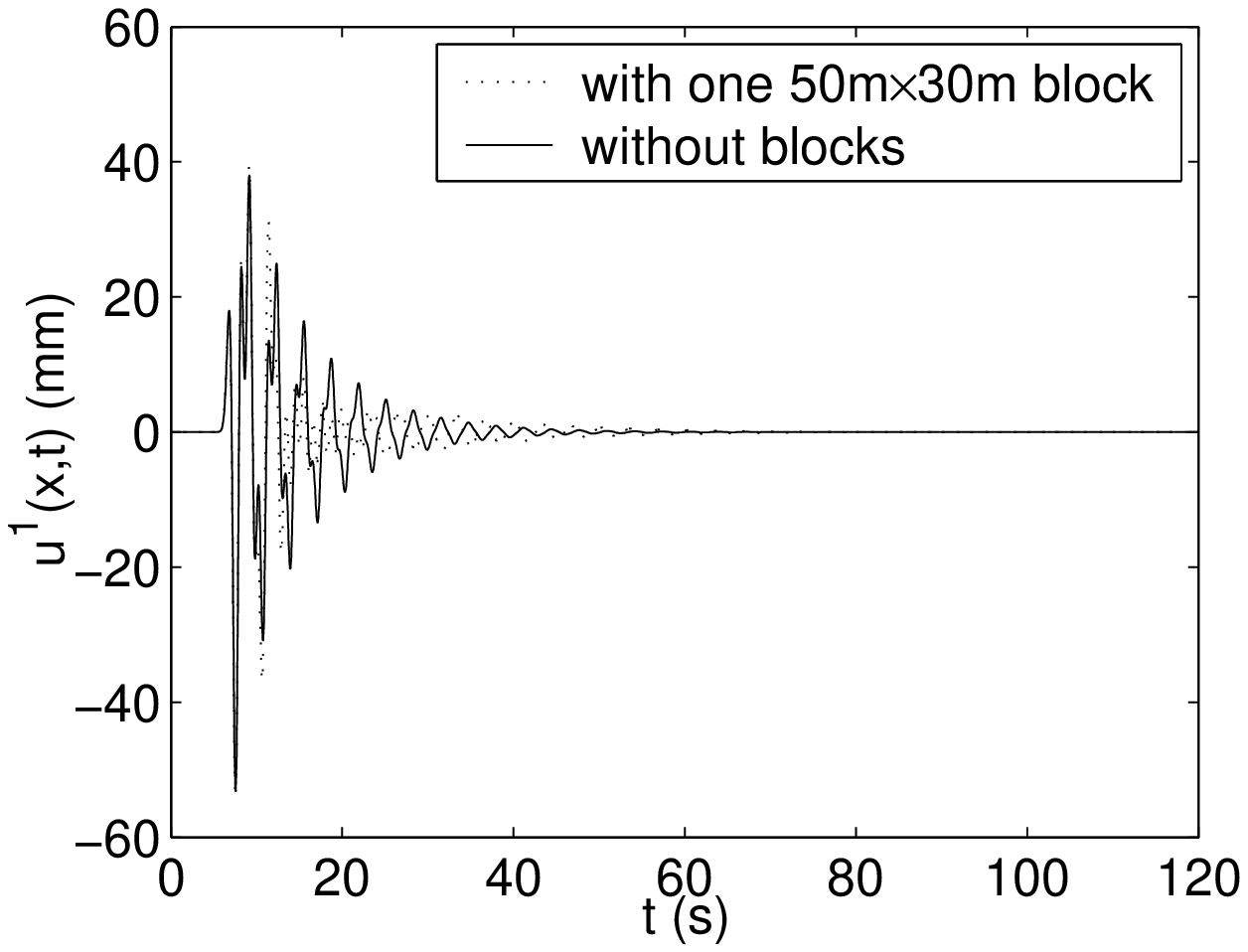}
\includegraphics[width=6.0cm] {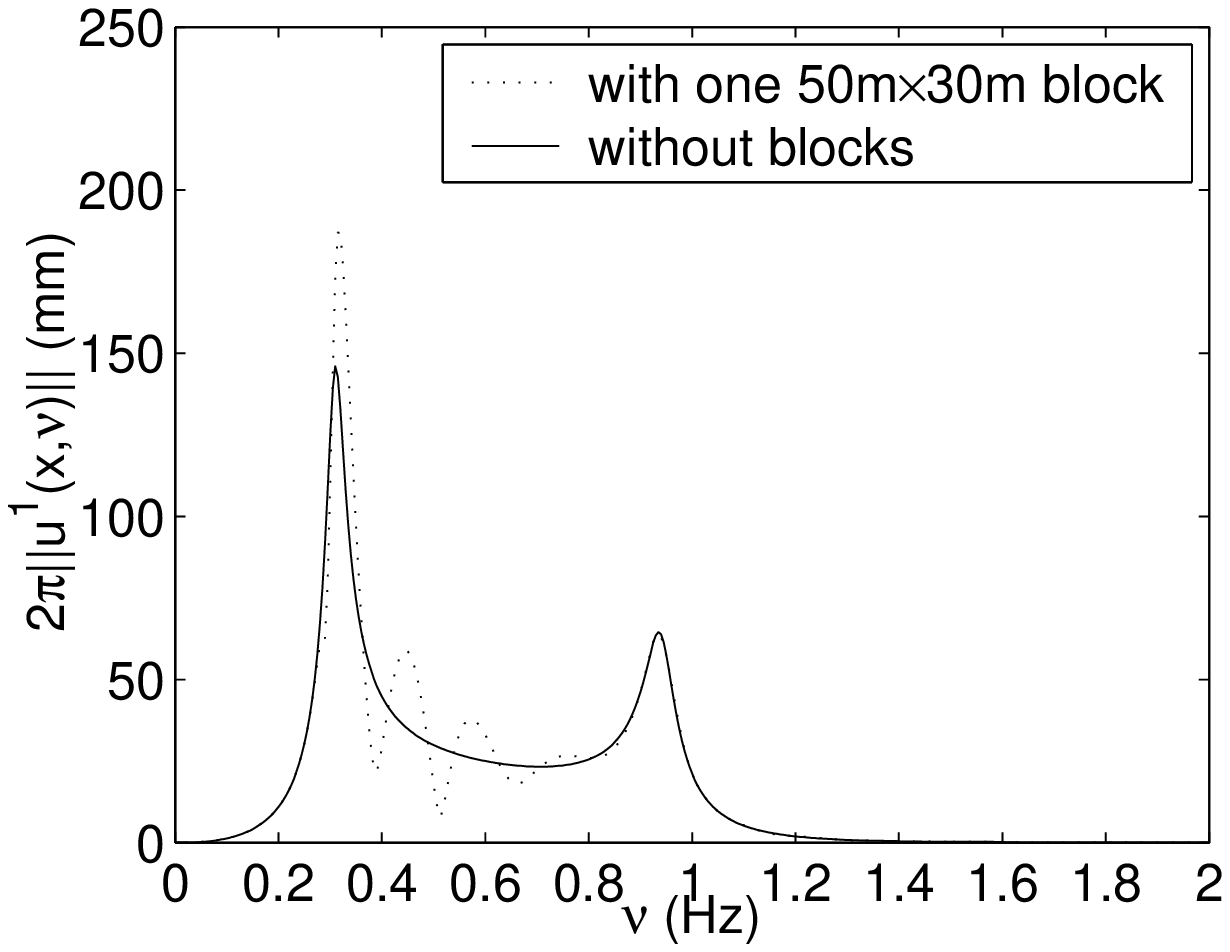}
\includegraphics[width=6.0cm] {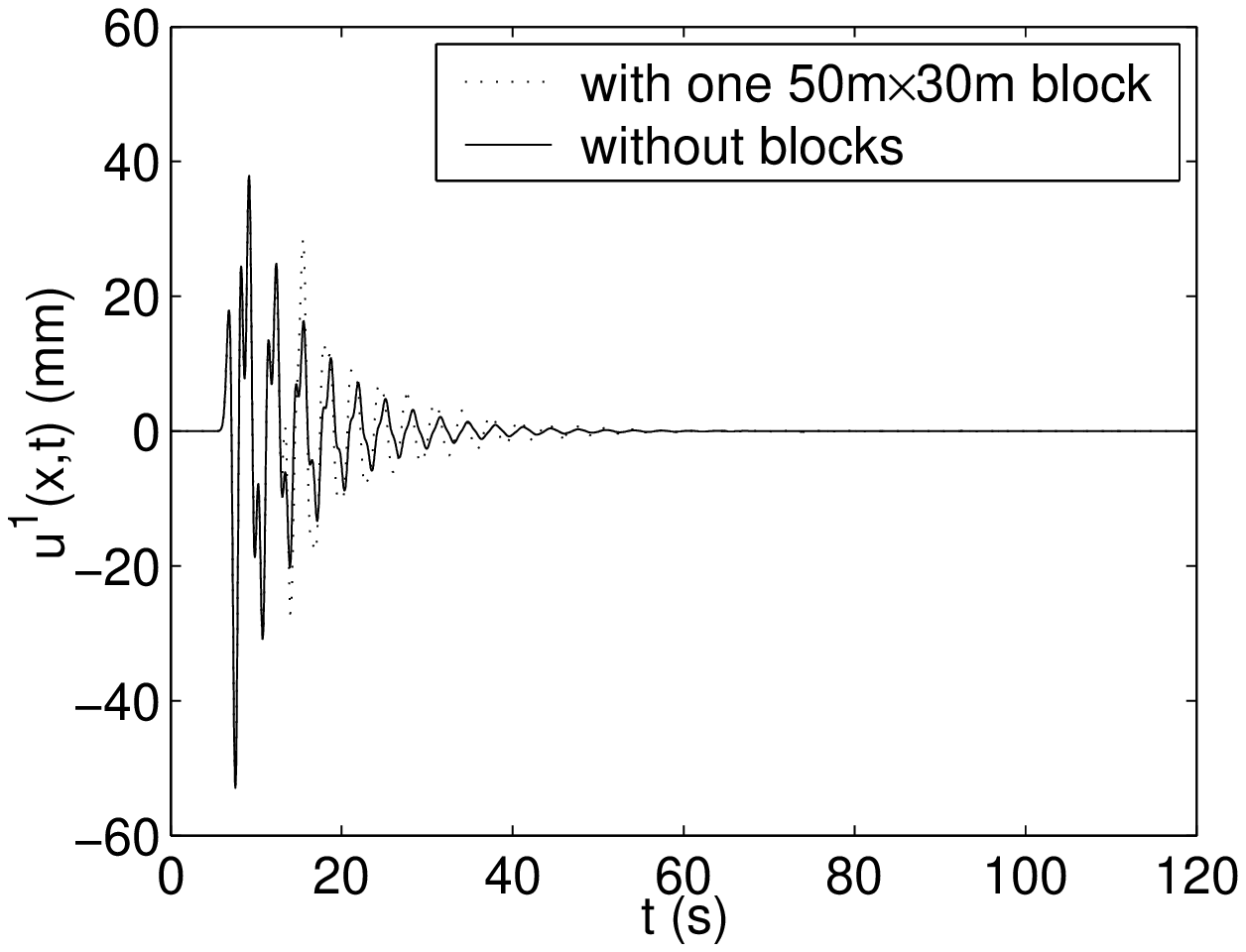}
\end{center}
\caption{Comparison of $2\pi$ times the spectrum (left panel) and
of the  time history (right panel) of the total displacement on
the ground in the absence of the block (solid curves) and in the
presence of the block (dashed curves) at various locations on the
ground: (from the top to the bottom) at 20m, 30m, 150, and 300m
from the center of a single $50m\times 30m$ block. The deep line
source is located at $\mathbf{x}^s=(0,3000m)$.}
\label{compspecttimebasxg}
\end{figure}

Fig. \ref{compspecttimebasxg} shows that  when the block is
present, the displacement field at a given location on the ground
is somewhat different from what it is in absence of the block,
even at a large (i.e. 300m) distance from the block, this being
particularly evident in the response spectra.  The duration and
cumulative amplitude of the displacement fields have a tendency of
increasing as one approaches the block, but not in monotonic
manner, as manifested by what happens at  $150m$.

This may be one of the causes of the spatial  variability of the
destructions noticed during the Michoacan earthquake. As shown by
the form of the matrix in section \ref{anotherquasisfb}, one can
expect one building to have an effect on its neigbors via the
waves traveling in the substructure and in particular in the layer
and along the ground. Thus,  the effect of one block on the other
can vary in a way that depends on the distance of their
separation. Nethertheless, the authors of \cite{wirgintsogkagroby}
did not find substantial differences in response for varying
building separations in the case of an idealized city composed of
ten blocks. It may be that in order for this effect to be
noticeable requires  either a  large distance of separation
between blocks or a single block.

\clearpage
\subsubsection{Displacement on the ground on one side of the block
for shallow line source solicitation}
\begin{figure}[ptb]
\begin{center}
\includegraphics[width=6.0cm] {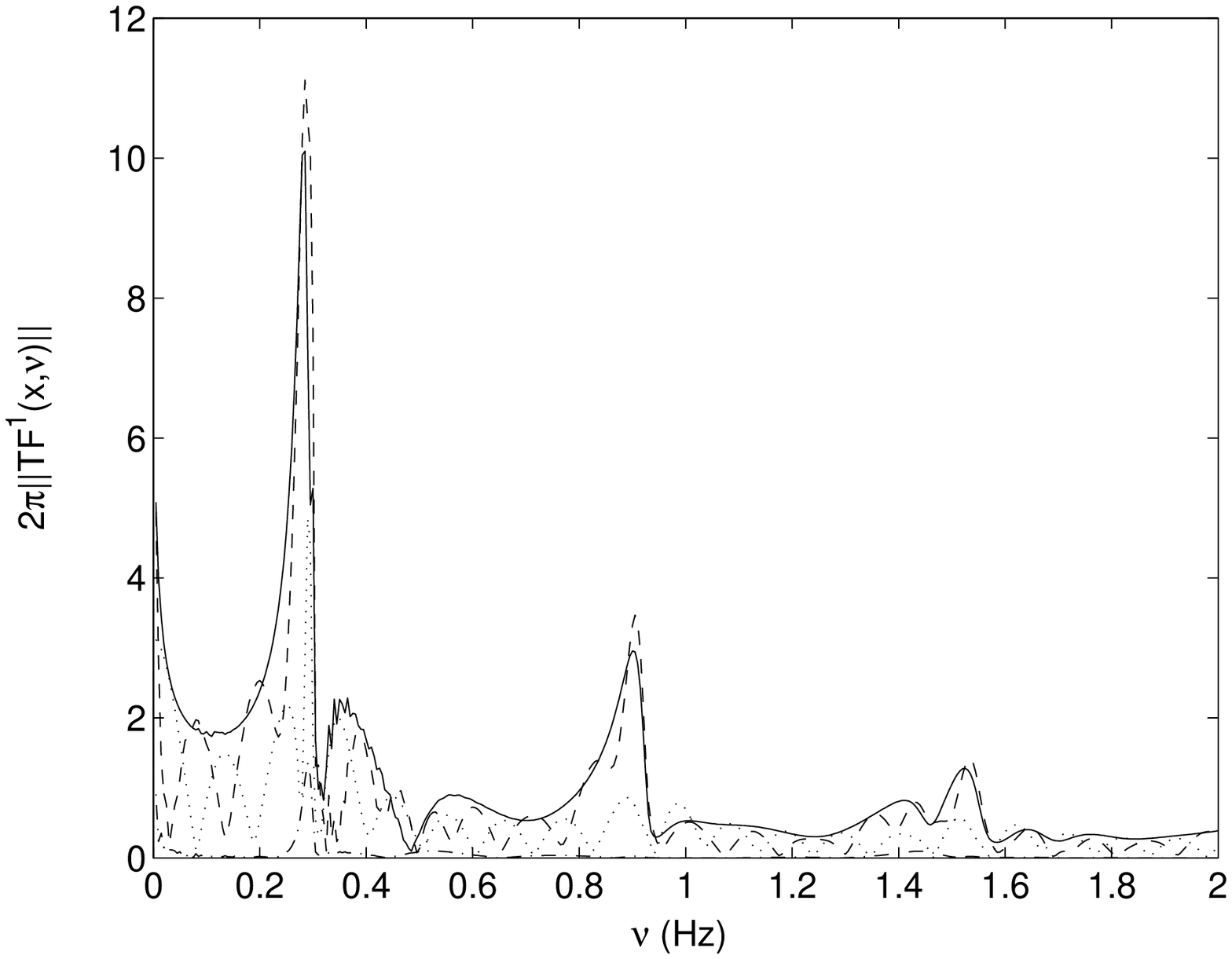}
\includegraphics[width=6.0cm] {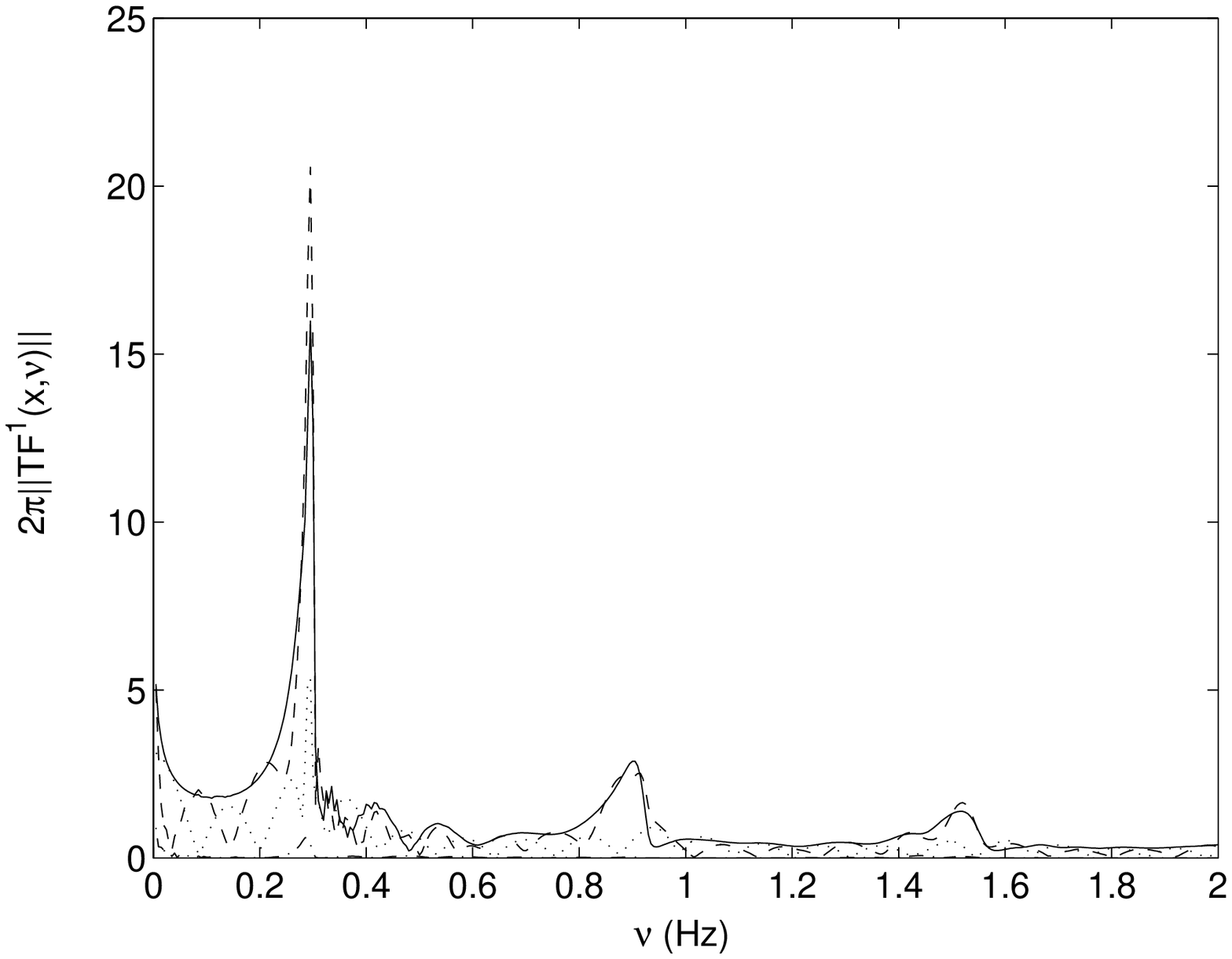}
\caption{$2\pi$ times the transfer function, solid curve, of the
total displacement on the ground at $\mathbf{x_{g}}=(150m,0m)$
(left panel), and $\mathbf{x_{g}}=(300m,0m)$ (right panel).  The
shallow source is located at $\mathbf{x^{s}}=(-3000m,100m)$. The
dotted curve represents the contribution along the interval
$\mathcal{I}_{1}$, the dashed curve represents the contribution
along the interval  $\mathcal{I}_{2}$, and the dotted-dashed curve
represents the contribution along the interval $\mathcal{I}_{3}$.}
\label{ftransgroundshallow}
\end{center}
\end{figure}

\begin{figure}[ptb]
\begin{center}
\includegraphics[width=6.0cm] {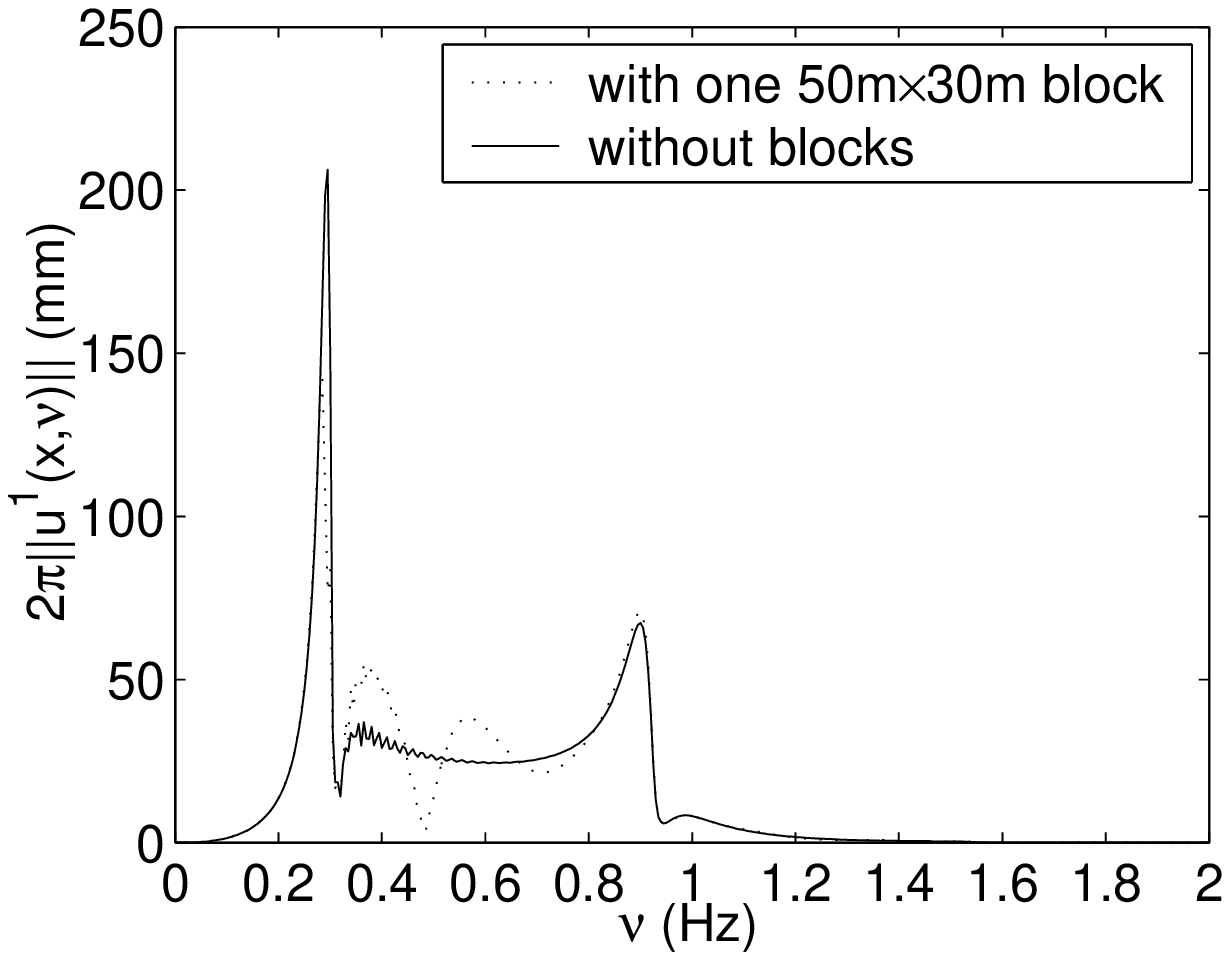}
\includegraphics[width=6.0cm] {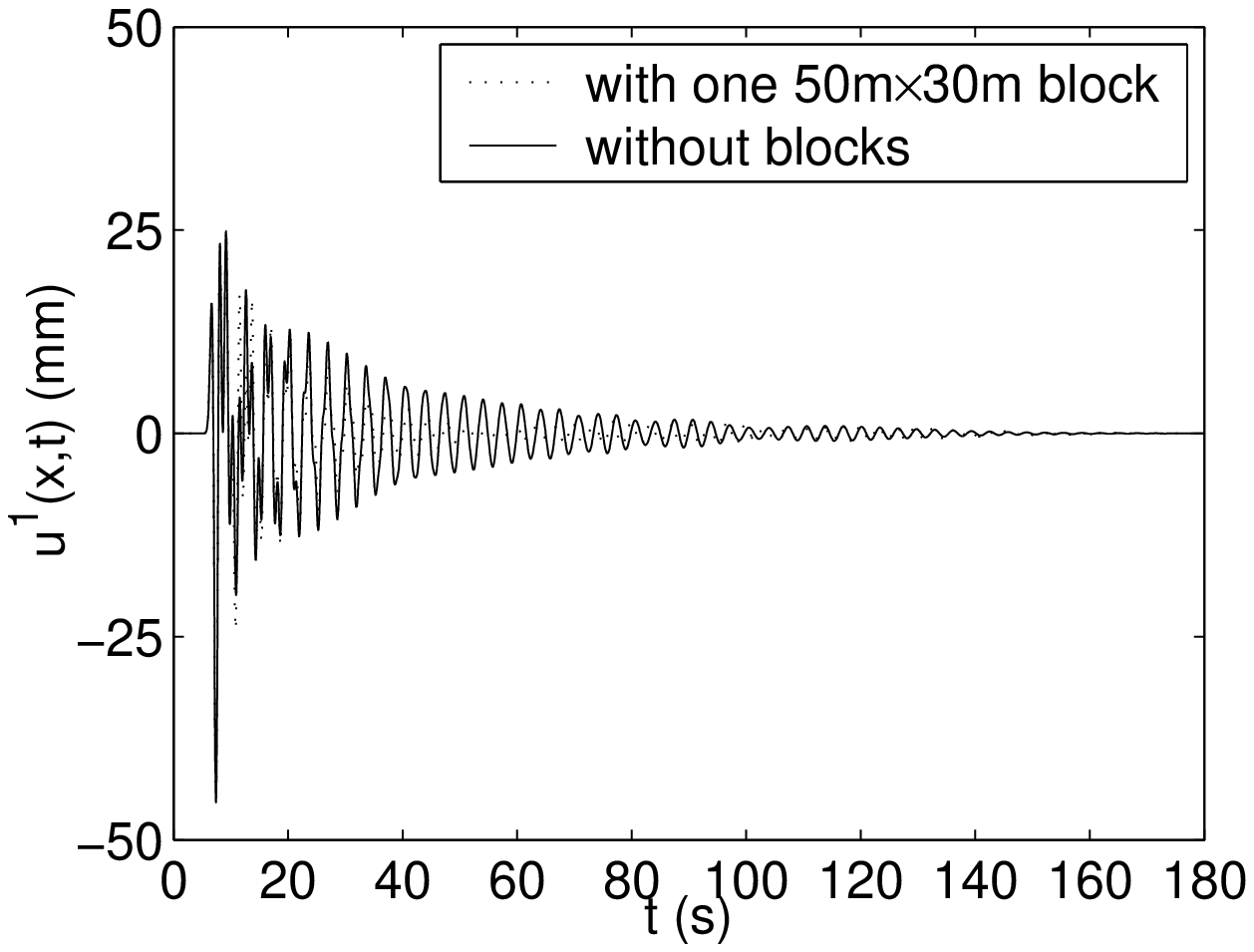}
\includegraphics[width=6.0cm] {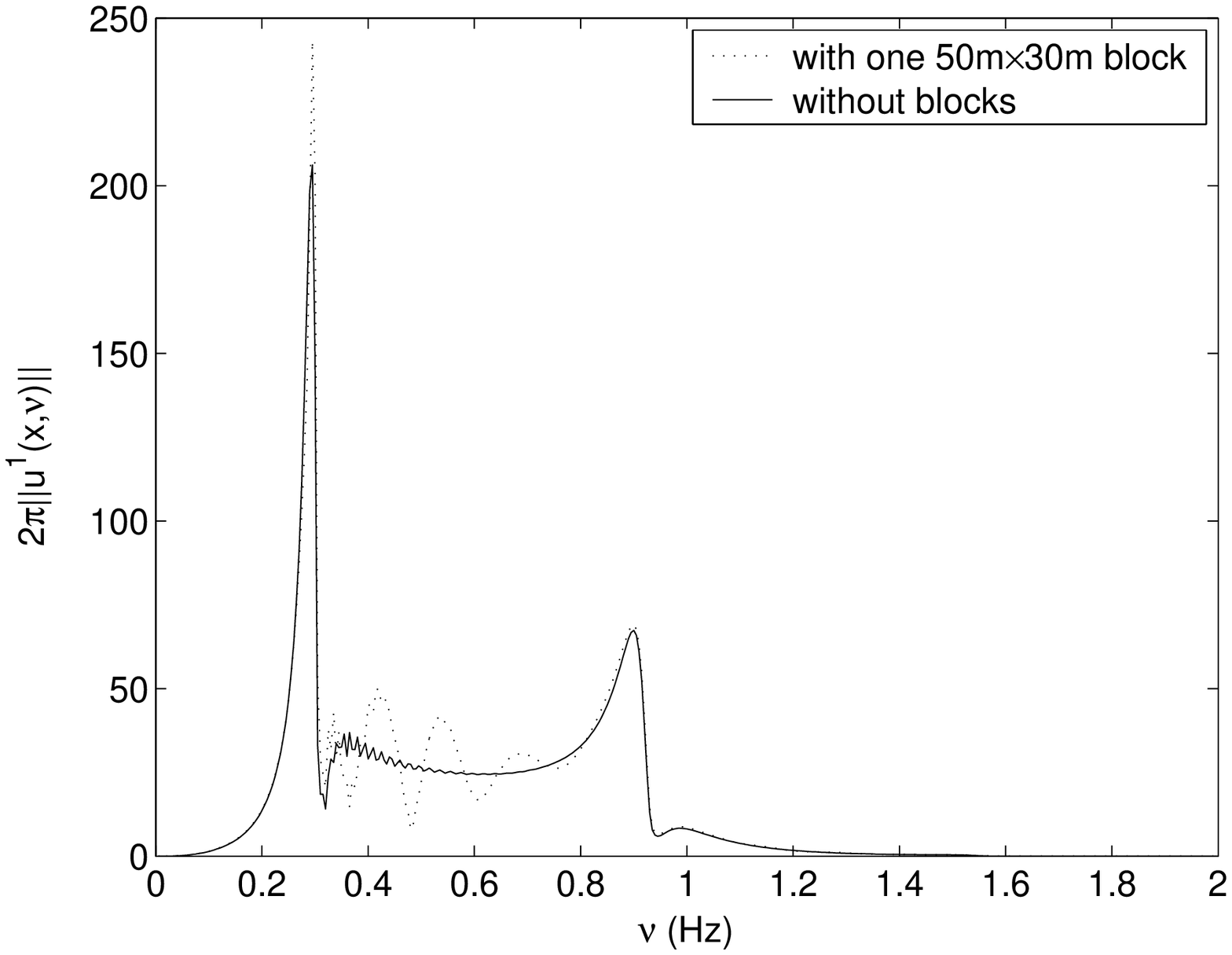}
\includegraphics[width=6.0cm] {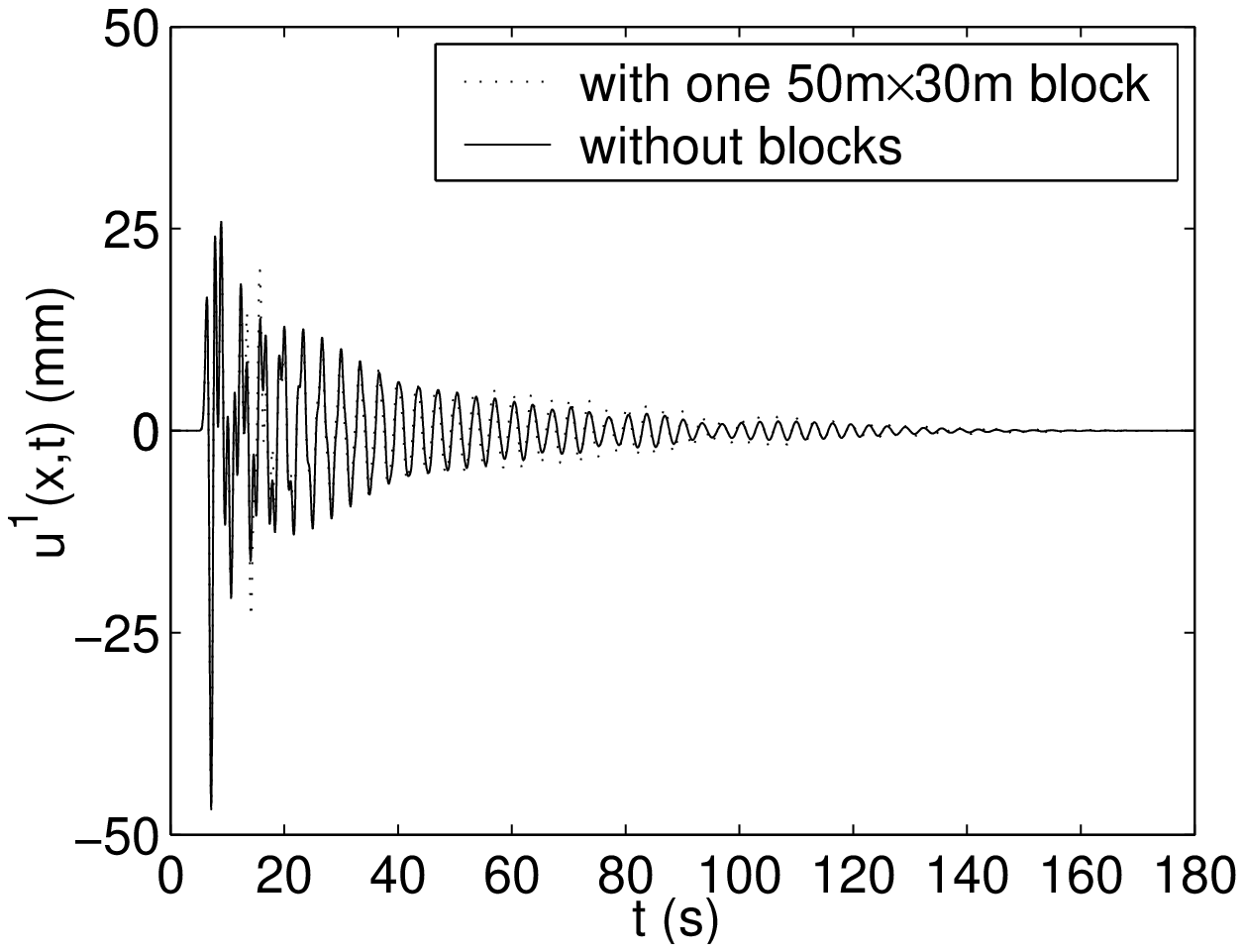}
\end{center}
\caption{Comparison of $2\pi$ times the spectrum (left panel)  and
of the time history (right panel) of the total displacement on the
ground in the absence of the block (solid curves) and in the
presence of the block (dashed curves) at various locations on the
ground: (from the top to the bottom) at 150 and 300m from the
center of a single $50m\times 30m$ block. The shallow source is
located at $\mathbf{x}^s=(-3000,100m)$.}
\label{compspecttimegroundshallow}
\end{figure}

When the seismic disturbance is delivered to the site by the wave
radiated from a {\it shallow line source} located at
$\mathbf{x^{s}}=(-3000m,100m)$, it should be recalled that, in the
absence of the block, the first peak (associated with the first
Love mode) is mainly composed of the component coming from the
integration over $\mathcal{I}_2$, whereas we observe in fig.
\ref{ftransgroundshallow}  that when the block is present, this
peak is composed mainly of two components, one from integration
over $\mathcal{I}_1$ and the other from integration over
$\mathcal{I}_2$. However, this double dependence of the first peak
finds no apparent translation in the time history of the
displacement on the ground, as seen in fig.
\ref{compspecttimegroundshallow}. In fact, we see in this figure
that there are no substantial differences in the responses on the
ground between the cases of the absence and presence of the block.
\subsubsection{Displacement in the substratum}\label{onebloc5030insub}
The focus here is on the displacement field in the subtratum when
the solicitation  is due to a deep source located at
$\mathbf{x}^{s}=(0,3000m)$.

\begin{figure}[ptb]
\begin{center}
\includegraphics[width=8cm] {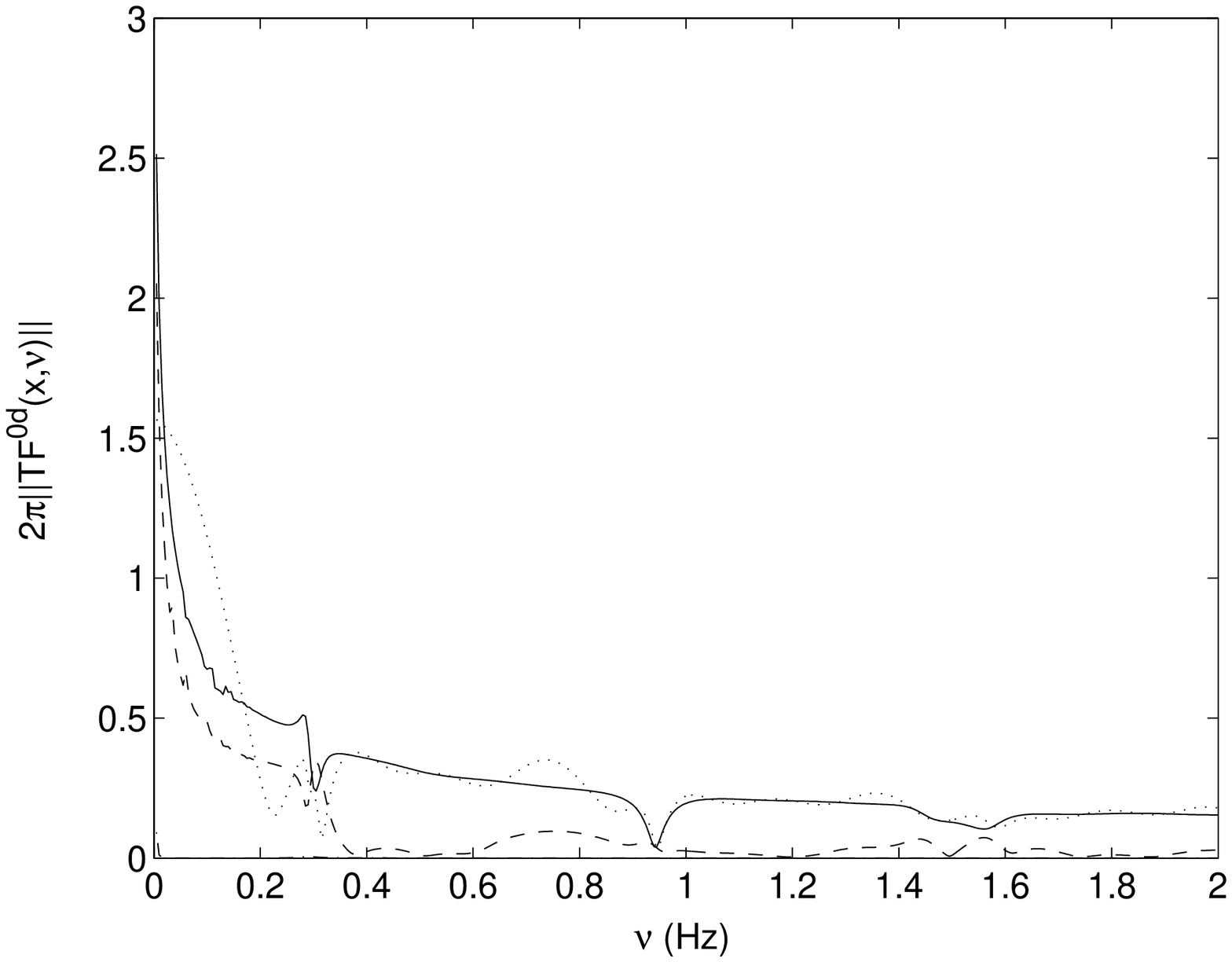}
\end{center}
\caption{The solid curve represents $2\pi$ times the transfer
function  of the diffracted displacement at the interface between
the substratum and the layer below the center of a single
$50m\times 30m$ block. The dotted curve represents the
contribution of the interval $\mathcal{I}_{1}$, the dashed curve
the contribution of the interval $\mathcal{I}_{2}$ and the
dot-dashed curve  the contribution of the interval
$\mathcal{I}_{3}$.} \label{transfud0}
\end{figure}

As shown in fig. \ref{transfud0}, the main  components of the
diffracted field (i.e. $u^{0}-u^{i}$) are those due to propagative
and evanescent waves in the substratum, the latter being
associated with a quasi-Love mode.
\begin{figure}[ptb]
\begin{center}
\includegraphics[width=6.0cm] {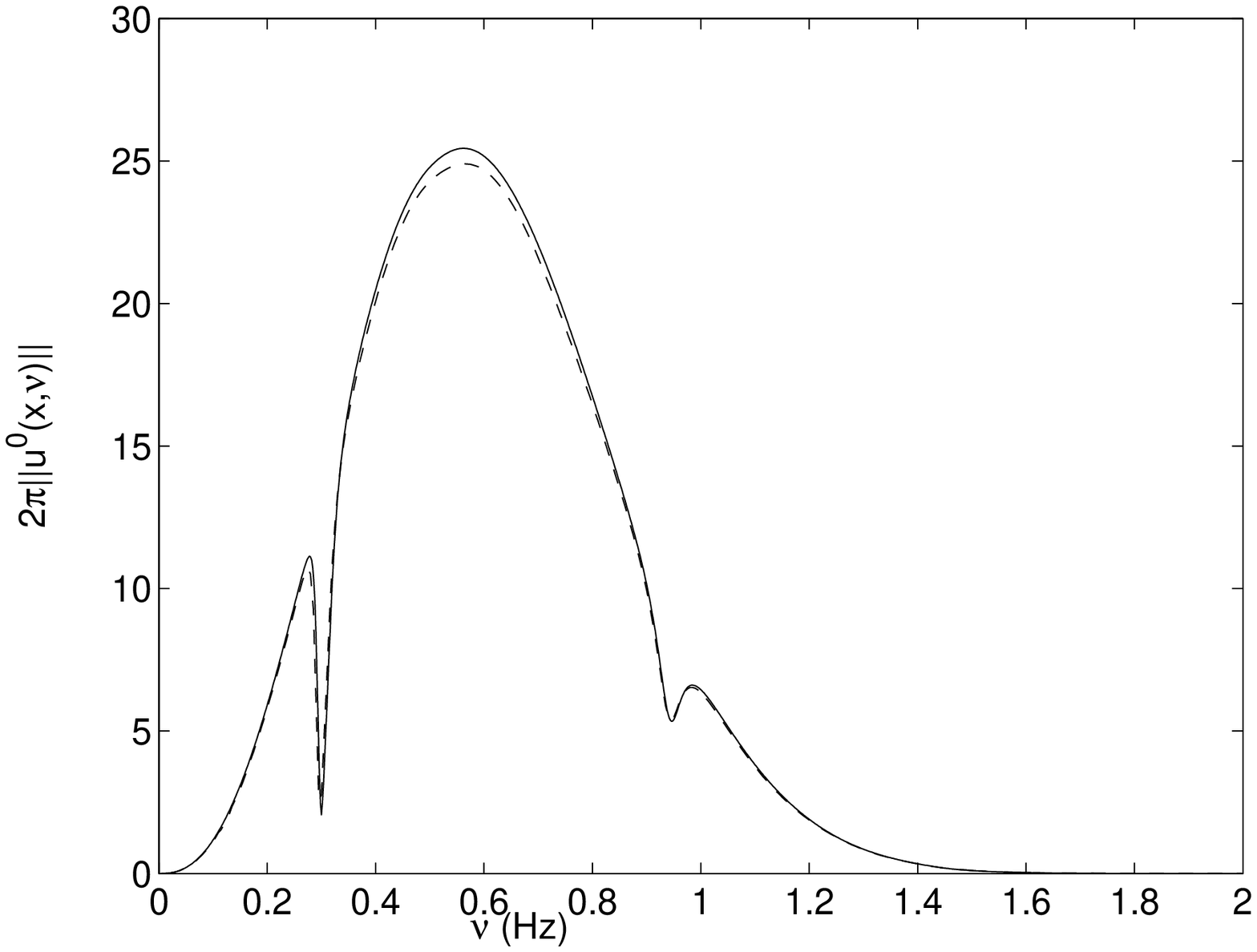}
\includegraphics[width=6.0cm] {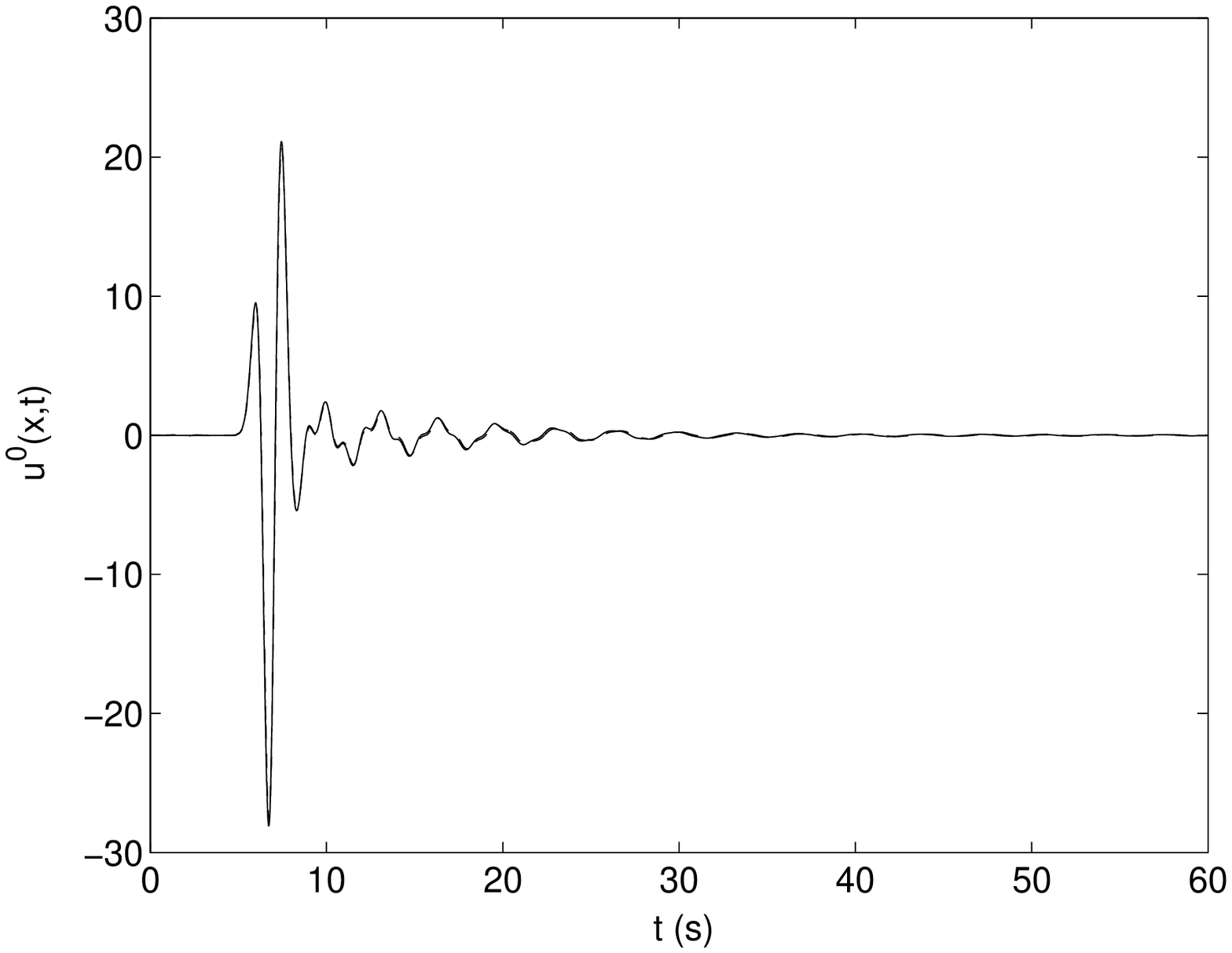}
\end{center}
\caption{$2\pi$ times the spectrum (left panel) and time history
(right panel)  of the total displacement at the interface between
the substratum and the layer below the center of a single
$50m\times 30m$ block. The dashed curves result from the
semi-analytical (mode-matching) computation, whereas the solid
curves result from the numerical (finite element) computation .
Note the splitting of the first peak.} \label{specttimesub0505030}
\end{figure}

The comparison of the results of the mode-matching and finite
element methods is made in
 fig. \ref{specttimesub0505030}.
The time history of the total displacement field is mainly
composed of the incident and  specularly-reflected fields.
Although a quasi-Love mode is excited, it hardly makes itself felt
in the substratum.
\section{Numerical results for two-block configurations in a Mexico City-like site}\label{twobloc}
Let us now consider the configuration involving {\it two blocks},
which is the simplest configuration for studying inter-block
coupling effects. The latter go by the name of
{\textit{structure-soil-structure interaction}}.

The  two blocks, situated in a Mexico City-like environment, are
located such that the center of the base segment of block 1 is
(0m,0m) and that  of block 2 is (-65m,0m).

The parameters of the site are: $\rho^{0}=2000$ kg/m$^{3}$,
$c^{0}$=600 m/s, $Q^{0}=\infty$, $\rho^{1}=1300$ kg/m$^{3}$,
$c^{1}$=60 m/s, $Q^{1}=30$, with the soft layer thickness being
$h=50$ m. The material constants of the blocks are: $\rho^{2}=325$
kg/m$^{3}$, $c^{2}$=100 m/s, $Q^{2}=100$.

The incident  cylindrical wave is radiated by a deep line source
located at (0m, 3000m).

Recall that the eigenfrequencies of the displacement-free base
block  are $\nu_{0m}^{FB}=\frac{c^{2}(2m+1)}{2b}$, and the Haskell
frequencies are $\nu_{m}^{HASK}=\frac{2m+1}{2}\frac{c^{1}}{2h}$,
wherein $m=0,1,2,...$. Thus, the Haskell frequencies are 0.3, 0.9,
1.5 Hz ,...

If the zeroth-order quasi-mode coefficient is relevant, the
dispersion relation of the configuration takes the from:
\begin{equation}
\left(\mathcal{F}_1^{(1)}-\mathcal{F}_2^{(1)} \right)\left(\mathcal{F}_1^{(2)}-
\mathcal{F}_2^{(2)} \right)-\mathcal{F}_2^{(12)}\mathcal{F}_2^{(21)}=\mathcal{F}=0
\end{equation}
wherein $\mathcal{F}_1^{(j)}-\mathcal{F}_2^{(j)}=0$, $j=1,2$ is
the dispersion relation  of the configuration with one block of
characteristics of the block $j$ (see eq.\ref{disrel1b}) and the
term
$\mathcal{F}_2^{(12)}\mathcal{F}_2^{(21)}=\left(\cos\left(k^{2}b_1\right)Q_{00}^{(12)}\right)
\left(\cos\left(k^{2}b_2\right)Q_{00}^{(21)}\right)$ accounts for
the coupling between the two blocks.
\subsection{Results relative to a two-block configuration consisting of a $50m\,\times\, 30m$ block and
a $40m\,\times\, 40m$ block}
Block 1 is $50m$ high and $30m$ wide and block 2 is  $40m$ high
and $40m$ wide. Their center-to-center separation  is $65m$.

The displacement-free base block eigenfrequencies are:
$\nu_{00}^{DFB}=0.5\mbox{, }1.5Hz...$ for block 1, and
$\nu_{00}^{DFB}=0.625\mbox{, }1.875Hz...$ for  block 2.

\begin{figure}[ptb]
\begin{center}
\includegraphics[width=8cm] {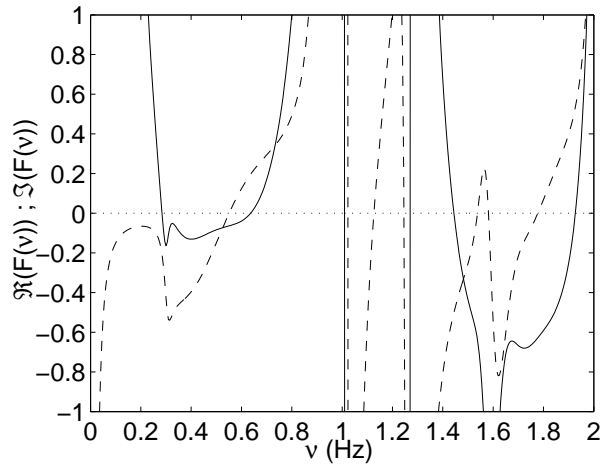}
\end{center}
\caption{Indications concerning the dispersion relation for a
system of two blocks, one with dimensions $50m\,\times\, 30m$ and
the other with dimensions $40m\,\times\, 40m$. The solid curve
represents $\Re(\mathcal{F})$ and the dashed curve
$|\Im(\mathcal{F})|$.} \label{reim50304040}
\end{figure}
Figure \ref{reim50304040} gives an indication of the frequency of
occurrence of the modes of the system. An eigenfrequency is a
frequency for which $\Re\left(\mathcal{F} \right)=0$. The
attenuation associated with a particular mode (at a frequency
$\nu^{\star}$) is given by
$\Im\left(\mathcal{F}(\nu^{\star})\right)$.

A quasi-Love mode is excited at $\nu\approx 0.3$. We could expect
two different quasi displacement-free base block modes for a
system with two non-identical blocks (at $\nu_{00}^{QDF}\approx
0.5$ and $0.625Hz$), but only one eigenfrequency, at $\approx
0.7Hz$, is found (actually this value is debatable, since the the
eigenfrequency might actually correspond to a minimum of
$\|\mathcal{F}\|$ rather than to a zero of $\mathcal{F}$). This
means that: i) the modes of a complex configuration are not the
union of the modes of the subsystems of which it is composed, ii)
the single quasi displacement-free block mode results from the
inter block coupling of the fields, i.e., the
(\textit{structure-soil-structure interaction}), iii) this mode is
a {\textit{coupled mode}} and is associated with a smaller
attenuation than either of the corresponding modes of the two
associated one-block configurations, iv) the correct expression of
the modes of the system is the one given in section
\ref{anotherquasisfb} involving the couplings between the blocks.
In particular, the coupling matrix $Q_{00}^{(ij)}$, $i\neq j$
cannot be neglected.

To distinguish this mode from the quasi displacement-free base
block mode for a single block that accounts only for the
structure-soil interaction, we  call it the
\textit{multi-displacement-free base block mode}.
\subsubsection{Response on the top and bottom segments
of the blocks}\label{twobloc50304040}
Now consider the response at the centers of the top segments of
the two blocks.
\begin{figure}[ptb]
\begin{center}
\includegraphics[width=6.0cm] {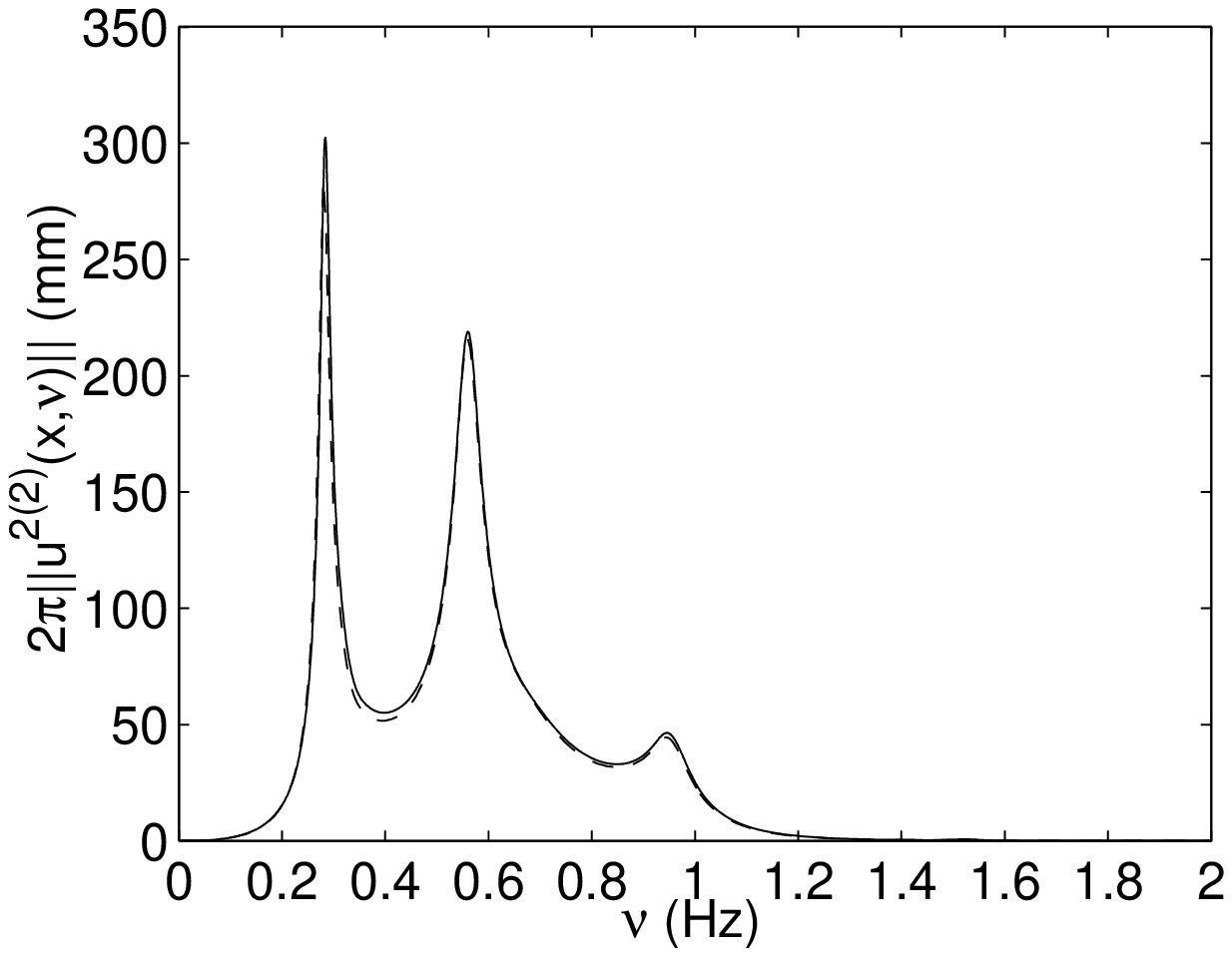}
\includegraphics[width=6.0cm] {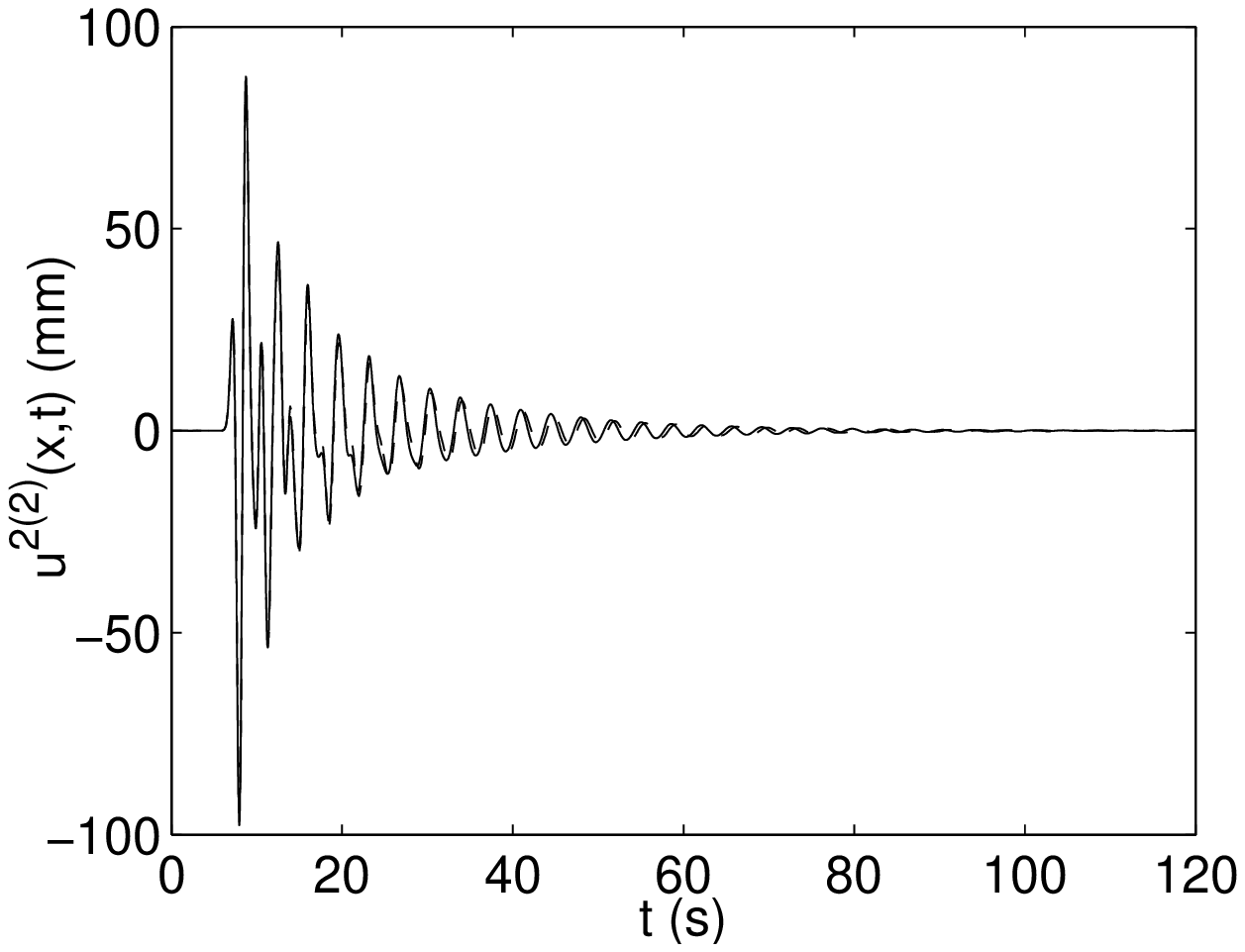}
\includegraphics[width=6.0cm] {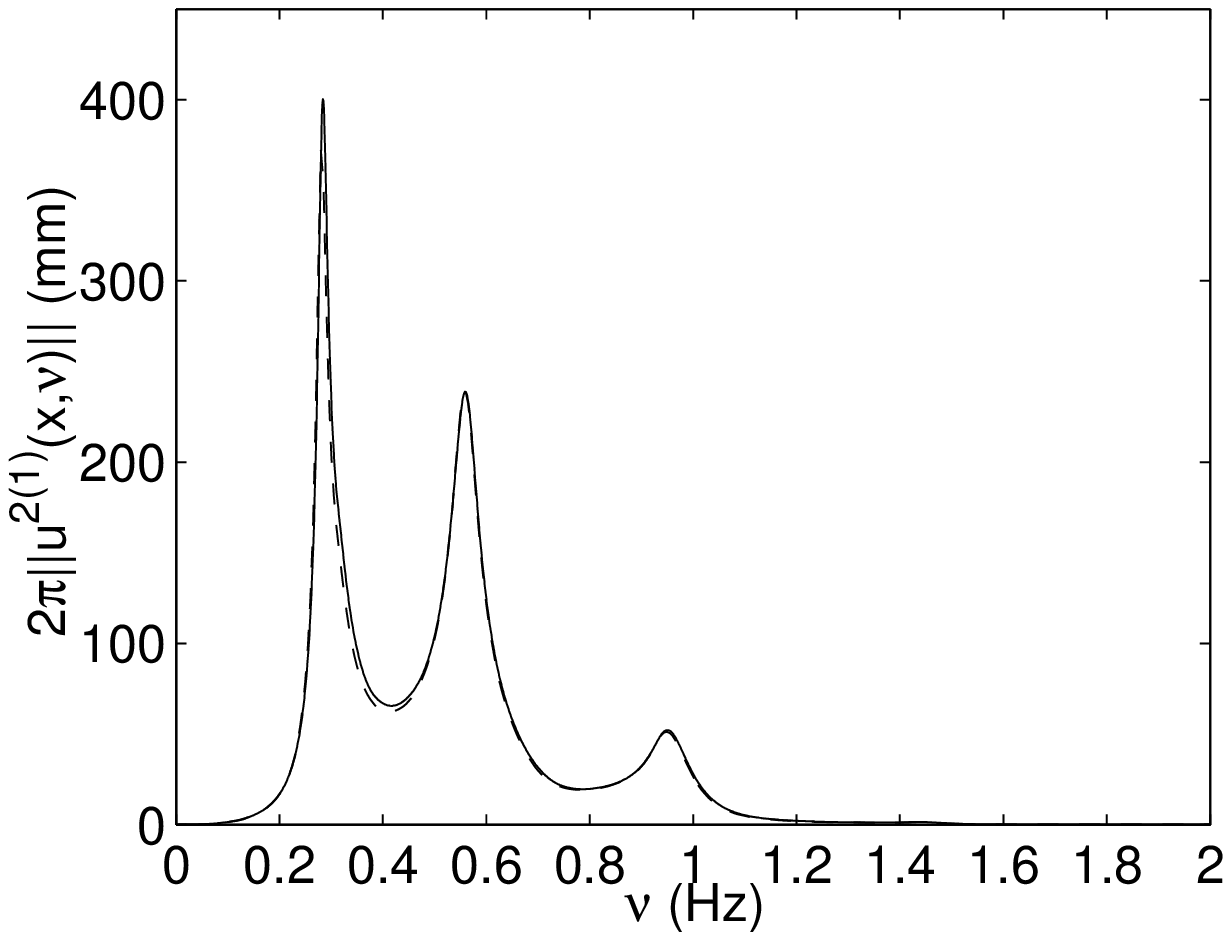}
\includegraphics[width=6.0cm] {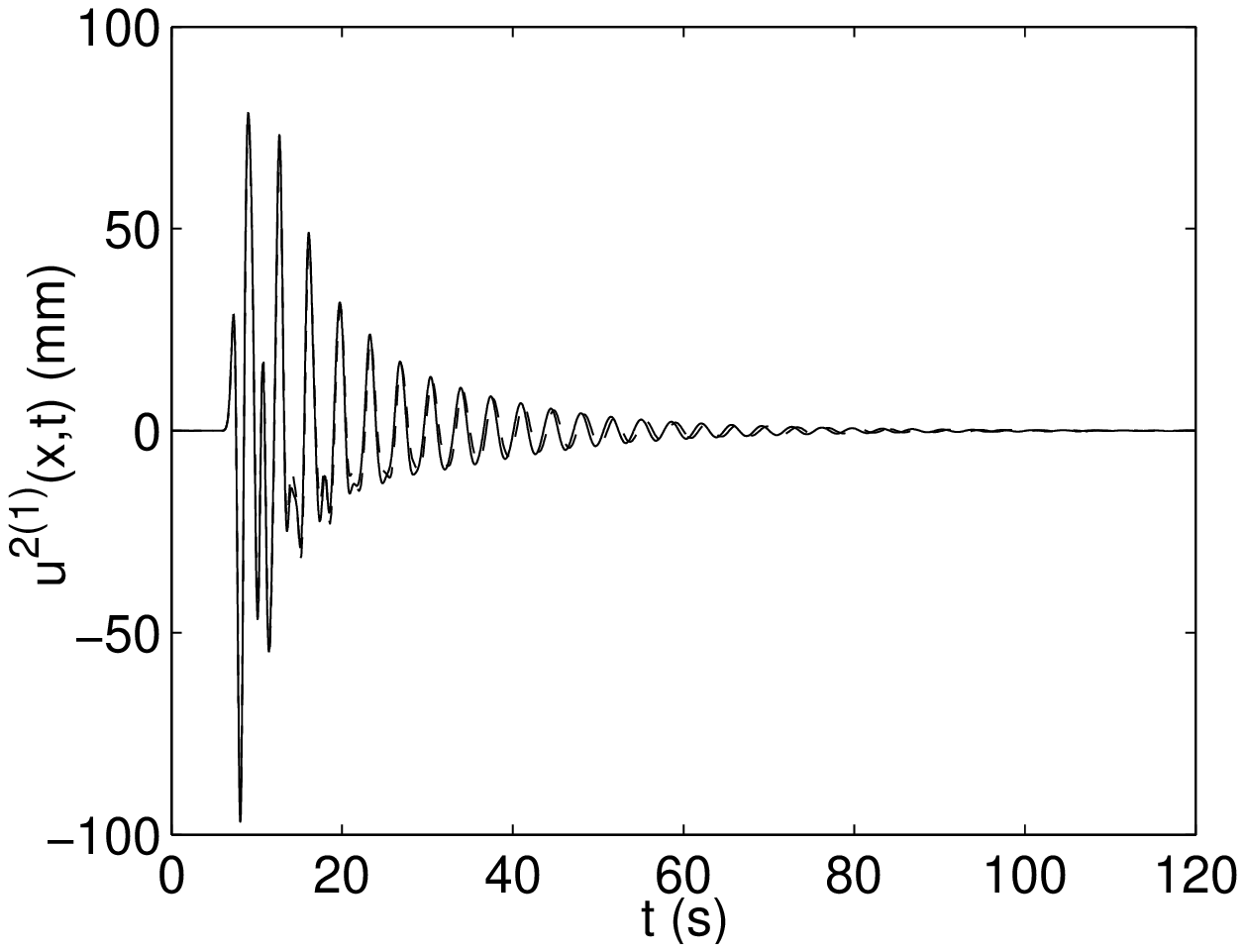}
\end{center}
\caption{$2\pi$ times the spectrum (left panels) and time
histories (right panels) of  the total displacement of a
configuration with two blocks for a deep seismic line source
located at $\mathbf{x}^s=(0,3000m)$. The top panels apply to the
field at the center of the top segment of block 2 ($40\times 40$
block). The lower  panels apply to the field at the center of the
top segment of block 1 ($50\times 30$ block). The dashed curves
result from a computation via the semi-analytical (quasi-mode)
method (taking account only of the zeroth-order quasi-mode in each
block), whereas the solid curves result from a finite element
computation.} \label{specttimesum50304040}
\end{figure}
The responses at these locations, as computed by the finite
element and mode matching methods (with account taken only of the
zeroth-order quasi-mode in each building), are seen in fig.
\ref{specttimesum50304040} to be almost identical in each block.
The peak at $\nu_{00}^{MDF}\approx 0.6Hz$, translates the
excitation of the multi-displacement-free base block mode. This
peak is sharper than the one encountered for only one block at
$\nu_{00}^{QDF}$ (due to excitation of the quasi displacement-free
base block mode), which fact is essentially due to its larger
amplitude. The important sharpness of this peak is also related to
the fact that the excited mode is a {\textit{coupled mode}}. This
suggests that the larger the number of blocks, the larger will be
the response at the resonance frequency of the
multi-displacement-free base block mode.

\begin{figure}[ptb]
\begin{center}
\includegraphics[width=6.0cm] {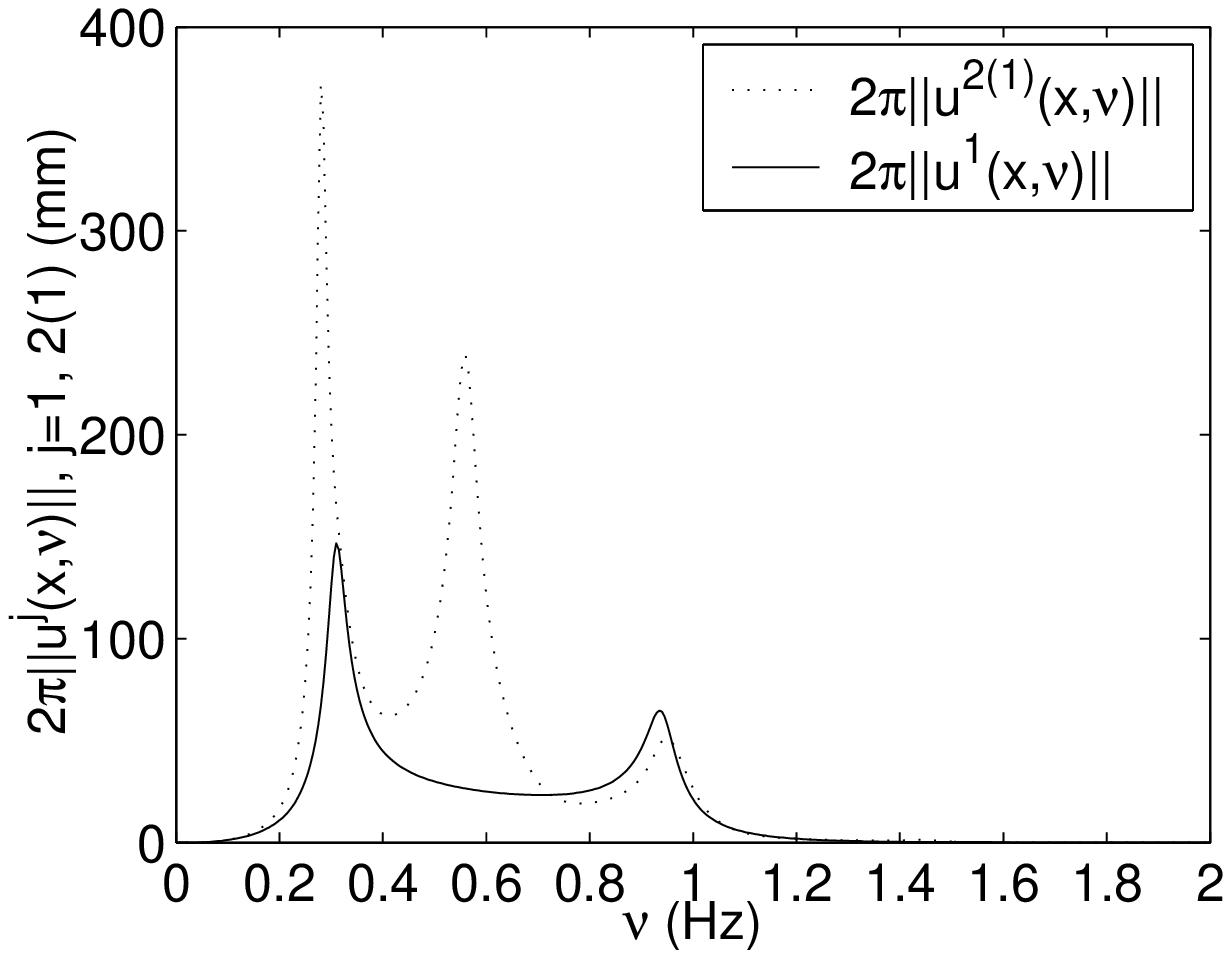}
\includegraphics[width=6.0cm] {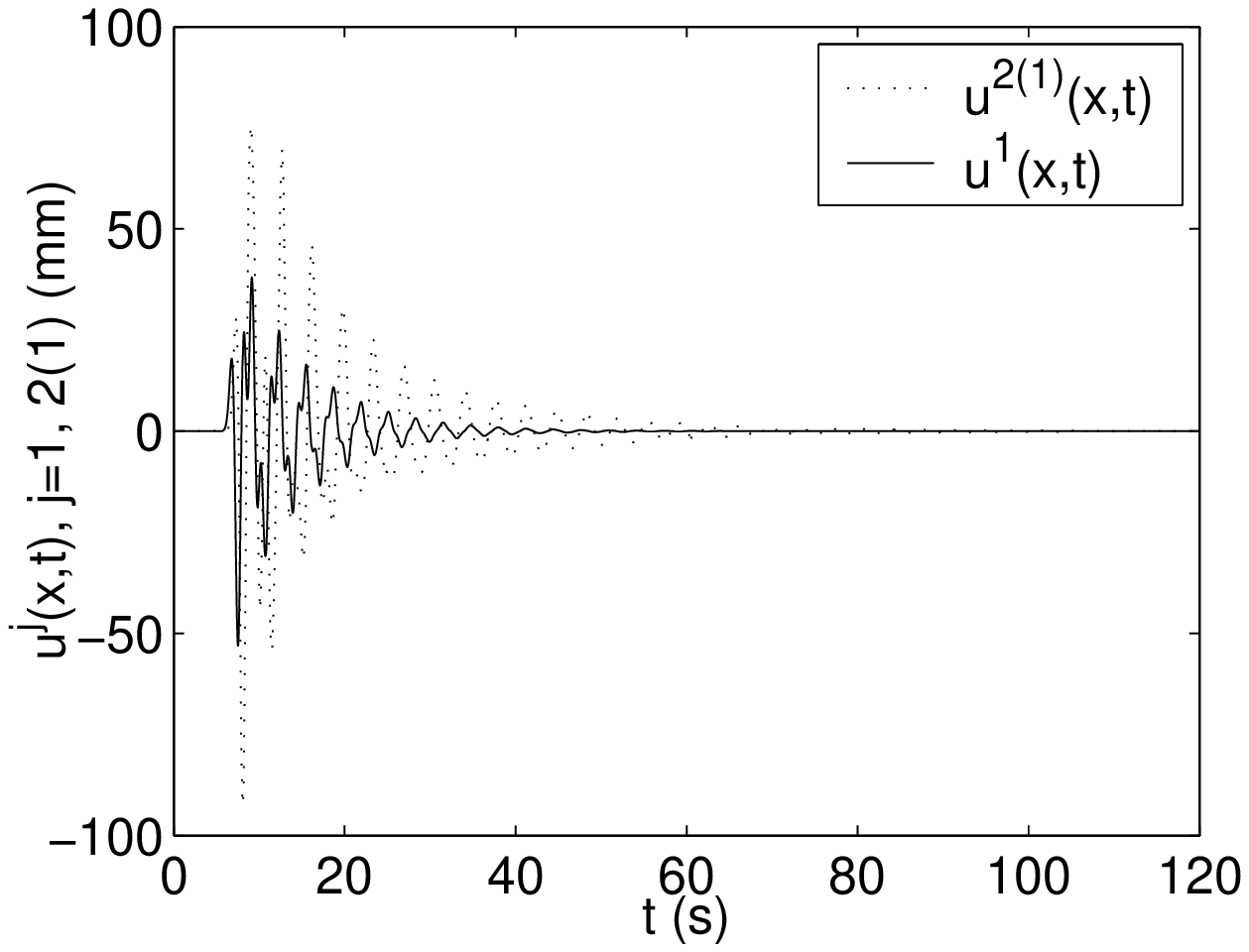}
\includegraphics[width=6.0cm] {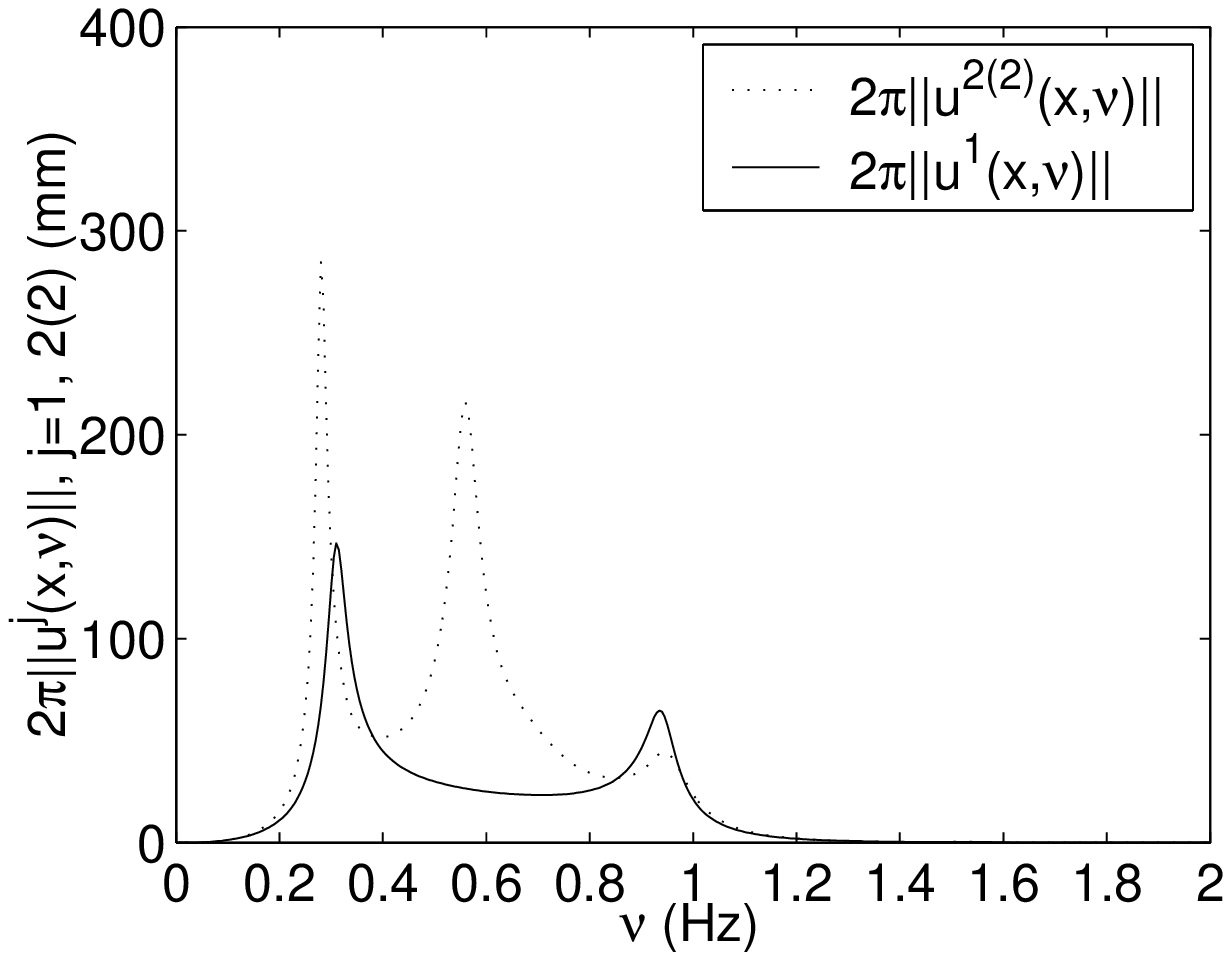}
\includegraphics[width=6.0cm] {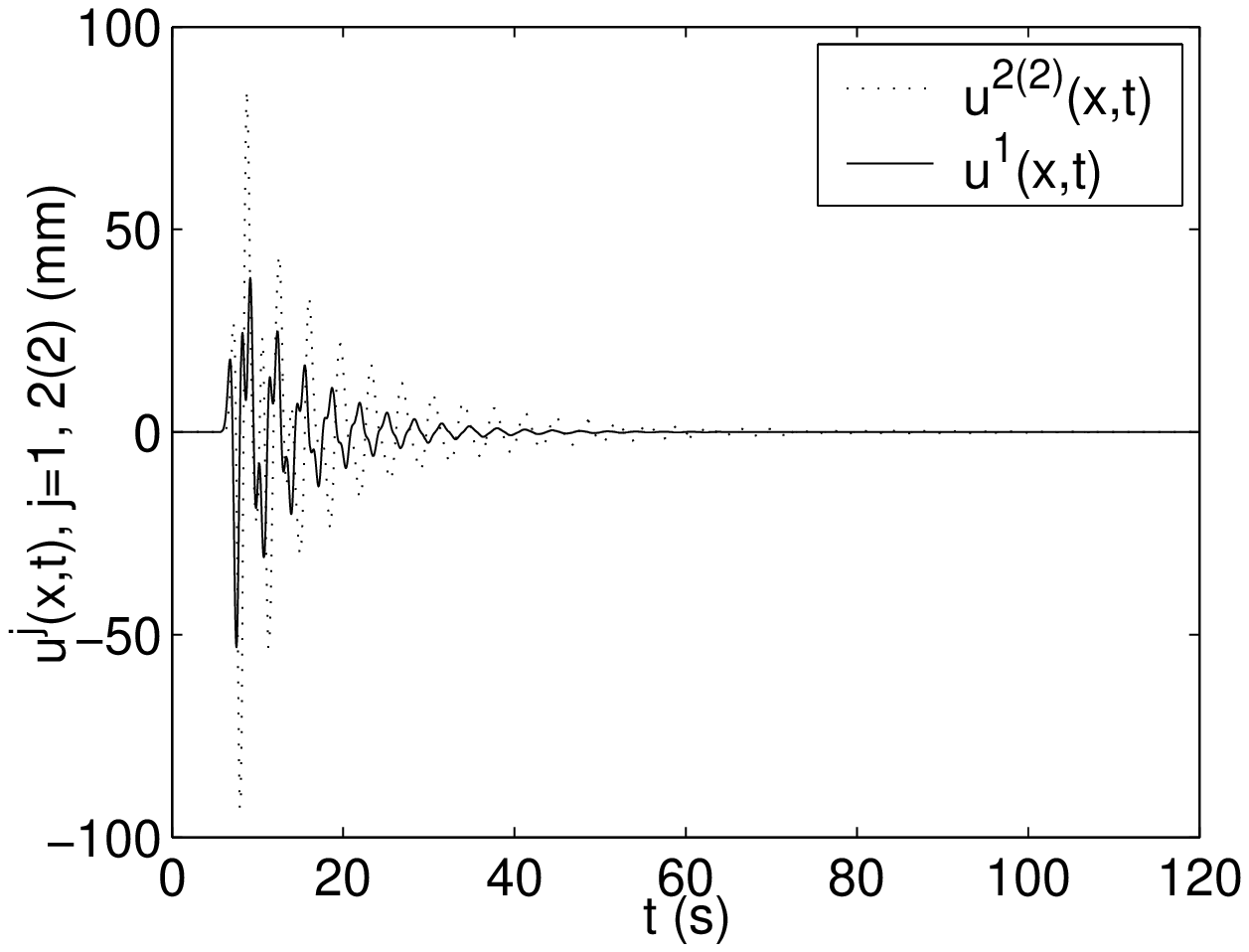}
\includegraphics[width=6.0cm] {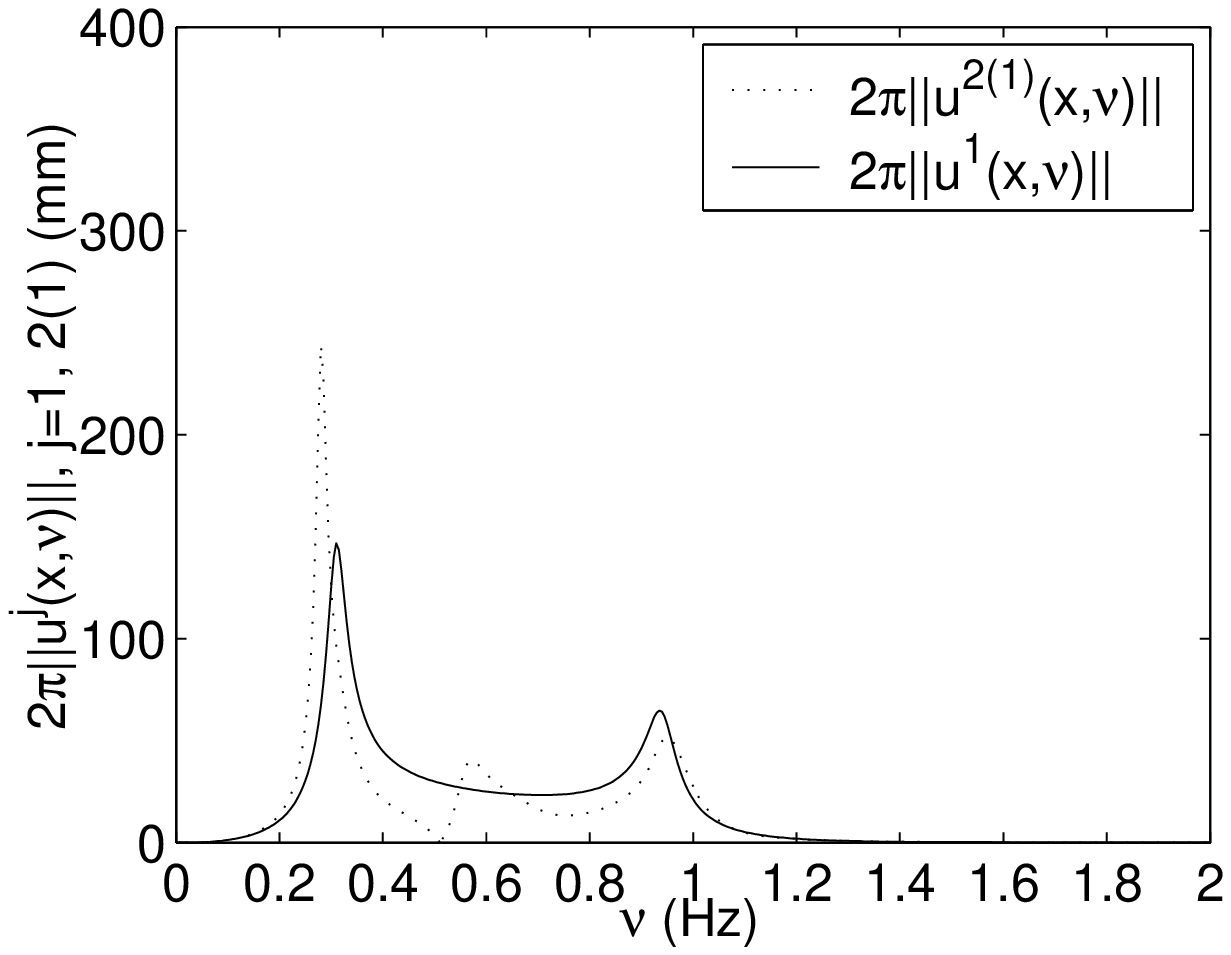}
\includegraphics[width=6.0cm] {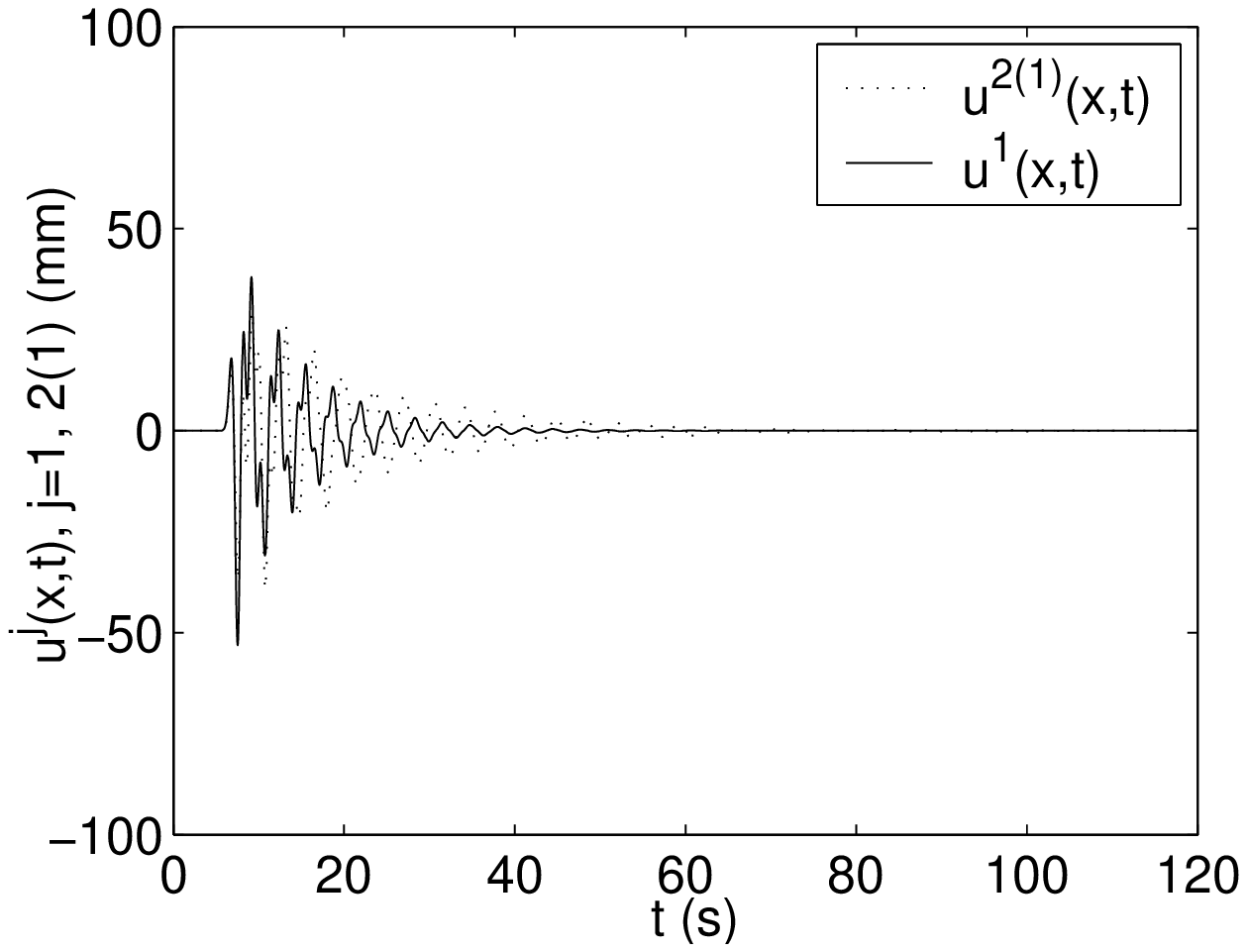}
\includegraphics[width=6.0cm] {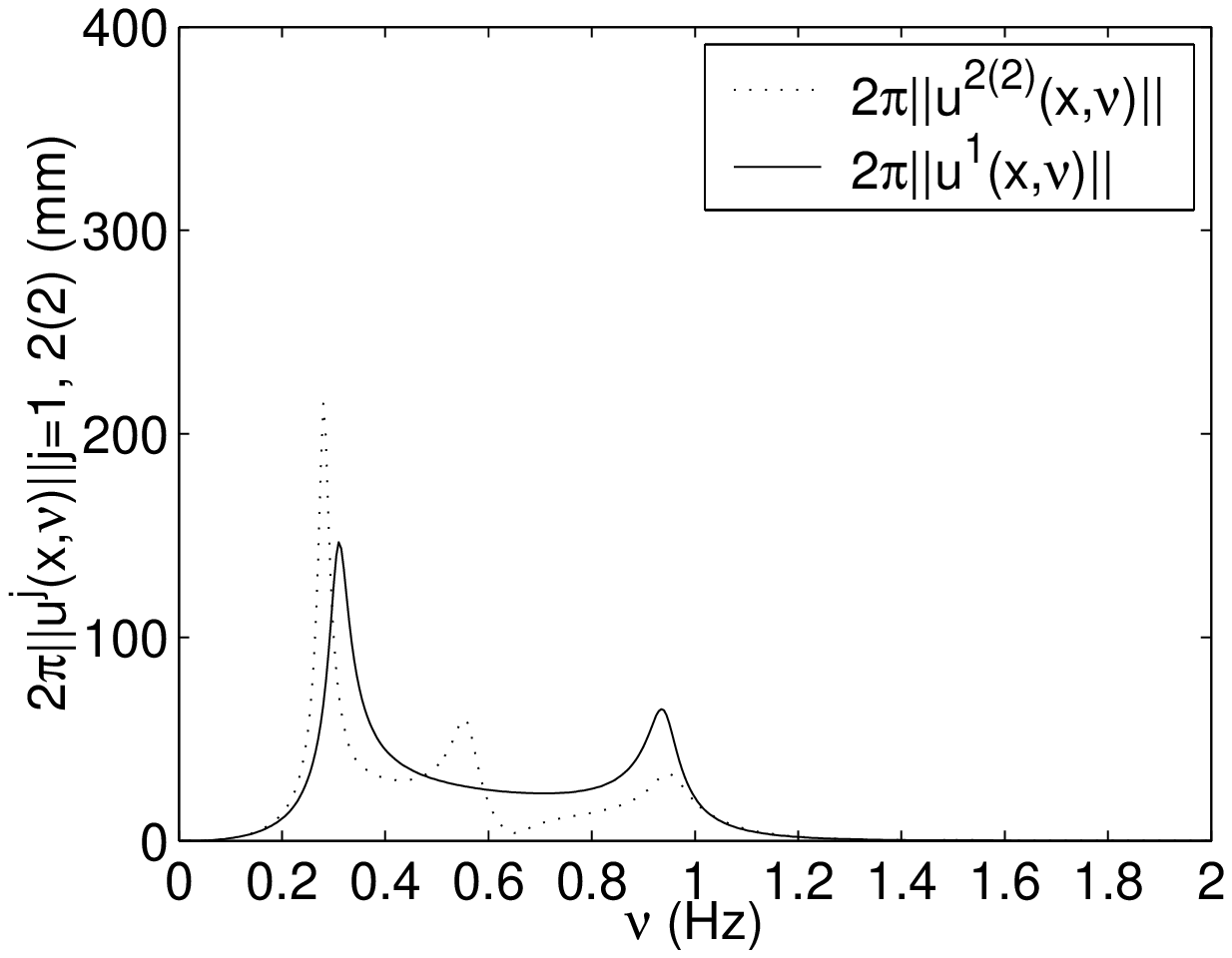}
\includegraphics[width=6.0cm] {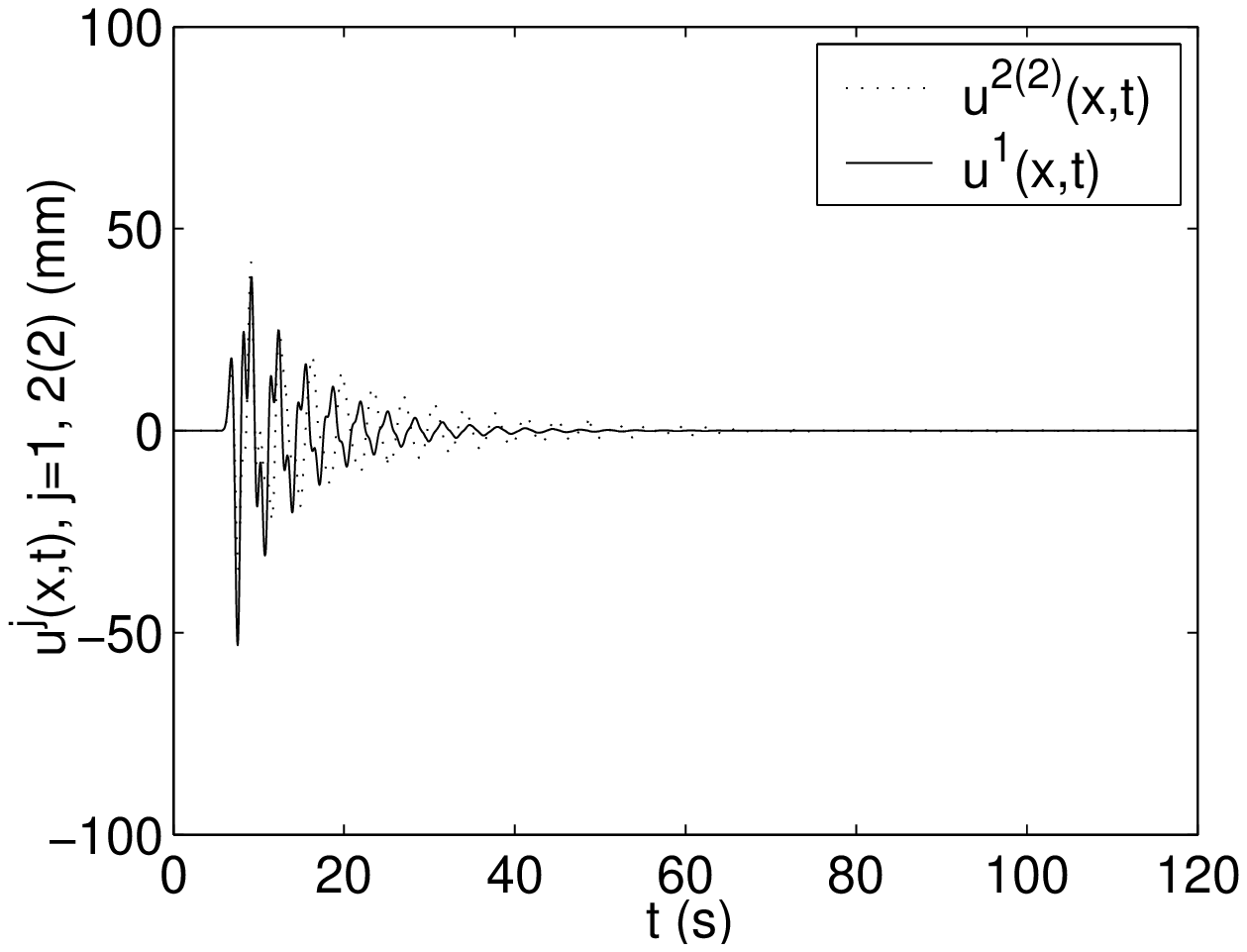}
\end{center}
\caption{Comparison of $2\pi$ times the  spectrum (left panels)
and
 time history (right panels) of the total displacement on the
ground in absence of blocks (solid curves) and in the presence of
blocks (dashed curves). From the top to bottom: at the center of
the summit segment of block $1$, at the center of the summit
segment of block $2$, at the center of the base segment of block
$1$, and at the center of the base segment of block $2$. The deep
line source is located at $\mathbf{x}^s=(0,-3000m)$.}
\label{compspecttime50304040}
\end{figure}

We compare in fig. \ref{compspecttime50304040} the results
obtained in (at the centers of the top and bottom segments of) the
two blocks with those on the ground (at the same locations as the
centers of bottom segments of the blocks) in the absence of these
blocks. In the time domain, a small increase of the duration and a
more substantial increase  of the peak and cumulative amplitudes
can be observed, in particular on the top segments of the two
blocks. These increases are more important than in the
configurations with a single  $40m\,\times \, 40m$ or $50m\,\times
\, 30m$ block (see figs. \ref{compspecttime4040} and
\ref{compspecttime5030}), and is also somewhat more important in
block 1 than in block 2, as  already noticed in the cases of
single blocks. The field vanishes at the center of the bottom
segments at the displacement-free base frequencies: $\nu=0.5Hz$
for block 1 and  $\nu=0.625Hz$ for block 2. This emphasizes the
fact that the multi displacement-free  block mode is different
from the quasi displacement-free block mode of the configuration
with only one block, and is a manifestation of geometrical
features.
\subsection{Results relative to a two-block configuration with two $40m\,\times\, 40m$ blocks}\label{twobloc4040}
The two blocks are $40 m$ high and $40 m$ wide and their
center-to-center separation is 65m.

The displacement-free base block eigenfrequencies of the two
blocks are 0.625, 1.875 Hz ...
\begin{figure}[ptb]
\begin{center}
\includegraphics[width=8cm] {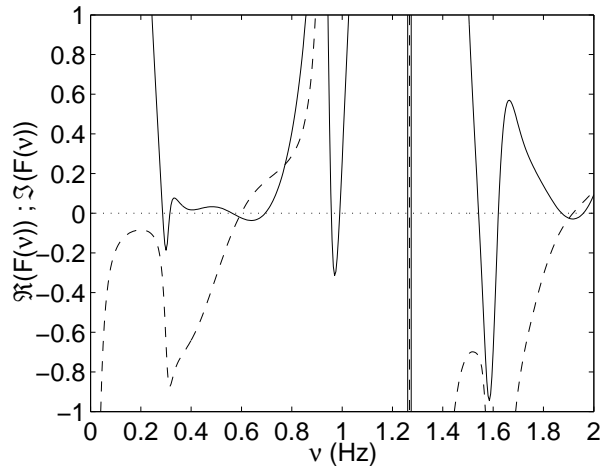}
\caption{Indications concerning the dispersion relation for a
configuration with  two $40m\,\times\, 40m$ blocks. The solid
curve represents $\Re(\mathcal{F})$ and the dashed curve
 $\Im(\mathcal{F})$.} \label{reim2b4040}
\end{center}
\end{figure}
Fig. \ref{reim2b4040} gives an indication  of the solutions of the
dispersion relation $\mathcal{F}=0$. The eigenfreqencies
corresponding to  $\Re\mathcal{F}=0$, are seen to occur at $0.3$,
$0.35$, $0.6$, $0.7$, $0.9$, $1.0$Hz... Some of them are close to
each other and can be gathered together into the groups:
$0.3-0.35$, $0.6-0.7$ and $0.9-1.0$ $Hz$. This suggests a
splitting associated with the lifting of the degeneracy of
eigenvalues.  Nevertheless, the attenuations of the modes at
$0.35$, $0.7$ and $0.9-1.0$ are very large, so that we can expect
them to be hardly excited (as will be confirmed in the following
results).
\subsubsection{Response on the top and bottom segments of the
blocks}
\begin{figure}[ptb]
\begin{center}
\includegraphics[width=6.0cm] {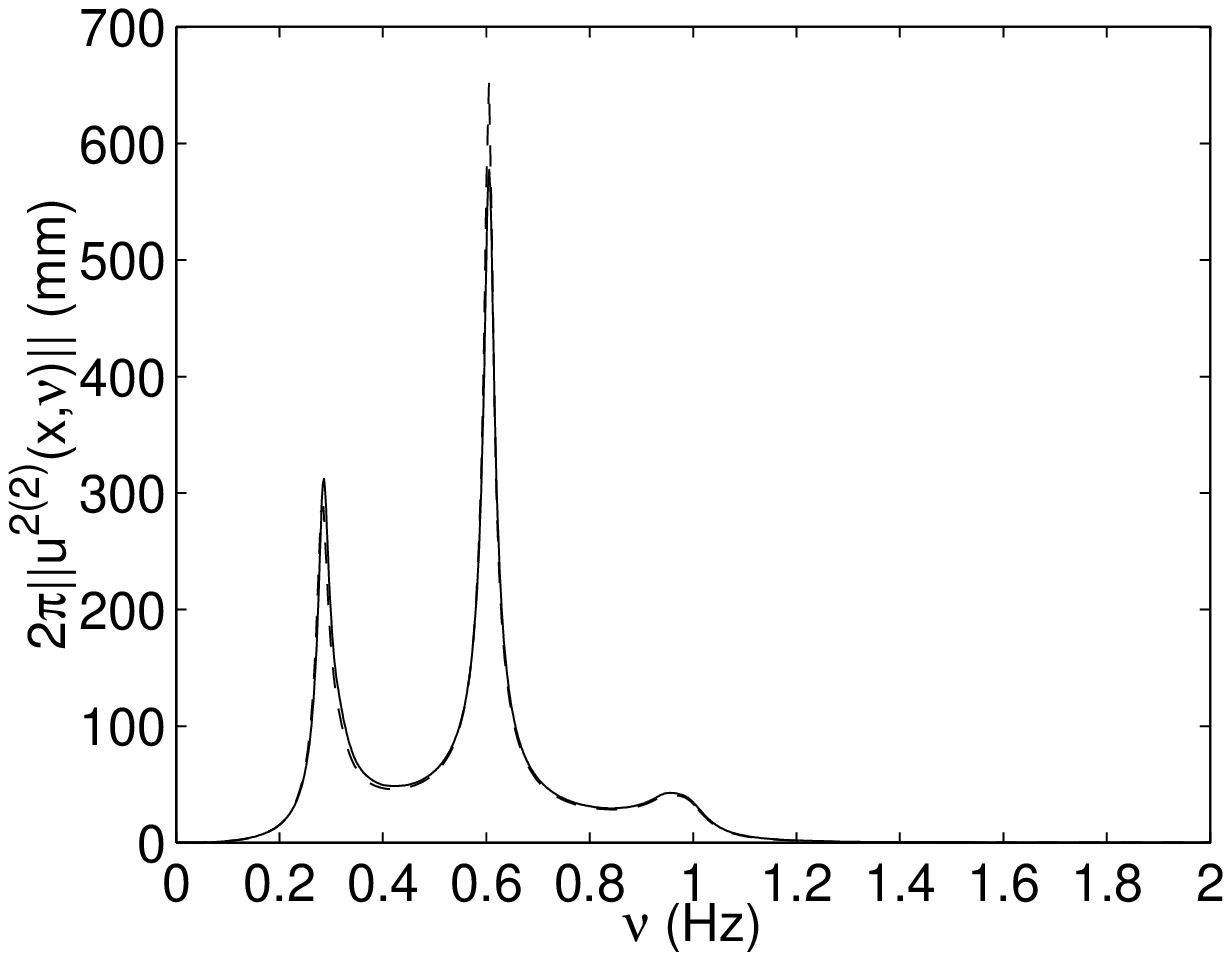}
\includegraphics[width=6.0cm] {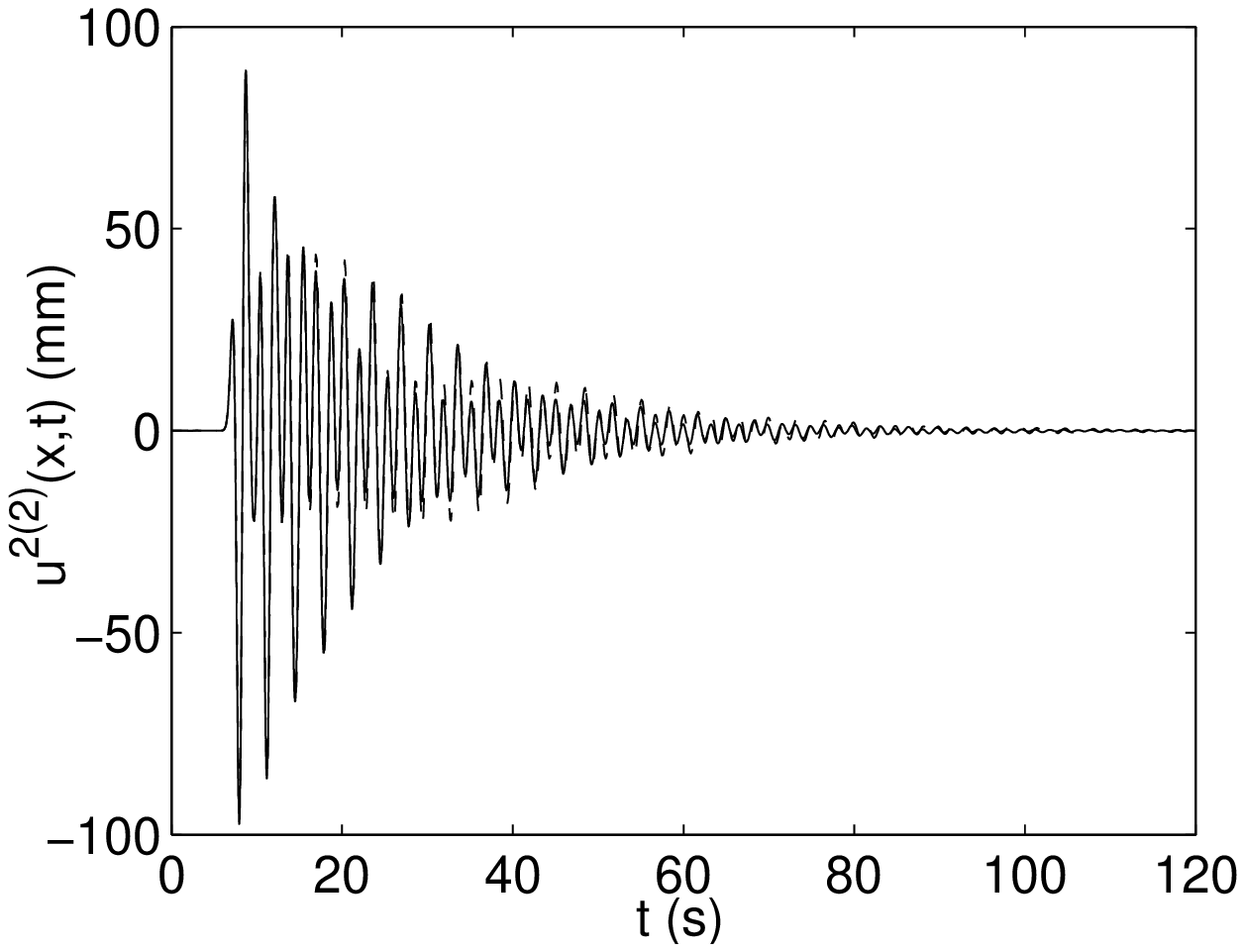}
\includegraphics[width=6.0cm] {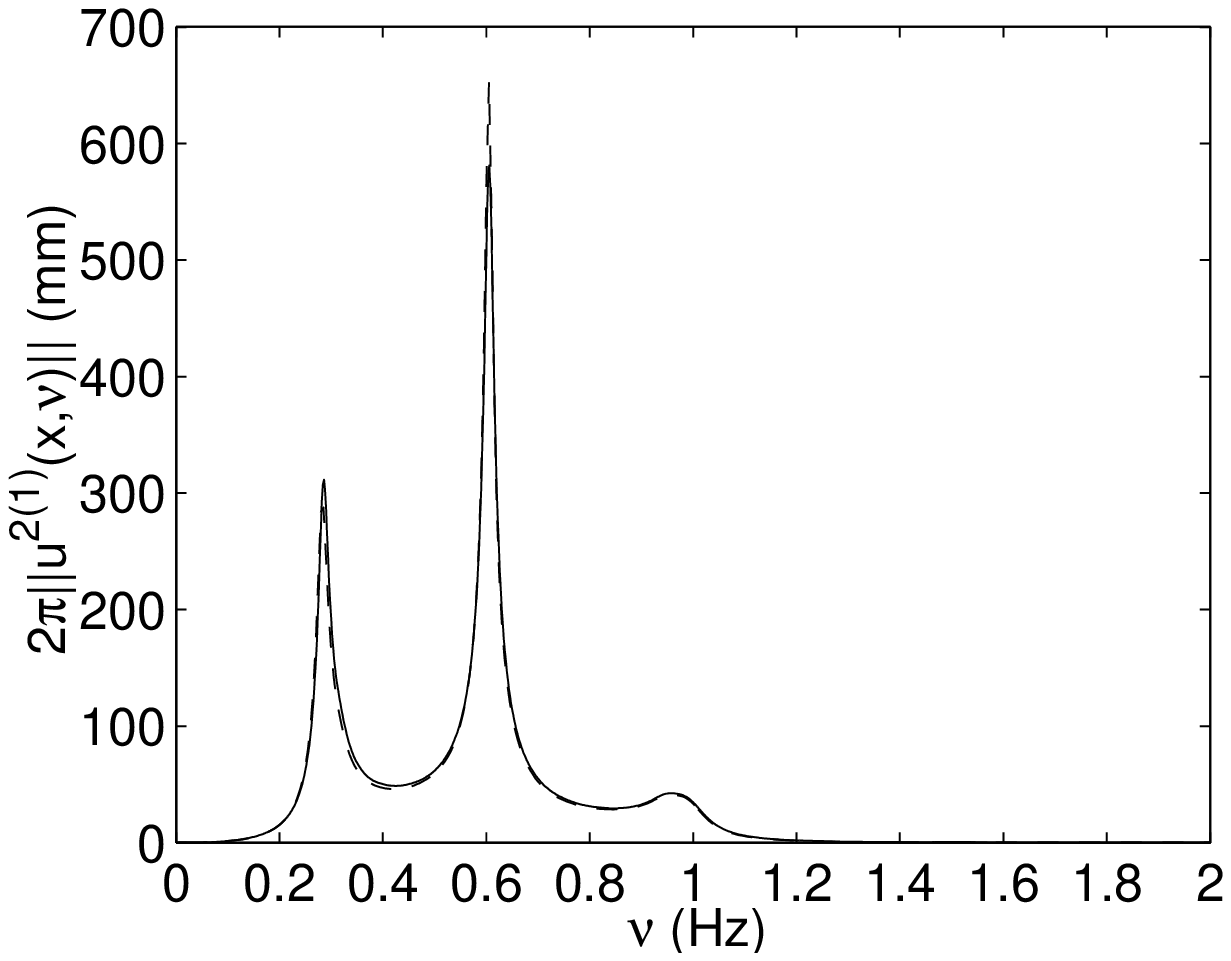}
\includegraphics[width=6.0cm] {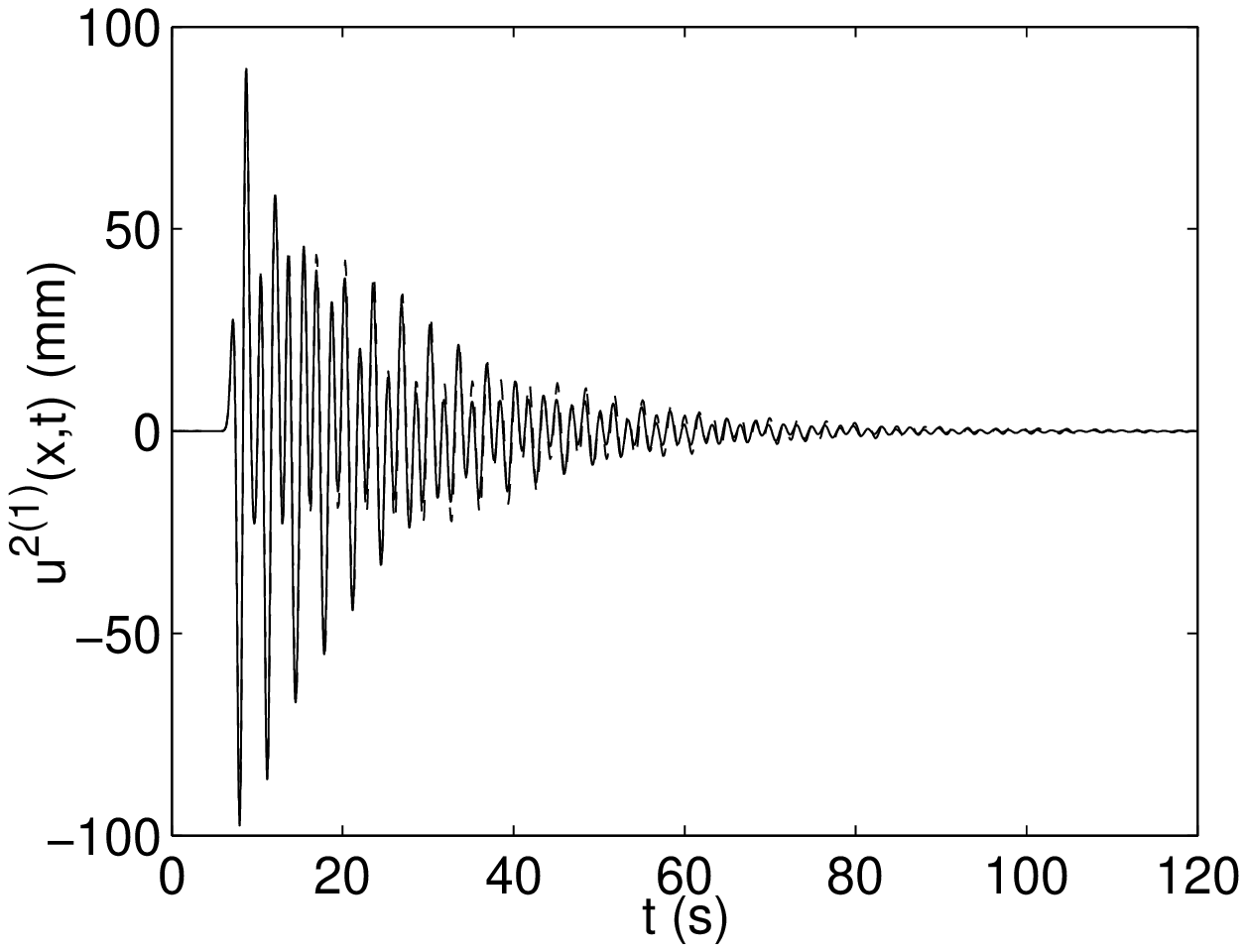}
\end{center}
\caption{$2\pi$ times the spectra (left panel) and time histories
(right panel) of the total displacement at the center of the
summit segment of block 2 (top panels), and at the center of the
summit segment of block 1 (bottom panels), as computed by the
mode-matching method (accounting only for the zeroth-order
quasi-mode) (dashed curves) and by the finite element method
(solid curves) for a configuration with  two $40m\,\times\, 40m$
blocks solicited by the cylindrical wave radiated by a deep line
source.} \label{specttimesum2b4040}
\end{figure}

In fig. \ref{specttimesum2b4040} we see that the results obtained
by the two computational methods pertaining to the response at the
center of the summit segments of the two $40m\,\times\, 40m$
blocks  are nearly identical. The amplitude of the
multi-displacement-free base block resonance peak is higher for
two identical blocks, than it was for two different blocks.  As
pointed out during the previous discussion of the dispersion
relation of the configuration, no splitting of the first and
second peaks is noticed because of the high attenuation of one of
each pair of split modes. On the contrary, the large response due
to the excitation of the other one of the pairs of split modes is
due to the small attenuation of these modes. Notice also that the
second (quasi displacement-free base mode) peaks in the spectra of
the figure are larger than the first peaks, in contrast to what
was obtained for two dissimilar blocks.

\begin{figure}[ptb]
\begin{center}
\includegraphics[width=6.0cm] {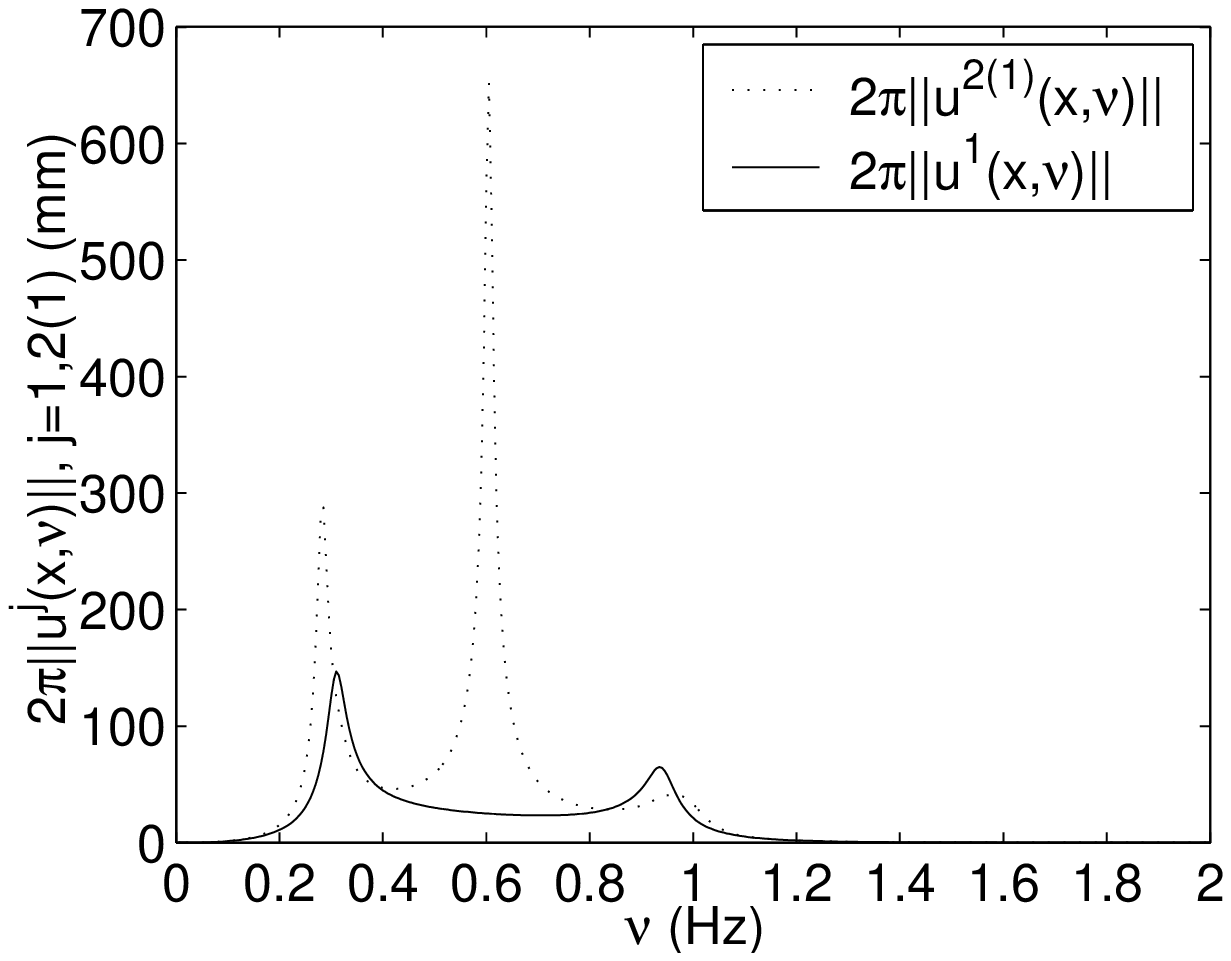}
\includegraphics[width=6.0cm] {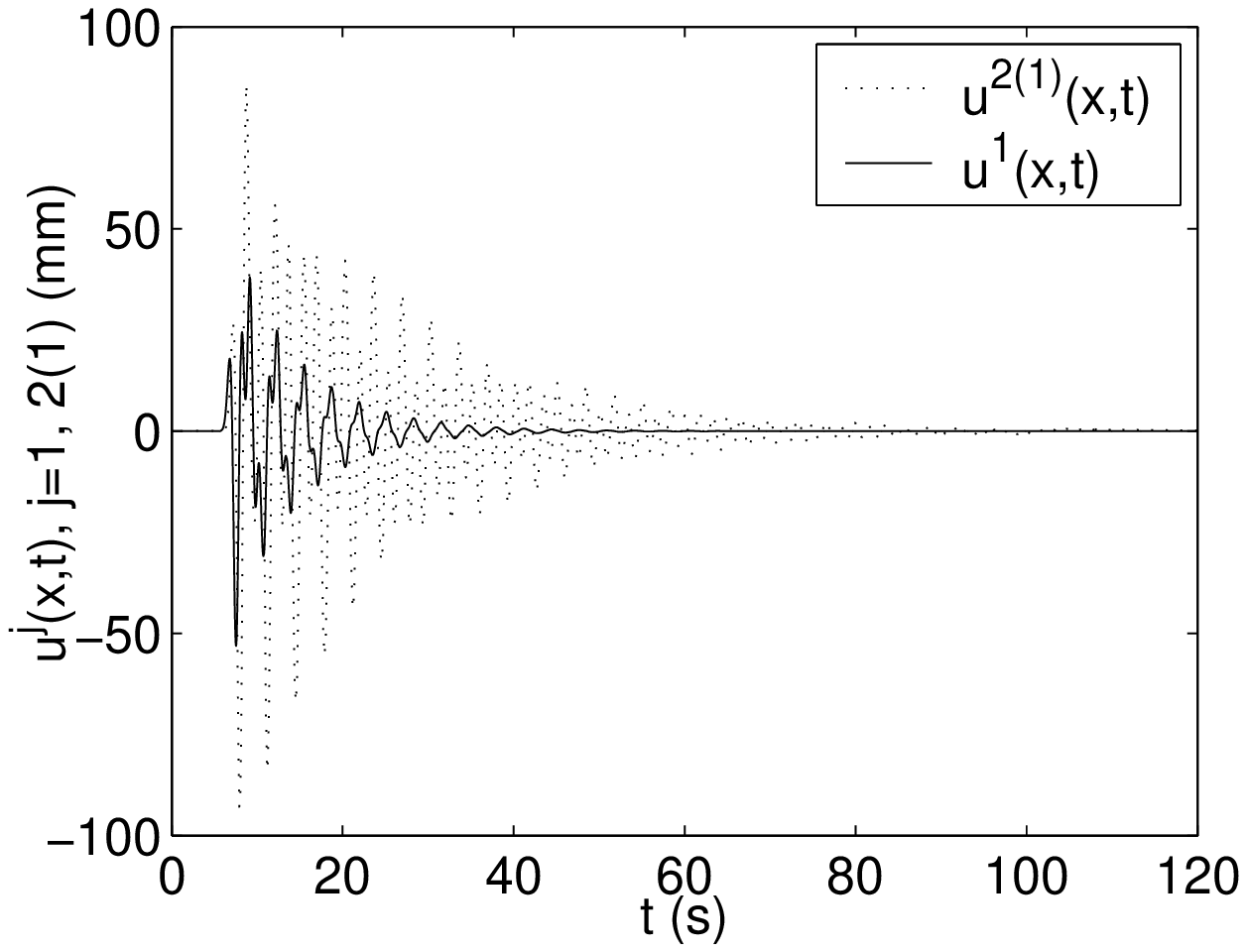}
\includegraphics[width=6.0cm] {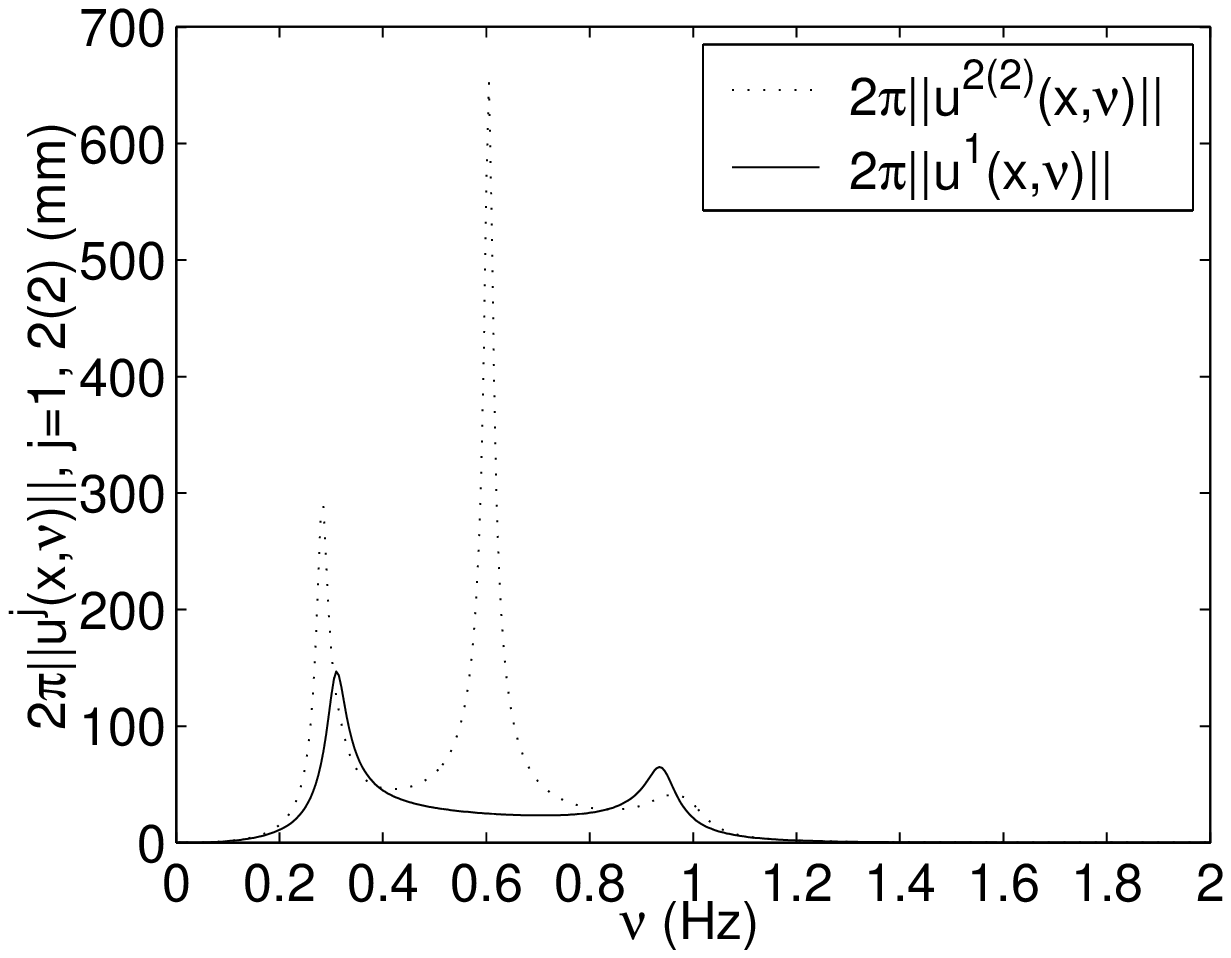}
\includegraphics[width=6.0cm] {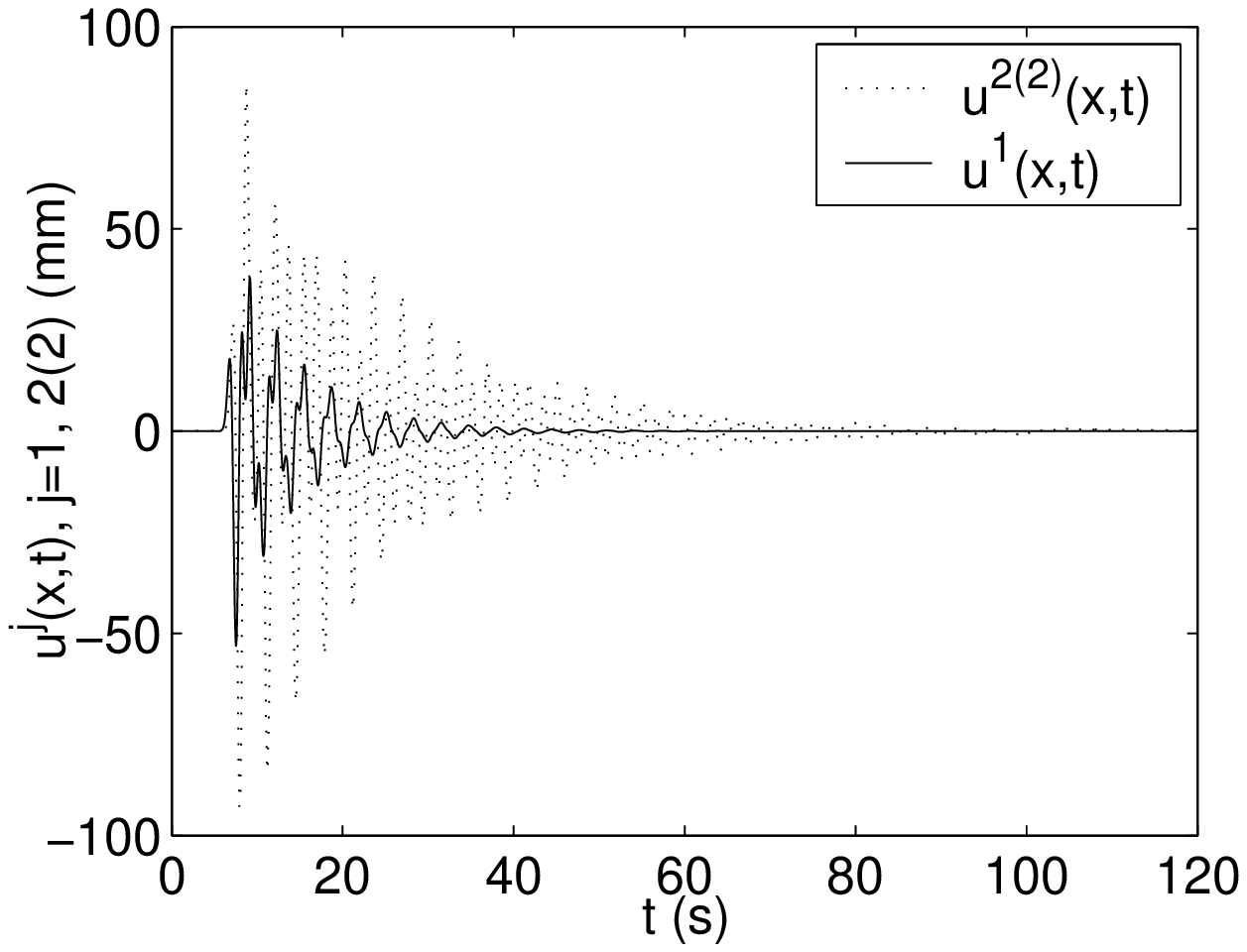}
\includegraphics[width=6.0cm] {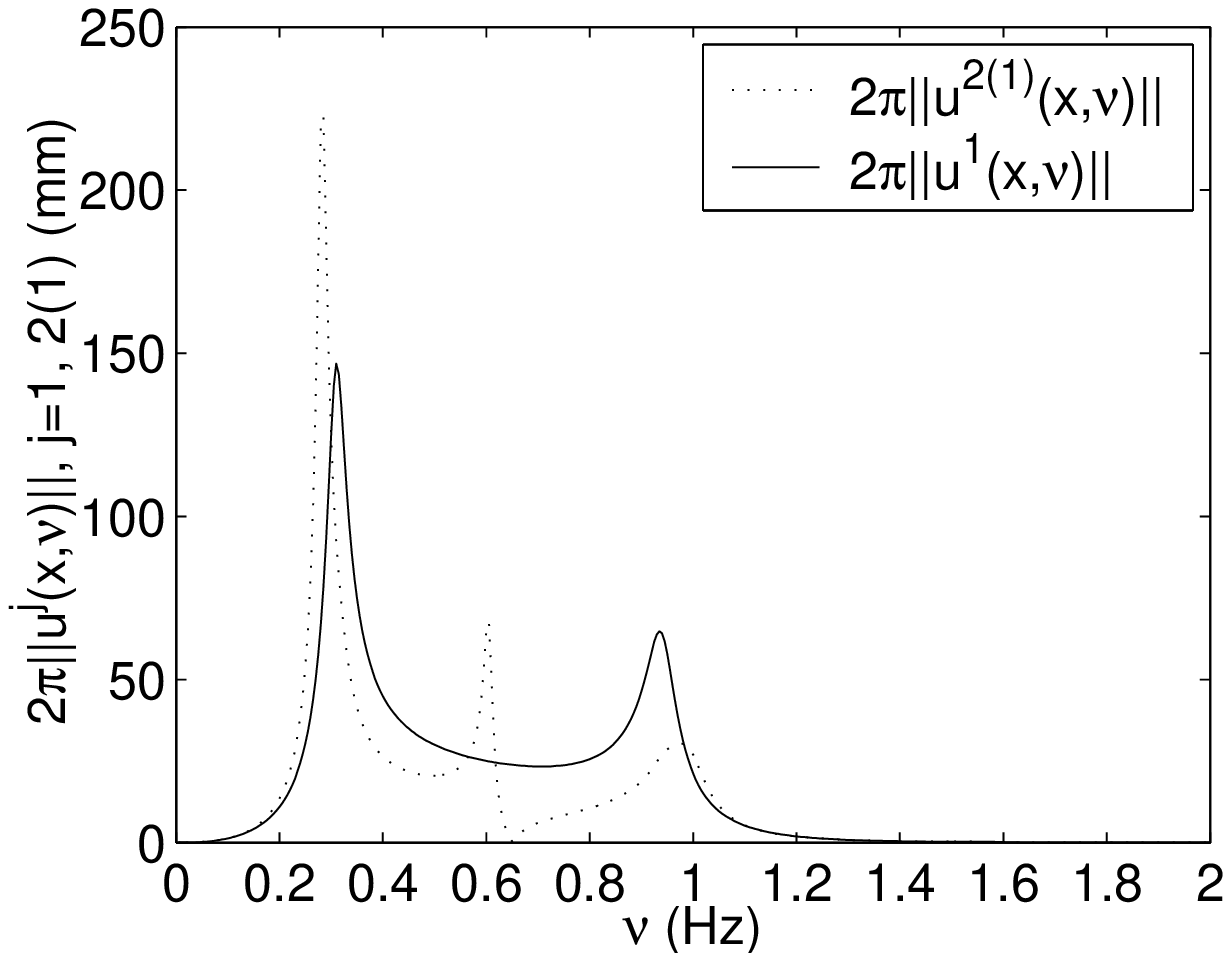}
\includegraphics[width=6.0cm] {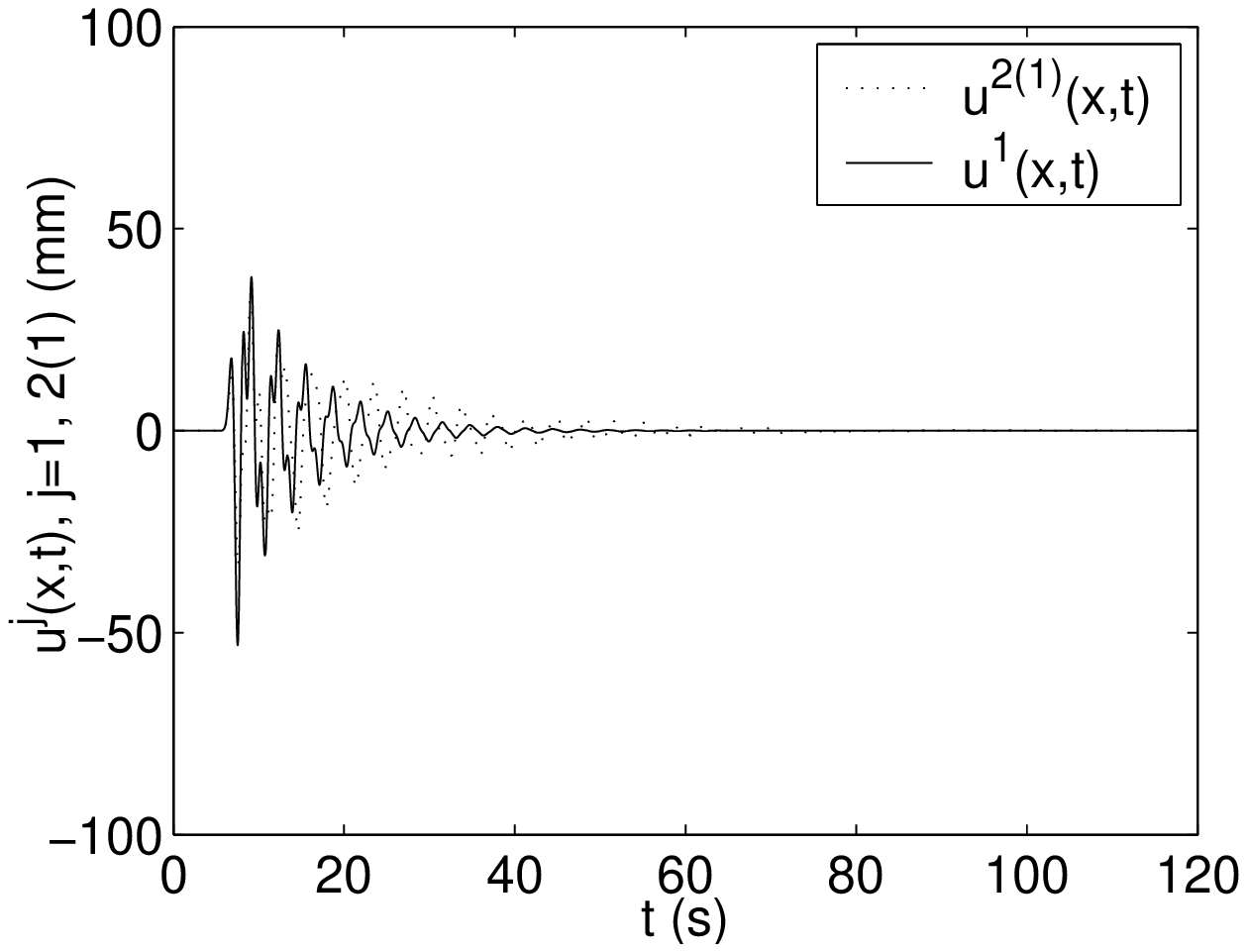}
\includegraphics[width=6.0cm] {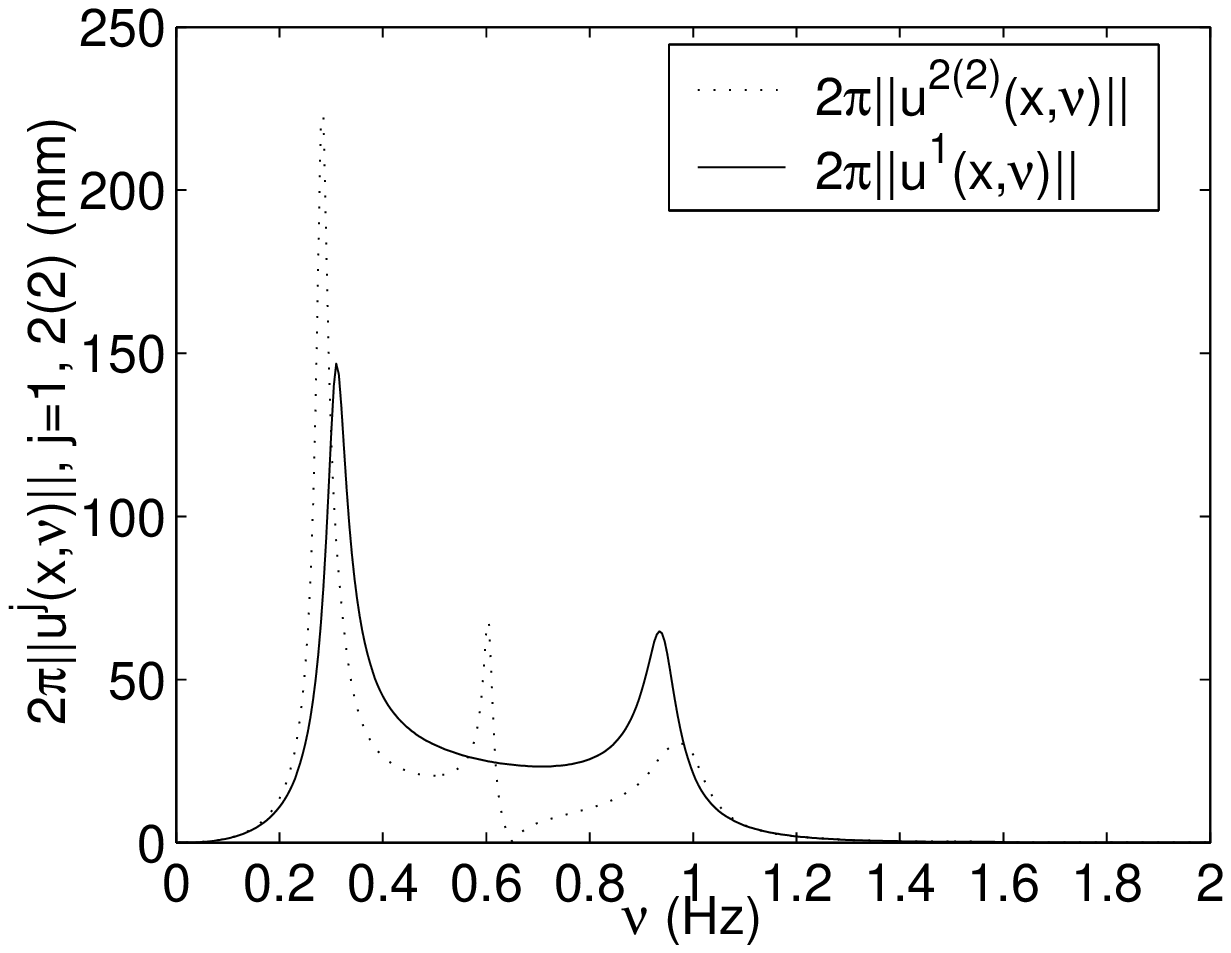}
\includegraphics[width=6.0cm] {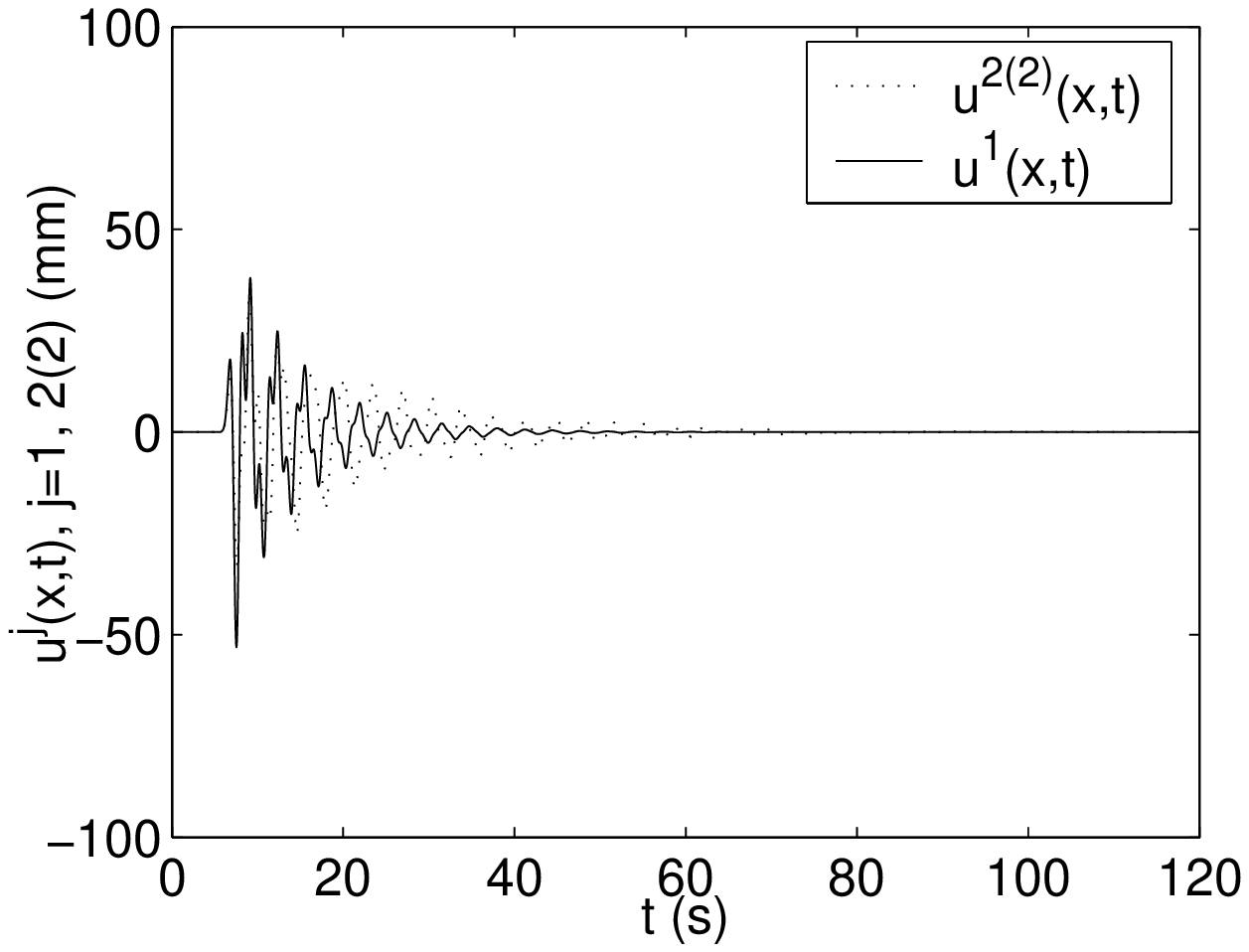}
\end{center}
\caption{Comparison of $2\pi$ times  the spectrum (left panels)
and of the time history (right panels) of the total displacement
on the ground in the absence of blocks (solid curves) and in the
presence of the blocks (dashed curves). From the top to the
bottom: at the center of the summit segment of block $1$, at the
center of the summit segment of  block $2$, at the center of the
base segment of block $1$, and at the center of the base segment
of  block $2$. Two $40m\,\times\, 40m$ blocks solicited by the
cylindrical wave radiated by a deep line source located at
$\mathbf{x}^s=(0,-3000m)$.} \label{compspecttime2b4040}
\end{figure}

In fig. \ref{compspecttime2b4040} we compare the displacement in
the two blocks (in particular, at the centers of  the summit and
base segments) with the displacement on the ground in the absence
of these blocks. The response of the two blocks are, in fact,
identical, due to the deep source being located on the vertical
dividing line between the two blocks. One observes an increase of
the duration and of the peak and cumulative amplitudes that is
larger than for two {\it different} blocks, in particular, on the
top segments. This increase seems to be  due to the much more
stronger response associated with the excitation of the multi
displacement-free base mode block. This hypothesis is backed up by
the fact that the said mode is more weakly excited on the base
segment and at the same time the time history is much closer to
that of the configuration with no blocks. Finally, the
displacement at the center of the base segments vanishes at the
frequency of occurrence of the displacement-free base mode of the
block (which is not a mode of the global configuration) and is
slightly different from the corresponding multi-displacement-free
block mode eigenfrequency.
\subsection{Results relative to a two-block configuration with two
$50m\, \times\, 30m$ blocks}\label{twobloc5030}
The blocks are $50 m$ high and $30 m$ wide and their
center-to-center separation is $65m$.

The displacement-free base block eigenfrequencies are $0.5$, $1.5
Hz$, .....
\begin{figure}[ptb]
\begin{center}
\includegraphics[width=8cm] {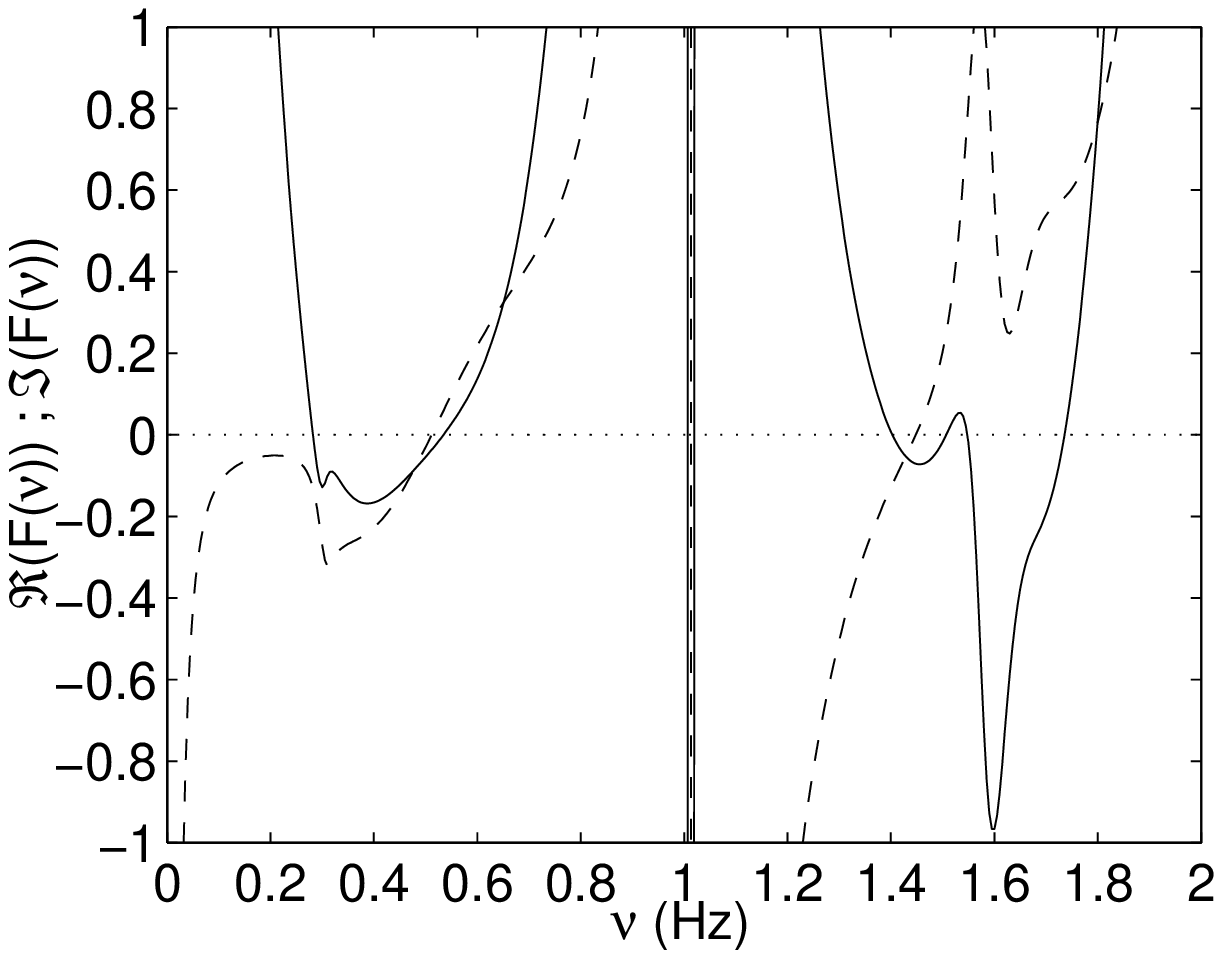}
\end{center}
\caption{Indications concerning the dispersion relation
$\mathcal{F}=0$. The solid curves represent $\Re(\mathcal{F})$ and
the  dashed curves  $\Im(\mathcal{F})$.} \label{reim2b5030}
\end{figure}
Fig. \ref{reim2b5030} gives an indication of the solution of the
dispersion relation. Eigenfreqencies (i.e., solutions of
$\Re\mathcal{F}=0$) are found at $0.3$, $0.55$, $1.0$, $1.4$...
The attenuations associated with the quasi-Love mode at $0.3Hz$
and the multi displacement-free base block mode at $0.55Hz$ are
rather small. The eigenfrequencies are not close to each other,
contrary to the case of two identical $40m\,\times 40m$ blocks.
\subsubsection{Response on the top and bottom segments of the blocks}
\begin{figure}[ptb]
\begin{center}
\includegraphics[width=6.0cm] {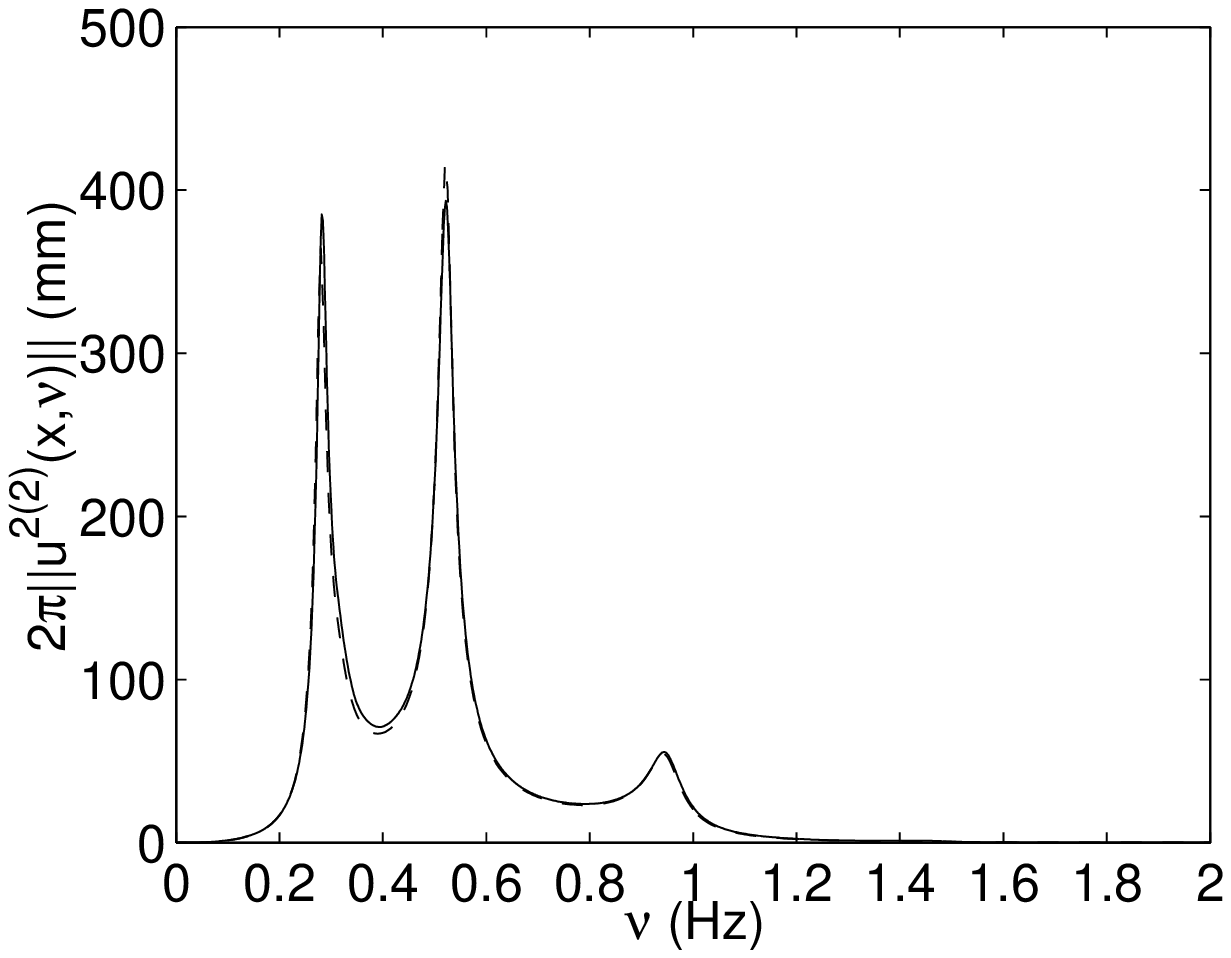}
\includegraphics[width=6.0cm] {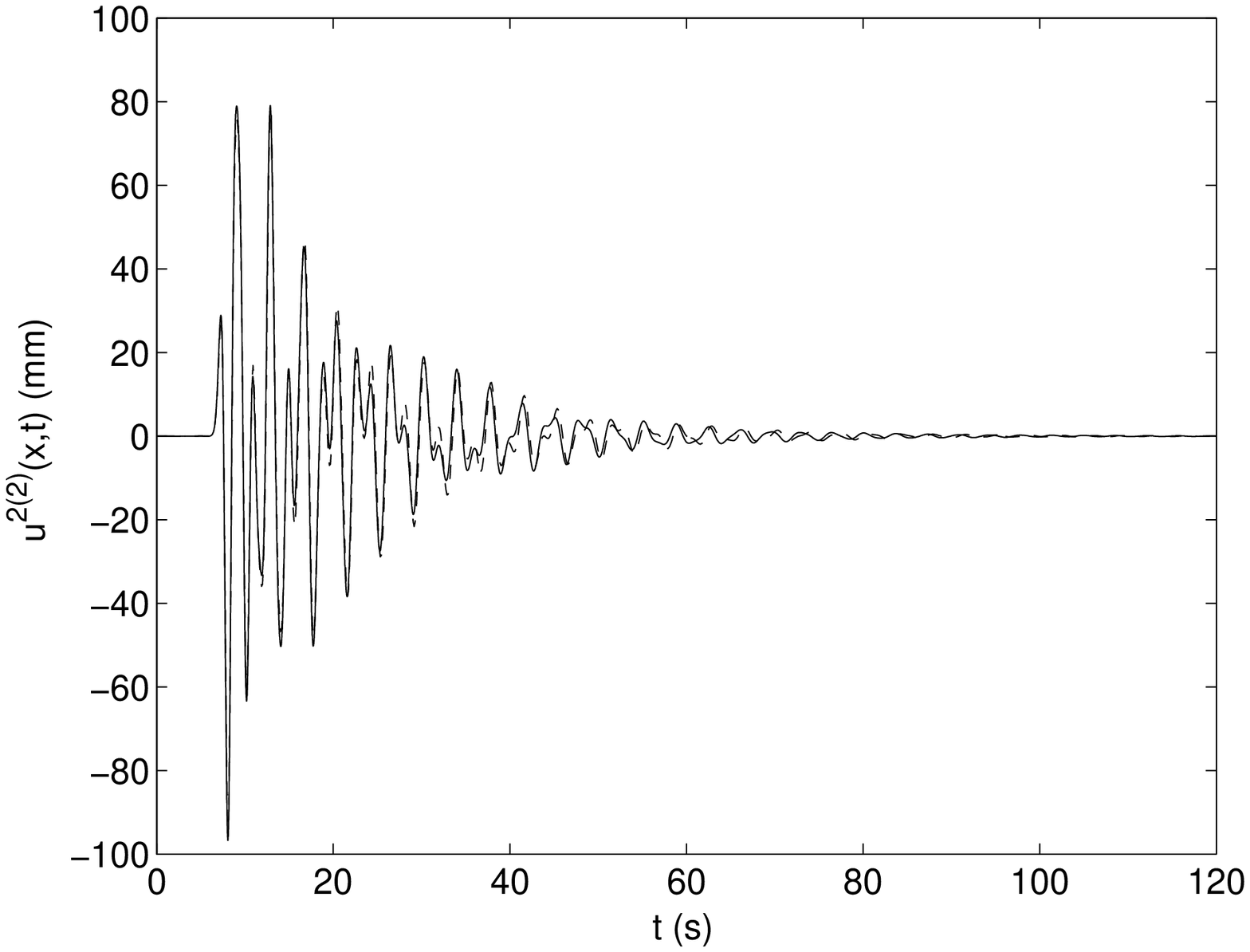}
\includegraphics[width=6.0cm] {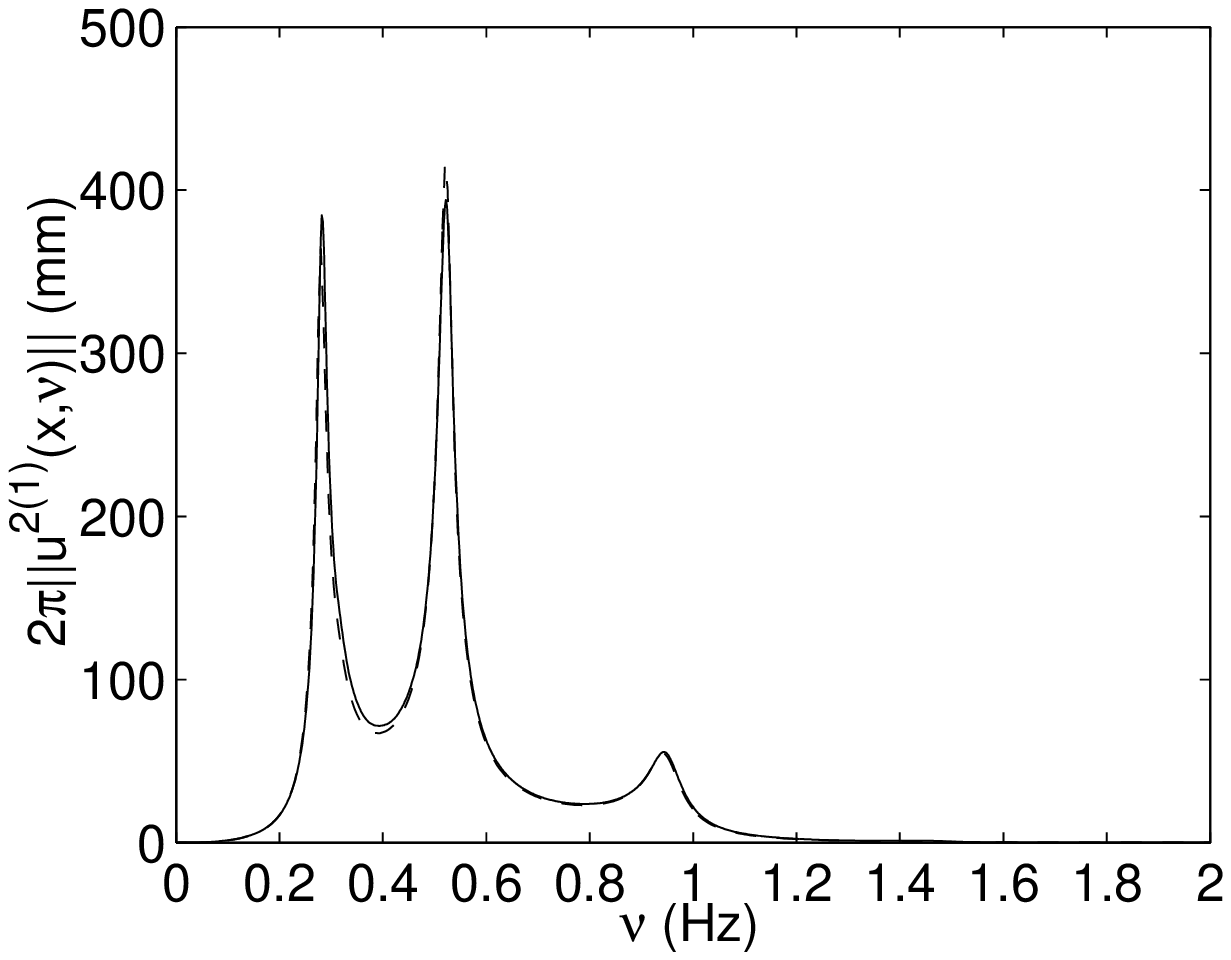}
\includegraphics[width=6.0cm] {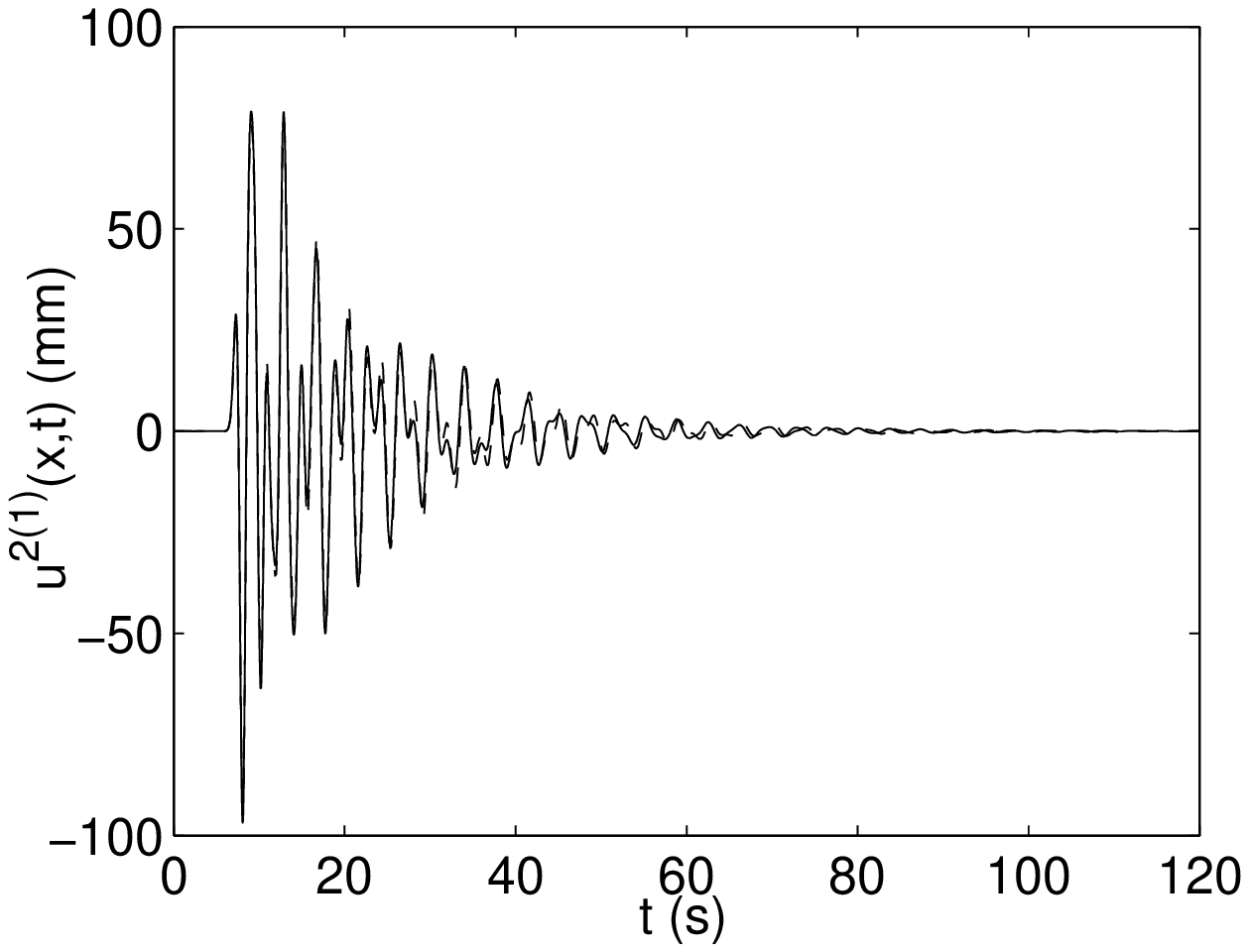}
\end{center}
\caption{$2\pi$ times the spectra (left panel) and time histories
(right panel) of the total displacement at the center of the
summit segment of block 2  (top panels), and of the block 1
(bottom panels), as computed by the mode-matching method (with
account taken only of the zeroth-order quasi-mode (dashed curves)
and the finite element method (solid curves).}
\label{specttimesum2b5030}
\end{figure}

The response at the centers of the summit segments of both blocks
computed by the finite element method is compared in fig.
\ref{specttimesum2b5030} to the corresponding response computed by
the mode matching method (with account taken only of the
zeroth-order quasi-mode. Once again, we observe these responses to
be almost identical. Moreover, the amplitude of the multi-
displacement-free base block resonance peak is observed (in the
same figure) to be higher for two identical blocks, than it was
for two different blocks.

\begin{figure}[ptb]
\begin{center}
\includegraphics[width=6.0cm] {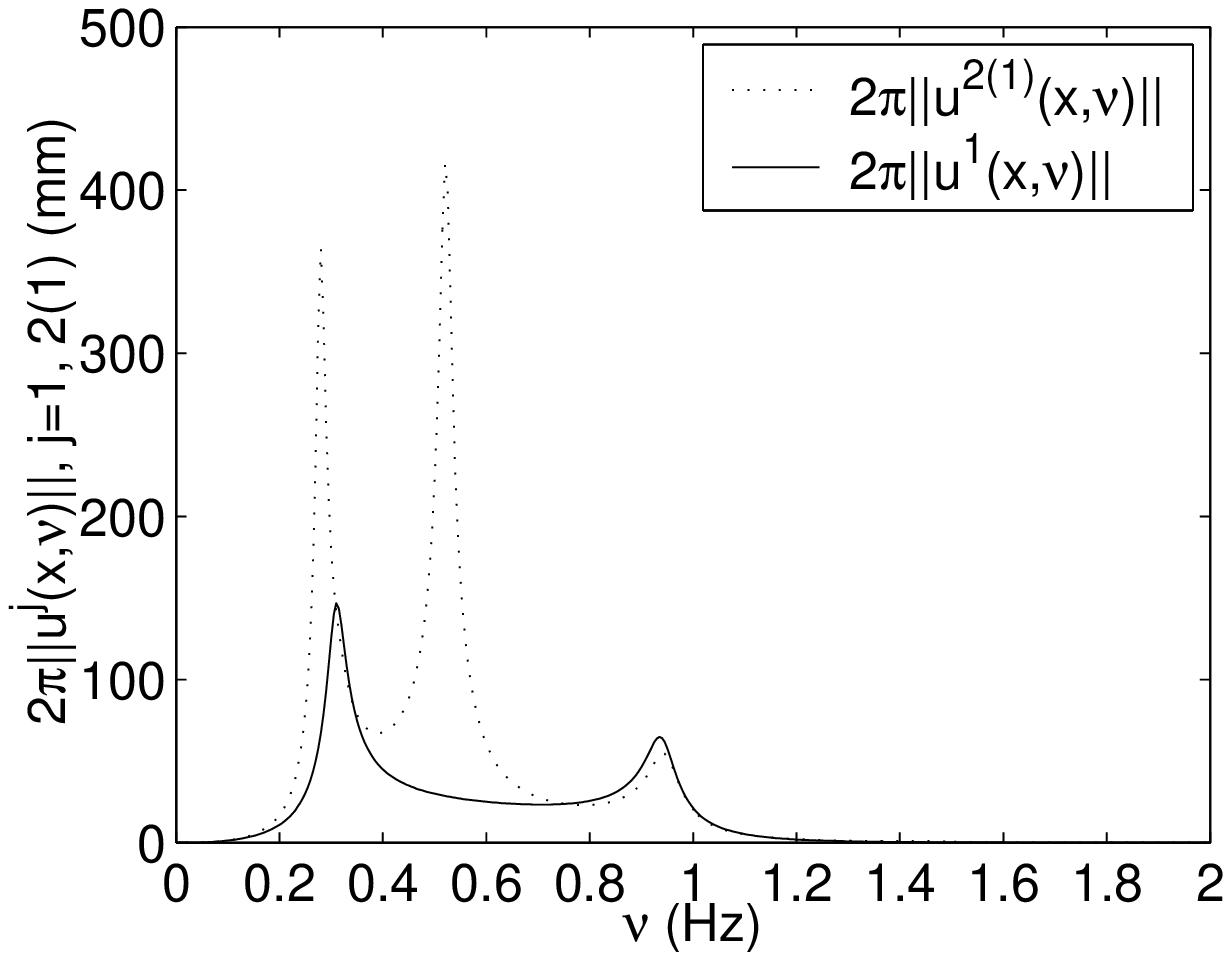}
\includegraphics[width=6.0cm] {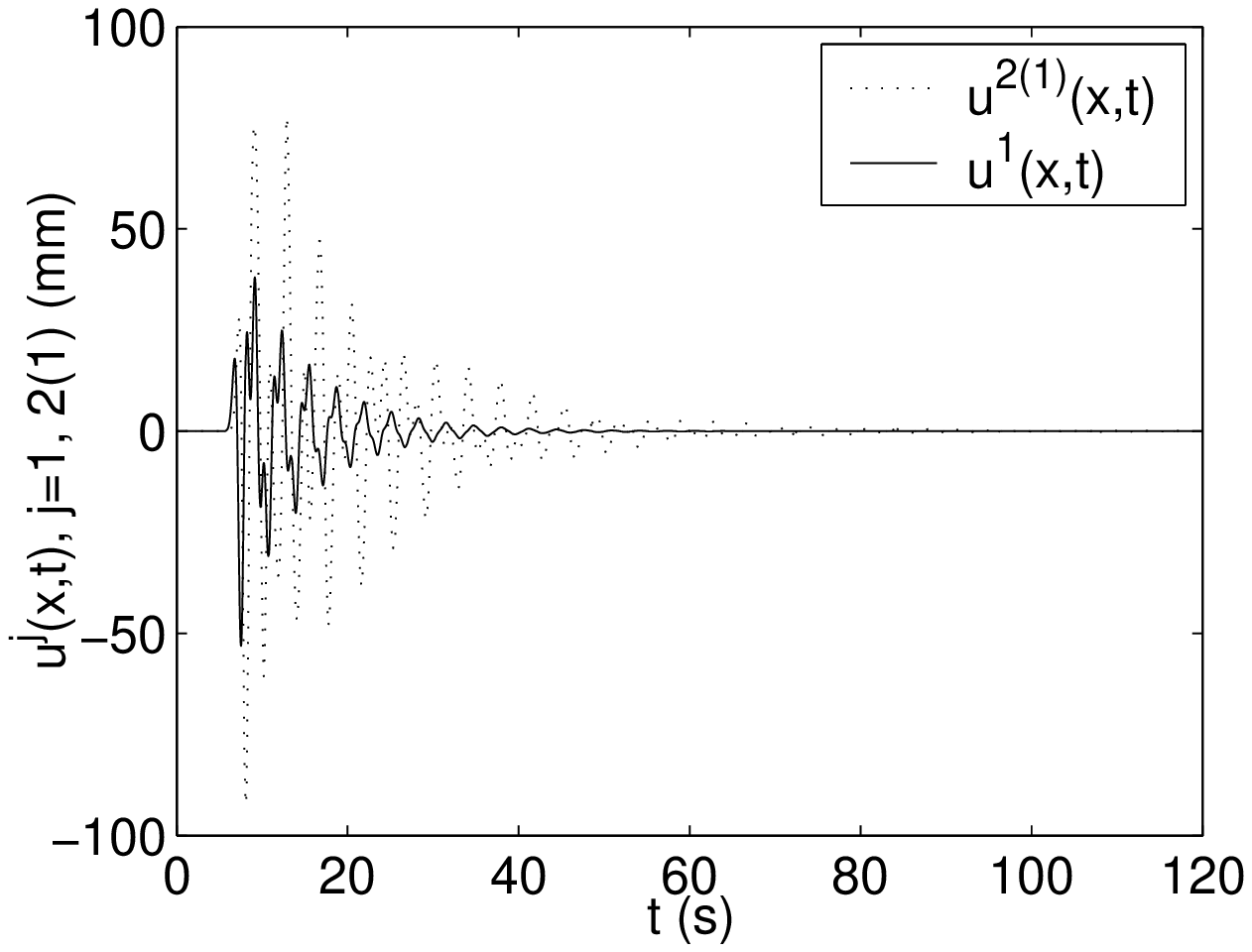}
\includegraphics[width=6.0cm] {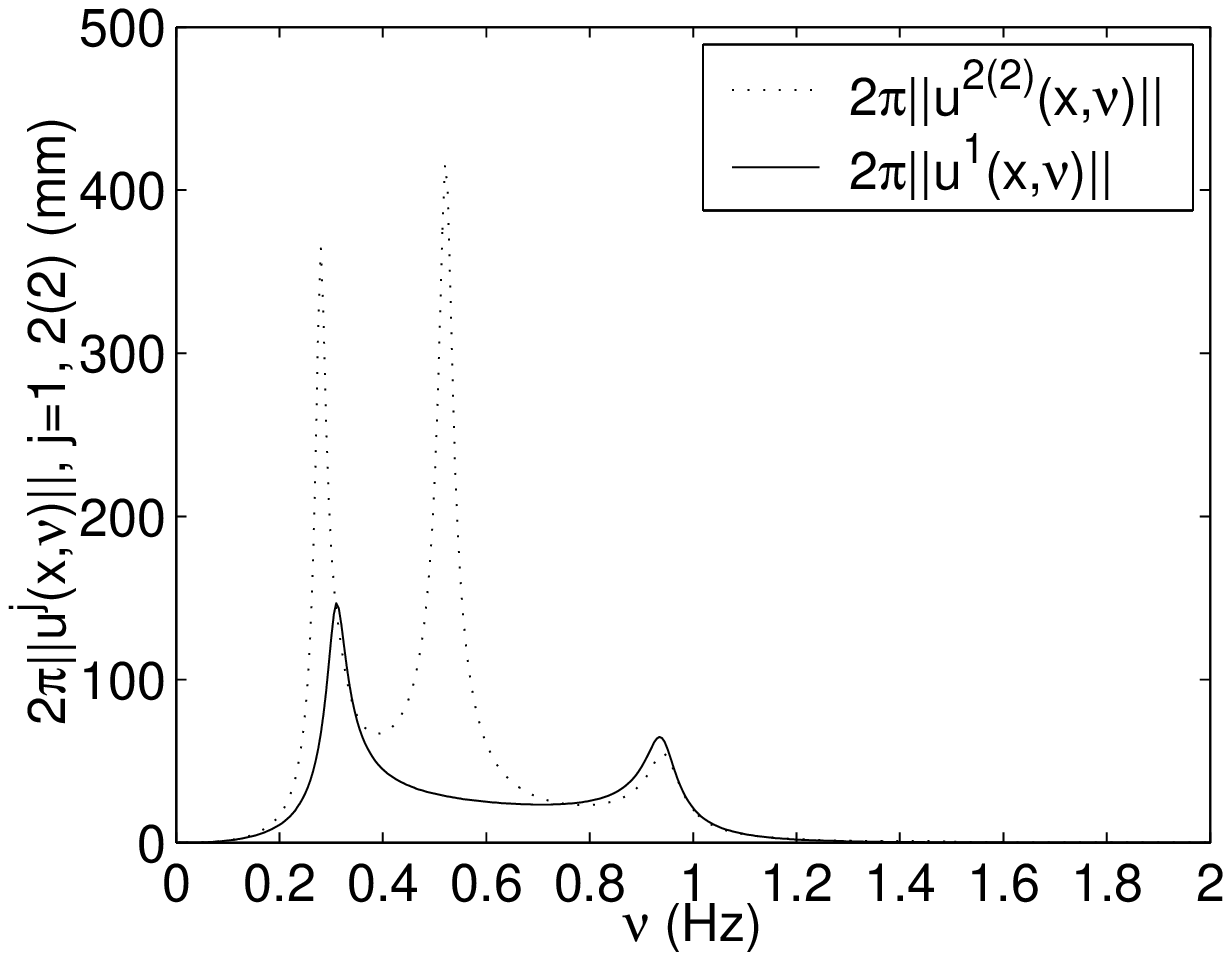}
\includegraphics[width=6.0cm] {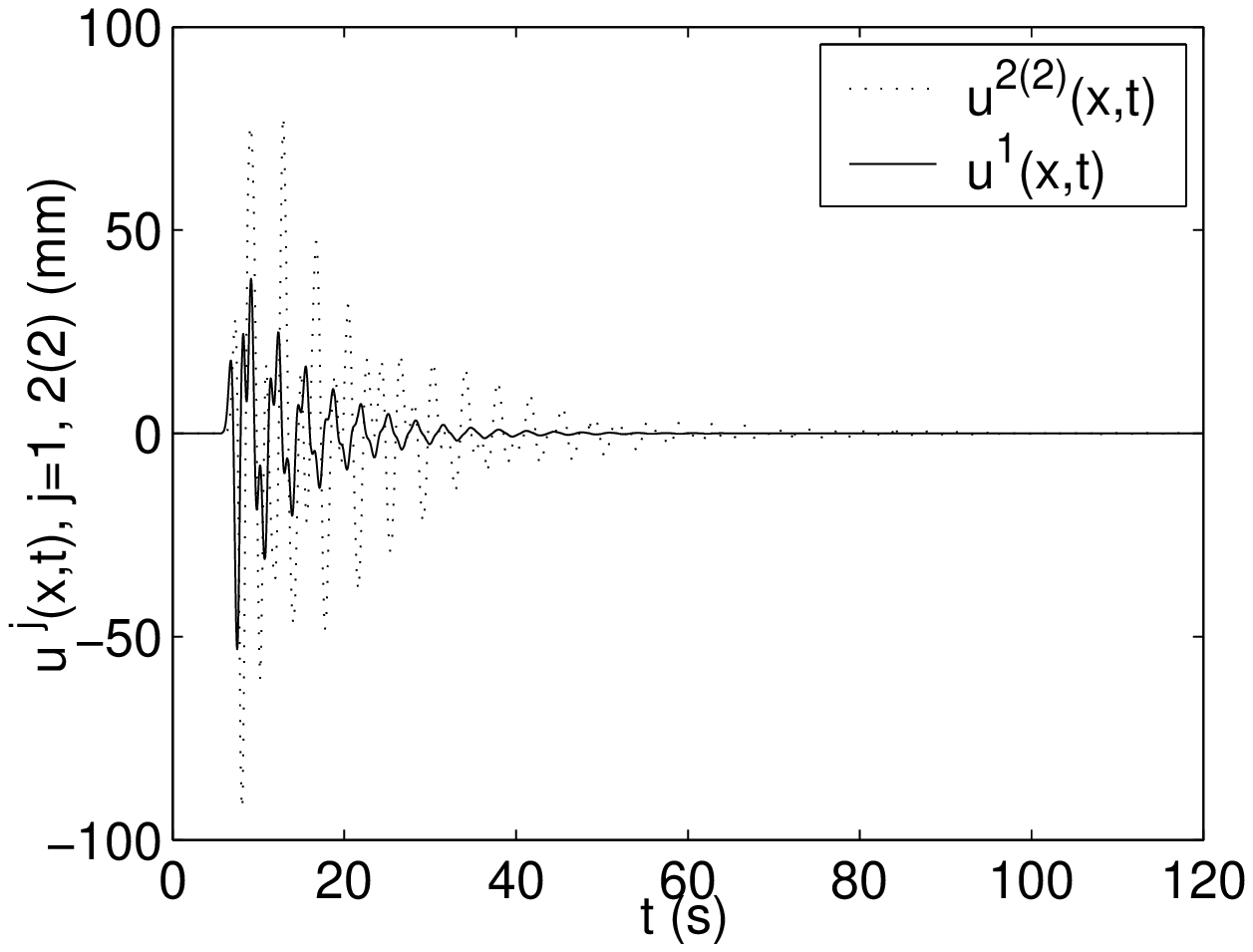}
\includegraphics[width=6.0cm] {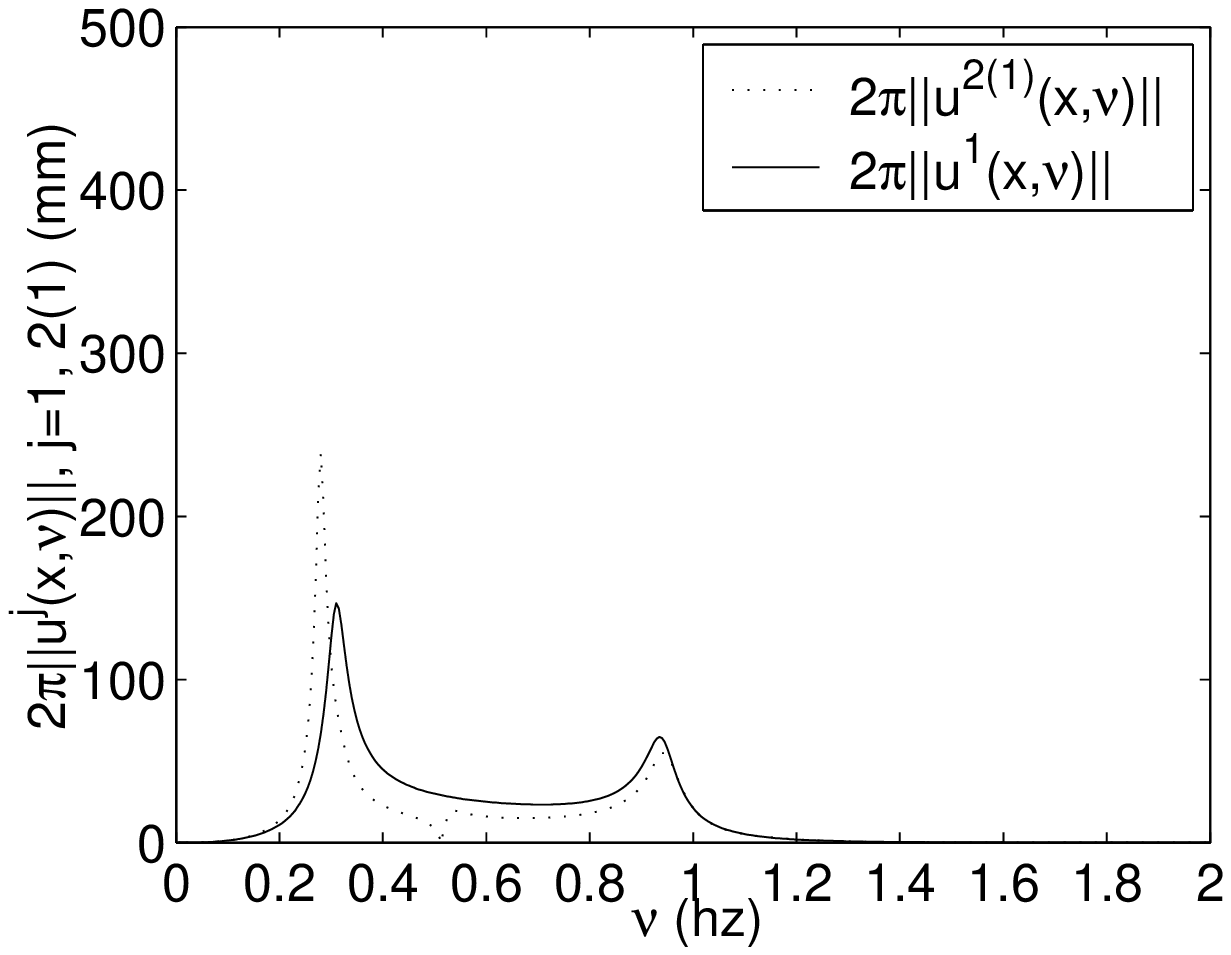}
\includegraphics[width=6.0cm] {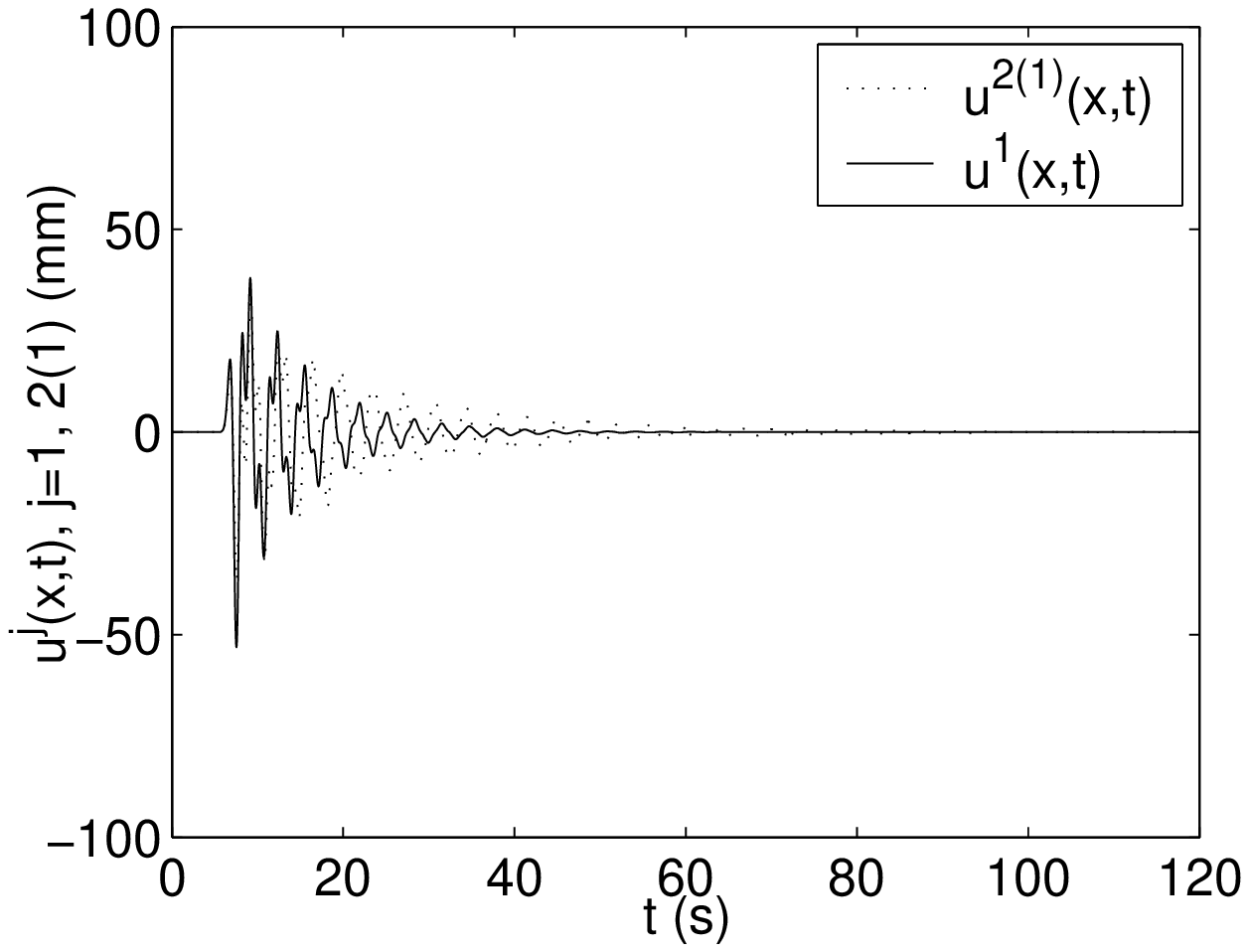}
\includegraphics[width=6.0cm] {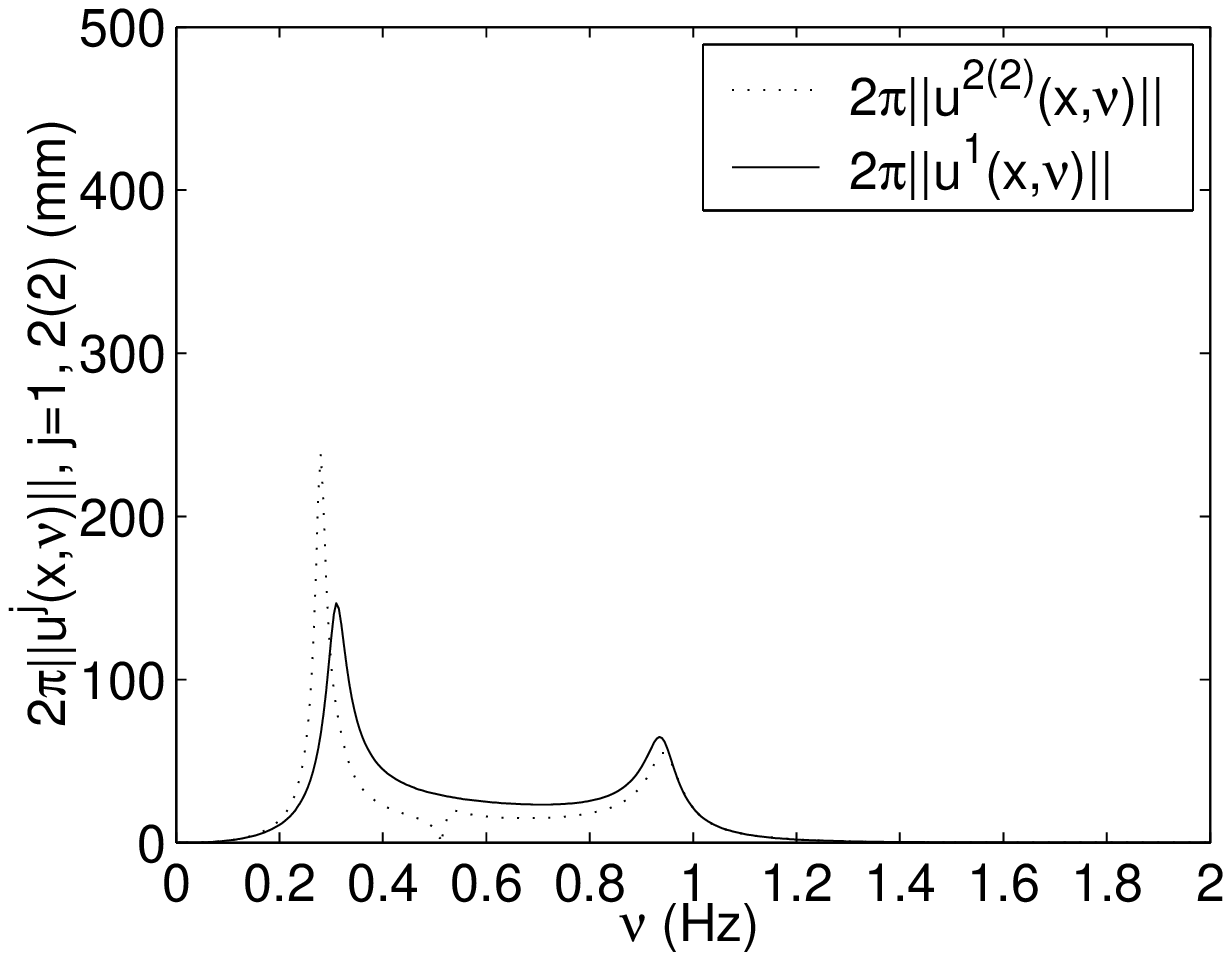}
\includegraphics[width=6.0cm] {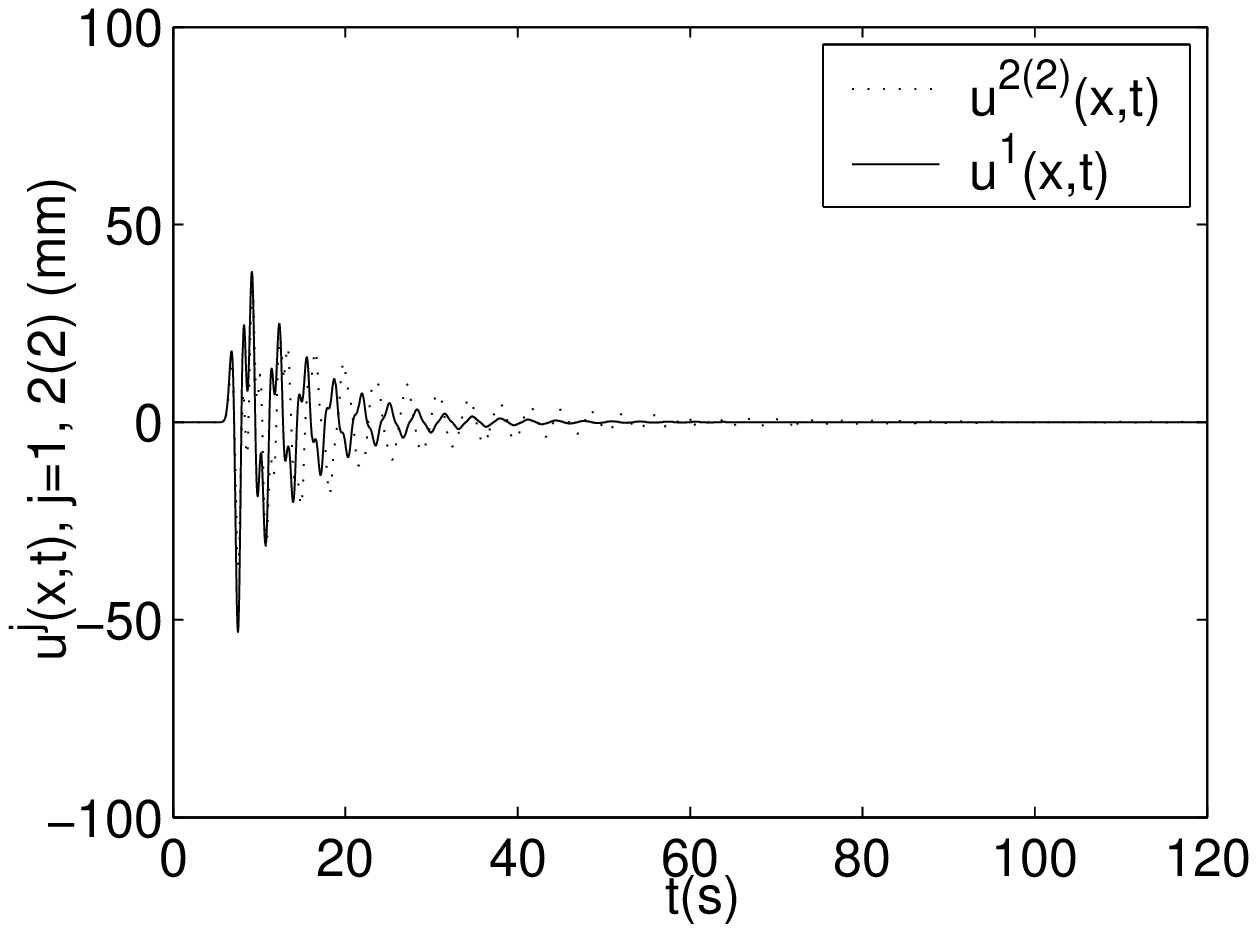}
\end{center}
\caption{Comparison of $2\pi$ times  the spectrum (left panels)
and of the time history (right panels) of the total displacement
on the ground in the absence of blocks (solid curves) and in the
presence of blocks (dashed curves). From the top to the bottom: at
the center of the summit segment of block $1$, at the center of
the summit segment of block $2$, at the center of the base segment
of block $1$, and at the center of the base segment of block $2$.
Two $50m\, \times\, 30m$ blocks solicited by the cylindrical wave
radiated by a deep line source located at
$\mathbf{x}^s=(0,-3000m)$.} \label{compspecttime2b5030}
\end{figure}

In fig. \ref{compspecttime2b5030} we  compare the displacements in
the two blocks (i.e., at the centers of the summit and base
segments) with the displacement on the ground (at the same points
as occupied by the centers of the base segments of the blocks) in
the absence of the blocks. The response of the two blocks are
identical for the previously-mentioned reasons. A much larger
increase of the duration and of the peak and cumulative amplitudes
(particularly on the top segments) is obtained for the two
identical blocks than for two different blocks. This increase
seems to be  due to the much stronger response at the frequency
corresponding to the excitation of  the multi displacement-free
base block  mode. Once again, the displacement at the center of
the base segments vanishes at a frequency corresponding to the
occurrence of the displacement-free base mode of the block.
\subsubsection{Response on the ground}
In order to get another grip on the  phenomena that are produced
when two blocks are present, we now focus on the displacement
field at points on the ground outside of the blocks.
\clearpage
\begin{figure}[ptb]
\begin{center}
\includegraphics[width=6.0cm] {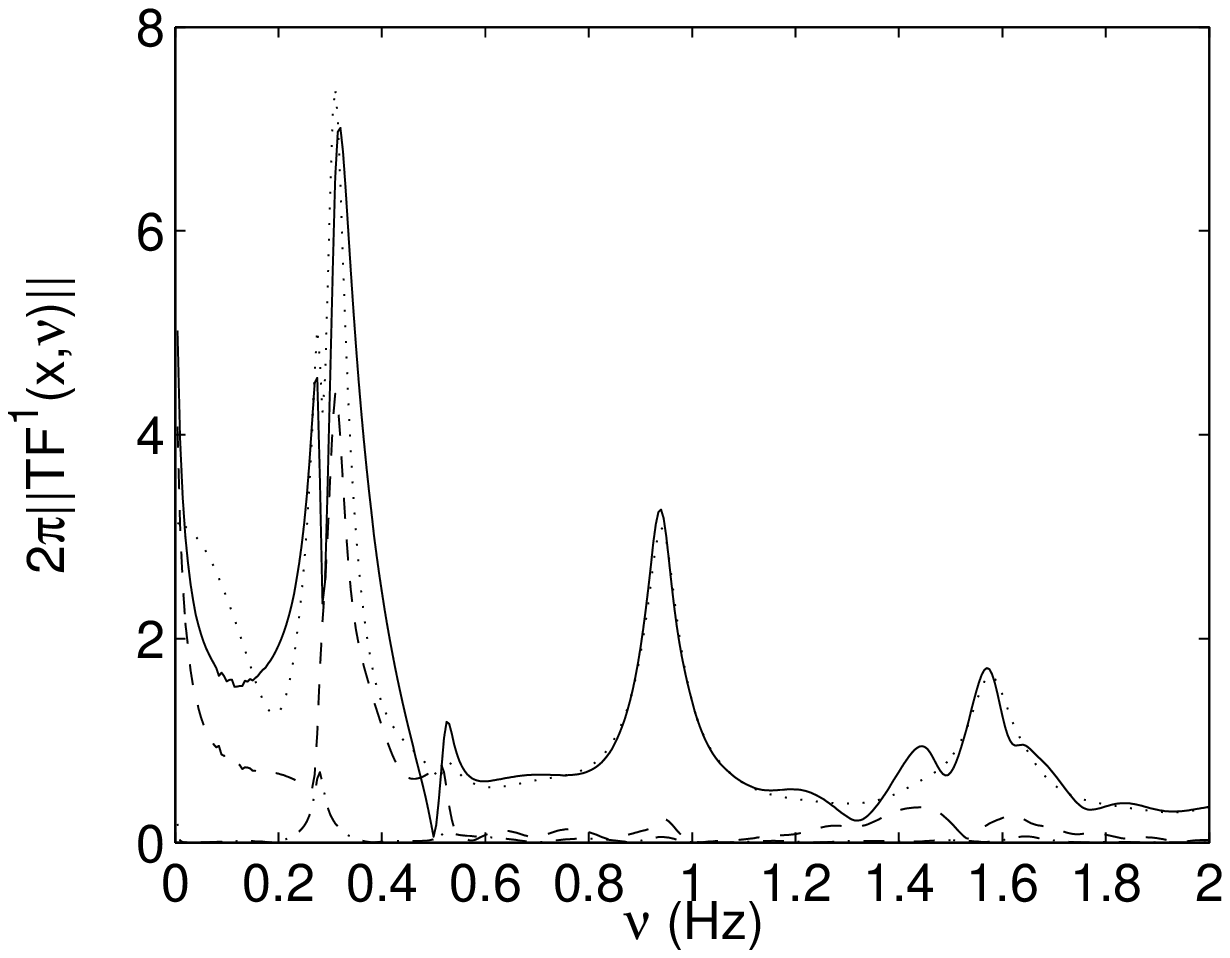}
\includegraphics[width=6.0cm] {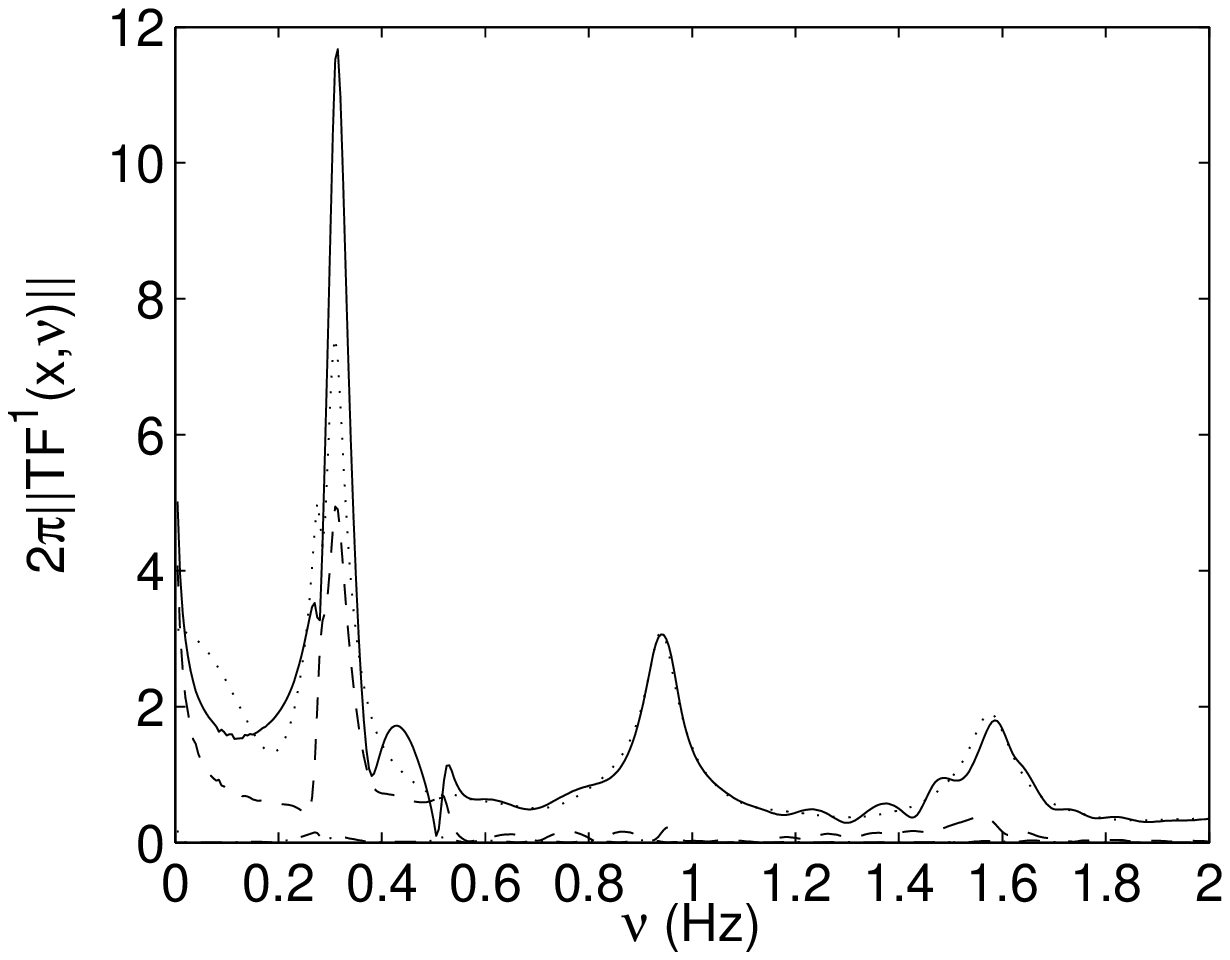}
\end{center}
\caption{$2\pi$ times the transfer function (solid curve) of the
total displacement on the ground at $\mathbf{x_{g}}=(150m,0m)$
(left panel),  and $\mathbf{x_{g}}=(300m,0m)$ (right panel) for a
two  $50m\, \times\, 30m$ block configuration solicited by the
cylindrical wave radiated by a deep line source located at
$\mathbf{x^{s}}=(0m,3000m)$. The dotted curve represents the
contribution along the interval $\mathcal{I}_{1}$, the dashed
curve  the contribution along the interval $\mathcal{I}_{2}$, and
the dotted-dashed curve  the contribution along the interval
$\mathcal{I}_{3}$.} \label{transfground2b}
\end{figure}

Fig. \ref{transfground2b} depicts  $2\pi$ times the transfer
function $u^{1}(\mathbf{x},\omega)/S(\omega)$ resulting from a
computation by the mode-matching method. The influence of two
blocks on the contribution of the various types of waves traveling
in the layer is close to the one we noticed when only one block is
present (see fig. \ref{ftransground}). The amplitude of the first
peak is more important than that of the one  in fig.
\ref{ftransground}.

\begin{figure}[ptb]
\begin{center}
\includegraphics[width=6.0cm] {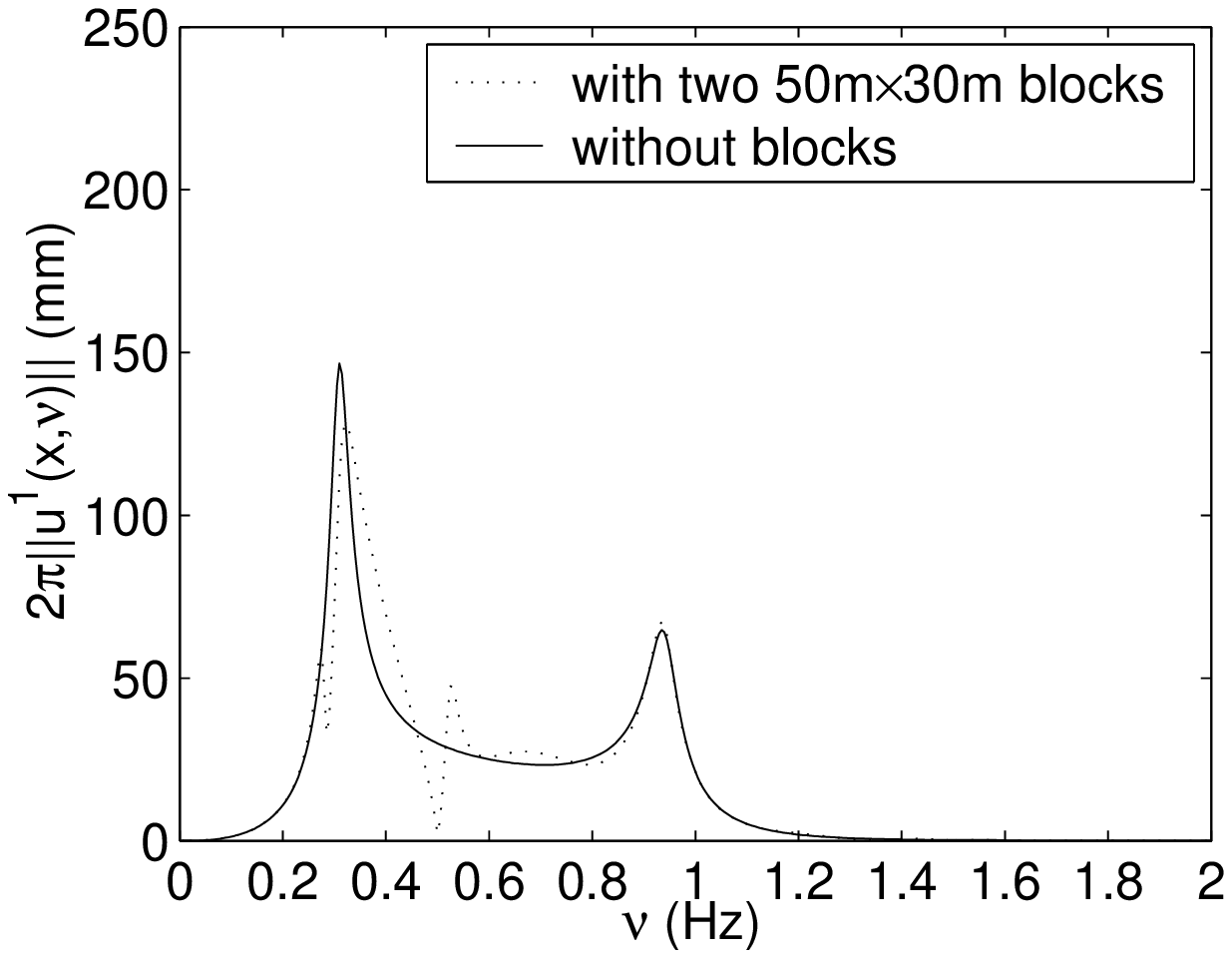}
\includegraphics[width=6.0cm] {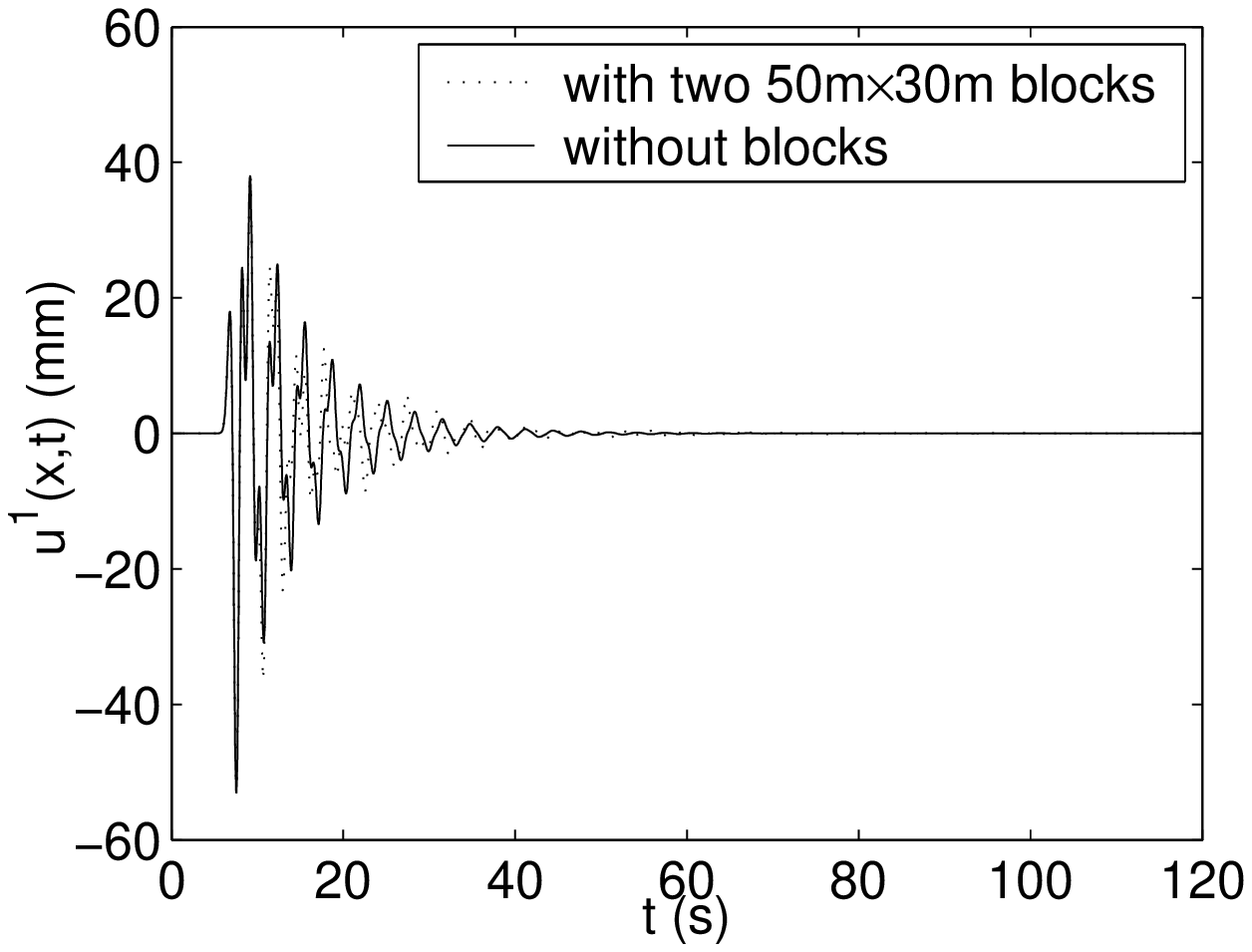}
\includegraphics[width=6.0cm] {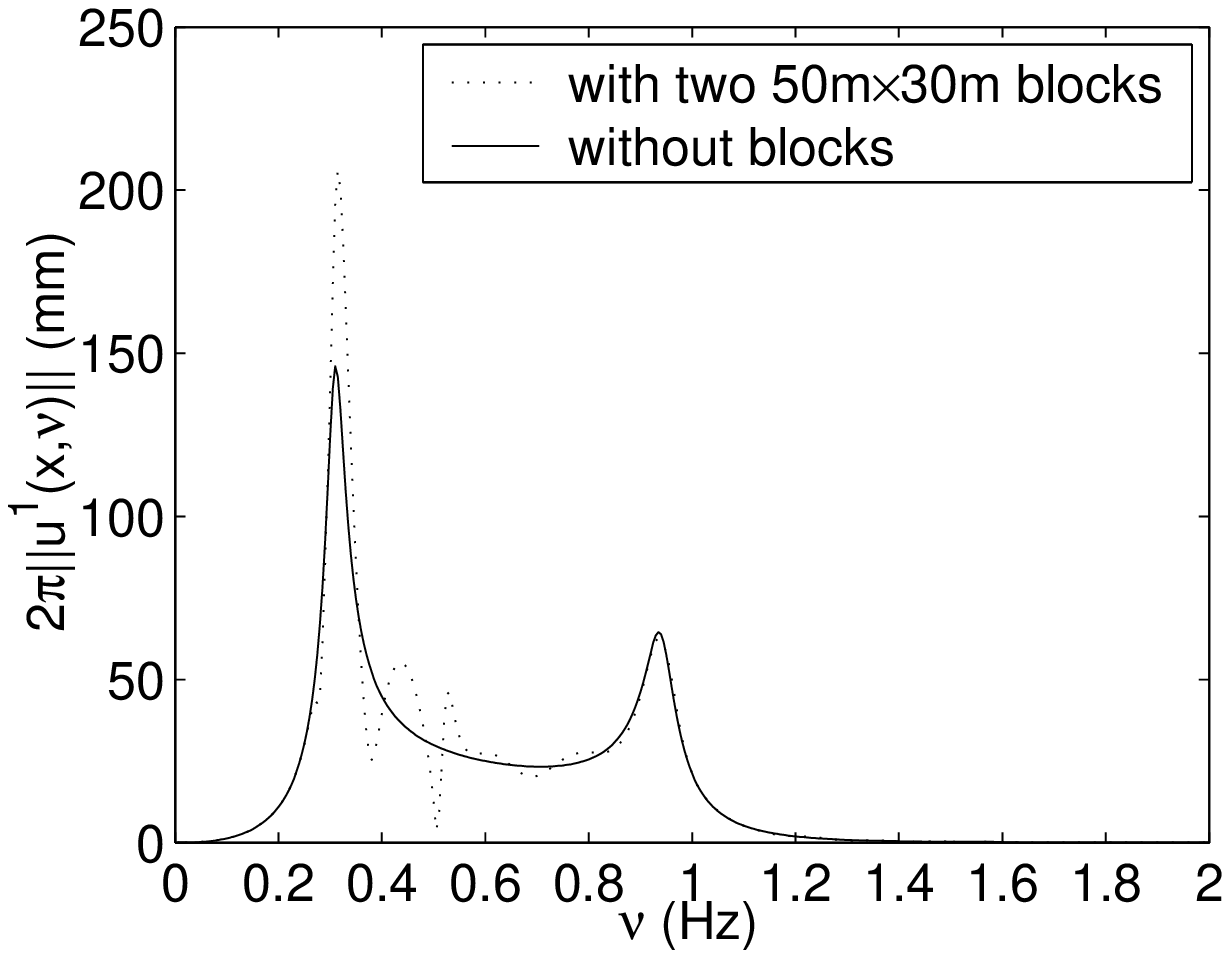}
\includegraphics[width=6.0cm] {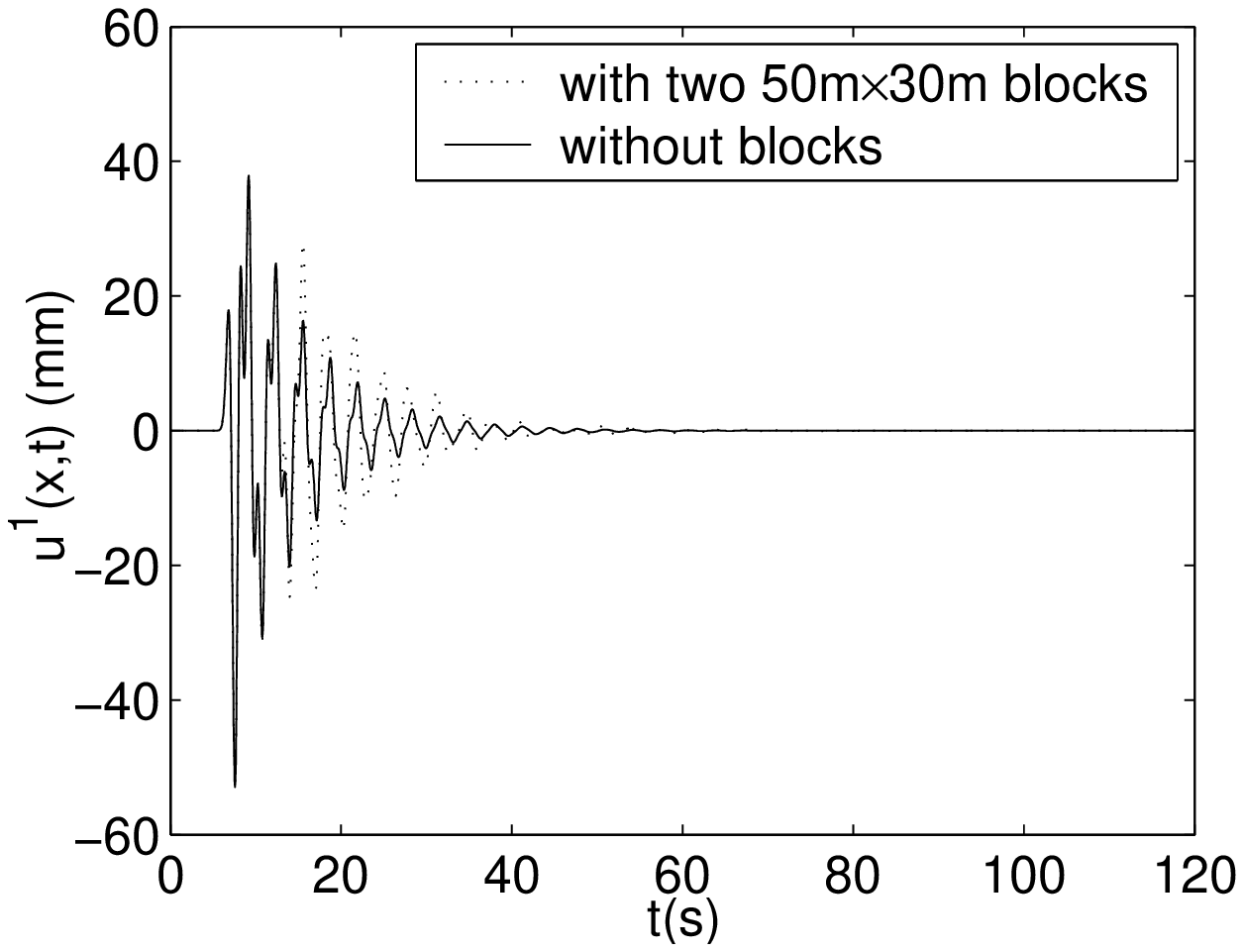}
\end{center}
\caption{Comparison of $2\pi$ times  the  spectrum (left panels)
and of the time history (right panels) of the total displacement
on the ground  at $\mathbf{x_{g}}=(150m,0m)$ (top panels),  and
$\mathbf{x_{g}}=(300m,0m)$ (bottom panels) in the absence of
blocks (solid curves) and in the presence of the blocks (dashed
curves). Case of two $50m\, \times\, 30m$ blocks solicited by the
cylindrical wave radiated by a deep line source located at
$\mathbf{x}^s=(0,-3000m)$} \label{compspecttimebase50304040}
\end{figure}

The comparison in fig. \ref{compspecttimebase50304040} between the
displacement in the presence of the blocks and in their absence
shows a small increase of the amplitude and of the duration in the
time domain, this depending non-linearly on the location on the
ground. In the frequency domain, the amplitude of the first peak
is larger than for a single block.
\section{Snapshots of the displacement fields for one- and two-block configurations in a Mexico City-like site}
\begin{figure}[ptb]
\begin{minipage}{5.2cm}
\centering{$t=12s$}
\end{minipage}\hfill
\begin{minipage}{5.2cm}
\centering{$t=18s$}
\end{minipage}\hfill
\begin{minipage}{5.2cm}
\centering{$t=24s$}
\end{minipage}\\[8pt]
\begin{minipage}{5.2cm}
\includegraphics[width=5.0cm] {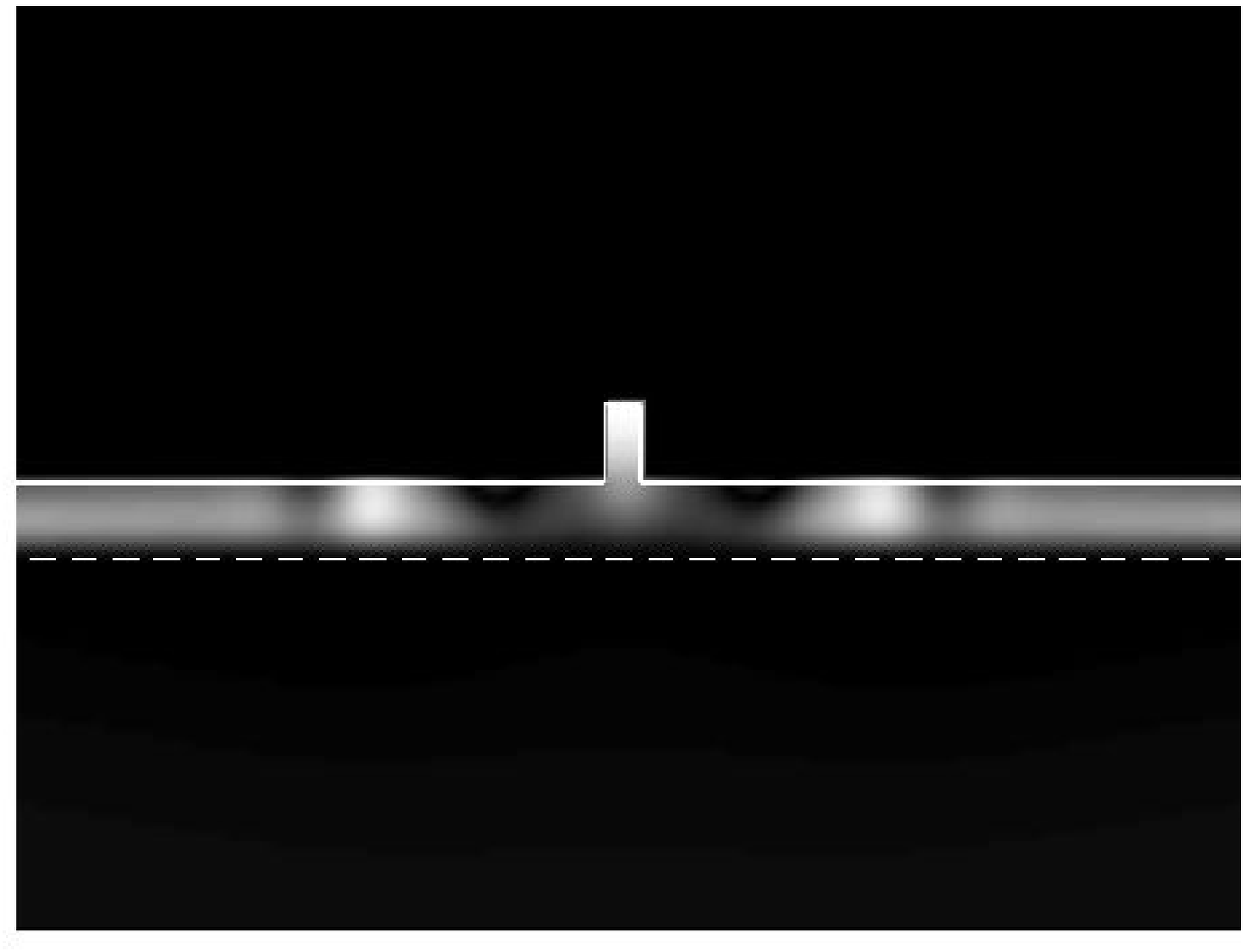}
\end{minipage}\hfill
\begin{minipage}{5.2cm}
\includegraphics[width=5.0cm] {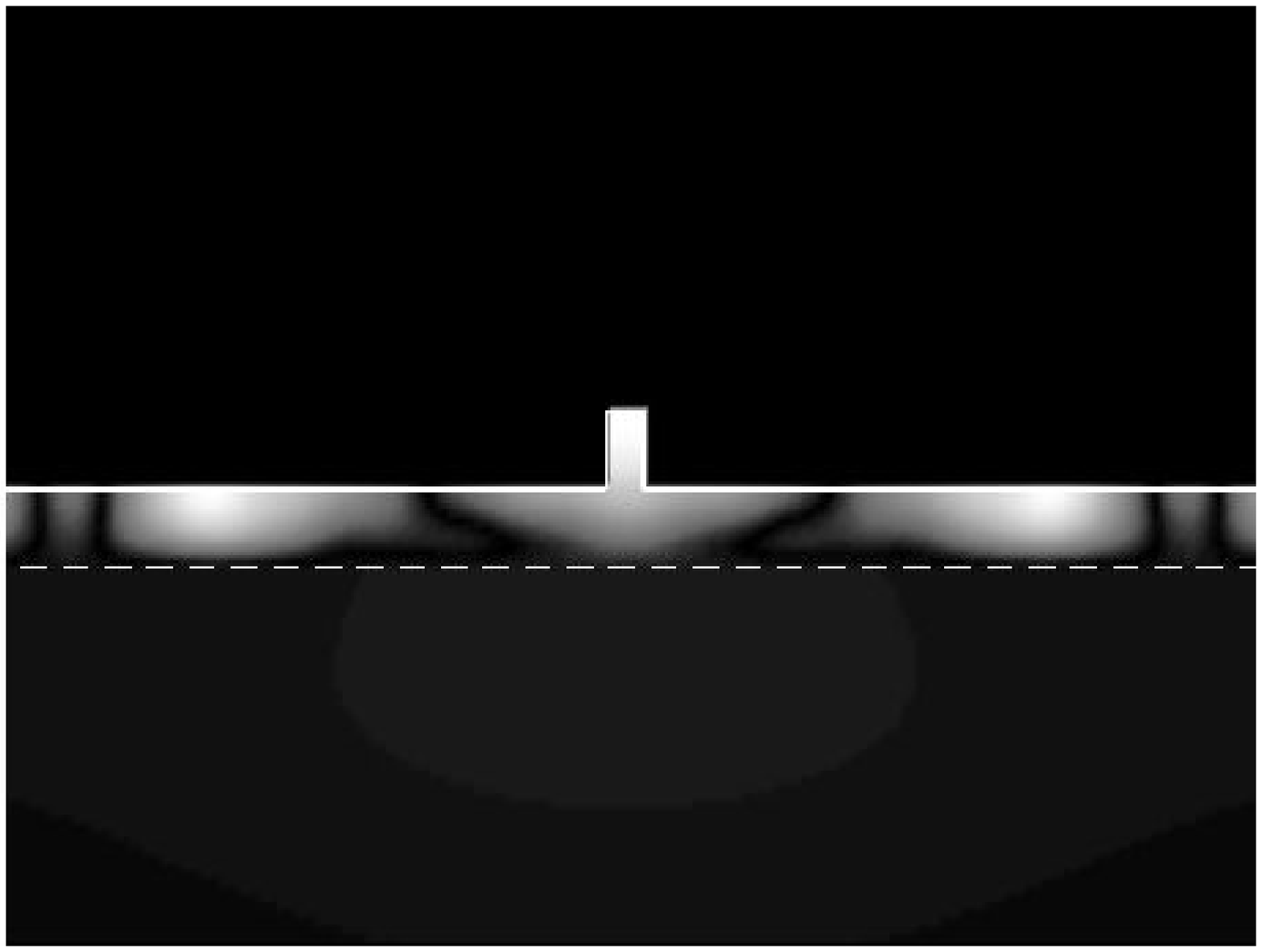}
\end{minipage}\hfill
\begin{minipage}{5.2cm}
\includegraphics[width=5.0cm] {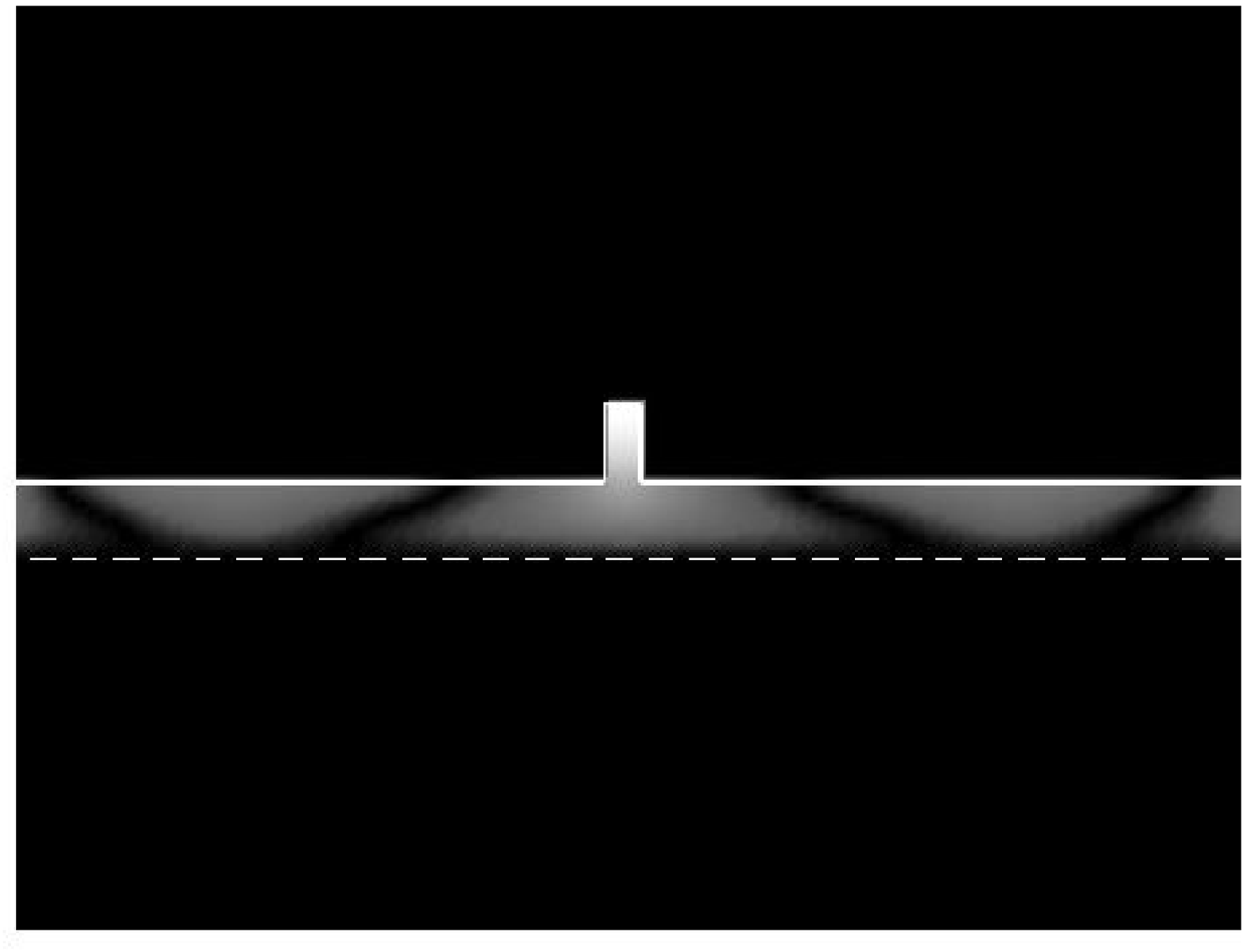}
\end{minipage}\\[12pt]
\begin{minipage}{5.2cm}
\centering{$t=30s$}
\end{minipage}\hfill
\begin{minipage}{5.2cm}
\centering{$t=36s$}
\end{minipage}\hfill
\begin{minipage}{5.2cm}
\centering{$t=42s$}
\end{minipage}\\[8pt]
\begin{minipage}{5.2cm}
\includegraphics[width=5.0cm] {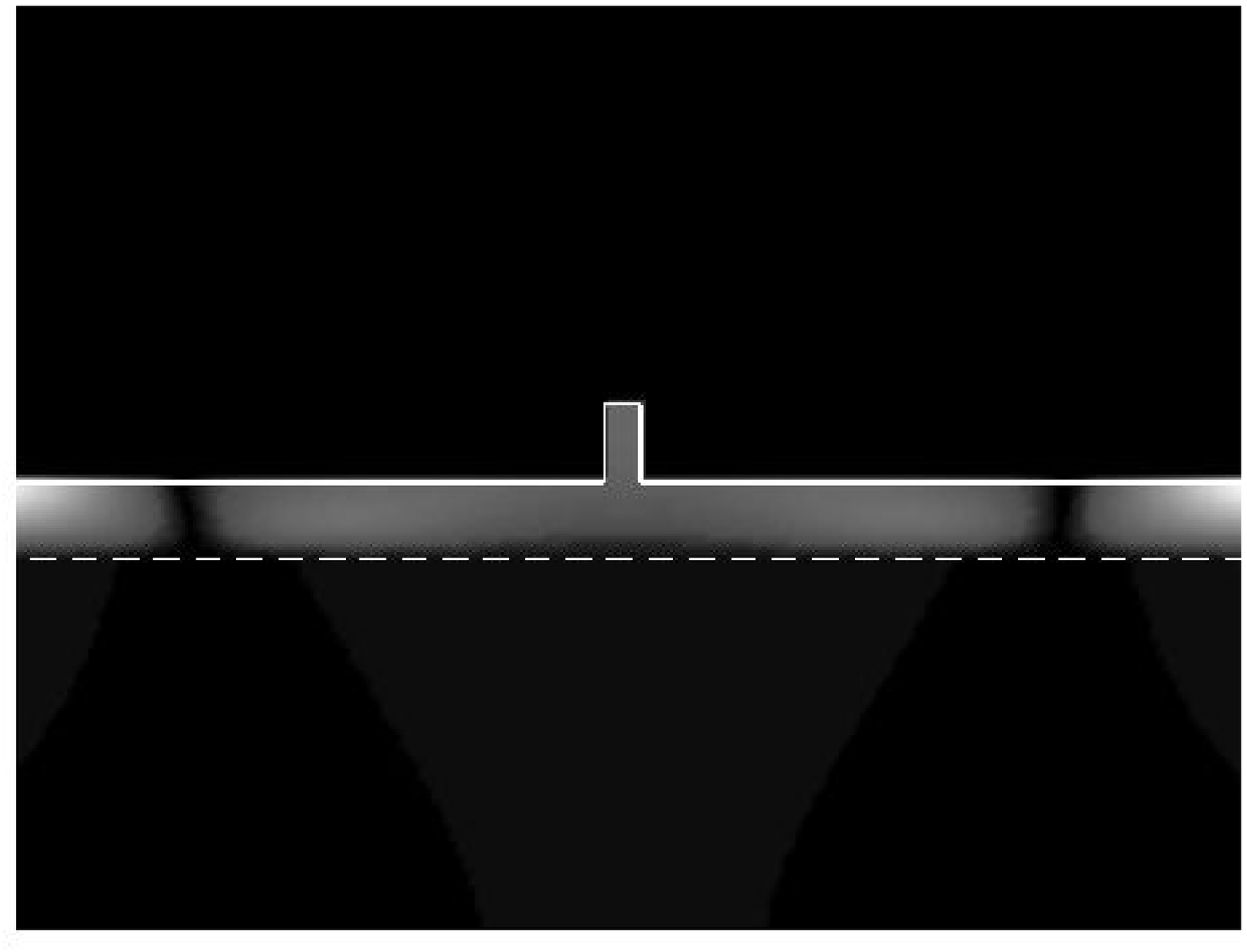}
\end{minipage}\hfill
\begin{minipage}{5.2cm}
\includegraphics[width=5.0cm] {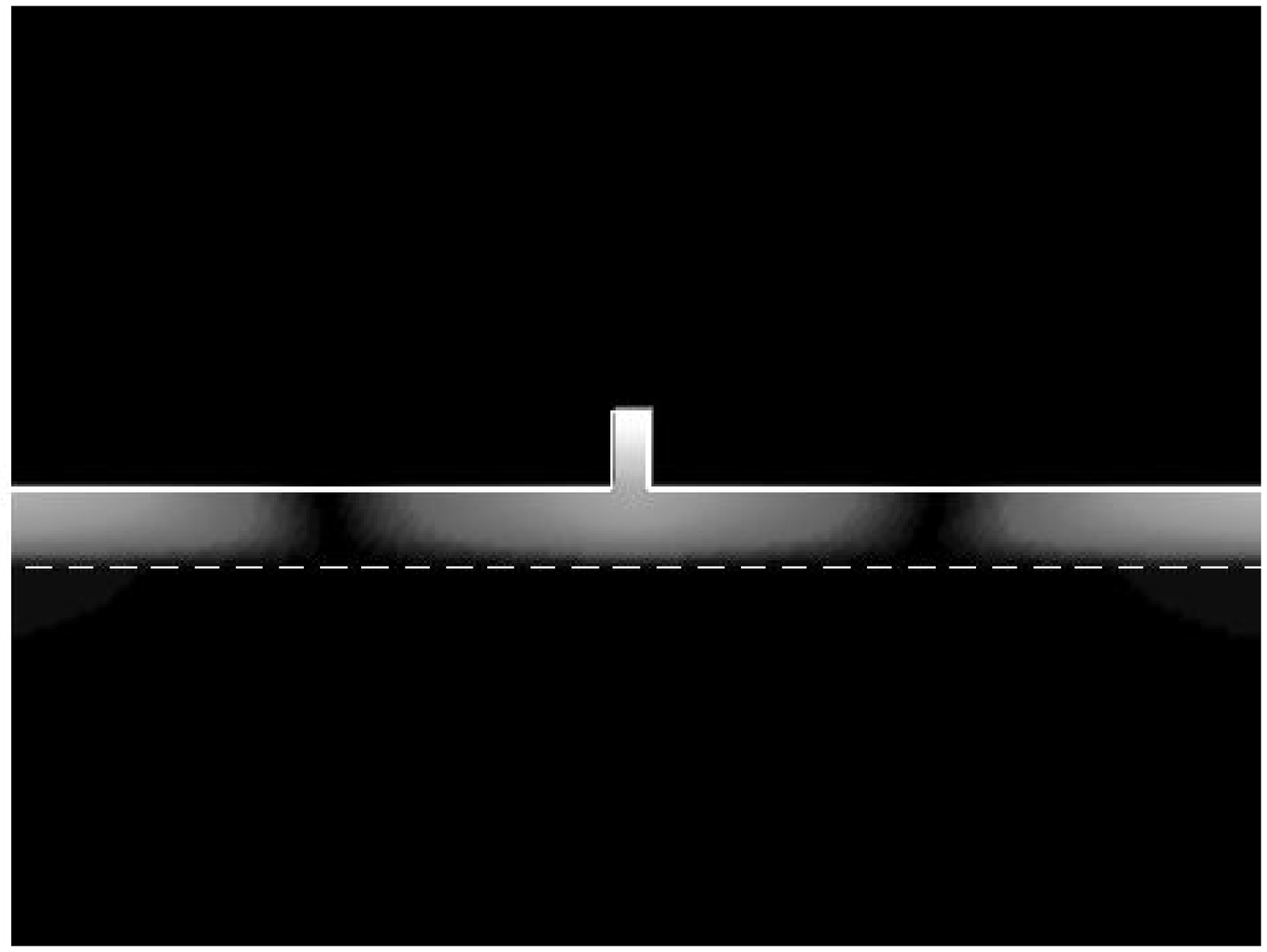}
\end{minipage}\hfill
\begin{minipage}{5.2cm}
\includegraphics[width=5.0cm] {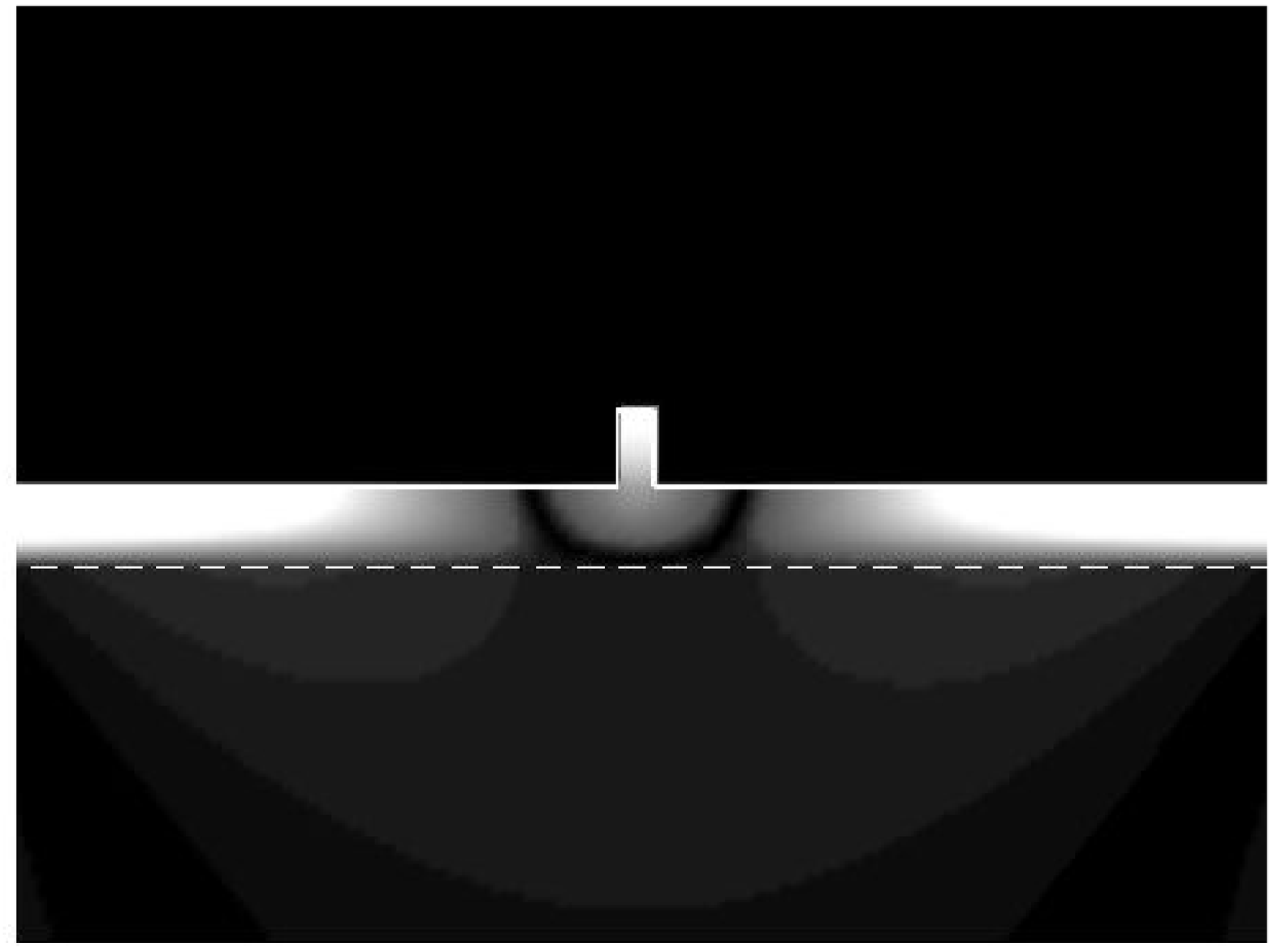}
\end{minipage}
\caption{Snapshots at various instants, $t=12s$, $t=18s$, $t=24s$,
$t=30s$, $t=36s$, $t=42s$, of the total displacement field   for
one $50m\, \times 30m$-block configuration, solicited by the
cylindrical wave radiated by a deep line source located at
$\mathbf{x}^s$=(0m,3000m).} \label{snap1}
\end{figure}

\begin{figure}[ptb]
\begin{minipage}{6.0cm}
\centering{$t=21s$}
\end{minipage}\hfill
\begin{minipage}{6.0cm}
\centering{$t=42s$}
\end{minipage}\\[8pt]
\begin{minipage}{6.5cm}
\includegraphics[width=6.0cm] {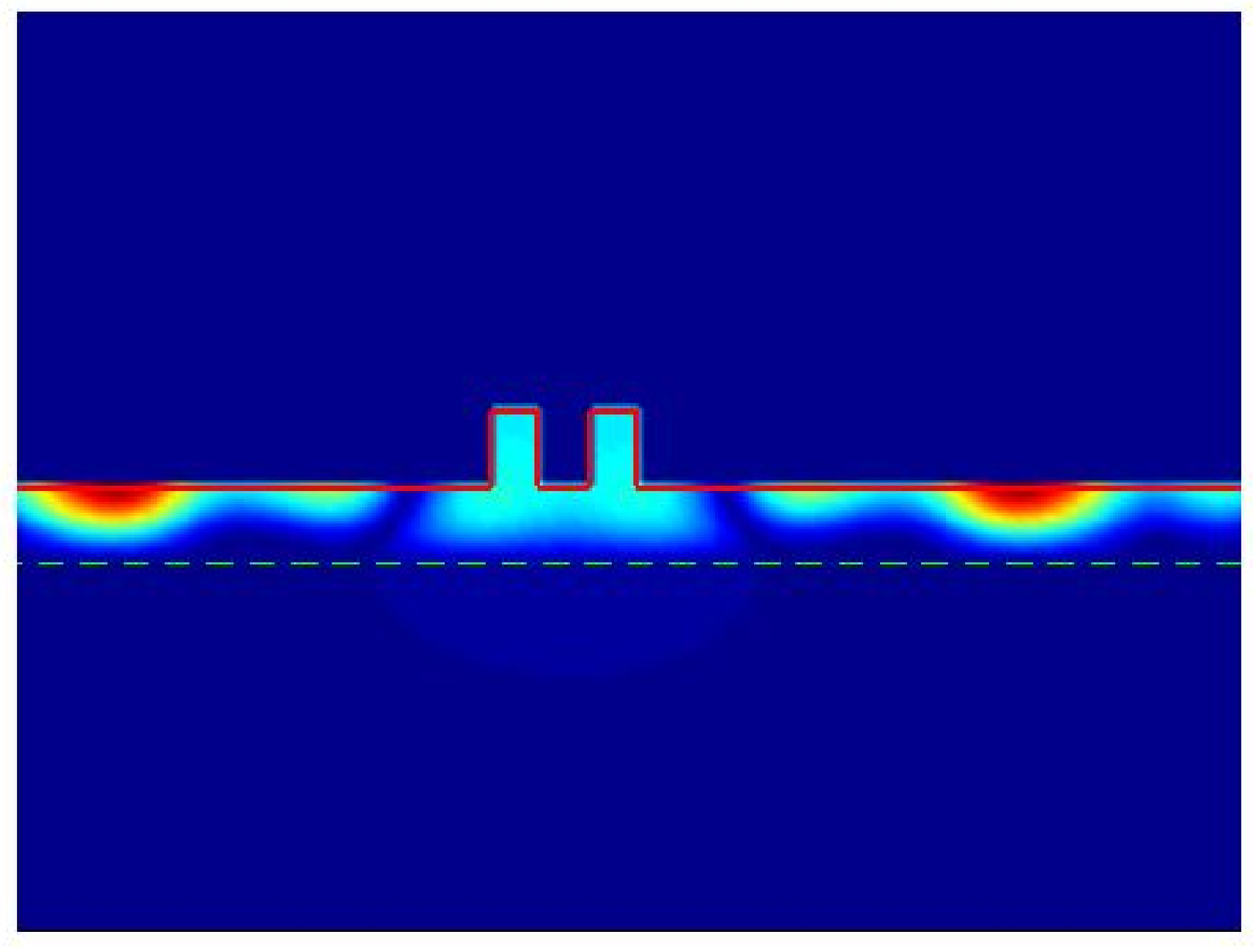}
\end{minipage}\hfill
\begin{minipage}{6.5cm}
\includegraphics[width=6.0cm] {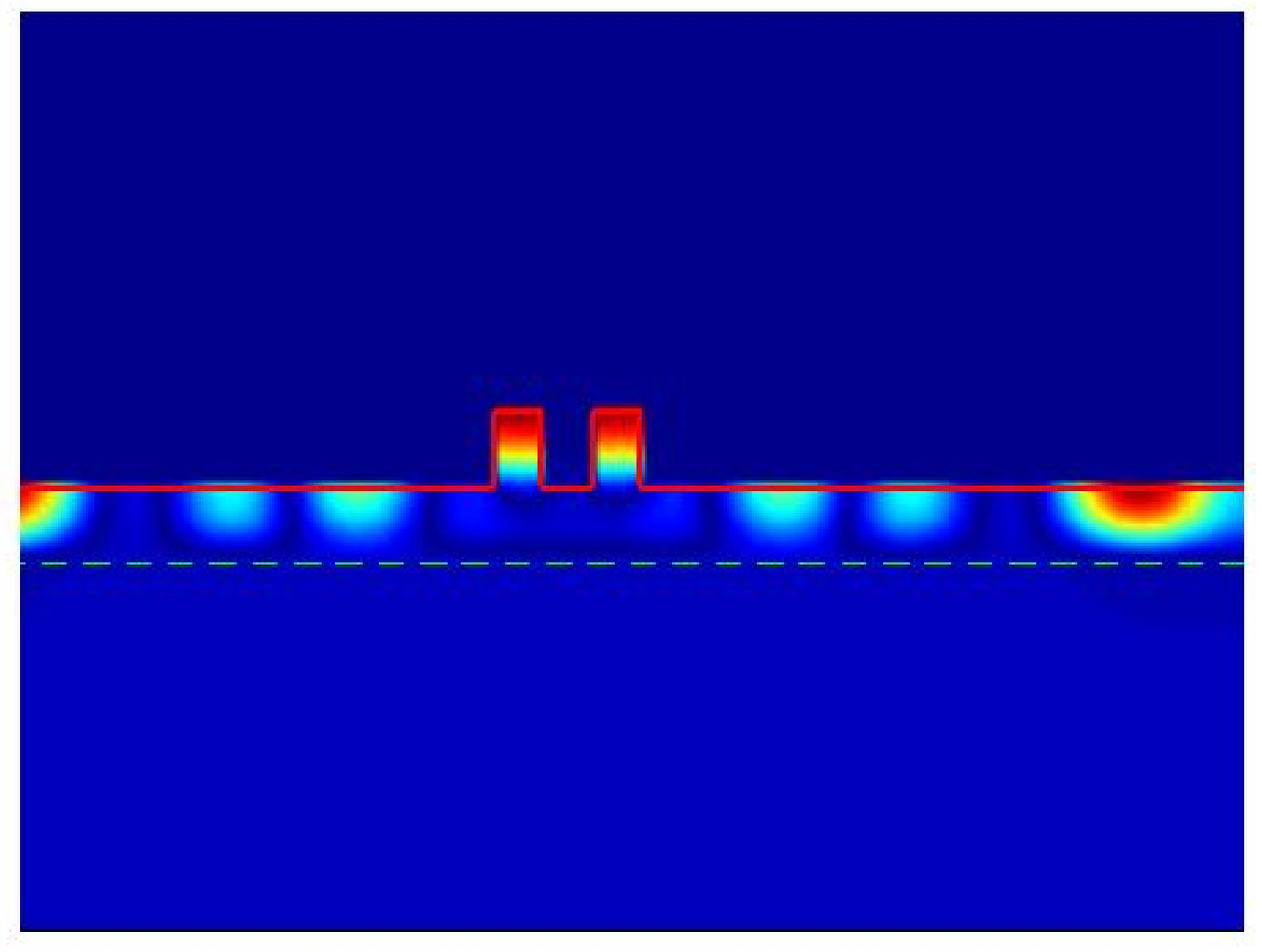}
\end{minipage}
\caption{Snapshots at $t=21s$ and $t=42s$ of the total
displacement field for  two identical $50m\, \times 30m$- block
configuration, solicited by the cylindrical wave radiated by a
deep line source located at $\mathbf{x}^s$=(0m,3000m).}
\label{snap2}
\end{figure}

In order to better visualize the excitation of the quasi-Love mode
due to the presence of one or two blocks in a Mexico City-like
site, we show, in figs. \ref{snap1} and \ref{snap2}, the snapshots
at various instants, of the displacement field, for one
$50m\,\times \,30m$ block, and two identical $50m\,\times \,30m$
blocks, respectively, in response to the cylindrical wave radiated
by a deep line source located at $\mathbf{x}^s=(0m,3000m)$.
We notice, in the layer regions of these figures,  waves that are
re-radiated from the base segments of the blocks and which evolve
into a field with a series of nodes and anti-nodes characteristic
of a sum of modes dominated by the quasi-Love modes. Coupling
between the two blocks (structure-soil-structure iteraction) is
also noticeable in fig. \ref{snap2} at $t=21s$.
\section{Conclusions and preview of the contents of the companion paper}
The response, to the cylindrical wave radiated by a line source
located
 in the substratum, of a finite set of
non-equally sized, non-equally spaced blocks, each block modeling
one, or a group of buildings, in welded contact with a soft layer
overlying a hard half space, was investigated in a theoretical
manner via the mode matching technique.

The capacity of this technique to account for the complex
phenomena provoked by the presence of blocks on the ground was
demonstrated by comparison of the numerical results to which it
leads to those obtained by a finite element method.

It was shown that the presence of blocks induces a modification of
the phenomena that are produced by the configuration without
blocks, or of a configuration of closed blocks disconnected from
the geophysical half-space. In particular, the blocks modify the
dispersion relation of what, in the absence of the blocks,
constitutes the Love modes.

Three different and complementary points of view were developed
(in the framework of the mode matching theory) concerning the
dispersion relation relative to the configuration with blocks. The
first emphasizes the role of the substructure (i.e. the
geophysical (flat ground/layer/substratum) structure ). The second
emphasizes the role of each particular block of the
superstructure. The third point of view emphasizes the couplings
of the fields in the superstructure with those in the
substructure.

The first two points of  views enable the definition of two new
types of
 modes relative to the configuration with blocks:
 the \textit{quasi-Love modes},
which are small perturbations of the Love mode (which can exist
when no blocks are present), and the  \textit{quasi displacement-
free base block modes}, which are small perturbations of the
displacement- free base block mode (which can exist when no
geophysical structure is present).

These two types of quasi-modes account for coupling between a
particular block of the superstructure with the substructure, but
not for the couplings between blocks (when more than one block is
present). Thus, the so-called \textit{soil-structure interaction}
is shown to be due to the excitation of quasi Love modes.

The third point of view emphasizes the coupling between blocks
(when more than one block is present). It was shown that this
coupling, manifested  by the existence of coupling matrices in the
expressions for the quasi mode amplitudes, is of the same form as
the coupling between one particular block and the substructure,
which underlines the fact that the coupling between blocks is
carried out via the substructure.

The study of the dispersion relation for the multi-block
configurations is very complex, but can be carried numerically.
This was done for a two-block configuration and showed that a
\textit {multi displacement-free base block mode} can be produced.
The latter  was shown to correspond to a coupled mode, constituted
by a combination of quasi displacement-free base block modes of
each block, which are no longer excited  as such in a
configuration with more than one block. The multi
displacement-free base block mode was shown to account for the
so-called \textit{structure-soil-structure interaction}.

It was underlined that the modes of a complete one- or multi-block
configuration do not  constitute  the reunion of the modes of the
individual component structures. Thus, the phenomena for a
complete $N$ block configuration are not the sum of the phenomena
for $N-1,N-2,..,0$ block configurations.

The excitation of these modes was then studied in the particular
case of one and two blocks. A common feature of the influence of
one/or more blocks is the excitation of the quasi-Love mode, which
occurs even for  solicitation by the waves radiated by a deep
source (recall that, for this type of solicitation, it is not
possible to excite ordinary Love modes in a flat ground (i.e., no
blocks)/soft horizontal layer/hard substratum configuration
\cite{grobyetwirgin2005}). The trace of quasi-Love mode excitation
in the frequency domain was shown to be: i) for a deep source, a
shift to lower frequency and an increase of the amplitude of the
first (lowest-frequency) peak of the response, and ii) for a
shallow source (the case in which  waves whose structure is close
to that of Love waves already exist in the layer and substratum in
the absence of the block(s)), an increase of the amplitude, and
little or no shift, of the first resonance peak.

As concerns deep sources, the change of the phenomena, from a
configuration without blocks (for which there exist only bulk
waves in the geophysical structure), to one with blocks (for which
there exist quasi-Love modes characterized by a field in the
substratum that is predominantly a surface wave in the substratum)
is a manifestation of  the so-called \textit{soil-structure
interaction}.

Quasi displacement-free base block modes are encountered only for
configurations with one block whereas multi displacement-free base
block modes are excited in configurations with two (and
presumably, more) blocks. Both of these modes are excited with
incident waves that are radiated either from deep or shallow
sources. The multi displacement-free base block mode is more
energetic and corresponds to a coupled mode. Its excitation is
tied up with the so-called \textit{structure-soil-structure
interaction}.

The modifications of frequency-domain response (and less so for
the time-domain response) were found to be fairly substantial,
even when only one block is present, at large distances from the
block on the ground.

Generally speaking, the peak and cumulative amplitudes and
duration of the time histories are larger for a one- and two-block
configuration than for  flat ground, especially at locations
within the block, but no spectacular effects, such as those
noticed during the Michoacan earthquake were found in the
numerical simulations, either for the one- or two-block "cities".

A configuration involving a larger number of blocks is difficult
to study theoretically in the framework of the mode-matching
formulation, which is why configurations with an infinite number
of blocks, each one of which is close to the average shape and
composition of the blocks of  typical
 cities, is investigated in the companion paper. It is also shown
 in this paper that, beyond of the order of ten identical blocks, the response
 of the laterally-finite city is very close to that of the city
 with an infinite number of blocks. More importantly, it is found
 in the companion paper,
 that the phenomena are quite different, for configurations with a large number of
 identical
blocks, from those for a small number of identical or
non-identical
 blocks, and that the said phenomena for a large number of blocks are
 evocative of those observed during earthquakes in urban areas
 such as Mexico City.

 This  indicates that it is probably not
 sufficient to carry out experiments and numerical simulations on an isolated building \cite{made95,gu00,to02} in
order to predict correctly the seismic response of the  building
when it is surrounded
 by other buildings (the situation of most buildings in a typical
 city).
\appendix
\section{An auxiliary problem: fields in
the layer and in the substratum when the seismic distrubance takes
the form of a ribbon source of width $w$ located in the
layer}\label{aux3}
The problem is to determine the total fields
$u^{1}(\mathbf{x},\omega)$ and $u^{0}(\mathbf{x},\omega)$, when
the configuration without blocks (i.e. involving only a soft layer
and a hard half space) is solicited by a horizontal ribbon source
of width $w$ centered at $\mathbf{x}^{s}$ in the soft layer.
\subsection{Boundary conditions and field representations}\label{auxbound}
The fields satisfy the continuity of displacement and  traction
conditions on $\Gamma_{h}$, i.e., (\ref{boundrad.5}) and
(\ref{boundrad.6}), the vanishing traction condition on
$\Gamma_{g}$ (\ref{boundrad.1}), and the outgoing wave (radiation)
condition.

In the layer and in the substratum the field representations are:
\begin{equation}\label{auxbound.1}
u^{1}(\mathbf{x},\omega)=u^{1i}(\mathbf{x},\omega)+
\int_{-\infty}^{\infty} \left(A^{1}(k_{1},\omega)
e^{-\mbox{i}k_{2}^{1}x_{2}}+B^{1}(k_{1},\omega)
e^{\mbox{i}k_{2}^{1}x_{2}} \right)e^{\mbox{i}k_{1}x_{1}}
\frac{dk_{1}}{k_{2}^{1}}~,
\end{equation}
and
\begin{equation}\label{auxbound.2}
u^{0}(\mathbf{x},\omega)=\int_{-\infty}^{\infty}
B^{0}(k_{1},\omega)
e^{\mbox{i}\left(k_{1}x_{1}+k_{2}^{0}x_{2}\right)}
\frac{dk_{1}}{k_{2}^{0}}~,
\end{equation}
respectively.
\subsection{The (incident) field radiated by a ribbon source of width $w$}\label{aux1}
By employing the Green's theorem, the incident field can be
written as:
\begin{equation}\label{aux1.1}
u^{i1}(\mathbf{x},\omega)=\int_{\mathbb{R}^{2}}
G(\|\mathbf{x}-\mathbf{y}\|,\omega)s^{i1}(\mathbf{y},\omega)d\varpi(\mathbf{y})
\mbox{; }\forall \mathbf{x},\in \mathbb{R}^{2}~,
\end{equation}
wherein $\mathbf{y}$ is a vector from the origin $O$ and pointing
to a generic point in the sagittal plane $(y_1,y_2)$,
$G(\|\mathbf{x}-\mathbf{y}\|,\omega)$ the 2D free-space  Green's
function, and $s^{i1}(\mathbf{y},\omega)$  the source density
function:
\begin{equation}\label{aux1.2}
s^{i1}(\mathbf{y},\omega)=\frac{S(\omega)}{w}\left[ H\left(
y_{1}-x_{1}^{s}+ \frac{w}{2}\right) -H\left( y_{1}-
x_{1}^{s}-\frac{w}{2}\right) \right] \delta\left(
y_{2}-x_{2}^{s}\right) ~,
\end{equation}
wherein $H(~.~)$ is the Heaviside  function and $\delta(~.~)$ the
Dirac delta distribution.

Introducing (\ref{aux1.2}) into (\ref{aux1.1}), enables the
incident field to be written as:
\begin{equation}
u^{i1}(\mathbf{x},\omega)=\frac{\mbox{i}}{4\pi}\int_{\infty}^{\infty}
S(\omega)
e^{\mbox{i}\left(k_{1}(x_{1}-x_{1}^{s})+k_{1}^{1}|x_{2}-x_{2}^{s}|
\right)}
\mbox{sinc}\left(k_{1}\frac{w}{2}\right)\frac{dk_{1}}{k_{2}^{1}}~,
\end{equation}
wherein sinc$(\zeta):=\frac{\sin\zeta}{\zeta}$.
\subsection{Expression of the fields in the the presence of a ribbon source of width w}
The introduction of the field representations (\ref{auxbound.1})
and (\ref{auxbound.2})  into the boundary conditions (with the
appropriate projection), leads, after the resolution of the
resulting linear system, to:
\begin{itemize}
\item for $x_{2}\geqq x_{2}^{s}$:
\begin{multline}
u^{1}(\mathbf{x},\omega)=\frac{\mbox{i}S(\omega)\mu^{1}}{2\pi
\mu^{0}} \int_{-\infty}^{\infty}
\frac{\cos\left(k_{2}^{1}\left(x_{2}^{s}-h\right)\right)+
\mbox{i}\frac{\mu^{0}k_{2}^{0}}{\mu^{1}k_{2}^{1}}\sin\left(k_{2}^{1}\left(x_{2}^{s}-h\right)\right)}
{\cos\left(k_{2}^{1}h\right)-\mbox{i}
\frac{\mu^{1}k_{2}^{1}}{\mu^{0}k_{2}^{0}}
\sin\left(k_{2}^{1}h\right)}\times \\
\cos\left(k_{2}^{1}x_{2}\right) e^{\mbox{i}k_{1}(x_{1}-x_{1}^{s})}
\mbox{sinc} \left( k_{1} \frac{w}{2}\right)
\frac{dk_{1}}{k_{2}^{0}}~,
\end{multline}
\item for $x_{2}\leqq x_{2}^{s}$:
\begin{multline}
u^{1}(\mathbf{x},\omega)=\frac{\mbox{i}S(\omega)\mu^{1}}{2\pi
\mu^{0}} \int_{-\infty}^{\infty}
\frac{\cos\left(k_{2}^{1}\left(x_{2}-h\right)\right)+
\mbox{i}\frac{\mu^{0}k_{2}^{0}}{\mu^{1}k_{2}^{1}}\sin\left(k_{2}^{1}\left(x_{2}-h\right)\right)}
{\cos\left(k_{2}^{1}h\right)-\mbox{i}
\frac{\mu^{1}k_{2}^{1}}{\mu^{0}k_{2}^{0}}
\sin\left(k_{2}^{1}h\right)}\\ \cos\left(k_{2}^{1}x_{2}^{s}\right)
e^{\mbox{i}k_{1}(x_{1}-x_{1}^{s})} \mbox{sinc} \left( k_{1}
\frac{w}{2}\right) \frac{dk_{1}}{k_{2}^{0}}~,
\end{multline}
\end{itemize}
and
\begin{multline}
u^{0}(\mathbf{x},\omega)=\frac{\mbox{i}S(\omega)
\mu^{1}}{2\pi\mu^{0}} \int_{-\infty}^{\infty}
\frac{e^{\mbox{i}(k_{1}(x_{1}-x_{1}^{s})+k_{2}^{0}(x_{2}-h))}}
{\cos\left(k_{2}^{1}h\right)-\mbox{i}
\frac{\mu^{1}k_{2}^{1}}{\mu^{0}k_{2}^{0}}
\sin\left(k_{2}^{1}h\right)}\mbox{sinc}\left(k_{1}\frac{w}{2}
\right) \cos\left(k_2^1 x_2^s \right)\frac{dk_1}{k_2^0}~.
\end{multline}


\begin{thebibliography}{77}

\bibitem{akla70} Aki K. \& Larner K.L., 1970, Surface motion of a
layered medium having an irregular interface due to incident plane
SH waves, \textit{J.Geophys.Res.}, 75, 933-954.

\bibitem{wirginetbard} Bard P.Y. \& Wirgin A., 1996, Effects of buildings on the
duration and amplitude of ground motion
in mexico city, \textit{Bull.Seism.Soc.Am.}, 86, 914-920.

\bibitem{becachejolyettsogka} B\'ecache E., Joly P. \& Tsogka C., 2001, Fictitious
domains, mixed finite elements and
perfectly matched layers for 2D elastic wave propagation, 9,
1175-1203, \textit{J.Comput.Acoust.}.

\bibitem{Boutinroussillon} Boutin C. \& Roussillon P., 2004, Assessement of the
urbanisation effect on seismic response,
\textit{Bull.Seism.Soc.Am.}, 94, 252-268.

\bibitem{Boutinrous} Boutin C. \& Roussillon P., 2006, Wave propagation in presence
of oscillators on the free surface,
\textit{Int.J.Engrg.Sci.}, 44, 180-204.

\bibitem{Cardenassato} Cardenas-Soto M. \& Chavez-Garcia F.J., 2003, Regional path
effect on seismic wave propagation
in central Mexico, \textit{Bull.Seism.Soc.Am.}, 93, 973-985.

\bibitem{cardenas} Cardenas-Soto M. \&  Chavez-Garcia F.J., 2006, Seismic
wavefield analysis in Mexico City using accelerometric arrays,
\textit{Abstracts of the First European Conference on Earthquake
Engineering and Seismology}, SSS, Geneva, 166.

\bibitem{chavezgarciabard} Chavez-Garcia F.J. \& Bard P.Y., 1994, Site effects in
Mexico-city height years after the
september 1985 Michoacan earthquakes,
\textit{SoilDyn.\&Earthq.Engrg.}, 13, 229-247.

\bibitem{chavezgarciasalazar} Chavez-Garcia F.J. \& Salazar L., 2002, Strong motion in
central Mexico: a model on data
analysis and simpler modeling, \textit{Bull.Seism.Soc.Am.}, 92,
3087-3101.

\bibitem{clouteau} Clouteau D. \& Aubry D., 2001, Modification of the ground
motion in dense urban areas,
\textit{J.Comput.Acoust.}, 9, 1-17.

\bibitem{collinoettsogka} Collino F. \& Tsogka C., 2001, Application of the PML
absorbing layer model to the linear
elastodynamic problem in anisotropic heterogeneous media, 66,
294-305, \textit{Geophys.}

\bibitem{doby} Doby R., Idriss I.M. \& Ng E., 1979, A reply,  \textit{Bull.Seism.Soc.Am.},
69, 2127-2128.

\bibitem{fah} Faeh D., Suhadolc P., Mueller St. \& Panza, G.F., 1994, A Hybrid method
for the estimation of
ground motion in sedimentary bains: quantitative modeling for
Mexico City, \textit{Bull.Seism.Soc.Am.}, 84, 1711-1797.

\bibitem{febi06} Fernandez-Ares A. \& Bielak J., 1973, Urban seismology: interaction
between earthquake ground motion
  and multiple buildings in urban regions, in {\it ESG 2006}, Bard P.-Y.,
  Chaljub E.,  Cornu C.,
Cotton F. \& Gu\'eguen P. (eds.), LCPC, Paris, 87-96.

\bibitem{flores} Flores J., Novaro O. \& Seligman T.H., 1987, Possible resonance
effect in the distribution of earthquake
damage in Mexico City, \textit{Nature}, 326.

\bibitem{Furumuraetkennet1998} Furumura T. \& Kennett B.L.N., 1998, On the nature
of regional seismic phase-III. The
influence of crustal heterogénéity on the wavefield for subduction
earthquakes: the 1989 Michoacan, and 1995 Copala, Guerro, Mexico
earthquake, \textit{Geophys.J.Intl.}, 135, 1060-1085.

\bibitem{grobythese} Groby J.P., 2005, Mod\'elisation de la propagation des ondes
\'elastiques g\'en\'er\'ees par un
s\'eisme proche ou \'eloign\'e à l'int\'erieur d'une ville,
\textit{PhD Thesis}, Universit\'e de la
M\'editerran\'ee-Aix-Marseille II.

\bibitem{grobyettsogka} Groby J.P. \& Tsogka C., 2006, A time domain method for
modeling viscoacoustic wave propagation, \textit{J.Comput.Acoust.}, 14, 201-236.

\bibitem{wirgintsogkagroby} Groby J.P., Tsogka C. \& Wirgin A., 2005,
Simulation of seismic response
in a city-like environment, \textit{Soil Dyn.\&Earthq.Engrg.}, 25,
487-504

\bibitem{grobyandwirgin2003} Groby J.-P. \& Wirgin A., 2004, On the causes
of anomalous seismic response in an urban site with a
two-component soft layer overlying a hard substratum, {\it EGU
1$^{st}$ General Assembly}, Nice.

\bibitem{grobyetwirgin2005} Groby J.P. \& Wirgin A., 2005, 2D ground motion
at a soft viscoelastic layer/hard
 substratum site in response to SH cylindrical sesimic waves radiated by deep
 and shallow line sources I. Theory,
 \textit{Geophys.J.Intl.}, 163, 165-191.

\bibitem{grobyetwirgin2005II} Groby J.P. \& Wirgin A., 2005, 2D ground
motion at a soft viscoelastic layer/hard
substratum site in response to SH cylindrical sesimic
waves radiated by deep and shallow line sources II. Numerical
results, \textit{Geophys.J.Intl.}, 163, 192-224.

\bibitem{gu00} Gu\'eguen P., 2000, Interaction sismique entre le sol et
le b\^ati: de l'interaction sol-structure \`a l'interaction
site-ville, \textit{Phd thesis}, Universit\'e Grenoble I,
Grenoble.

\bibitem{Guegen} Gu\'eguen P., Bard, P.Y. \& Ch\'avez-Garcia F.J., 2002,
Site-City seismic interaction in Mexico city-like
environments: an analytical study, \textit{Bull.Seism.Soc.Am.},
92, 794-811.

\bibitem{haghshenas} Haghshenas E.,  Jafari M.,  Bard P.-Y.,
Moradi A.S. \& Hatzfeld D., 2006, Preliminary results of site
effects assessment in the city of Tabriz (Iran) using earthquakes
recording, in {\it ESG 2006}, Bard P.-Y., Chaljub E.,  Cornu C.,
Cotton F. \& Gu\'eguen P. (eds.), LCPC, Paris, 993-1001.

\bibitem{hill}  Hill N.R. \& Levander A.R., 1984, Resonances of low-velocity layers with lateral variations,
\textit{Bull.Seism.Soc.Am.}, 74, 521-537.

\bibitem{iwata} Iwata T.,  Kagawa T.  Petukhin A. \&  Onishi
Y., 2006, Basin and crustal structure model for strong ground
motion simulation in Kinki, Japan, in {\it ESG 2006}, Bard P.-Y.,
Chaljub E.,  Cornu C., Cotton F. \& Gu\'eguen P. (eds.), LCPC,
Paris, 435-442.

\bibitem{jennings} Jennings P.C., 1970, Distant motions from a building vibration test,
\textit{Bull.Seism.Soc.Am.}, 60, 2037-2043.

\bibitem{jebi73} Jennings P.C. \& Bielak J., 1973, Dynamics of building-soil interaction,
\textit{Bull.Seism.Soc.Am.}, 63, 9-48.

\bibitem{kj79} Kjartansson E., 1979, Constant Q wave propagation and attenuation,
\textit{J.Geophys.Res.}, 84,4737-4748.

\bibitem{levander} Levander A.R \& Hill N.R., 1984, P-SV resonances in
irregular low-velocity surface layer,
 \textit{Bull.Seism.Soc.Am.}, 75, 847-864.

\bibitem{lombert} Lombaert G., Clouteau D., Ishizawa O. \& Mezher N., 2004,
The city-site effect : a fuzzy substructure
approach and numerical simulations, In Doolin et al. ed., Proc.
\textit{11th Intl.Conf. on Soil Dyn.\&Earthq.Engrg.}, vol2, 68-74.

\bibitem{luco} Luco J.E. \& Contesse L., 1993, Dynamic structure-soil-structure
interaction, \textit{Bull.Seism.Soc.Am.}, 62, 449-462.

\bibitem{maeda} Maeda N., Nakajima Y., Matsuda I. \& Abeki N., 2006,
Evaluation of seismic amplification characteristics at a
university campus with complicated relief of basement, in {\it ESG
2006}, Bard P.-Y., Chaljub E.,  Cornu C., Cotton F. \& Gu\'eguen
P. (eds.), LCPC, Paris, 443-451.

\bibitem{made95} Manos G.C., Demosthenous M., Triamataki M., Yasin
B. and Skalkos P., 1995, Construction and instrumentation of a 5
storey masonry infilled RC building at the Volvi-Thessaloniki
Euro-Seistest site: correlation of measured and numerically
predicted dynamic properties, in {\it Proceedings of the Third
International Conference on Earthquake Engineering}, Amman,
Jordan, 2, 449-462.

\bibitem{mateos} Mateos J.L., Flores J., Novaro O., Seilgman T.H.
\& Alvarez-Costado J.M., 1993,
Resonant response models
for the valleys of Mexico - II The trapping of horizontal P waves,
\textit{Geophys.J.Intl.},113, 449-462.

\bibitem{MF53} Morse P.M. \& Feshbach H., 1953, {\it Methods of Theoretical
Physics}, Mc Graw-Hill, New York, 1953.

\bibitem{Perezrocha} Perez-Rocha L.E., Sanchez-Sesma F.J. \& Reinoso E., 1991,
 Three-Dimensional site effects in
mexico-city: evidence from the accelerometric network observation
and theorical results, In Proc. \textit{4th Intl. Conf. on
Seismics Zonation}, volume II, 327-334.

\bibitem{rial89} Rial J.A., 1989, Seismic wave resonances in 3-D
sedimentary basins, \textit{Geophys.J.Int.}, 99, 81-90.

\bibitem{ro06} Roussillon P., 2006, Interaction sol-structure et
interaction site-ville: aspects fondamentaux et mod\'elisation,
\textit{Phd thesis}, INSA, Lyon.

\bibitem{savage} Savage B.K., 2004, Regional Seismic wavefield propagation, \textit{PhD Thesis},
California Institute of Technology.

\bibitem{semblat} Semblat J.-F., Duval A.-M. \& Dangla P., 2000, Numerical analysis of
seismic wave amplification in
Nice (France) comparison with experiments, \textit{Soil
Dyn.\&Earthq.Engrg.}, 19, 347-362.

\bibitem{semblat2003} Semblat J.F., Gu\'egen P., Kham M., Bard P.Y.
\& Duval A.M., 2003, Site-city interaction at local and global
scales, In Proc. \textit{12th European Conf. on Earthq.Engrg.},
paper no. 807.

\bibitem{shaw} Shaw D.E., 1979, Comment on "Duration characteristics of horizontal components
of strong-motion earthqauke records",
\textit{Bull.Seism.Soc.Am.}, 69, 2125-2126.

\bibitem{to02} Todorovska M.I., 2002, Full-scale experimental
studies of soil-structure interaction, {\it ISET J.Earthqu.Tech}., 39,
139-165.

\bibitem{trifunac} Trifunac M.D., 1972, Interaction of a shear wall with the soil for
incident  plane SH wave,
\textit{Bull.Seism.Soc.Am.}, 62, 63-83.

\bibitem{wirgintsogka} Tsogka C. \& Wirgin A., 2003, Simulation of seismic response in
an idealized city, \textit{Soil Dynam.Earthquake Engrg.}, 23,
391-402.

\bibitem{wirginandgroby2006} Wirgin A. \& Groby J.P., 2006, Amplification and increased
duration of earthquake motion on uneven stress-free ground,
https://hal.ccsd.cnrs.fr/ccsd-00076746,
http://fr.arxiv.org/abs/physics/0605239.

\bibitem{wirgingroby} Wirgin A. \& Groby J.-P., 2006, Amplification and
increased duration of earthquake motion on uneven stress-free
ground, in {\it ESG 2006}, Bard P.-Y., Chaljub E.,  Cornu C.,
Cotton F. \& Gu\'eguen P. (eds.), LCPC, Paris, 559-568.

\bibitem{wongtrif} Wong H.L. \& Trifunac M.D., 1974, Surface motion of sediment-elliptical
alluvial valley for incident plane SH waves,
\textit{Bull.Seism.Soc.Am.}, 64, 1389-1408.
%
\end{thebibliography}
\end{document}